\documentclass[11pt,a4paper]{article}
\pdfoutput=1

\usepackage{jheppub}
\usepackage[utf8]{inputenc}
\usepackage{amsfonts}
\usepackage{amsmath}
\usepackage{dsfont}
\usepackage{color}
\usepackage{graphicx}
\usepackage[normalem]{ulem}
\usepackage{verbatim}
\usepackage{subcaption}
\usepackage{feynmp-auto}
\usepackage{tikz}
\usepackage{braket}
\usepackage{simplewick}
\usepackage{pifont}
\usepackage{slashed}
\usepackage{bm}

\newcommand{\dd}{\mathrm{d}}
\newcommand{\mM}{\mathcal{M}}
\newcommand{\Sp}{\mathbf{Sp}}

\newcommand{\red}[1]{\textcolor{red}{#1}}

\newcommand{\eg}{\textit{e.g.}\ }
\newcommand{\ie}{\textit{i.e.}\ }

\newcommand{\viz}{\textit{viz.}\ }
\renewcommand{\eqref}[1]{Eq.~(\ref{#1})}
\newcommand{\scdot}{\!\cdot\!}

\newcommand{\X}{\scriptsize{\ding{53}}}
\newcommand{\tabref}[1]{Tab.~\ref{#1}}
\newcommand{\secref}[1]{Sec.\,\,\ref{#1}}  

\def\fmfovalblob#1#2#3{\fmfcmd{input vovalblob; vovalblob ((#1), (#2), \fmfpfx{#3});}}

\def\fmfrectangle#1#2#3{\fmfcmd{input vrectangle; vrectangle ((#1), (#2), \fmfpfx{#3});}}

\allowdisplaybreaks

\title{Multi-Emission Kernels for\\ Parton Branching Algorithms}
\author[a,d]{Maximilian L\"oschner\footnote{M.L.\ would like to dedicate this work to his father.},}
\author[b,c,d]{Simon Pl\"atzer}
\author[a]{and Emma Simpson Dore}
\affiliation[a]{Institute for Theoretical Physics, Karlsruhe Institute of
  Technology, Wolfgang-Gaede-Strasse 1, 76131 Karlsruhe, Germany }
\affiliation[b]{Institute of Physics, NAWI Graz, University of Graz, Universit\"atsplatz 5, A-8010 Graz, Austria}
\affiliation[c]{Particle Physics, Faculty of Physics, University of Vienna, Boltzmanngasse 5, A-1090 Wien, Austria}
\affiliation[d]{Erwin Schr\"odinger Institute for Mathematics and
  Physics, University of Vienna, Boltzmanngasse 9, A-1090 Wien}

\emailAdd{maximilian.loeschner@kit.edu}
\emailAdd{simon.plaetzer@uni-graz.at}
\emailAdd{e.simpsondore@gmail.com}

\preprint{\begin{flushright}KA-TP-22-2021\\P3H-21-076\\UWTHPH-2020-07\\MCnet-21-14\end{flushright}}

\abstract{We introduce a general framework to construct multi-emission
  kernels for parton branching algorithms at the amplitude level and
  across different soft and collinear limits.
  We highlight the connection of kinematic parameterizations and
  recoil schemes to the underlying power counting, and discuss in
  detail how soft radiation can be partitioned in between different
  collinear configurations beyond the single-emission picture
  underpinning the traditional dipole and angular ordering
  approaches. Our work is a vital cornerstone to build parton
  branching algorithms which include multiply-unresolved emissions in
  a fully differential way, and our construction can
  also be used to obtain splitting functions for probabilistic
  algorithms or other cross-section level objects such as subtraction
  terms.}

\begin{document}

\maketitle

\section{Introduction}

A detailed description of QCD dynamics at current and future collider
experiments is vital to discriminate Standard Model background from
signals of New Physics using the most advanced methods ranging from
traditional event shape variables to jet substructure methods to
machine learning techniques. The comparison to detailed theory
predictions is vital and a large range of analytic predictions up to
the most versatile Monte Carlo event generators are used in comparison
to experimental data. Specifically, Monte Carlo event generators have
recently seen a tremendous development regarding the simulation of the
hard scattering process, and the attention has shifted to the core,
parton shower, component among other developments on non-perturbative
modelling. The question of how the accuracy of parton branching
algorithms can be quantified, and how they can eventually be extended
beyond their current approximations -- both regarding resummed
perturbation theory in Quantum Chromodynamics (QCD), as well as beyond
the customary limit of a large number of colour charges $N$ -- has
received significant attention which we will briefly review in this
introduction to motivate the present work.

Parton branching algorithms have long been used in the context of
analytic resummation \cite{Catani:1989ne}, including up to
next-to-leading logarithmic (NLL) accuracy for a large class of event
shape variables \cite{Catani:1992ua} and properties of hard objects
created in hadron-hadron collisions \cite{Catani:1990rr}. These
techniques notably exploit the coherence properties of QCD and the
structure of the observable such that angular ordering can be used
upon azimuthal averaging to simplify the resummation procedure. They
have also been the key ingredient behind designing the Herwig
angular-ordered parton shower algorithm
\cite{Marchesini:1987cf,Gieseke:2003rz,Bahr:2008pv,Bellm:2019zci},
which is thus capable of describing global event shape variables at
NLL accuracy. Due to ambiguities in how recoil of the emissions is
distributed at the end of the evolution, however, failures of this
naive expectation may be introduced and need careful investigation
\cite{Bewick:2019rbu}. Even in the case of NLL accurate coherent
branching evolution, a number of shortcomings are present in these
algorithms, and have motivated additional work, notably spin and
colour correlations \cite{Richardson:2018pvo} as well as mass effects
\cite{Cormier:2018tog}.

The biggest shortcoming, however, is the inability of coherent
branching algorithms to describe observables which are sensitive to a
change in colour structure from subsequent emissions beyond the
dynamics accounted for by QCD coherence, and this applies already in
the large-$N$ limit. Crucially, it is the lack of a dipole-type
picture which disallows coherent branching from describing these
correlations, and hence there is no chance for such an evolution to
account for any of the logarithmic enhancements in non-global
observables. It is clear though, that non-global observables
\cite{Dasgupta:2001sh, Becher:2016mmh, Martinez:2018ffw} should be
setting the level of complexity which needs to be addressed for the
plethora of sophisticated analysis methods employed at the currently
operating hadron colliders, and projected to be used even at future
$e^+e^-$ colliders. Note however that coherent branching is in fact
able to describe the exact-$N$ structure of global event shapes at
least at leading logarithmic (LL) level.

Dipole shower algorithms \cite{Gleisberg:2008ta, Platzer:2011bc,
  Dinsdale:2007mf, Giele:2013ema}, on the other hand, do offer the
possibility to properly resum non-global effects at leading colour
(and in fact have been used, in a specialized form, for specifically
this task \cite{Dasgupta:2001sh, Becher:2016mmh}). Despite the fact
that they have been used to try and extend parton branching beyond the
leading order \cite{Dulat:2018vuy}, their initial accuracy remained
unclear and has been proven to be wrong at various levels for global
observables
\cite{Dasgupta:2018nvj,Dasgupta:2020fwr,Forshaw:2020wrq}. Recent work
has therefore focused on trying to combine the best of dipole and
coherent branching algorithms such as to maintain the accuracy for
global event shapes, and a proper (\ie at least leading-colour and
leading logarithmic accurate) description of non-global observables
\cite{Hamilton:2020rcu,Holguin:2020joq}. Within this context it turned
out that recoils can cause severe accuracy problems, and that there is
a delicate interplay between ordering, recoil and the question of how
the soft behaviour is distributed across the partonic systems
competing to emit the next parton, a mechanism typically referred to
as `partitioning'. Notice that a dipole-type picture, or extensions
thereof, is also required to have accurate control over recoils
distributed emission by emission, something which is extremely
important when trying to combine with fixed-order QCD calculations,
either through matching or merging.  Most of the time, the development
of parton showers within this context has been closely tied to the
development of fixed-order subtraction terms reflecting the
singularity structure of QCD {\it cross sections} for the emission of
one, or possibly two, unresolved partons.

The shortcomings of existing parton shower algorithms, and possible
improvements, have been addressed in \cite{Dasgupta:2018nvj} by
checking how well they reproduce the soft radiation pattern for two
subsequent emissions, while \cite{Forshaw:2020wrq} has been taking a
different point of view in starting from formulating parton branching
algorithms at the amplitude level \cite{Forshaw:2019ver}. This is a
theoretical framework which addresses the iterative build-up of an
amplitude (and its conjugate) describing the emission of multiple
unresolved partons, and is thus the prime framework to analyse how
cross sections can be described in the limit of a large number of
unresolved partons. A systematic expansion of the resulting
expressions around the large-$N$ limit, for example, then leads to
commonly used parton branching algorithms which can be extended beyond
the large-$N$ limit.  The same formalism can also be used to derive
further limiting cases such as coherent branching, which has
ultimately led to the proposal of an improved dipole shower
algorithm. We stress that only within such a framework, and similar
approaches pursued by Nagy and Soper \cite{Nagy:2020rmk}, is it
possible to analyse the entire set of correlations in between multiple
emissions, and to systematically obtain an approximation of the
iterations, rather than iterating a single-emission approximation of
the cross section. The parton branching at amplitude level was
designed around the soft gluon evolution algorithm presented earlier,
and included the hard-collinear contributions by extending the dipole
structure of the leading order soft evolution. Work is underway to
extend this algorithm to the next order, including an expansion around
the colour-diagonal contributions \cite{Platzer:2020lbr} corresponding
to the $d'$-expansions advocated in \cite{DeAngelis:2020rvq}.

In this work we focus on extending the parton branching in amplitude
level algorithms to account for more than singly-unresolved limits,
concentrating on corrections due to iterating the emission of two or
more simultaneously unresolved partons. We work at a finite number of
colours, and include the full spin dependence aiming to establish
factorization at the level of the 'density operator', \ie the
amplitude and its conjugate, for which we can devise the definition of
splitting kernels.  These encompass all enhanced configurations of
both soft, collinear, or any combination of unresolved limits. We also
highlight the connection of the underlying power counting to the way
recoil is distributed and how kinematics are parametrized.

This paper is structured as follows: In Sec.~\ref{sec:Factorization}
we will outline the general framework in which we establish the
iterative factorization of the density operator. We will highlight the
connection to the light-cone gauge we employed, as well as the
distribution of recoil of emissions among the hard jet axes, which
allows us to establish a diagrammatic framework in which the splitting
kernels can be calculated. After the projection onto different
collinearly enhanced contributions, which we refer to as partitioning,
we will be able to establish a systematic power counting and a final
calculational recipe for the emission operators. In
Sec.~\ref{sec:Paritioning}, the partitioning is set up in an
algorithmic manner and is suited to be carried out to an in principle
arbitrary number of emissions from a hard process.

In Sec.~\ref{sec:Mapping} we detail one particular instance of a
suitable momentum mapping and discuss its properties in the unresolved
limit, as well as how it can be used to construct an actual parton
branching algorithm concerning the real emission contributions.  In
combination with singling out collinear limits, and the proper choice
of a backward vector $n$, we can determine the splitting kernels.  We
discuss a few examples in Sec.~\ref{sec:Kernels}, also addressing how
the underlying amplitude can be taken on-shell from the final state
considered (before emissions are added) and which power suppressed
effects we neglect in doing so.

In Sec.~\ref{sec:Applications} we conclude by discussing applications
of our formalism, and how our framework can be generalized to the case
of virtual corrections. The determination of a full set of QCD
splitting kernels is devoted to a future publication, however we do
discuss several examples, as well as the possibility to link our
formalism to a partitioning of the known singular behaviour of two
emissions, and a systematic removal of overlap between the soft and
collinear limits. A number of technical details and material for
future reference is devoted to several appendices, which do not need
to be considered to understand the main findings of our current work.

\section{Construction of splitting kernels}
\label{sec:Factorization}

While traditional parton branching algorithms use probabilistic
paradigms based on the factorization of cross sections, our aim is to
address parton branching algorithms at the amplitude level, as
recently theoretically developed and implemented in
\cite{Martinez:2018ffw,Forshaw:2019ver,Forshaw:2020wrq,DeAngelis:2020rvq}. Only
within such an approach is it possible to iteratively approximate
amplitudes with many legs and to derive Markovian algorithms which
then might be able to perform multiple emissions at the probabilistic
level; iterating approximations of a fixed-order cross section will
yield the right singularity structure, but otherwise not guarantee the
right resummation properties. In the present work, we focus on the
systematic construction of real emission splitting kernels within such
a branching algorithm, addressing the simultaneous emission of more
than one unresolved parton. The goal of the present work is to not
perform this specifically in one unresolved limit, but to obtain a
combination of splitting operators which smoothly cover the different
singular limits. Some aspects of virtual corrections have recently
also been addressed in \cite{Platzer:2020lbr} and will be
combined with the present effort in future work. We can of course use
our results to obtain splitting kernels which can also serve as
subtraction terms in fixed-order corrections, though this is not our
primary goal.

To set the notation, we consider scattering amplitudes $|{\cal
  M}\rangle$ to be vectors in colour and spin space (see
\cite{Catani:1996vz} for an outline of this formalism), from which
cross sections originate as
\begin{equation}
  \label{eqs:crosssec}
  \sigma[u] = \sum_n \int {\rm Tr} \left[|{\cal M}(1,...,n)\rangle
    \langle{\cal M}(1,...,n)|\right] u(1,...,n) {\rm d}\phi(1,...,n|Q) \ ,
\end{equation}
where ${\rm d}\phi(1,...,n|Q)$ commonly refers to the $n$-parton phase
space given a total momentum $Q$ and the arguments of the amplitude
vector and phase space measure refer to both parton momenta, as well
as other degrees of freedom needed to identify a particular final
state. $u$ is a generic observable with the same convention on
arguments understood, \ie $u(1,...,n) = u(p_1,...,p_n)$ for
specific momenta. In the present work we also only consider massless
partons. The trace in \eqref{eqs:crosssec} refers to summing over
the colour and spin degrees of freedom in the amplitude and its
conjugate, seen as an operator in colour and spin space, the cross
section density operator
\begin{equation}
  {\mathbf A}_n = |{\cal M}(1,...,n)\rangle \langle{\cal M}(1,...,n)| \ ,
\end{equation}
which will be the central object of our investigation. The trace can
typically be expressed using the completeness of external wave
functions, and we shall decompose the respective numerator of a cut
propagator in such a way that we can find a similar decomposition of
internal lines, with suitably parametrized kinematics. With this in
mind, we will then formulate a diagrammatic construction of the relevant
splitting kernels, directly involving a power counting to identify the
leading contributions of interest. Working in a physical gauge with
gauge fixing $C^a = n\cdot A^a$, $n^2=0$, we in particular note that
we can express the propagator numerators for cut gluon lines of
on-shell momentum $q$ as
\begin{equation}
  d_{\mu\nu}(q) = -\eta^{\mu\nu} + \frac{n^\mu q^\mu + n^\nu q^\mu}{n\cdot q} \ ,
\end{equation}
such that we can use
\begin{equation} \label{eq:projector-gluon}
  d_{\mu\nu}(q) = d_{\mu\rho}(q)\ P^{\rho\sigma}(p) \ d_{\sigma\nu}(q),
  \qquad P^{\rho\sigma}(p) = d^{\rho\sigma}(p) \ ,
\end{equation}
which holds for $q^2=0$, while for quark lines we can employ
\begin{equation}\label{eq:projector-fermion}
  \slashed{q} =
  \sqrt{\frac{n\cdot q}{n\cdot p}}\ \frac{ n\cdot p}{n\cdot q} \slashed{q} \
  \slashed{P}(p)\ \slashed{q} \frac{n\cdot p}{n\cdot q}\ \sqrt{\frac{n\cdot q}{n\cdot p}},
  \qquad \slashed{P}(p) = \frac{\slashed{n}}{2n\cdot p}  \ ,
\end{equation}
where $p$ is the forward component we associate to the external
momentum $q$ in a decomposition
\begin{equation}
  q^\mu = \alpha\ p^\mu + \beta\ n^\mu + q_\perp^\mu \ ,\qquad q_\perp\cdot n
  = q_\perp\cdot p = 0 \ .
\end{equation}
The reason for the awkward decomposition of the quark numerator as in
\eqref{eq:projector-fermion} is to associate the square-root factors
with the vertex the quark couples into (or leave it as an explicit
factor if such a coupling is not considered), while the factors
$n\cdot p/n\cdot q$ provide a convenient normalization in the power
counting such that quark and gluon lines have a uniform scaling. The
role of soft quarks in the factorization can then be tracked
separately from more general considerations on factorization of the
amplitude. This is discussed in Sec.~\ref{sec:decompositions-intro}.
In a more abstract notation we thus define
\begin{align}\label{eq:projectors-general}
	\mathbf{P}(q) \equiv
	\begin{cases}
		P^{\rho \sigma}(p) = d^{\rho \sigma} (p), & \text{for gluons}, \\
		\slashed{P}(p) = \frac{\slashed{n}}{2n \cdot p}, & \text{for quarks},
	\end{cases}
\end{align}
where it should always be clear which forward momentum $p$ we
associate to a certain external momentum $q$, subject to a
decomposition of the density operator into different collinearly
singular configurations to be discussed in Sec.~\ref{sec:decompositions-counting}.  Moreover, we
can use polarisation sums to represent these operators, \ie
\begin{align}
	d^{\mu\nu}(p) &= \epsilon^\mu_+ (p,n) \epsilon ^\nu_- (p,n) + (\mu \leftrightarrow \nu), \\
	\slashed{n} &= \sum\limits_\lambda u_\lambda (n) \bar{u}_\lambda (n),
\end{align}
with
\begin{align}
	\epsilon_{\pm}^2 = 0, \quad \epsilon_{\pm}\cdot \epsilon_{\mp} = -1, \quad \epsilon_{\pm} \cdot p = \epsilon_\pm \cdot n = 0.
\end{align}
Also note that for
internal lines carrying a sum of momenta, we can decompose the
respective numerators via
\begin{eqnarray}
	\label{eqs:linearity}
	d^{\mu\nu}(q_I+q_J) &=& \frac{n\cdot q_I}{n\cdot q_I + n\cdot q_J}\times d^{\mu\nu}(q_I)
	+ (I \leftrightarrow J), \\ \nonumber
	\slashed{q}_I+\slashed{q}_J &=& 
	\sqrt{\frac{n\cdot q_I}{n\cdot p_i}}
	\left( \frac{n\cdot p_i}{n\cdot q_I} \slashed{q}_I\right) \sqrt{\frac{n\cdot q_I}{n\cdot p_i}}
	+ (I
	\leftrightarrow J),
\end{eqnarray}
where $q_I$ and $q_J$ themselves can be sums of momenta in distinct
sets of partons $I$ and $J$. Moreover, the square root factors in the
second line are of the same use as in \eqref{eq:projector-fermion}.
Similarly, we can decompose the three-gluon vertex due to its
linearity in the momenta, \ie
\begin{fmffile}{soft-three-vertex-decomp}
	\fmfset{thin}{.7pt}
	\fmfset{arrow_len}{2.5mm}
	\fmfset{curly_len}{1.8mm}
	\fmfset{dash_len}{1.5mm}
	\begin{eqnarray}\label{eq:linearity-three-gluon-vertex}
		\begin{gathered}
			\vspace{-4pt}
			\begin{fmfgraph*}(50,50)
				\fmfbottom{l,m,r}
				\fmftop{t}
				\fmf{phantom}{m,v,t}
				\fmffreeze
				\fmf{curly}{r,v,l}
				\fmf{curly}{t,v}
				\fmfv{label=\small{$q_J$},label.angle=0}{t}
				\fmfv{label=\small{$q_I$},label.angle=-90}{r}
				\fmfv{label=\small{$q_I+q_J$},label.angle=-90}{l}
			\end{fmfgraph*}
		\end{gathered}
		\quad  = \quad
		\begin{gathered}
			\vspace{-4pt}
			\begin{fmfgraph*}(50,50)
				\fmfbottom{l,m,r}
				\fmftop{t}
				\fmf{phantom}{m,v,t}
				\fmffreeze
				\fmf{curly}{r,v,l}
				\fmf{curly}{t,v}
				\fmfv{label=\small{$0$},label.angle=0}{t}
				\fmfv{label=\small{$q_I$},label.angle=-90}{r}
				\fmfv{label=\small{$q_I$},label.angle=-90}{l}
			\end{fmfgraph*}
		\end{gathered}
		\quad + \quad 
		\begin{gathered}
			\vspace{-4pt}
			\begin{fmfgraph*}(50,50)
				\fmfbottom{l,m,r}
				\fmftop{t}
				\fmf{phantom}{m,v,t}
				\fmffreeze
				\fmf{curly}{r,v,l}
				\fmf{curly}{t,v}
				\fmfv{label=\small{$q_J$},label.angle=0}{t}
				\fmfv{label=\small{$0$},label.angle=-90}{r}
				\fmfv{label=\small{$q_J$},label.angle=-90}{l}
			\end{fmfgraph*}
		\end{gathered}\; .
	\end{eqnarray}
\end{fmffile}%
 If now all of the partons in
the set $J$ become soft, then only the first terms in
\eqref{eqs:linearity} and \eqref{eq:linearity-three-gluon-vertex} will
contribute in a leading soft limit. This paves the way to establish a
universal set of rules to decompose the cross section with many
external, unresolved, legs as follows: Using
\eqref{eq:projectors-general}, we can write
\begin{equation}\label{eq:trace-amplitude}
  {\rm Tr} \left[{\mathbf A}_n\right] = {\rm Tr}_c
  \left[\tilde{{\mathbf A}}_n\times {\mathbf P}_n  \right]  \ ,
\end{equation}
where ${\rm Tr}_c$ now solely refers to a trace in colour space,
$\tilde{{\mathbf A}}_n$ is the density operator obtained from
${\mathbf A}_n$ by removing all external wave functions.  The sum over
external polarisations is implicitly carried out by associating the
normalized completeness relations of external wave functions to each
external line, \ie applying a factor $d^{\mu\nu}(q)$ for each
external gluon line and a factor of $(\slashed{q}\ n\cdot p / n\cdot
q)\times \sqrt{n\cdot q / n\cdot p}$ to each external fermion line,
together with the insertion of the respective operator ${\mathbf P}$.
This setup allows us to treat internal and external lines for quarks
and gluons alike on the same footing and devise universal power
counting rules in the following parts of this section.  Those will
eventually be used to dissect numerator structures of emission
amplitudes in terms of their soft or collinear scaling and allows for
a precise association of which parts of an amplitude contribute in
singular limits at a given power.  The results of these considerations
for one and two emissions are given in
Sec.~\ref{sec:single-emission-splitting-kernel} and
Sec.~\ref{sec:two-emission-splitting-kernel} respectively.  Also note
that it is in principle possible to include a polarized measurement
into the definition of the ${\mathbf P}$ operator by appropriately
restricting the sums of external wave functions, though this is beyond
the scope of the present work.

Starting from the amplitude, we consider a certain subset of diagrams
contributing to additional emissions from a general amplitude $|{\cal
  M}\rangle$, which we consider to be a vector in spin and colour
space. In particular, we are interested in the emission of several, $k\ge
1$ additional particles, for which we consider the diagrams
factorizing in an $(n+k)$-particle amplitude in terms of emission
diagrams and an underlying, $n$-particle hard amplitude as
\begin{multline}
|{\cal M}_{n+k}(1,...,{n+k})\rangle =
\sum_{p=1}^k \sum_{r\in S_{n,p,k}}
\mathbf{Sp}_{(r_{11}|...|r_{1\ell_1})}...\mathbf{Sp}_{(r_{p1}|...|r_{p\ell_p})}\\
|\tilde{{\cal M}}_{n}(1,...,{(r_{11}|...|r_{1\ell_1})},...,
{(r_{p1}|...|r_{p\ell_p})},...,{n+k} )\rangle,
\end{multline}
or in a diagrammatic representation
\vspace{5pt}
\begin{fmffile}{master-diagram-splitting}
	\fmfset{thin}{.7pt}
	\fmfset{dot_len}{1.2mm}
	\fmfset{dot_size}{5}
	\fmfset{arrow_len}{2.5mm}
	\fmfset{wiggly_len}{2.2mm}
	\begin{align}\label{master-diagram-splitting}
	\begin{gathered}
	\begin{fmfgraph*}(80,60)
	\fmfleft{l}
	\fmfright{rd,rm,ru}
	\fmf{phantom,tension=4}{l,v0}
	\fmf{plain}{v0,vdu,ru}
	\fmf{plain}{v0,vdd,rd}
	\fmfv{label=\tiny{${1}$},label.angle=30}{ru}
	\fmfv{label=\tiny{${n+k}$},label.angle=-30}{rd}
	\fmffreeze
	\fmf{dots,left=.2,width=1}{vdu,vdd}
	\fmfv{decor.shape=circle,decor.filled=empty,decor.size=20,label=$\mM$,label.dist=0}{v0}
	\end{fmfgraph*}
	\end{gathered}
	\quad \xrightarrow{\makebox[1.6cm]{\tiny\text{singular terms}}}
	\quad
	\sum\limits_{p=1}^k \sum\limits_{r\in S_{n,p,k}}
	\begin{gathered}
	\begin{fmfgraph*}(80,80)
	\fmfleft{l}
	\fmfright{r4,r3,rm,r2,r1}
	\fmf{phantom,tension=5}{l,v0}
	\fmf{phantom,tension=2}{v0,v0u,v1}
	\fmf{plain}{v1,v1u,r1}
	\fmf{plain}{v1,v1d,r2}
	\fmf{phantom,tension=2}{v0,v0d,v2}
	\fmf{plain}{v2,v2u,r3}
	\fmf{plain}{v2,v2d,r4}
	\fmfv{label=\tiny{${r_{11}}$},l.a=45}{r1}
	\fmfv{label=\tiny{${r_{1l_1}}$},l.a=20}{r2}
	\fmfv{label=\tiny{${r_{p1}}$},l.a=-20}{r3}
	\fmfv{label=\tiny{${r_{pl_p}}$},l.a=-45}{r4}
	\fmffreeze
	\fmf{dots,left=.2,width=1}{v1,v2}
	\fmf{dots,left=.2,width=1}{v1u,v1d}
	\fmf{dots,left=.2,width=1}{v2u,v2d}
	\fmf{dbl_plain,tension=2,background=white,rubout}{v0,v1}
	\fmf{dbl_plain,tension=2,background=white,rubout}{v0,v2}
	\fmfv{decor.shape=circle,decor.filled=empty,decor.size=20,label=$\tilde{\mM}$,label.dist=.00}{v0}
	\fmfv{decor.shape=circle,decor.filled=empty,decor.size=13,label=\tiny{$\mathbf{Sp}$},label.dist=0}{v1}
	\fmfv{decor.shape=circle,decor.filled=empty,decor.size=13,label=\tiny{$\mathbf{Sp}$},label.dist=0}{v2}
	\end{fmfgraph*}
	\end{gathered}
	\end{align}
\end{fmffile}
$S_{n,p,k}$ is the set of possible splitting assignments involving $p$
emitters and $k$ emissions out of $n+k$ external particles, and
\begin{equation}
  {(r_{i1}|...|r_{i\ell_i})} = {r_{i1}}+...+{r_{i\ell_i}}
\end{equation}
are the off-shell momenta of the branching internal lines, where each
of the $r_{ik}$ denotes a certain external momentum $q_j$ involved in
the splitting process. We are interested in the unresolved limits of
the $k$ emitted partons, in which all of the off-shell momenta
${(r_{i1}|...|r_{i\ell_i})}$ will become on-shell and the splitting
amplitudes are expected to factorize in a universal manner, within
which the underlying amplitude $\tilde{\mathcal{M}}$ in fact becomes an on-shell
amplitude. The leading singular behaviour of the cross section is then
obtained by considering the square of the amplitude, or equivalently a
density-operator type object, such that none of the partons
participating in an unresolved limit are connected to an internal
line. To be more precise we can write
\begin{multline}\label{eqs:density-operator-general}
|{\cal M}_{n+k}(1,...,{n+k})\rangle\langle {\cal M}_{n+k}(1,...,{n+k})| = \\
\sum_{p=1}^k \sum_{\bar{p}=1}^k \sum_{r\in S_{n,p,k}}\sum_{\bar{r}\in S_{n,\bar{p},k}} 
\mathbf{Sp}_{(r_1)}...\mathbf{Sp}_{(r_p)}
  |{\cal M}_{n}(1,...,{(r_1)},...,
  {(r_p)},...,{n+k} )\rangle\\ \langle {\cal M}_{n}(1,...,{(\bar{r}_1)},...,
  {(\bar{r}_{\bar{p}})},...,{n+k} ) | \mathbf{Sp}^\dagger_{(\bar{r}_1)}...
  \mathbf{Sp}^\dagger_{(\bar{r}_{\bar{p}})} \ \times \hat\Delta^{r}_{\bar{r}}\ +\\
  \text{(subleading)}
\end{multline}
where we have introduced short-hands for the splitting labels, $(r_i) =
(r_{i1}|...|r_{i\ell_i})$ and similar.  The tensor
$\hat\Delta^{r}_{\bar{r}}$ ensures that we only include those diagram topologies
which give rise to leading singularities.  See
Appendix~\ref{sec:delta-tensor} for its explicit construction.
This extraction of leading singularities is valid as long as we work
in a physical gauge.  The diagrammatic representation of one such leading
singular contribution to the density operator is:
\bigskip
	
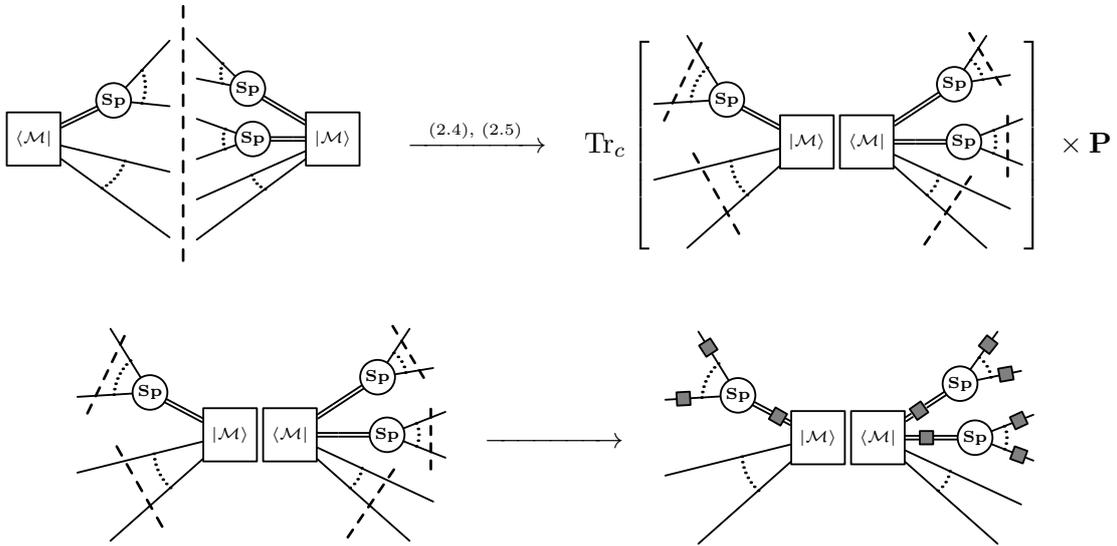
\begin{fmffile}{splitting-general-algorithm-squared}
	\fmfset{thin}{.7pt}
	\fmfset{dot_len}{.8mm}
	\fmfset{dot_size}{5}
	\fmfset{arrow_len}{2.5mm}
	\fmfset{wiggly_len}{2.2mm}
	\begin{align*}
	\begin{gathered}
	\begin{fmfgraph*}(140,90)
	\fmfleft{l4,l3,l2,l1}
	\fmfright{r6,r5,r4,r3,r2,r1}
	\fmf{phantom}{l1,ml1,mu,mr1,r1}
	\fmf{phantom}{l4,ml4,md,mr6,r6}
	\fmf{phantom,tension=4.5}{ml1,mr1}
	\fmf{phantom,tension=4.5}{ml4,mr6}
	\fmf{phantom,tension=-0.25}{mu,md}
	\fmf{phantom}{ml1,ml2,ml3,ml4}
	\fmf{phantom}{mr1,mr2,mr3,mr4,mr5,mr6}
	\fmf{phantom}{l1,vl,l4}
	\fmf{phantom}{r1,vr,r6}
	\fmffreeze
	\fmf{dbl_plain,tension=1.4,background=white}{vl,vdl1,vl1}
	\fmf{plain}{vl1,vl11,ml1}
	\fmf{plain}{vl1,vl12,ml2}
	\fmf{plain}{vl,vl2,vl22,ml3}
	\fmf{plain}{vl,vl3,ml4}
	\fmffreeze
	\fmf{dots,left=.2,width=1}{vl11,vl12}
	\fmf{dots,left=.2,width=1}{vl22,vl3}
	\fmf{dashes,width=1}{mu,md}
	\fmfv{decor.shape=circle,decor.filled=empty,decor.size=13,label=\tiny{$\mathbf{Sp}$},label.dist=0}{vl1}
	\fmf{dbl_plain,tension=1.2,background=white}{vr,vdr1,vr1}
	\fmf{plain}{vr1,vr11,mr1}
	\fmf{plain}{vr1,vr12,mr2}
	\fmf{dbl_plain,tension=1.4,background=white}{vr,vdr2,vr2}
	\fmf{plain}{vr2,vr21,mr3}
	\fmf{plain}{vr2,vr22,mr4}
	\fmf{plain}{vr,vr3,vr32,vr33,vr35,mr5}
	\fmf{plain}{vr,vr4,mr6}
	\fmffreeze
	\fmf{dots,right=.2,width=1}{vr11,vr12}
	\fmf{dots,right=.2,width=1}{vr21,vr22}
	\fmf{dots,right=.2,width=1}{vr33,vr4}
	\fmfv{decor.shape=circle,decor.filled=empty,decor.size=13,label=\tiny{$\mathbf{Sp}$},label.dist=0}{vr1}
	\fmfv{decor.shape=circle,decor.filled=empty,decor.size=13,label=\tiny{$\mathbf{Sp}$},label.dist=0}{vr2}
	\fmfv{decor.shape=square,decor.filled=empty,decor.size=20,label=\tiny{$\bra{\mM}$},label.dist=0}{vl}
	\fmfv{decor.shape=square,decor.filled=empty,decor.size=20,label=\tiny{$\ket{\mM}$},label.dist=0}{vr}
	\end{fmfgraph*}
	\end{gathered}
	\quad \xrightarrow{\makebox[1.6cm]{\tiny(2.4), (2.5)}}
	\quad
	{\rm Tr}_c \left[
	\begin{gathered}
	\begin{fmfgraph*}(140,80)
	\fmfleft{l4,l3,l2,l1}
	\fmfright{r6,r5,r4,r3,r2,r1}
	\fmf{phantom}{l1,vl,l4}
	\fmf{phantom}{r1,vr,r6}
	\fmf{phantom,tension=4}{vl,vr}
	\fmffreeze
	\fmf{dbl_plain,tension=0.7,background=white}{vl,vl1}
	\fmf{plain}{vl1,vl11,l1}
	\fmf{plain}{vl1,vl12,l2}
	\fmf{plain}{vl,vl2,l3}
	\fmf{plain}{vl,vl3,l4}
	\fmffreeze
	\fmf{dots,right=.2,width=1}{vl11,vl12}
	\fmf{dots,right=.2,width=1}{vl2,vl3}
	\fmf{phantom}{l4,lc4,lc41,rc61,rc6,r6}
	\fmf{phantom}{l1,lc1,vl11}
	\fmf{phantom}{l2,lc2,lc21,lc3}
	\fmf{phantom}{l3,lc3,vl1}
	\fmf{dashes,width=1,tension=0.2}{lc4,lc3}
	\fmf{dashes,width=1,tension=-0.2}{lc1,lc2}
	\fmfv{decor.shape=circle,decor.filled=empty,decor.size=13,label=\tiny{$\mathbf{Sp}$},label.dist=0}{vl1}
	\fmf{dbl_plain,tension=1,background=white}{vr,vdr1,vr1}
	\fmf{plain}{vr1,vr11,r1}
	\fmf{plain}{vr1,vr12,r2}
	\fmf{dbl_plain,tension=1.2,background=white}{vr,vdr2,vr2}
	\fmf{plain}{vr2,vr21,r3}
	\fmf{plain}{vr2,vr22,r4}
	\fmf{plain}{vr,vr3,r5}
	\fmf{plain}{vr,vr4,r6}
	\fmffreeze
	\fmf{dots,left=.2,width=1}{vr11,vr12}
	\fmf{dots,left=.2,width=1}{vr21,vr22}
	\fmf{dots,left=.2,width=1}{vr3,vr4}
	\fmf{phantom}{r1,rc1,vr11}
	\fmf{phantom}{r2,rc2,vr12}
	\fmf{phantom}{r3,rc3,vr21}
	\fmf{phantom}{r4,rc4,vr22}
	\fmf{phantom}{r5,rc51,rc5,vr2}
	\fmf{dashes,width=1,tension=0.1}{rc6,rc5}
	\fmf{dashes,width=1,tension=-0.5}{rc4,rc3}
	\fmf{dashes,width=1,tension=-0.5}{rc1,rc2}
	\fmfv{decor.shape=circle,decor.filled=empty,decor.size=13,label=\tiny{$\mathbf{Sp}$},label.dist=0}{vr1}
	\fmfv{decor.shape=circle,decor.filled=empty,decor.size=13,label=\tiny{$\mathbf{Sp}$},label.dist=0}{vr2}
	\fmfv{decor.shape=square,decor.filled=empty,decor.size=20,label=\tiny{$\ket{\mM}$},label.dist=0}{vl}
	\fmfv{decor.shape=square,decor.filled=empty,decor.size=20,label=\tiny{$\bra{\mM}$},label.dist=0}{vr}
	\end{fmfgraph*}
	\end{gathered}
	\right] \ \times  {\mathbf P}
	\end{align*}
        \end{fmffile}%

\begin{fmffile}{density-operator-boxes}
	\fmfset{thin}{.7pt}
	\fmfset{dot_len}{.8mm}
	\fmfset{dot_size}{5}
	\fmfset{arrow_len}{2.5mm}
	\fmfset{wiggly_len}{2.2mm}
	\begin{align*}
	\begin{gathered}
	\begin{fmfgraph*}(140,80)
	\fmfleft{l4,l3,l2,l1}
	\fmfright{r6,r5,r4,r3,r2,r1}
	\fmf{phantom}{l1,vl,l4}
	\fmf{phantom}{r1,vr,r6}
	\fmf{phantom,tension=4}{vl,vr}
	\fmffreeze
	\fmf{dbl_plain,tension=0.7,background=white}{vl,vl1}
	\fmf{plain}{vl1,vl11,l1}
	\fmf{plain}{vl1,vl12,l2}
	\fmf{plain}{vl,vl2,l3}
	\fmf{plain}{vl,vl3,l4}
	\fmffreeze
	\fmf{dots,right=.2,width=1}{vl11,vl12}
	\fmf{dots,right=.2,width=1}{vl2,vl3}
	\fmf{phantom}{l4,lc4,lc41,rc61,rc6,r6}
	\fmf{phantom}{l1,lc1,vl11}
	\fmf{phantom}{l2,lc2,lc21,lc3}
	\fmf{phantom}{l3,lc3,vl1}
	\fmf{dashes,width=1,tension=0.2}{lc4,lc3}
	\fmf{dashes,width=1,tension=-0.2}{lc1,lc2}
	\fmfv{decor.shape=circle,decor.filled=empty,decor.size=13,label=\tiny{$\mathbf{Sp}$},label.dist=0}{vl1}
	\fmf{dbl_plain,tension=1,background=white}{vr,vdr1,vr1}
	\fmf{plain}{vr1,vr11,r1}
	\fmf{plain}{vr1,vr12,r2}
	\fmf{dbl_plain,tension=1.2,background=white}{vr,vdr2,vr2}
	\fmf{plain}{vr2,vr21,r3}
	\fmf{plain}{vr2,vr22,r4}
	\fmf{plain}{vr,vr3,r5}
	\fmf{plain}{vr,vr4,r6}
	\fmffreeze
	\fmf{dots,left=.2,width=1}{vr11,vr12}
	\fmf{dots,left=.2,width=1}{vr21,vr22}
	\fmf{dots,left=.2,width=1}{vr3,vr4}
	\fmf{phantom}{r1,rc1,vr11}
	\fmf{phantom}{r2,rc2,vr12}
	\fmf{phantom}{r3,rc3,vr21}
	\fmf{phantom}{r4,rc4,vr22}
	\fmf{phantom}{r5,rc51,rc5,vr2}
	\fmf{dashes,width=1,tension=0.1}{rc6,rc5}
	\fmf{dashes,width=1,tension=-0.5}{rc4,rc3}
	\fmf{dashes,width=1,tension=-0.5}{rc1,rc2}
	\fmfv{decor.shape=circle,decor.filled=empty,decor.size=13,label=\tiny{$\mathbf{Sp}$},label.dist=0}{vr1}
	\fmfv{decor.shape=circle,decor.filled=empty,decor.size=13,label=\tiny{$\mathbf{Sp}$},label.dist=0}{vr2}
	\fmfv{decor.shape=square,decor.filled=empty,decor.size=20,label=\tiny{$\ket{\mM}$},label.dist=0}{vl}
	\fmfv{decor.shape=square,decor.filled=empty,decor.size=20,label=\tiny{$\bra{\mM}$},label.dist=0}{vr}
	\end{fmfgraph*}
	\end{gathered}
	\quad \xrightarrow{\makebox[1.6cm]{}}
	\quad
	\begin{gathered}
	\begin{fmfgraph*}(140,80)
	\fmfleft{l4,l3,l2,l1}
	\fmfright{r6,r5,r4,r3,r2,r1}
	\fmf{phantom}{l1,vl,l4}
	\fmf{phantom}{r1,vr,r6}
	\fmf{phantom,tension=4}{vl,vr}
	\fmffreeze
	\fmf{dbl_plain,tension=0.7,background=white}{vl,vl1}
	\fmf{plain}{vl1,vl11,l1}
	\fmf{plain}{vl1,vl12,l2}
	\fmf{plain}{vl,vl2,l3}
	\fmf{plain}{vl,vl3,l4}
	\fmffreeze
	\fmf{dots,right=.2,width=1}{vl11,vl12}
	\fmf{dots,right=.2,width=1}{vl2,vl3}
	\fmf{phantom}{vl1,vl11,bl1,l1}
	\fmf{phantom}{vl1,vl12,bl2,l2}
	\fmf{phantom}{vl,bl3,vl1}
	\fmfv{decor.shape=square,decor.filled=50,decor.angle=33,decor.size=5}{bl1}
	\fmfv{decor.shape=square,decor.filled=50,decor.angle=3,decor.size=5}{bl2}
	\fmfv{decor.shape=square,decor.filled=50,decor.angle=60,decor.size=5}{bl3}
	\fmfv{decor.shape=circle,decor.filled=empty,decor.size=13,label=\tiny{$\mathbf{Sp}$},label.dist=0}{vl1}
	\fmf{dbl_plain,tension=1.2,background=white}{vr,vdr1,vr1}
	\fmf{plain}{vr1,vr11,r1}
	\fmf{plain}{vr1,vr12,r2}
	\fmf{dbl_plain,tension=1.2,background=white}{vr,vdr2,vr2}
	\fmf{plain}{vr2,vr21,r3}
	\fmf{plain}{vr2,vr22,r4}
	\fmf{plain}{vr,vr3,r5}
	\fmf{plain}{vr,vr4,r6}
	\fmffreeze
	\fmf{dots,left=.2,width=1}{vr11,vr12}
	\fmf{dots,left=.2,width=1}{vr21,vr22}
	\fmf{dots,left=.2,width=1}{vr3,vr4}
	\fmf{phantom}{vr1,vr11,br1,r1}
	\fmf{phantom}{vr1,vr12,br2,r2}
	\fmf{phantom}{vr2,vr21,br3,r3}
	\fmf{phantom}{vr2,vr22,br4,r4}
	\fmfv{decor.shape=square,decor.filled=50,decor.angle=52,decor.size=5}{br1}
	\fmfv{decor.shape=square,decor.filled=50,decor.angle=10,decor.size=5}{br2}
	\fmfv{decor.shape=square,decor.filled=50,decor.angle=35,decor.size=5}{vdr1}
	\fmfv{decor.shape=square,decor.filled=50,decor.angle=25,decor.size=5}{br3}
	\fmfv{decor.shape=square,decor.filled=50,decor.angle=70,decor.size=5}{br4}
	\fmfv{decor.shape=square,decor.filled=50,decor.angle=0,decor.size=5}{vdr2}
	\fmfv{decor.shape=circle,decor.filled=empty,decor.size=13,label=\tiny{$\mathbf{Sp}$},label.dist=0}{vr1}
	\fmfv{decor.shape=circle,decor.filled=empty,decor.size=13,label=\tiny{$\mathbf{Sp}$},label.dist=0}{vr2}
	\fmfv{decor.shape=square,decor.filled=empty,decor.size=20,label=\tiny{$\ket{\mM}$},label.dist=0}{vl}
	\fmfv{decor.shape=square,decor.filled=empty,decor.size=20,label=\tiny{$\bra{\mM}$},label.dist=0}{vr}
	\end{fmfgraph*}
	\end{gathered}
	\end{align*}
\end{fmffile}%
This procedure needs to be understood as follows: First we apply our
identities to decompose the square of the matrix element into the
operators ${\mathbf P}$ and a set of cut lines, where a dashed cut
line indicates a use of
$$
\slashed{q} \frac{\ n\cdot p}{n\cdot q}\quad \text{or}\quad d^{\mu\nu}(q)
$$
for each on-shell external quark or gluon line,
respectively. Internal lines carrying a sum of different momenta will
be decomposed as indicated by Eqs.~(\ref{eqs:linearity}), and a factor
of $\sqrt{n\cdot q_a/n\cdot p_a}\sqrt{n\cdot q_b/n\cdot p_b}$ will be
applied to each quark-gluon vertex with incoming fermion momentum
$q_a$ (with forward momentum $p_a$) and each outgoing fermion momentum
$q_b$ (with forward momentum $p_b$).  We then carry out the
decomposition of an amplitude's numerator in terms of its forward (along
jet directions $p_i^\mu$), backward (along the gauge vector $n^\mu$)
and transverse ($n_\perp^\mu$) components, indicated by the grey boxes
in the last diagram of the workflow depicted above.  Forward
directions are chosen with respect to jets emerging from a hard
interaction, such that we can properly trace collinear and/or soft
configurations through several emitted partons in the diagram, as well
as iterating kernels describing a particular unresolved configuration,
however a universal factorization onto an on-shell density operator
will generally only happen in very specific configurations of
unresolved emissions. To this extend we shall re-interpret
Eq.~(\ref{eqs:density-operator-general}) by writing it as a sum over
different topologies of diagrams in the amplitude and conjugate
amplitude, which we label $\rho$ and $\bar{\rho}$, respectively:
\begin{equation}\label{eqs:diagrams}
  \tilde{{\mathbf A}}_{n+k} =
  \sum_{\rho,\bar{\rho}} \widetilde{\mathbf{Sp}}^{(k)}_{n,\rho}\ {\mathbf B}_{n,n+k,\rho,\bar{\rho}}\
  \widetilde{\mathbf{Sp}}^{(k),\dagger}_{n,\bar{\rho}} \ ,
\end{equation}
where ${\mathbf B}_{n,n+k,\rho,\bar{\rho}}$ is obtained by removing
the splitting diagrams labelled by $\rho$ and $\bar{\rho}$, and still
constitutes the outer product of two amplitudes with different
compositions of on-shell and off-shell external lines.

Crucially, the
off-shell lines at which we will attempt to factor off emission
subdiagrams will be nearly on-shell in the unresolved configurations,
yet we need to restore overall energy momentum conservation. In our
approach we choose to do this by a global Lorentz transformation
combined with a scaling of each of the final state momenta, such that
momenta $q_i$ after emission, and with momentum conservation
maintained, $\sum_i q_i = Q$ are parametrised as
\begin{equation}
  \label{eq:globalrecoil}
  q_i^\mu = \frac{1}{\hat{\alpha}} \Lambda^\mu {}_\nu \hat{q}_i^\nu \ ,
\end{equation}
where the $\hat{q}_i$ are decomposed into the jet directions and do
not obey momentum conservation, $\sum_i \hat{q}_i = Q+ N$. The scaling
factor $\hat{\alpha}$ then simply encodes the correction of the different
mass shells $(Q+N)^2$ and $Q^2$. The recoil momentum $N$ can thus
easily be absorbed through the Lorentz transformation. The advantage
of such an approach is that recoil effects are now completely removed
from considerations about factorizing the amplitude, as we can use
Lorentz invariance and the known mass dimension of the amplitude to
remove this effect. More precisely,
\begin{equation}
    \label{eq:amplitudeglobalrecoil}
    |{\cal M}(q_1,...,q_n)\rangle = \frac{1}{\hat{\alpha}^{2n-4}}
    |{\cal M}(\hat{q}_1,...,\hat{q}_n)\rangle \ .
\end{equation}
In the following we will, unless stated otherwise, therefore treat the
transformed and untransformed momenta as equivalent. Notice that for
additional, massive particles involved, we can easily extend such a
recoil scheme which will not change the argument given here. Notably,
when working in a physical gauge, as we choose below, the gauge vector
$n$ will also have to be transformed by the recoil prescription in
order for the argument to stay valid. We will link this to how the
recoil transformation should be iterated in
Sec.~\ref{sec:decompositions-counting}.

An expansion around a certain limit in which the off-shell lines
become on-shell in \eqref{eqs:diagrams} is possible only if we can
expose this particular limit by a weighting factor, ensuring that no
other singular limit can contribute by virtue of either the weighting
(or partitioning) factor we introduce or the physical gauge we
choose. We therefore write \eqref{eqs:diagrams} as
\begin{multline}\label{eqs:paritioning-origin}
  \tilde{{\mathbf A}}_{n+k} =
  \sum_{\rho}\left[
  \sum_{\lambda}\left(\frac{w_{\rho,\lambda;\rho}}{w_{\rho,\lambda}}
  \widetilde{\mathbf{Sp}}^{(k)}_{n,\rho}\ {\mathbf B}_{n,n+k,\rho,\lambda}\
  \widetilde{\mathbf{Sp}}^{(k),\dagger}_{n,\lambda} +
  \frac{w_{\lambda,\rho;\rho}}{w_{\lambda,\rho}}
  \widetilde{\mathbf{Sp}}^{(k)}_{n,\lambda}\ {\mathbf B}_{n,n+k,\lambda,\rho}\
  \widetilde{\mathbf{Sp}}^{(k),\dagger}_{n,\rho}
  \right)\right. +\\\left.
  \sum_{\lambda\ne \rho}\sum_{\sigma\ne \rho} \frac{w_{\sigma,\lambda;\rho}}{w_{\sigma,\lambda}}
  \widetilde{\mathbf{Sp}}^{(k)}_{n,\sigma}\ {\mathbf B}_{n,n+k,\sigma,\lambda}\
  \widetilde{\mathbf{Sp}}^{(k),\dagger}_{n,\lambda}
  \right] \ ,
\end{multline}
with $w_{\rho,\bar{\rho}} = \sum_\sigma w_{\rho,\bar{\rho};\sigma}$
and where $w_{\rho,\bar{\rho};\sigma}$ ensures that a diagram with
propagator structures identified by $\rho$ and $\bar{\rho}$ contains a
collinear enhancement solely as dictated by the topology $\sigma$, but
not by those identified through $\rho$ and $\bar{\rho}$ (though all of
these might contain common configurations of collinear enhancement).
An index $\rho$ can be visualized as part of a set of distinct
emission patterns, which can in turn be mapped to ordered tuples of
the set of all partons involved in an $(n+k)$-splitting.  This
procedure will allow us to uniquely identify one parametrization of
the external momenta in terms of forward and backward directions which
we can then use to set up a systematic power counting in each singular
limit.

As an example to illustrate the contributions in
\eqref{eqs:paritioning-origin}, we examine it for fixed $\rho =
(ijk)$, meaning a splitting with a hard line $i$ and two sequential
emissions $j$ and $k$ off of it.  Then the first line in
\eqref{eqs:paritioning-origin} represents all two emission topologies
containing this emission pattern, such as the self energy $E^{(1)}$
for $\lambda = \rho$ or topology $B^{(1)}$ for $\lambda=(ij)(lk)$ and
their conjugates (see App.~\ref{app:two-emission-diagrams} for a list
of topologies).  The third term represents all topologies that do not
contain this emission pattern, such as the $X$-topology.  Each
contribution is multiplied by their respective weighting factor which
cancels all singular factors not belonging to the $(ijk)$-collinear
sector.  These are precisely the partitioning factors of
Sec.~\ref{sec:Paritioning}.  In this way of a decomposition, we are
able to collect all possible contributions to a specific splitting
pattern.  The collection of such terms for fixed $\rho$ is what we
call \emph{splitting kernels} which are defined in
Sec.~\ref{sec:Kernels}.

\subsection{Decomposition of amplitudes in a physical gauge}
\label{sec:decompositions-intro}

Once we have identified a certain splitting process by the emission
subdiagrams contributing in the amplitude factor, \ie one term
of fixed $\rho$ in \eqref{eqs:paritioning-origin}, we decompose all
of the momenta involved in each splitting $(r_i)$ as
\begin{equation}\label{eq:Sudakov-decomp-single}
  r_{ik}^\mu = z_{ik} \, p_i^\mu +
  \frac{p_{\perp,ik}^2}{z_{ik}\ 2 p_i\cdot n}\, n^\mu + k_{\perp,ik}^\mu \ ,
\end{equation}
where $k_{\perp,ik}^2=-p_{\perp,ik}^2$, $p_i\cdot k_{\perp,ik} =
n\cdot k_{\perp,ik}=0$ and $p_i$ is the momentum of a progenitor of
the splitting. An off-shell internal line $I$ within a splitting
subdiagram $(r_i)$ in the amplitude carries momenta of the form
\begin{align}\label{eq:Sudakov-decomp-combined}
  q_I^\mu = \sum_{k\in I} r_{ik} = z_I \, p_i^\mu +
  \frac{S_I + p_{\perp,I}^2}{ z_I\ 2 p_i\scdot n}\,  n^\mu + k_{\perp,I}^\mu \ ,
\end{align}
where $q_I^2=S_I>0$ and the transverse components as well as momentum
fractions are additive
\begin{equation}
 k_{\perp,I}^\mu = \sum_{k\in I}  k_{\perp,ik}^\mu,\qquad z_I = \sum_{k\in I} z_{ik},
\end{equation}
while
\begin{equation}
  S_I = \Big(\sum_{k\in I} z_{ik}\Big) \sum_{l\in I} \frac{p_{\perp,il}^2}{z_{il}} +
  \sum_{k,l\in I} k_{\perp,ik}\cdot k_{\perp,il}, \qquad p_{\perp,I}^2 =
  \Big(\sum\limits_{k \in I} k_{\perp,ik}^\mu\Big)^2,
\end{equation}
such that
\begin{equation}
  \beta_I \equiv \frac{S_I + p_{\perp,I}^2}{2 z_I\ p_i\scdot n} =\frac{1}{2p_i\cdot n} \sum_{k\in I} \frac{p_{\perp,ik}^2}{z_{ik}},
\end{equation}
consistent with momentum conservation.  Notice that, if all lines
combining into $I$ are collinear w.r.t.\ $p_i$, meaning
$p_{\perp,ik}\to \lambda p_{\perp,ik}$, we obtain $S_I\sim \lambda^2$;
if all lines are soft, $p_{\perp,ik}\to \lambda p_{\perp,ik}$ and
$z_{ik}\to \lambda z_{ik}$, we obtain the same.  However, if one line
$k$ is hard collinear in the sense that it has an $z_{ik} = {\cal
  O}(1)$, we find $S_I\sim \lambda^2$ only in the limit of all lines
becoming collinear, while $S_I\sim \lambda$ in the combined
soft/collinear limit. For the backward component this implies that it
is ${\cal O}(\lambda)$ for all soft, or a soft/collinear combination,
and ${\cal O}(\lambda^2)$ for an all-collinear configuration.  An
overview of the line- and vertex-scalings in all possible collinear
and soft settings is shown in Sec.~\ref{sec:scaling-of-lines}.

We can use \eqref{eq:Sudakov-decomp-combined} for decomposing the
scaled fermion numerators $(n\cdot p_i/n\cdot q_I) \slashed{q}_I$ as
\begin{fmffile}{fermion-lines-sudakov-scaling}
	\fmfset{thin}{.7pt}
	\fmfset{arrow_len}{1.8mm}
	\fmfset{curly_len}{1.8mm}
	\fmfset{dash_len}{1.5mm}
	\begin{subequations}\label{eq:power-counting-lines-fermion}
	\begin{align}
	\begin{gathered}
	\vspace{-4pt}
	\begin{fmfgraph*}(50,30)
	\fmfleft{l}
	\fmfright{r}
	\fmf{phantom}{l,vbox,r}
	\fmffreeze
	\fmf{fermion}{l,vbox,r}
	\fmfv{decor.shape=circle,decor.filled=empty,decor.size=2}{l}
	\fmfv{decor.shape=circle,decor.filled=empty,decor.size=2}{r}
	\fmfv{decor.shape=square,decor.filled=empty,decor.size=8}{vbox}
	\end{fmfgraph*}
	\end{gathered}
	& \; = \slashed{p}_i\ , 
	\\
	\begin{gathered}
	\vspace{-4pt}
	\begin{fmfgraph*}(50,30)
	\fmfleft{l}
	\fmfright{r}
	\fmf{phantom}{l,vbox,r}
	\fmffreeze
	\fmf{fermion}{l,vbox,r}
	\fmfv{decor.shape=circle,decor.filled=empty,decor.size=2}{l}
	\fmfv{decor.shape=circle,decor.filled=empty,decor.size=2}{r}
	\fmfv{decor.shape=square,decor.filled=full,decor.size=8}{vbox}
	\end{fmfgraph*}
	\end{gathered}
	& \; = \frac{S_I + p_{\perp,I}^2}{2 z_I^2 \, p_i \scdot n}\slashed{n} = \frac{1}{z_I}\sum_{k\in I} \frac{-k_{\perp,ik}^2}{z_{ik} 2 p_i\cdot n} \slashed{n}\ ,
	\\
	\begin{gathered}
	\vspace{-4pt}
	\begin{fmfgraph*}(50,30)
	\fmfleft{l}
	\fmfright{r}
	\fmf{phantom}{r,vbox,l}
	\fmffreeze
	\fmf{fermion}{l,vbox,r}
	\fmfv{decor.shape=circle,decor.filled=empty,decor.size=2}{l}
	\fmfv{decor.shape=circle,decor.filled=empty,decor.size=2}{r}
	\fmfv{decor.shape=square,decor.filled=empty,decor.size=8,label=\tiny{$\perp$},label.dist=0}{vbox}
	\end{fmfgraph*}
	\end{gathered}
	& \; = \frac{\slashed{k}_{\perp,I}}{z_I} = \frac{\sum_{k\in I} \slashed{k}_{\perp, ik}}{\sum_{k\in I} z_{ik}}\ ,
	\end{align}
	\end{subequations}	
\end{fmffile}%

and for gluon lines $d^{\mu\nu}(q_I)$ as
\begin{fmffile}{gluon-lines-sudakov-scaling}
	\fmfset{thin}{.7pt}
	\fmfset{arrow_len}{1.8mm}
	\fmfset{curly_len}{1.8mm}
	\fmfset{dash_len}{1.5mm}
	\begin{subequations}\label{eq:power-counting-lines-gluon}
		\begin{align}
		\begin{gathered}
		\vspace{-4pt}
		\begin{fmfgraph*}(50,30)
		\fmfleft{l}
		\fmfright{r}
		\fmf{phantom}{l,vbox,r}
		\fmffreeze
		\fmf{curly}{l,vbox,r}
		\fmfv{decor.shape=circle,decor.filled=empty,decor.size=2}{l}
		\fmfv{decor.shape=circle,decor.filled=empty,decor.size=2}{r}
		\fmfv{decor.shape=square,decor.filled=empty,decor.size=8}{vbox}
		\end{fmfgraph*}
		\end{gathered}
		& \; = d^{\mu \nu} (p_i)\ , 
		\\
		\begin{gathered}
		\vspace{-4pt}
		\begin{fmfgraph*}(50,30)
		\fmfleft{l}
		\fmfright{r}
		\fmf{phantom}{l,vbox,r}
		\fmffreeze
		\fmf{curly}{l,vbox,r}
		\fmfv{decor.shape=circle,decor.filled=empty,decor.size=2}{l}
		\fmfv{decor.shape=circle,decor.filled=empty,decor.size=2}{r}
		\fmfv{decor.shape=square,decor.filled=full,decor.size=8}{vbox}
		\end{fmfgraph*}
		\end{gathered}
		& \; = \frac{S_I + p_{\perp,I}^2}{ (z_I \, p_i \scdot n)^2}\, n^\mu n^\nu =   \frac{1}{z_I}\sum_{k\in I} \frac{-k_{\perp,ik}^2}{z_{ik} (p_i \scdot n)^2}\, n^\mu n^\nu\ ,
		\\
		\begin{gathered}
		\vspace{-4pt}
		\begin{fmfgraph*}(50,30)
		\fmfleft{l}
		\fmfright{r}
		\fmf{phantom}{r,vbox,l}
		\fmffreeze
		\fmf{curly}{l,vbox,r}
		\fmfv{decor.shape=circle,decor.filled=empty,decor.size=2}{l}
		\fmfv{decor.shape=circle,decor.filled=empty,decor.size=2}{r}
		\fmfv{decor.shape=square,decor.filled=empty,decor.size=8,label=\tiny{$\perp$},label.dist=0}{vbox}
		\end{fmfgraph*}
		\end{gathered}
		& \; = \frac{k_{\perp,I}^\mu n^\nu + n^\mu k_{\perp,I}^\nu}{z_I \,p_i \scdot n}=\frac{\sum_{k\in I}k_{\perp,ik}^\mu n^\nu + \sum_{k\in I}n^\mu k_{\perp,ik}^\nu}{\sum_{k\in I}z_{ik} \,p_i \scdot n} \ .
		\end{align}
	\end{subequations}	
\end{fmffile}%
These decompositions hold both for the numerators we obtain from the
sum over external wave functions, and internal lines.  Comparing
\eqref{eq:power-counting-lines-fermion} with
\eqref{eq:power-counting-lines-gluon} shows that the components in
these decompositions give the same soft and collinear scaling
behaviour for quarks and gluons.  To this extent we can 'colour' all
internal lines in the amplitude and the conjugate amplitude, as well
as the external lines by the different contributions in the
decomposition above.  Note that it suffices both for external on-shell
gluon and quark lines to use the projectors with only the forward
momentum component $p_i$ as an argument, due to
\begin{subequations}
  \begin{align}
    \slashed{r}_{ik} \, \frac{\slashed{n}}{2 n \scdot p_i} \,
    \slashed{r}_{ik} &= z_{ik} \slashed{r}_{ik},
    \\
    d^{\mu \rho}
    (r_{ik}) \, d_{\rho \sigma} (p_i) \, d^{\sigma \nu} (r_{ik}) &=
    d^{\mu \nu} (r_{ik}).
  \end{align}
\end{subequations}
Moreover, we find the following rules for connecting on-shell lines
via the projectors, \viz
\begin{fmffile}{external-line-rules}
	\fmfset{thin}{.7pt}
	\fmfset{arrow_len}{1.8mm}
	\fmfset{curly_len}{1.8mm}
	\fmfset{dash_len}{1.5mm}
	\begin{subequations}\label{eq:line-rules}
		\begin{align}
			\; &\mathbf{P}(p_i) \;
			\begin{gathered}
				\vspace{-4pt}
				\begin{fmfgraph*}(30,30)
					\fmfleft{l}
					\fmfright{r}
					\fmf{phantom}{l,vbox,r}
					\fmffreeze
					\fmf{plain}{l,vbox,r}
					\fmfv{decor.shape=circle,decor.filled=empty,decor.size=2}{r}
					\fmfv{decor.shape=square,decor.filled=full,decor.size=8}{vbox}
				\end{fmfgraph*}
			\end{gathered}
			\;= \;
			0, 
			\label{eq:line-rules-black}
			\\
			\begin{gathered}
				\vspace{-4pt}
				\begin{fmfgraph*}(30,30)
					\fmfleft{l}
					\fmfright{r}
					\fmf{phantom}{l,vbox,r}
					\fmffreeze
					\fmf{plain}{l,vbox,r}
					\fmfv{decor.shape=circle,decor.filled=empty,decor.size=2}{l}
					\fmfv{decor.shape=square,decor.filled=empty,decor.size=8}{vbox}
				\end{fmfgraph*}
			\end{gathered}
			\; &\mathbf{P}(p_i) \;
			\begin{gathered}
				\vspace{-4pt}
				\begin{fmfgraph*}(30,30)
					\fmfleft{l}
					\fmfright{r}
					\fmf{phantom}{l,vbox,r}
					\fmffreeze
					\fmf{plain}{l,vbox,r}
					\fmfv{decor.shape=circle,decor.filled=empty,decor.size=2}{r}
					\fmfv{decor.shape=square,decor.filled=empty,decor.size=8}{vbox}
				\end{fmfgraph*}
			\end{gathered}
			\;= \;
			\begin{gathered}
				\vspace{-4pt}
				\begin{fmfgraph*}(30,30)
					\fmfleft{l}
					\fmfright{r}
					\fmf{phantom}{l,vbox,r}
					\fmffreeze
					\fmf{plain}{l,vbox,r}
					\fmfv{decor.shape=circle,decor.filled=empty,decor.size=2}{l}
					\fmfv{decor.shape=circle,decor.filled=empty,decor.size=2}{r}
					\fmfv{decor.shape=square,decor.filled=empty,decor.size=8}{vbox}
				\end{fmfgraph*}
			\end{gathered}\;,
		\label{eq:line-rules-white}
		\\
			\begin{gathered}
				\vspace{-4pt}
				\begin{fmfgraph*}(30,30)
					\fmfleft{l}
					\fmfright{r}
					\fmf{phantom}{l,vbox,r}
					\fmffreeze
					\fmf{plain}{l,vbox,r}
					\fmfv{decor.shape=circle,decor.filled=empty,decor.size=2}{l}
					\fmfv{decor.shape=square,decor.filled=empty,decor.size=8,label=\tiny{$\perp$},label.dist=0}{vbox}
				\end{fmfgraph*}
			\end{gathered}
			\; &\mathbf{P}(p_i) \;
			\begin{gathered}
				\vspace{-4pt}
				\begin{fmfgraph*}(30,30)
					\fmfleft{l}
					\fmfright{r}
					\fmf{phantom}{l,vbox,r}
					\fmffreeze
					\fmf{plain}{l,vbox,r}
					\fmfv{decor.shape=circle,decor.filled=empty,decor.size=2}{r}
					\fmfv{decor.shape=square,decor.filled=empty,decor.size=8,label=\tiny{$\perp$},label.dist=0}{vbox}
				\end{fmfgraph*}
			\end{gathered}
			\;= \;
			\begin{gathered}
				\vspace{-4pt}
				\begin{fmfgraph*}(30,30)
					\fmfleft{l}
					\fmfright{r}
					\fmf{phantom}{l,vbox,r}
					\fmffreeze
					\fmf{plain}{l,vbox,r}
					\fmfv{decor.shape=circle,decor.filled=empty,decor.size=2}{l}
					\fmfv{decor.shape=circle,decor.filled=empty,decor.size=2}{r}
					\fmfv{decor.shape=square,decor.filled=full,decor.size=8,label=\tiny{$\perp$},label.dist=0}{vbox}
				\end{fmfgraph*}
			\end{gathered}\;,
		\label{eq:line-rules-perp-perp}
			\\
			\begin{gathered}
				\vspace{-4pt}
				\begin{fmfgraph*}(30,30)
					\fmfleft{l}
					\fmfright{r}
					\fmf{phantom}{l,vbox,r}
					\fmffreeze
					\fmf{plain}{l,vbox,r}
					\fmfv{decor.shape=circle,decor.filled=empty,decor.size=2}{l}
					\fmfv{decor.shape=square,decor.filled=empty,decor.size=8,label=\tiny{$\perp$},label.dist=0}{vbox}
				\end{fmfgraph*}
			\end{gathered}
			\; &\mathbf{P}(p_i) \;
			\begin{gathered}
				\vspace{-4pt}
				\begin{fmfgraph*}(30,30)
					\fmfleft{l}
					\fmfright{r}
					\fmf{phantom}{l,vbox,r}
					\fmffreeze
					\fmf{plain}{l,vbox,r}
					\fmfv{decor.shape=circle,decor.filled=empty,decor.size=2}{r}
					\fmfv{decor.shape=square,decor.filled=empty,decor.size=8}{vbox}
				\end{fmfgraph*}
			\end{gathered}
		\; + \;
			\begin{gathered}
				\vspace{-4pt}
				\begin{fmfgraph*}(30,30)
					\fmfleft{l}
					\fmfright{r}
					\fmf{phantom}{l,vbox,r}
					\fmffreeze
					\fmf{plain}{l,vbox,r}
					\fmfv{decor.shape=circle,decor.filled=empty,decor.size=2}{l}
					\fmfv{decor.shape=square,decor.filled=empty,decor.size=8}{vbox}
				\end{fmfgraph*}
			\end{gathered}
			\; \mathbf{P}(p_i) \;
			\begin{gathered}
				\vspace{-4pt}
				\begin{fmfgraph*}(30,30)
					\fmfleft{l}
					\fmfright{r}
					\fmf{phantom}{l,vbox,r}
					\fmffreeze
					\fmf{plain}{l,vbox,r}
					\fmfv{decor.shape=circle,decor.filled=empty,decor.size=2}{r}
					\fmfv{decor.shape=square,decor.filled=empty,decor.size=8,label=\tiny{$\perp$},label.dist=0}{vbox}
				\end{fmfgraph*}
			\end{gathered}
			\;= \;
			\begin{gathered}
				\vspace{-4pt}
				\begin{fmfgraph*}(30,30)
					\fmfleft{l}
					\fmfright{r}
					\fmf{phantom}{l,vbox,r}
					\fmffreeze
					\fmf{plain}{l,vbox,r}
					\fmfv{decor.shape=circle,decor.filled=empty,decor.size=2}{l}
					\fmfv{decor.shape=circle,decor.filled=empty,decor.size=2}{r}
					\fmfv{decor.shape=square,decor.filled=empty,decor.size=8,label=\tiny{$\perp$},label.dist=0}{vbox}
				\end{fmfgraph*}
			\end{gathered}\;,
		\end{align}
	\end{subequations}	
\end{fmffile}
which hold both for quarks and gluons.  Due to
\eqref{eq:line-rules-black} and \eqref{eq:line-rules-perp-perp}, we
note that the backwards ($n$) components on external lines do not need
to be taken into account in the decomposition of an amplitude.  Via
these rules, we are able to translate our density operator-like
discussion to the one of cut diagrams or vice versa.

Let us first discuss the vertex structures coupling these different
internal lines. We find that, making use of the fact that $n_\mu
d^{\mu\nu}=0$, the quark gluon vertex and three gluon vertex can
effectively be decomposed into two components depending only on the
longitudinal and only on the transverse components,
\begin{fmffile}{definition-col-perp-vertices}
	\fmfset{thin}{.7pt}
	\fmfset{arrow_len}{2.5mm}
	\fmfset{curly_len}{1.8mm}
	\fmfset{dash_len}{1.5mm}
	\begin{subequations}\label{eq:vertex-rules}
	\begin{eqnarray}
	\begin{gathered}
	\vspace{-4pt}
	\begin{fmfgraph*}(50,50)
	\fmfbottom{l,m,r}
	\fmftop{t}
	\fmf{phantom}{m,v,t}
	\fmffreeze
	\fmf{fermion}{l,v,r}
	\fmf{curly}{t,v}
	\fmfv{label=\small{$k$},label.angle=0}{t}
	\fmfv{label=\small{$i$}}{r}
	\fmfv{label=\small{$j$}}{l}
	\fmfv{decor.shape=circle,decor.filled=empty,decor.size=16,label=\small{$\parallel$},label.dist=0}{v}
	\end{fmfgraph*}
	\end{gathered}
	& \quad = \ \sqrt{z_iz_j} \ &
	\begin{gathered}
	\vspace{-4pt}
	\begin{fmfgraph*}(50,50)
	\fmfbottom{l,m,r}
	\fmftop{t}
	\fmf{phantom}{m,v,t}
	\fmffreeze
	\fmf{fermion}{l,v,r}
	\fmf{curly}{t,v}
	\fmfv{label=\small{$z_k p_k$},label.angle=0}{t}
	\fmfv{label=\small{$z_i p_i$},label.angle=-90}{r}
	\fmfv{label=\small{$z_j p_j$},label.angle=-90}{l}
	\end{fmfgraph*}
	\end{gathered}
	\\[10pt]
	\begin{gathered}
	\vspace{-4pt}
	\begin{fmfgraph*}(50,50)
	\fmfbottom{l,m,r}
	\fmftop{t}
	\fmf{phantom}{m,v,t}
	\fmffreeze
	\fmf{curly}{r,v,l}
	\fmf{curly}{t,v}
	\fmfv{label=\small{$i$},label.angle=0}{t}
	\fmfv{label=\small{$j$}}{r}
	\fmfv{label=\small{$k$}}{l}
	\fmfv{decor.shape=circle,decor.filled=empty,decor.size=16,label=\small{$\parallel$},label.dist=0}{v}
	\end{fmfgraph*}
	\end{gathered}
	& \quad = \quad &
	\begin{gathered}
	\vspace{-4pt}
	\begin{fmfgraph*}(50,50)
	\fmfbottom{l,m,r}
	\fmftop{t}
	\fmf{phantom}{m,v,t}
	\fmffreeze
	\fmf{curly}{r,v,l}
	\fmf{curly}{t,v}
	\fmfv{label=\small{$z_i p_i$},label.angle=0}{t}
	\fmfv{label=\small{$z_j p_j$},label.angle=-90}{r}
	\fmfv{label=\small{$z_k p_k$},label.angle=-90}{l}
	\end{fmfgraph*}
	\end{gathered}
	\label{eq:three-gluon-col-vertex}
	\\[10pt]
	\begin{gathered}
	\vspace{-4pt}
	\begin{fmfgraph*}(50,50)
	\fmfbottom{l,m,r}
	\fmftop{t}
	\fmf{phantom}{m,v,t}
	\fmffreeze
	\fmf{curly}{r,v,l}
	\fmf{curly}{t,v}
	\fmfv{label=\small{$i$},label.angle=0}{t}
	\fmfv{label=\small{$j$}}{r}
	\fmfv{label=\small{$k$}}{l}
	\fmfv{decor.shape=circle,decor.filled=empty,decor.size=16,label=\small{$\perp$},label.dist=0}{v}
	\end{fmfgraph*}
	\end{gathered}
	& \quad = \quad &
	\begin{gathered}
	\vspace{-4pt}
	\begin{fmfgraph*}(50,50)
	\fmfbottom{l,m,r}
	\fmftop{t}
	\fmf{phantom}{m,v,t}
	\fmffreeze
	\fmf{curly}{r,v,l}
	\fmf{curly}{t,v}
	\fmfv{label=\small{$k_{\perp i}$},label.angle=0}{t}
	\fmfv{label=\small{$k_{\perp j}$},label.angle=-90}{r}
	\fmfv{label=\small{$k_{\perp k}$},label.angle=-90}{l}
	\end{fmfgraph*}
	\end{gathered}
	\end{eqnarray}
	\end{subequations}
\end{fmffile}%
We also find that certain combinations of lines always vanish,
irrespective of the progenitor momentum and/or the off-shellness of the
lines. In particular this applies to the cases in which more than one
line with a black square connects to a vertex, indicating that a
backward component can only propagate upon exchanging at least some
transverse or longitudinal momentum with another parton,
\begin{fmffile}{boxed-vertices}
	\fmfset{thin}{.7pt}
	\fmfset{arrow_len}{2.5mm}
	\fmfset{curly_len}{1.8mm}
	\fmfset{dash_len}{1.5mm}
	\begin{align}\label{eq:vanishing-vertices-1}
	\begin{gathered}
	\vspace{-4pt}
	\begin{fmfgraph*}(40,40)
	\fmftop{r,m,l}
	\fmfbottom{b}
	\fmf{phantom}{m,v,b}
	\fmffreeze
	\fmf{plain}{l,v,r}
	\fmf{plain}{b,v}
	\fmf{phantom}{b,bb,v}
	\fmf{phantom}{l,lb,v}
	\fmf{phantom}{r,rb,v}
	\fmfv{decor.shape=square,decor.filled=full,decor.size=6}{bb}
	\fmfv{decor.shape=square,decor.filled=full,decor.size=6}{lb}
	\fmfv{decor.shape=square,decor.filled=full,decor.size=6}{rb}
	\end{fmfgraph*}
	\end{gathered}
	\; = \; 0,\quad
	\begin{gathered}
	\vspace{-4pt}
	\begin{fmfgraph*}(40,40)
	\fmftop{r,m,l}
	\fmfbottom{b}
	\fmf{phantom}{m,v,b}
	\fmffreeze
	\fmf{plain}{l,v,r}
	\fmf{plain}{b,v}
	\fmf{phantom}{b,bb,v}
	\fmf{phantom}{l,lb,v}
	\fmf{phantom}{r,rb,v}
	\fmfv{decor.shape=square,decor.filled=empty,decor.size=6}{bb}
	\fmfv{decor.shape=square,decor.filled=full,decor.size=6}{lb}
	\fmfv{decor.shape=square,decor.filled=full,decor.size=6}{rb}
	\end{fmfgraph*}
	\end{gathered}
	\; = \; 0, \quad
	\begin{gathered}
	\vspace{-4pt}
	\begin{fmfgraph*}(40,40)
	\fmftop{r,m,l}
	\fmfbottom{b}
	\fmf{phantom}{m,v,b}
	\fmffreeze
	\fmf{plain}{l,v,r}
	\fmf{plain}{b,v}
	\fmf{phantom}{b,bb,v}
	\fmf{phantom}{l,lb,v}
	\fmf{phantom}{r,rb,v}
	\fmfv{decor.shape=square,decor.filled=empty,decor.size=6,label=\scalebox{.5}{$\perp$},label.dist=0}{bb}
	\fmfv{decor.shape=square,decor.filled=full,decor.size=6}{lb}
	\fmfv{decor.shape=square,decor.filled=full,decor.size=6}{rb}
	\end{fmfgraph*}
	\end{gathered}
	\; = \; 0.
	\end{align}
\end{fmffile}%

both for quark and gluon insertions. The following vertices also
vanish but only in the gluon case:
\begin{fmffile}{boxed-gluon-vertices}
	\fmfset{thin}{.7pt}
	\fmfset{arrow_len}{2.5mm}
	\fmfset{curly_len}{1.5mm}
	\fmfset{dash_len}{1.5mm}
	\begin{align}\label{eq:vanishing-vertices}
	\begin{gathered}
	\vspace{-4pt}
	\begin{fmfgraph*}(50,60)
	\fmfbottom{l,m,r}
	\fmftop{t}
	\fmf{phantom}{m,v,t}
	\fmffreeze
	\fmf{phantom}{r,rb,rb2,v,lb2,lb,l}
	\fmf{phantom}{t,tb,tb2,tb3,v}
	\fmf{curly}{r,rb,v,lb,l}
	\fmf{curly}{t,tb,v}
	\fmfv{decor.shape=square,decor.filled=full,decor.size=6}{tb}
	\fmfv{decor.shape=square,decor.filled=empty,decor.size=6}{lb}
	\fmfv{decor.shape=square,decor.filled=empty,decor.size=6}{rb}
	\fmfv{decor.shape=circle,decor.filled=empty,decor.size=16,label=\small{$\perp$},label.dist=0}{v}
	\end{fmfgraph*}
	\end{gathered}
	\; = \; 0, \quad
	\begin{gathered}
	\vspace{-4pt}
	\begin{fmfgraph*}(50,60)
	\fmfbottom{l,m,r}
	\fmftop{t}
	\fmf{phantom}{m,v,t}
	\fmffreeze
	\fmf{phantom}{r,rb,rb2,v,lb2,lb,l}
	\fmf{phantom}{t,tb,tb2,tb3,v}
	\fmf{curly}{r,rb,v,lb,l}
	\fmf{curly}{t,tb,v}
	\fmfv{decor.shape=square,decor.filled=full,decor.size=6}{tb}
	\fmfv{decor.shape=square,decor.filled=empty,decor.size=6}{lb}
	\fmfv{decor.shape=square,decor.filled=empty,decor.size=6,label=\scalebox{.5}{$\perp$},label.dist=0}{rb}
	\fmfv{decor.shape=circle,decor.filled=empty,decor.size=16,label=\small{$\perp$},label.dist=0}{v}
	\end{fmfgraph*}
	\end{gathered}
	\; = \; 0, \quad
	\begin{gathered}
	\vspace{-4pt}
	\begin{fmfgraph*}(50,60)
	\fmfbottom{l,m,r}
	\fmftop{t}
	\fmf{phantom}{m,v,t}
	\fmffreeze
	\fmf{phantom}{r,rb,rb2,v,lb2,lb,l}
	\fmf{phantom}{t,tb,tb2,tb3,v}
	\fmf{curly}{r,rb,v,lb,l}
	\fmf{curly}{t,tb,v}
	\fmfv{decor.shape=square,decor.filled=full,decor.size=6}{tb}
	\fmfv{decor.shape=square,decor.filled=empty,decor.size=6,label=\scalebox{.5}{$\perp$},label.dist=0}{lb}
	\fmfv{decor.shape=square,decor.filled=empty,decor.size=6,label=\scalebox{.5}{$\perp$},label.dist=0}{rb}
	\fmfv{decor.shape=circle,decor.filled=empty,decor.size=16,label=\small{$\perp$},label.dist=0}{v}
	\end{fmfgraph*}
	\end{gathered}
	\; = \; 0.
	\end{align}
\end{fmffile}%

Further simplifications occur if all lines which are coupled together
share the same forward direction; in this case we also have that
vertices coupling three forward components vanish identically, if the
participating momenta have been decomposed with respect to the same
forward direction $p_i$.

\subsection{Decomposing around singular limits}
\label{sec:decompositions-counting}

Let us first focus on those terms in \eqref{eqs:paritioning-origin}
for which we can uniquely assign a splitting configuration to the
diagram in the amplitude or conjugate amplitude, \ie the terms
in the first line of \eqref{eqs:paritioning-origin}.
\begin{fmffile}{notation}
	\fmfset{thin}{.7pt}
	\fmfset{dot_len}{.8mm}
	\fmfset{dot_size}{5}
	\fmfset{arrow_len}{2.5mm}
	\fmfset{wiggly_len}{2.2mm}
	\setlength{\belowdisplayskip}{0pt}
	\begin{align*}
	\begin{gathered}
	\begin{fmfgraph*}(140,80)
	\fmfleft{l6,l5,l4,l3,l2,l1}
	\fmfright{r6,r5,r4,r3,r2,r1}
	\fmf{phantom}{l1,vl,l6}
	\fmf{phantom}{r1,vr,r6}
	\fmf{phantom,tension=4}{vl,vr}
	\fmffreeze
	\fmf{dbl_plain,tension=1,background=white}{vl,vl1}
	\fmf{plain}{vl1,vl11,l1}
	\fmf{plain}{vl1,vl12,l2}
	\fmf{plain}{vl1,vl13,l3}
	\fmf{plain}{vl,vl2,l4}
	\fmf{plain}{vl,vl3,l5}
	\fmffreeze
	\fmf{dots,right=.2,width=1}{vl2,vl3}
	\fmf{phantom}{vl1,vl11,bl1,l1}
	\fmf{phantom}{vl1,vl12,bl2,l2}
	\fmf{phantom}{vl1,vl13,bl3,l3}
	\fmf{phantom}{vl,bl4,vl1}
	\fmfv{decor.shape=square,decor.filled=50,decor.angle=28,decor.size=5}{bl1}
	\fmfv{decor.shape=square,decor.filled=50,decor.angle=60,decor.size=5}{bl2}
	\fmfv{decor.shape=square,decor.filled=50,decor.angle=15,decor.size=5}{bl3}
	\fmfv{decor.shape=square,decor.filled=50,decor.angle=63,decor.size=5}{bl4}
	\fmfv{decor.shape=circle,decor.filled=empty,decor.size=13,label=\tiny{$\mathbf{Sp}$},label.dist=0}{vl1}
	\fmf{dbl_plain,tension=1.2,background=white}{vr,vdr1,vr1}
	\fmf{plain}{vr1,vr11,r1}
	\fmf{plain}{vr1,vr12,r2}
	\fmf{dbl_plain,tension=1.4,background=white}{vr,vdr2,vr2}
	\fmf{plain}{vr2,vr21,r3}
	\fmf{plain}{vr2,vr22,r4}
	\fmf{plain}{vr,vr3,r5}
	\fmf{plain}{vr,vr4,r6}
	\fmffreeze
	\fmf{dots,left=.2,width=1}{vr3,vr4}
	\fmf{phantom}{vr1,vr11,br1,r1}
	\fmf{phantom}{vr1,vr12,br2,r2}
	\fmf{phantom}{vr2,vr21,br3,r3}
	\fmf{phantom}{vr2,vr22,br4,r4}
	\fmfv{decor.shape=square,decor.filled=50,decor.angle=52,decor.size=5}{br1}
	\fmfv{decor.shape=square,decor.filled=50,decor.angle=10,decor.size=5}{br2}
	\fmfv{decor.shape=square,decor.filled=50,decor.angle=22,decor.size=5}{br3}
	\fmfv{decor.shape=square,decor.filled=50,decor.angle=-22,decor.size=5}{br4}
	\fmfv{decor.shape=square,decor.filled=50,decor.angle=35,decor.size=5}{vdr1}
	\fmfv{decor.shape=square,decor.filled=50,decor.angle=0,decor.size=5}{vdr2}
	\fmfv{decor.shape=circle,decor.filled=empty,decor.size=13,label=\tiny{$\mathbf{Sp}$},label.dist=0}{vr1}
	\fmfv{decor.shape=circle,decor.filled=empty,decor.size=13,label=\tiny{$\mathbf{Sp}$},label.dist=0}{vr2}
	\fmfv{decor.shape=square,decor.filled=empty,decor.size=20,label=\tiny{$\ket{\mM}$},label.dist=0}{vl}
	\fmfv{decor.shape=square,decor.filled=empty,decor.size=20,label=\tiny{$\bra{\mM}$},label.dist=0}{vr}
	\fmfv{label=\tiny{splitter lines},label.dist=2mm,l.a=0}{vl11}
	\fmfv{label=\tiny{conjugate splitter lines},label.dist=3mm,l.a=10}{vr11}
	\fmfv{label=\tiny{interferer lines},label.dist=3.5mm,l.a=-15}{vr21}
	\fmfv{label=\tiny{1},label.dist=1.5mm,l.a=160}{l1}
	\fmfv{label=\tiny{2},label.dist=1.5mm,l.a=160}{l2}
	\fmfv{label=\tiny{3},label.dist=1.5mm,l.a=160}{l3}
	\fmfv{label=\tiny{4},label.dist=1.5mm,l.a=160}{l4}
	\fmfv{label=\tiny{1},label.dist=1.5mm,l.a=30}{r1}
	\fmfv{label=\tiny{2},label.dist=1.5mm,l.a=-20}{r2}
	\fmfv{label=\tiny{3},label.dist=1.5mm,l.a=22}{r3}
	\fmfv{label=\tiny{4},label.dist=1.5mm,l.a=-10}{r4}
	\end{fmfgraph*}
	\end{gathered}
	\end{align*}
	
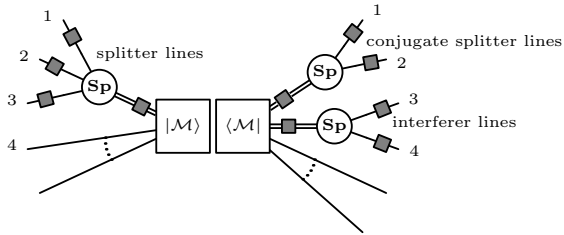
\captionof{figure}{Illustration of the terminology used to describe the different parts of a splitting topology.}
	\label{fig:notation}
        \end{fmffile}%
\medskip
        
To illustrate our terminology let us focus on a topology like the one
depicted in Fig.~\ref{fig:notation}, where we have identified a
certain set of splitting amplitudes in the amplitude side (the left
hand part of the diagram) which we refer to as `splitter lines', and
possibly different combinations of splitting amplitudes in the
conjugate (right hand) side of the density operator diagram. Splittings
of this kind, which solely involve momenta which are attached to a
single splitting blob on the left hand side are referred to as
`conjugate splitting lines', while otherwise we refer to them as
`interferer lines'. Fig.~\ref{fig:notation} contains an example of our
notation.
\begin{fmffile}{collinear-directions}
	\fmfset{thin}{.7pt}
	\fmfset{dot_len}{.8mm}
	\fmfset{dot_size}{5}
	\fmfset{arrow_len}{2.5mm}
	\fmfset{wiggly_len}{2.2mm}
	\setlength{\belowdisplayskip}{0pt}
	\begin{align*}
	\begin{gathered}
	\begin{fmfgraph*}(140,80)
	\fmfleft{l6,l5,l4,l3,l2,l1}
	\fmfright{r7,r6,r5,r4,r3,r2,r1}
	\fmf{phantom}{l1,vl,l6}
	\fmf{phantom}{r1,vr,r7}
	\fmf{phantom,tension=4}{vl,vr}
	\fmffreeze
	\fmf{dbl_plain,tension=1.2,background=white,fore=(1,,0.6,,0)}{vl,vdl1,vl1}
	\fmf{plain,fore=(1,,0.6,,0)}{vl1,vl11,l1}
	\fmf{plain,fore=(1,,0.6,,0)}{vl1,vl12,l2}
	\fmf{dbl_plain,tension=1.4,background=white,fore=blue}{vl,vdl2,vl2}
	\fmf{plain,fore=blue}{vl2,vl21,l3}
	\fmf{plain,fore=blue}{vl2,vl22,l4}
	\fmf{plain,fore=red}{vl,vl5,l5}
	\fmf{plain,fore=red}{vl,vl6,l6}
	\fmffreeze
	\fmf{dots,right=.2,width=1}{vl11,vl12}
	\fmf{dots,right=.2,width=1}{vl21,vl22}
	\fmf{dots,right=.2,width=1}{vl5,vl6}
	\fmf{phantom}{vl1,vl11,bl1,l1}
	\fmf{phantom}{vl1,vl12,bl2,l2}
	\fmf{phantom}{vl1,vl21,bl3,l3}
	\fmf{phantom}{vl1,vl22,bl4,l4}
	\fmfv{decor.shape=square,decor.filled=50,decor.angle=35,decor.size=5}{bl1}
	\fmfv{decor.shape=square,decor.filled=50,decor.angle=77,decor.size=5}{bl2}
	\fmfv{decor.shape=square,decor.filled=50,decor.angle=65,decor.size=5}{bl3}
	\fmfv{decor.shape=square,decor.filled=50,decor.angle=20,decor.size=5}{bl4}
	\fmfv{decor.shape=square,decor.filled=50,decor.angle=55,decor.size=5}{vdl1}
	\fmfv{decor.shape=square,decor.filled=50,decor.angle=0,decor.size=5}{vdl2}
	\fmfv{decor.shape=circle,decor.filled=empty,decor.size=13,fore=(1,,0.6,,0),label=\tiny{$\mathbf{Sp}$},label.dist=0}{vl1}
	\fmfv{decor.shape=circle,decor.filled=empty,decor.size=13,fore=blue,label=\tiny{$\mathbf{Sp}$},label.dist=0}{vl2}
	\fmf{dbl_plain,tension=1.2,background=white,fore=(1,,0.6,,0)}{vr,vdr1,vr1}
	\fmf{plain,fore=(1,,0.6,,0)}{vr1,vr11,r1}
	\fmf{plain,fore=blue}{vr1,vr12,r2}
	\fmf{dbl_plain,tension=1.4,background=white,fore=blue}{vr,vdr2,vr2}
	\fmf{plain,fore=blue}{vr2,vr21,r3}
	\fmf{plain,fore=(1,,0.6,,0)}{vr2,vr22,r4}
	\fmf{dbl_plain,tension=1.9,background=white,fore=red}{vr,vdr3,vr3}
	\fmf{plain,fore=blue}{vr3,vr31,r5}
	\fmf{plain,fore=(1,,0.6,,0)}{vr3,vr32,r6}
	\fmf{plain,fore=red}{vr3,vr35,r7}
	\fmffreeze
	\fmf{dots,left=.2,width=1}{vr11,vr12}
	\fmf{dots,left=.2,width=1}{vr21,vr22}
	\fmf{dots,left=.2,width=1}{vr31,vr32,vr35}
	\fmf{phantom}{vr1,vr11,br1,r1}
	\fmf{phantom}{vr1,vr12,br2,r2}
	\fmf{phantom}{vr2,vr21,br3,r3}
	\fmf{phantom}{vr2,vr22,br4,r4}
	\fmf{phantom}{vr3,vr31,br5,r5}
	\fmf{phantom}{vr3,vr32,br6,r6}
	\fmf{phantom}{vr3,vr35,br7,r7}
	\fmfv{decor.shape=square,decor.filled=50,decor.angle=50,decor.size=5}{br1}
	\fmfv{decor.shape=square,decor.filled=50,decor.angle=10,decor.size=5}{br2}
	\fmfv{decor.shape=square,decor.filled=50,decor.angle=22,decor.size=5}{br3}
	\fmfv{decor.shape=square,decor.filled=50,decor.angle=-10,decor.size=5}{br4}
	\fmfv{decor.shape=square,decor.filled=50,decor.angle=5,decor.size=5}{br5}
	\fmfv{decor.shape=square,decor.filled=50,decor.angle=-28,decor.size=5}{br6}
	\fmfv{decor.shape=square,decor.filled=50,decor.angle=-60,decor.size=5}{br7}
	\fmfv{decor.shape=square,decor.filled=50,decor.angle=35,decor.size=5}{vdr1}
	\fmfv{decor.shape=square,decor.filled=50,decor.angle=10,decor.size=5}{vdr2}
	\fmfv{decor.shape=square,decor.filled=50,decor.angle=-30,decor.size=5}{vdr3}
	\fmfv{decor.shape=circle,decor.filled=empty,decor.size=13,fore=(1,,0.6,,0),label=\tiny{$\mathbf{Sp}$},label.dist=0}{vr1}
	\fmfv{decor.shape=circle,decor.filled=empty,decor.size=13,fore=blue,label=\tiny{$\mathbf{Sp}$},label.dist=0}{vr2}
	\fmfv{decor.shape=circle,decor.filled=empty,decor.size=13,fore=red,label=\tiny{$\mathbf{Sp}$},label.dist=0}{vr3}
	\fmfv{decor.shape=square,decor.filled=empty,decor.size=20,label=\tiny{$\ket{\mM}$},label.dist=0}{vl}
	\fmfv{decor.shape=square,decor.filled=empty,decor.size=20,label=\tiny{$\bra{\mM}$},label.dist=0}{vr}
	\fmfv{label=\tiny{$p_1$},label.dist=2mm,l.a=70}{vdl1}
	\fmfv{label=\tiny{$p_2$},label.dist=1.8mm,l.a=130}{vdl2}
	\fmfv{label=\tiny{$p_3$},label.dist=1.8mm,l.a=-90}{vdr3}
	\end{fmfgraph*}
	\end{gathered}
	\end{align*}
	
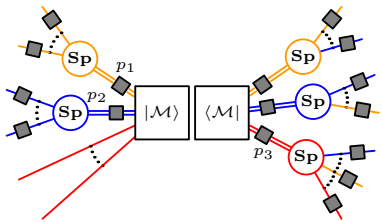
\captionof{figure}{Illustration of the collinear directions}
	\label{fig:coll-directions}
        \end{fmffile}%
  \medskip

Our power counting in the singular limits will be organized around the
assumption that we can prevent those momenta attached to a splitter
line from developing collinear singularities with respect to any other
than their progenitor direction.  This can be imposed by the
restrictions of a certain observable which \eg is requiring a fixed
number of jets, or by an algorithmic construction of a `partitioning',
an example of which we will outline in Sec.~\ref{sec:Paritioning} for
one and two emissions.  We illustrate this idea in
Fig.~\ref{fig:coll-directions}, where double lines represent different
hard directions and the colours signify which lines are connected in a
corresponding cut-diagram.  Using the weighting factors $\omega_{\rho
  \bar{\rho}; \sigma}$ of \eqref{eqs:paritioning-origin}, one can
extract the collinear behaviour to some subset of hard lines, \eg the
yellow ones in Fig.~\ref{fig:coll-directions}, via the choice of the
respective splitting configuration $\sigma$.  For a fixed-order
analysis such a partitioning can serve as the basis to construct
subtraction terms, or we can view it in an integrated way to obtain
evolution kernels for a full-fledged parton branching algorithm. A
further assumption we make is that each `splitter line' has one hard
momentum flowing through it, and that each `conjugate splitter' and
`interferer line' do so as well. We therefore consider to decompose
all momenta which are assigned to the `splitter' lines to be
decomposed with one progenitor momentum per splitter line, both in the
amplitude as well as its conjugate, \ie we shall {\it not} decompose
the momenta attaching to interferer lines in terms of the hard
momentum defining the interferer direction. Conversely, the hard
momentum in the interferer direction shall be parametrized taking into
account the interferer direction. Couplings between different
collinear sectors, hence, only need to be considered for conjugate
splitter lines, or for vertices coupling lines from different splitter
lines.

Let us stress that for conjugate splitter or interferer lines which
have internal lines from different collinear sectors we can employ
\eqref{eqs:linearity} and \eqref{eq:linearity-three-gluon-vertex}
to decouple the different sectors. Such a line, say one carrying
$q_I+q_J$, will then split into one with a numerator according to
$q_I$ and one according to $q_J$, and the net enhancement to be
studied is the partitioning's effect along with the combination
\begin{equation}\nonumber
  \frac{z_I}{z_I+z_J (n\cdot p_j/n\cdot p_i)} \times \frac{1}{(q_I+q_J)^2}
  \times \text{(partitioning/observable)} \ ,
\end{equation}
and a similar one for exchanging $I$ with $J$. If the partitioning
and/or the observable thus guarantees that there will not be a
singularity arising from $q_I$ becoming collinear to $q_J$, the above
configuration is finite also in the limit of $q_I$ becoming soft,
$z_I\to 0$ as long as a hard momentum flow asserts $z_J={\cal
  O}(1)$. Notice that we still might need to consider those
contributions in expanding to a certain power of $\lambda$, however in
the present work we shall only be concerned with extracting the
singular behaviour.

The contributions entering in the second line of
\eqref{eqs:paritioning-origin} are slightly more complicated in the
sense that there is no one-to-one correspondence between a collinear
sector and a sub-diagram. 
However, since the partitioning factor $w_{\lambda,\sigma;\rho}$
guarantees that there will only be a collinear enhancement as dictated by the
topology $\rho$ and this topology being excluded from the ones summed over,
emission lines in  $\lambda$ and $\sigma$ are forced to become
soft for the contribution to be enhanced.
Then, our discussion on the
decomposition of internal lines applies, and the coupling within
common collinear sectors applies otherwise.
Therefore, we can think of these
contributions consisting of interferer lines in both the amplitude and
conjugate amplitude.

\subsection{Iterating the recoil transformation}
\label{sec:recoil-iteration}

Our discussion on the power counting can also be applied to iterating
emission kernels in a Markovian manner, as long as we make sure that
the recoil transformation does not change the general
decomposition. To be precise, we consider, in a $k+1$st emission
step, a decomposition\footnote{We suppress the label for the
	collinear sector in this discussion.}
\begin{equation}\label{eq:qkplus1-iteration}
	q_{(k+1)}^{\mu} = \frac{1}{\hat{\alpha}_{(k+1)}}(\Lambda_{(k+1)})^\mu {}_\nu
	\left(z_{(k+1)} p^{\nu}_{(k)} + \beta_{(k+1)} n_{(k+1)}^{\nu} +
	k_{\perp,(k+1)}^\nu\right) \, ,
\end{equation}
where the forward direction has been changed due to the previous
emission and itself admits a decomposition (we do here not distinguish
what branching it belonged to)
\begin{equation}\label{eq:pk-iteration}
	p_{(k)}^{\mu} = \frac{1}{\hat{\alpha}_{(k)}}(\Lambda_{(k)})^\mu {}_\nu
	\left(z_{(k)} p^{\nu}_{(k-1)} + \beta_{(k)} n_{(k)}^{\nu} +
	k_{\perp,(k)}^\nu\right) \, .
\end{equation}
It is therefore tempting to redefine the backward direction as
\begin{equation}
	\label{eqs:nredef}
	n^\mu_{(k+1)} = \frac{1}{\hat{\alpha}_{(k)}}
	(\Lambda_{(k)})^\mu {}_\nu \, n^\nu_{(k)},
\end{equation}
which leads to
\begin{align}
	q_{(k+1)}^{\mu} = \frac{1}{\hat{\alpha}_{(k+1)} \hat{\alpha}_{(k)} }(\Lambda_{(k+1)})^\mu {}_\nu (\Lambda_{(k)})^\nu {}_\rho
	&\left[z_{(k+1)} z_{(k)} p^{\rho}_{(k-1)} 
	+ \left(\beta_{(k+1)} + z_{(k+1)} \beta_{(k)} \right) n_{(k)}^{\rho} \right. \nonumber \\
	& {}+ \left.z_{(k+1)} k_{\perp,(k)} +\hat{\alpha}_{(k)} (\Lambda_{(k)}^{-1})^\rho {}_\sigma \, k_{\perp,(k+1)}^\sigma \right] \, ,
\end{align}
However, this might actually not be sufficient to guarantee
that the mapping iterates the momentum parametrization in the way that
would not change our general power counting arguments.
In particular
this happens if, for an emission off of a hard line, we allow a
transverse component assigned through the emission process. In this
case, iterating the above to the second emission will generate a
transverse momentum which, despite being still orthogonal to the
backward direction chosen initially by virtue of \eqref{eqs:nredef},
will not be orthogonal to the forward direction anymore, \ie 
\begin{equation}
	p_{(k-1)} \cdot (\Lambda^{-1}_{(k)} k_{\perp,(k+1)}) = -\frac{\hat{\alpha}_{(k)} }{z_{(k)}} k_{\perp,(k)} \cdot  (\Lambda^{-1}_{(k)} k_{\perp,(k+1)}) \neq 0,
\end{equation}
potentially introducing azimuthal correlations between the emissions
and spoiling our power counting arguments.  As already mentioned, this
can only be avoided by not introducing transverse momentum components
for hard lines in the first place, which we will call ``unbalanced
mapping'', or by parametrizing two or more emissions in one step.  In
Sec.~\ref{sec:Mapping}, we explore mappings of this kind in detail. We
believe that this discussion intimately connects to the findings one
of the authors and others have been highlighting in \cite{Forshaw:2020wrq,Gieseke:2003rz,Dasgupta:2018nvj,Dasgupta:2020fwr}. We
therefore do not limit the discussion of our mappings to include
balance of transverse recoil, and leave this open to an algorithmic
choice which can, however, crucially impact the accuracy of algorithms
build on top of our calculations.  Similar remarks apply to the
ordering variable, which should then be chosen, in line with the
partitioning which has been used to separate different collinear
sectors, to not spill the configurations we have been discussing
within our power counting arguments. We also stress that the above way
of re-defining the backward (gauge) vector is crucial to establish
that the recoil transformation factors out of the amplitude in the
same homogeneous way as discussed earlier, since the amplitude
additionally depends on the backward direction $n$.  In the presence
of non-coloured (or coloured, massive) objects, the recoil scheme
above can easily be generalized and still be implemented by virtue of
a single Lorentz transformation; however the amplitude will not easily
satisfy a scaling property since the massive momenta cannot be
rescaled in order to remain on their definite, non-vanishing mass
shell. However, in the singular regions, the scaling factor
$\hat{\alpha}$ will in any case tend to unity such that no additional
complication should arise in this case.

Depending on how the transverse recoil has been chosen, different
diagrams then contribute after decomposing the internal lines into the
Sudakov decomposition and applying the vertex rules above. We will
highlight a few examples below, and defer a more detailed analysis to
Sec.~\ref{sec:Kernels}, where we also discuss the relation to splitting functions
and the soft limit, as well as the strategy employed for the dipole
subtraction terms in \cite{Catani:1996vz}.

\subsection{Local backward direction}\label{sec:local-backward}
Provided we have decomposed the momenta into a component towards a certain hard
direction, but with an arbitrary backward direction,
\begin{equation}
	q^\mu = \tilde{z}\, p^\mu + \frac{\tilde{p}_\perp^2}{\tilde{z} \, 2p\scdot\tilde{n}} \tilde{n}^\mu + \tilde{k}_\perp \ ,
\end{equation}
with $\tilde{k}_\perp\cdot p = \tilde{k}_\perp\cdot \tilde{n} = 0$,
$\tilde{k}_\perp^2 = -\tilde{p}_\perp^2$ and $p\cdot \tilde{n}>0$ as
usual, we can uniquely relate this to our global parametrization with
respect to $p$, $n$ and an appropriate transverse momentum $k_\perp$
as
\begin{equation}
	p_\perp^2 = \tilde{p}_\perp^2 R \ , \qquad
	z = \tilde{z} R \ , \qquad 
	R = 1 + \frac{1}{\tilde{z}} \frac{n\cdot \tilde{k}_\perp}{n\cdot p} 
	+ \frac{\tilde{p}_\perp^2}{2\tilde{z}^2} \frac{n\cdot \tilde{n}}{n\cdot p\ p\cdot \tilde{n}} \ ,
\end{equation}
and
\begin{equation}\label{eq:kperp-redef-local-ni}
	k_\perp^\mu = \frac{\tilde{p}_\perp^2}{2\tilde{z}}
	\left(\frac{\tilde{n}^\mu}{\tilde{n}\cdot p} - \frac{n^\mu}{n\cdot
		p}\right) +\tilde{z}(1-R)p^\mu + \tilde{k}_\perp^\mu \ .
\end{equation}
Hence the leading scaling in the soft and collinear limits is the same
in between the different parametrizations: the $R$ factor is of ${\cal
  O}(1)$ in each of the soft and collinear limits, and the transverse
momentum has the same leading scaling in both parametrizations, as
well. The above result will also allows us to formulate local recoil
schemes in which we choose a backward direction per collinear
configuration, and to properly link this parametrization to our power
counting rules.

\subsection{Scaling of individual internal lines and vertices}
\label{sec:scaling-of-lines}

In order to analyse how the individual diagrams scale in the relevant
limits, we need to consider external lines carrying a hard momentum
(which we refer to as `resolved' in the sense of a strong ordering),
internal lines with a hard momentum and a momentum composed of soft,
collinear or soft and collinear momenta. On top of this we consider
`unresolved' lines, which can become arbitrarily soft and/or collinear
and for which both external as well as internal lines deliver the same
scaling. It is clear that this distinction is only making sense if we
separate the contributions in such a way that we need to consider
collinear singular configurations only with respect to one line
carrying a hard momentum, and soft contributions otherwise. Achieving
such a separation will be discussed in
Sec.~\ref{sec:partitioning-algorithm}, where we introduce one example
of partitioning factors allowing for such a separation.

From the general form of the Sudakov decomposition
\eqref{eq:Sudakov-decomp-combined} we then conclude that we need to
assign scaling factors as outlined in Tables~\ref{tab:power-counting1}
and \ref{tab:power-counting2}, respectively, where we refer to the
different momentum flows as:
\begin{itemize}
\item h = external line with all momenta hard, possibly perturbed by a
  transverse momentum with respect to the hard direction, though we
 also consider the case in which the original direction is kept;
\item h+c = a sum of hard and collinear momenta;
\item h+s = a sum of hard and soft momenta; and 
\item h+c+s = a sum of hard, collinear and soft momenta, for which,
  owing to the lower scaling power, it mostly is the soft limit which
  determines the properties.
\end{itemize}
For lines which are allowed to become arbitrarily unresolved, we refer to
\begin{itemize}
\item s = a sum of purely soft momenta, or a single soft momentum; 
\item c = a sum of purely collinear momenta, or a single collinear momentum; and
\item s+c = a combination of soft and collinear momenta.
\end{itemize}
Notice that the lines with the white square need to be considered as
${\cal O}(1)$ in any of the unresolved limits.

\begin{table}[h]
	\begin{center}
		\begin{tabular}{c|cccc}
		& h & h+c & h+s & h+c+s   \\
			\hline
				\begin{minipage}{3cm}\begin{fmffile}{power-counting1}
	\fmfset{thin}{.7pt}
	\fmfset{arrow_len}{1.8mm}
	\fmfset{curly_len}{1.8mm}
	\fmfset{dash_len}{1.5mm}
	\begin{align*}
	\begin{gathered}
	\vspace{-4pt}
	\begin{fmfgraph*}(50,20)
	\fmfleft{l}
	\fmfright{r}
	\fmf{phantom}{l,vbox,r}
	\fmffreeze
	\fmf{plain}{l,vbox,r}
	\fmfv{decor.shape=circle,decor.filled=empty,decor.size=2}{l}
	\fmfv{decor.shape=circle,decor.filled=empty,decor.size=2}{r}
	\fmfv{decor.shape=square,decor.filled=empty,label=\tiny{$\perp$},label.dist=0,decor.size=8}{vbox}
	\end{fmfgraph*}
	\end{gathered}
	\end{align*} 
	\end{fmffile}\end{minipage} & $\lambda$ & $\lambda$ & $\lambda$ & $\lambda$	\\
			 & $1$	& $\lambda$ & $\lambda$ & $\lambda$	\\
			 \hline
			 	\begin{minipage}{3cm}\begin{fmffile}{power-counting2}
	\fmfset{thin}{.7pt}
	\fmfset{arrow_len}{1.8mm}
	\fmfset{curly_len}{1.8mm}
	\fmfset{dash_len}{1.5mm}
	\begin{align*}
	\begin{gathered}
	\vspace{-4pt}
	\begin{fmfgraph*}(50,20)
	\fmfleft{l}
	\fmfright{r}
	\fmf{phantom}{l,vbox,r}
	\fmffreeze
	\fmf{plain}{l,vbox,r}
	\fmfv{decor.shape=circle,decor.filled=empty,decor.size=2}{l}
	\fmfv{decor.shape=circle,decor.filled=empty,decor.size=2}{r}
	\fmfv{decor.shape=square,decor.filled=full,decor.size=8}{vbox}
	\end{fmfgraph*}
	\end{gathered}
	\end{align*} 
	\end{fmffile}\end{minipage}  & $\lambda^2$	& $\lambda^2$ & $\lambda$ & $\lambda$	\\
			 & $1$ & $\lambda^2$ & $\lambda$ & $\lambda$	\\
		\end{tabular}
	\end{center}
	\caption{Power counting for hard line with (upper row in each
          entry) and without $k_\perp$ recoil (lower row in each entry).}
	\label{tab:power-counting1}
\end{table}

In \tabref{tab:power-counting1}, we show the scaling of hard lines (emitters or spectators) in all possible combinations of hard, soft and collinear momenta shown above.
The first rows in each block correspond to the case where the $k_\perp$
recoil is included in the emitter mapping.
The second rows correspond to the
case without a $k_\perp$ component which applies to spectator lines
in general and can apply to emitters depending on the mapping used.
Both of these mapping types are discussed in
\secref{sec:Mapping}. The hard lines are signified by their horizontal orientation.
In \tabref{tab:power-counting2}, vertical lines are used to refer to unresolved (soft or collinear) lines and the scaling is shown for the two different `box' contributions.
For both the hard and unresolved lines, the white
box contributions do not have a scaling.
Notice that we need to
complete this picture by propagator factors which scale as
$1/\lambda^2$ in the hard+collinear configuration, $1/\lambda$ in the
hard+soft configuration as well as $1/\lambda$ in the
hard+collinear+soft configuration. Completely unresolved propagators
scale as $1/\lambda^2$ in a collinear or soft configuration, and as
$1/\lambda$ in a soft+collinear configuration.

\begin{table}[h]
	\begin{center}
		\begin{tabular}{c|ccc}
			 & s & c & s+c  \\
			\hline
			\begin{minipage}{3cm}\begin{fmffile}{power-counting3}
	\fmfset{thin}{.7pt}
	\fmfset{arrow_len}{1.8mm}
	\fmfset{curly_len}{1.8mm}
	\fmfset{dash_len}{1.5mm}
	\begin{align*}
	\begin{gathered}
	\vspace{4pt}
	\begin{fmfgraph*}(50,35)
	\fmfbottom{b}
	\fmftop{t}
	\fmf{phantom}{t,vbox,b}
	\fmffreeze
	\fmf{plain}{t,vbox,b}
	\fmfv{decor.shape=circle,decor.filled=empty,decor.size=2}{t}
	\fmfv{decor.shape=circle,decor.filled=empty,decor.size=2}{b}
	\fmfv{decor.shape=square,decor.filled=empty,label=\tiny{$\perp$},label.dist=0,decor.size=8}{vbox}
	\end{fmfgraph*}
	\end{gathered}
	\end{align*} 
	\end{fmffile}\end{minipage} 
	& $1$ & $\lambda$ & $\lambda$ 	\\
	\hline
			\begin{minipage}{3cm}\begin{fmffile}{power-counting4}
	\fmfset{thin}{.7pt}
	\fmfset{arrow_len}{1.8mm}
	\fmfset{curly_len}{1.8mm}
	\fmfset{dash_len}{1.5mm}
	\begin{align*}
	\begin{gathered}
	\vspace{4pt}
	\begin{fmfgraph*}(50,35)
	\fmfbottom{b}
	\fmftop{t}
	\fmf{phantom}{t,vbox,b}
	\fmffreeze
	\fmf{plain}{t,vbox,b}
	\fmfv{decor.shape=circle,decor.filled=empty,decor.size=2}{t}
	\fmfv{decor.shape=circle,decor.filled=empty,decor.size=2}{b}
	\fmfv{decor.shape=square,decor.filled=full,decor.size=8}{vbox}
	\end{fmfgraph*}
	\end{gathered}
	\end{align*} 
	\end{fmffile}\end{minipage}
	& $1$	& $\lambda^2$ & $\lambda$ \\
		\end{tabular}
	\end{center}
	\caption{Power counting for unresolved lines.}
	\label{tab:power-counting2}
\end{table}

\section{Partitioning}
\label{sec:Paritioning}

An intrinsic assumption of our power counting is that we envisage to
employ a partitioning of the soft behaviour into mutually exclusive
collinear limits. This is important in order to organize a given
observable into deviations from an ideal jet topology, but it will
also give us a key to parametrize a kinematic mapping suited to
collinear branching along a given direction and is thus of crucial
importance to set up a parton shower algorithm. In this section we
generalize both of the typically exploited partitionings to the
multi-emission case. With the case of multiple emissions there are
often multiple collinear combinations which can also be of different
orders \ie double or triple collinear. Whether a diagram contributes
to a collinear limit or not is determined by the presence of internal
propagators going on-shell in the respective limit. For $k$ emissions,
each diagram carries $2k$ internal propagators which can independently
become singular in different collinear limits.  It is the goal of our
algorithm to partition these propagator factors into a set of
splitting kernels, such that each kernel is only singular in the
\emph{one} collinear configuration it addresses.  The partitioning
factors themselves only contain soft singularities and for a given
diagram add up to one across all collinear configurations.  After
partitioning, a specific momentum mapping can be applied to the
amplitudes which contribute to a kernel, as the emitter and emission
momenta can be labelled.

\subsection{Basic purpose}\label{sec:basic-example-partitioning}

The approach we use is best explained in a simple example of a
rational function $f$ singular in two independent variables $x_1$ and
$x_2$.
\begin{align}
	f(x_1,x_2) = \frac{n(x_1,x_2)}{S_1(x_1,x_2) S_2(x_1,x_2)},
\end{align}
where $n(x_1,x_2)$ is a polynomial.  We are firstly interested in the
leading collinear singular behaviour for some specific configuration
which corresponds to one of the $x_i \rightarrow 0$.  Here, this is
parametrized by $S_1\rightarrow 0$ for $x_1\rightarrow 0$ while $
S_2\neq 0$ and vice versa for $x_2\rightarrow 0$.  By defining the
partitioning factors
\begin{align}
	\mathbb{P}^{(f)}_{(x_1)} = \frac{S_2}{S_1+S_2}, \quad \mathbb{P}^{(f)}_{(x_2)} = \frac{S_1}{S_1+S_2}, 
\end{align}
we can decompose $f$ into
\begin{align}\label{eq:partitioning-ex-1}
	f = \left[\mathbb{P}^{(f)}_{(x_1)} + \mathbb{P}^{(f)}_{(x_2)} \right] f = \frac{n(x_1,x_2) }{S_1(S_1+S_2)} + \frac{n(x_1,x_2) }{S_2(S_1+S_2)}.
\end{align}
This allows us to define the splitting kernels
\begin{align}\label{kernel-example}
	\mathbb{U}_{(x_i)}  = \mathbb{P}^{(f)}_{(x_i)} f
	= \frac{1}{S_i} \frac{n(x_1,x_2)}{S_1+S_2}.
\end{align}
What we have achieved is a decomposition into objects which are
singular solely in one of the variables. These objects will later be
identified with so called splitting kernels.  Moreover, the second
factor on the right hand side of \eqref{kernel-example} is
non-singular in any single $x_i\rightarrow 0$ collinear limit and
therefore only contains soft singularities.
Note that in this simple case, it is possible to set $x_i=0$ in the
numerator polynomial $n(x_1,x_2)$ without loss of information for the
leading singular behaviour.  Nevertheless, this is not possible in
more complicated cases where the $x_i$ can appear to some power $>1$
in the denominator.  Lastly, by only using the singular variables
$x_i$ in the partitioning, we make sure that we smoothly approach the
original collinear singular behaviour of the function when both
variables go to zero simultaneously.  In the actual partitioning
algorithm, this corresponds to keeping the correct soft-singular
behaviour.

A second interesting option for partitioning is given by means of subtractions
instead of partitioning factors of \eqref{eq:partitioning-ex-1}.  It
reads
\begin{equation}
	f = \frac{1}{2}\left[f -\Delta_2 + \Delta_1\right] + (1 \leftrightarrow 2),
\end{equation}
where 
\begin{equation}
	\Delta_1 = \frac{1}{S_1 \, S_2\vert_{x_1 \to 0} }, \quad \Delta_2 = \frac{1}{S_2 \, S_1\vert_{x_2 \to 0} }.
\end{equation}
Here we use the fact that $S_1(x_2 \to 0)$ and $S_2(x_1 \to 0)$ are non-zero.
In this case, a
splitting kernel can be defined as
\begin{equation}
	\mathbb{U}_{(x_1)} = \frac{1}{2}\left[\frac{1}{S_1 S_2} - \Delta_2 + \Delta_1\right].
\end{equation}
This again gives a function which shows no singularity when $x_2
\rightarrow 0$, but reproduces the original singular behaviour when
$x_1\rightarrow 0$.  We discuss this type of partitioning in
Sec.~\ref{sec:AOPartitioning}.

\subsection{Partial fractioning partitioning algorithm}\label{sec:partitioning-algorithm}

As a first non-trivial example for our algorithm, we present the
approach for the case of two emissions in this section. Firstly, it is
expedient to define the various limits we are interested in.  These
can be represented by the set of collinear configurations $\mathbf{C}$
for a given number of partons and emissions.  In the two emission
case, this set reads
\begin{equation}
\mathbf{C} = \{ (i\parallel j \parallel k), (i \parallel j \parallel l), \dots, (i \parallel j)(k \parallel l),\dots \},
\end{equation}
where the notation $(i \parallel j)$ stands for two different external
partons with momenta $q_i$ and $q_j$ becoming collinear.  It consists
of all configurations contributing to a triple collinear limit, \ie
all triplets and pairs-pairs one can build from the set of external
partons.  
Next, we define
the set $\mathbf{C}^{(d)}$ which contains only the configurations in which
diagram $d$ can become singular.  As an example, we take the two
emission interference diagram $A^{(1)}$ (see
Fig.~\ref{fig:triplet-triplet-diagrams} in
App.~\ref{app:two-emission-diagrams}), of which the propagator factors
$\mathcal{P}(A_1)$ are given by
\begin{align}\label{eq:propagators-A1}
\mathcal{P}(A_1) = \frac{1}{S_{ij} \, S_{ijk} \, S_{kl} \, S_{jkl}}.
\end{align}
Here, we have used the notation
\begin{align}
S_{ij} = S(q_i, q_j) \equiv (q_i + q_j)^2.
\end{align}
The singular collinear configurations and respective factors are collected in Table~\ref{tab:singular-factors-A1}.
Then, $\mathbf{C}^{(A_1)}$ is represented by the first column of the table.
\begin{table}[h]
	\begin{center}
		\begin{tabular}{c|cc}
			Configuration & Vanishing & Non-vanishing  \\
			\hline
			$(i \parallel j \parallel k)$ & $S_{ij}S_{ijk}$ & $ S_{kl} S_{jkl} $	\\
			$(i \parallel j \parallel l)$ & $S_{ij}$	& $ S_{ijk}S_{kl}S_{jkl} $	\\
			$(i \parallel k \parallel l)$ & $S_{kl}$	& $ S_{ij}S_{ijk}S_{jkl} $	\\
			$(j \parallel k \parallel l)$ & $S_{kl} S_{jkl}$ & $ S_{ij} S_{ijk} $	\\
			$(i \parallel j), (k \parallel l)$ & $S_{ij} S_{kl}$ & $ S_{ijk} S_{jkl} $  \\
		\end{tabular}
	\end{center}
	\caption{Singular configurations and the relevant propagator factors for the two emission diagram $A_1$.}
	\label{tab:singular-factors-A1}
\end{table}
Then we can define $\mathbf{S}_c^d$, the set of vanishing S-invariants
contained in diagram $d$ for configuration $c$, which corresponds to
the second column in Table~\ref{tab:singular-factors-A1}.  The
partitioning factors for some singular configuration $c$ and diagram
$d$ can then generally be defined as
\begin{align}\label{partitioning-definition}
\mathbb{P}_c^{(d)} \equiv \frac{ F_c^{(d)} }{\mathbb{F}^{(d)} },
\end{align}
where the cancelling factors $F_c^{(d)}$ are given by
\begin{align} \label{cancelling-factors}
F^{(d)}_{c} = \Big( \prod\limits_{S_{c'} \in \mathbf{S}_c^{(d)}} S_{c'} \, \mathcal{P}(\mathcal{A}^{(d)}) \Big)^{-1} \varsigma^{p},
\end{align}
and
\begin{equation}\label{sum-cancelling-factors}
\mathbb{F}^{(d)} \equiv \sum\limits_{c \in \mathbf{C}^{(d)}} F_c^{(d)}.
\end{equation}
The cancelling factors $F^{(d)}_c$ correspond to the last column in Table~\ref{tab:singular-factors-A1}.
Lastly, we have introduced the scale $\varsigma^p$ in order to keep
the partitioning factors dimensionless.
The power $p$ needs to be
chosen such that all cancelling factors carry the same mass
dimension.\footnote{Note that this scale could be avoided by only
partitioning into the \emph{leading} singular behaviours. The caveat
is that there are diagrams for which it is \emph{a priori} unclear
which of the singular structures will give leading contributions to
some kernel. Therefore, it is safer in this sense to simply partition
into all possible singular structures.} 

In the following, we assume that $i$ and $l$ are hard partons.  For
topology $A^{(1)}$, this provokes that we only need to deal with the three
leading singular configurations of
Table~\ref{tab:singular-factors-A1}.  Its cancelling factors are
\begin{subequations}
	\begin{eqnarray}
	F^{(A^{(1)})}_{(ijk)} &=& \left(S_{ij}S_{ijk} A^{(1)} \right)^{-1} = S_{kl} S_{jkl}, \\
	F^{(A^{(1)})}_{(jkl)} &=& \left(S_{kl}S_{jkl} A^{(1)} \right)^{-1} =  S_{ij} S_{ijk}, \\
	F^{(A^{(1)})}_{(ij)(kl)} &=& \left(S_{ij}S_{kl} A^{(1)} \right)^{-1} = S_{ijk} S_{jkl},
	\end{eqnarray}
\end{subequations}
which corresponds to the entries of the third column of Table~\ref{tab:singular-factors-A1}. 
Then, the amplitude's fully partitioned propagator is given by
\begin{align} \label{eq:A1-partitioned}
\sum\limits_{c\in \mathbf{C}_{A^{(1)}}} \mathbb{P}^{(A^{(1)})}_c \mathcal{P}( A^{(1)}) 
= \frac{1}{S_{kl}S_{jkl} + S_{ij}S_{ijk} + S_{ijk}S_{jkl}  } \times \left(\frac{1}{S_{ij}S_{ijk}} +  \frac{1}{S_{kl}S_{jkl}} + \frac{1}{S_{ij}S_{kl}}  \right).
\end{align}
The first,second and third term will give leading singular
contributions to splitting kernels $\mathbb{U}_{(ijk)}$,
$\mathbb{U}_{(jkl)}$ and $\mathbb{U}_{(ij)(kl)}$,  respectively.

Note that topologies with only one hard line, \ie the self-energy like ones need no partitioning due to the absence of a recoiler that the emissions could become collinear to.

In \tabref{tab:scaling-two-emission}, we show the collinear and soft scalings for the two-emission topologies of App.~\ref{app:two-emission-diagrams} together with their fractional partitioning factors for the $(i\parallel j \parallel k)$ configuration.
\begin{table}[h]
	\begin{center}
		\begin{tabular}{c|cccc}
			& CC & CS & SC & SS  \\
			\hline
			$ A^{(1)} $ & $1/\lambda^4$ & $1/\lambda^4$ & $1/\lambda^2$ & $1/\lambda^4$\\
			$ A^{(2)} $ & $1/\lambda^4$ & $1/\lambda^3$ & $1/\lambda^3$ & $1/\lambda^4$ \\
			$ A^{(3)} $ & $1/\lambda^4$ & $1/\lambda^3$ & $1/\lambda^2$ & $1/\lambda^4$ \\
			$ A^{(4)} $ & $1/\lambda^6$ & $1/\lambda^4$& $1/\lambda^3$ & $1/\lambda^5$ \\
			$ A^{(5)} $ & $1/\lambda^6$ & $1/\lambda^3$ & $1/\lambda^3$  & $1/\lambda^6$ \\
			$ B^{(1)} $ & $1/\lambda^6$ & $1/\lambda^6$& $1/\lambda^3$ & $1/\lambda^4$ \\
			$ B^{(2)} $ & $1/\lambda^6$ & $1/\lambda^4$ & $1/\lambda^5$ & $1/\lambda^4$ \\
			$ B^{(3)} $ & $1/\lambda^6$ & $1/\lambda^5$ & $1/\lambda^3$ & $1/\lambda^5$ \\
			$ B^{(4)} $ & $1/\lambda^6$ & $1/\lambda^6$& $1/\lambda^3$ & $1/\lambda^4$ \\
			$ B^{(5)} $ & $1/\lambda^6$ & $1/\lambda^4$ & $1/\lambda^5$ & $1/\lambda^4$ \\
			$ B^{(6)} $ & $1/\lambda^6$ & $1/\lambda^5$ & $1/\lambda^3$ & $1/\lambda^5$ \\
			$ X^{(1)} $ & $1/\lambda^4$ & $1/\lambda^4$ & $1/\lambda^4$ & $1/\lambda^4$ \\
			$ X^{(2)} $ & $1/\lambda^4$ & $1/\lambda^4$ & $1/\lambda^2$ & $1/\lambda^2$ \\
			$ E^{(1)} $ & $1/\lambda^8$& $1/\lambda^6$ & $1/\lambda^4$ & $1/\lambda^4$ \\
			$ E^{(2)} $ & $1/\lambda^8$ & $1/\lambda^5$ &$1/\lambda^5$ & $1/\lambda^4$  \\
			$ E^{(3)} $ & $1/\lambda^8$& $1/\lambda^5$ & $1/\lambda^4$& $1/\lambda^5$ \\
			$ E^{(4)} $ & $1/\lambda^8$& $1/\lambda^5$ & $1/\lambda^4$& $1/\lambda^5$ \\
			$ E^{(5)} $ & $1/\lambda^8$& $1/\lambda^4$ & $1/\lambda^4$& $1/\lambda^6$ \\			
		\end{tabular}
	\end{center}
	\caption{Scaling for propagator times partitioning factor of two emission single emitter topologies when partitioned to $(i\parallel j \parallel k)$. Here, `CC' refers to the triple collinear limit where $i||j||k$, `CS' refers to $(i \parallel j)$ with soft $k$, `SC' to $(i \parallel k)$ and $j$ soft and `SS' is the double soft limit where both $j$ and $k$ are soft.}
	\label{tab:scaling-two-emission}
\end{table}
Eventually, these scalings need to be combined with the respective numerators scalings discussed in Sec.~\ref{sec:scaling-of-lines} in order to determine which topology contributes to a splitting kernel at a given power in $\lambda$.
We show the resulting table in Sec.~\ref{sec:Applications}.
	
The partitioning factors as defined in \eqref{partitioning-definition}
fulfil two important requirements as already mentioned in
Sec.~\ref{sec:basic-example-partitioning}., \viz
\begin{enumerate}
	\item Non-singular behaviour in any collinear limit,
	\item Soft-collinear and purely soft limits are reproduced correctly.
\end{enumerate}
The first property is guaranteed by the fact that the sum in
\eqref{sum-cancelling-factors} does not vanish in \emph{any} collinear
limit.  One can show this fact by realising that for all given
collinear configurations $c \in \mathbf{C}^{(d)}$, the partitioning
factors $F^{(d)}_{c} \neq 0$ per definition.  An example is
$c=(i||j)(k||l)$ for which $F^{(A^{(1)})}_{c_1} = \varsigma \, S_{ijk}
S_{jkl} \neq 0$ in said configuration.  It can be that $F^{(d)}_{c}$
vanishes in another configuration $c'$, but not all of them
simultaneously.  The reason is that the in the case where partitioning
is carried out, there are necessarily recoiler partons present leading
to the fact that not all propagator factors of the respective diagram
vanish simultaneously.  In all other configurations $c
\notin \mathbf{C}^{(d)}$, we know that $(A^{(d)})^{-1} \neq 0$ and
therefore, neither the partitioning factors $F^{(A_d)}_c$ can vanish
here.  Thus, there is no configuration where all $F^{(A_d)}_{c} = 0$
simultaneously for a given diagram $d$.

Secondly, the property of the correct scaling behaviour in soft- and
soft-collinear limits is guaranteed by the fact that for a given
diagram, we only use the $S$-invariants appearing in the respective
diagram to construct the partitioning factors, which form a partition
of unity.  Therefore, there is no spurious finite remainder in
$\mathbb{F}^{(d)}$ that could spoil the approach of the original soft
singularity.

We stress that the analysis in this section has focused on the case
that we have identified certain partons as those carrying the hard
momentum. Additional combinatorics are thus needed if several partons
in the same jet could carry the hard momentum, but in an algorithmic
implementation, see Sec.~\ref{sec:Applications}, we will be able to
decide at each branching which of the daughter particles is to carry
the hard momentum, and whether that particular configuration will then
contribute to the leading behaviour, \eg our power counting directly
implies that a hard quark transitioning into a soft quark and a hard
gluon will be immediately suppressed by the $\sqrt{z_i z_j}$ factor we
associate to the effective quark-gluon vertex.

\subsection{Subtraction (angular ordered) partitioning algorithm}
\label{sec:AOPartitioning}

An alternative approach to partitioning is to isolate and subtract
collinear divergencies in a way that a single collinear divergence
remains in the diagram just considered. For one emission, this
procedure leads to angular ordering, essentially using
\begin{equation}
	\label{eq:angular-ordering}
	\frac{1}{S_{ij}\, S_{jk}} = \frac{1}{S_{ik}} \frac{1}{2 E_j^2}
	\left(\frac{n_i\cdot n_k}{n_i\cdot n_j\, n_j\cdot n_k} -
	\frac{1}{n_j\cdot n_k} + \frac{1}{n_i\cdot n_j}\right) +
	(i\leftrightarrow k) \ ,
\end{equation}
where we have decomposed the momenta into energies and directions,
$p_j = E_j\, n_j$;
in this case, any measure of energy could have been
used, \ie in general we consider $E_i = T\cdot p_i$ with some
timelike vector $T$. In the above expression, the first term is
singular only in the $(i||j)$ configuration, while the second one is
singular in the complementary $(k||j)$ limit.
In this section we show
how this paradigm can be generalized to a larger number of emissions
and we find that the resulting partitioning shows the same scaling
behaviour as the partial fractioning variant, thus allowing to use the same power counting.

The basic idea of the subtraction partitioning now starts from a
slight re-write of the angular ordering logic; to this extent rewrite
\eqref{eq:angular-ordering} as
\begin{equation}
	\frac{1}{S_{ij}\, S_{jk}} = \frac{1}{2}\left(\frac{1}{S_{ij}S_{jk}} -
	\frac{E_k}{E_j\, S_{ik}} \frac{1}{S_{jk}} +
	\frac{E_i}{E_j\, S_{ik}}\frac{1}{S_{ij}}\right) + (i\leftrightarrow
	k) \ .
\end{equation}
The second term in the parentheses has been constructed to subtract
off the $1/S_{jk}$ divergence upon replacing $1/S_{ij}$ by its
limiting expression for $(j||k)$, \ie
\begin{equation}
	S_{ij} \xrightarrow{(j\parallel k)} E_i E_j \, n_i \scdot n_k = \frac{E_j}{ E_k} S_{ik} 
\end{equation}
and the second term adds back what
has been subtracted in the term with $i$ and $k$ interchanged.
This decomposition allows to define the partitioned propagator
\begin{equation}
	{\mathbb P}_{(i \parallel j)}\left[\frac{1}{S_{ij}\, S_{jk}}\right] = 
	\frac{1}{2} \left(\frac{1}{S_{ij}\, S_{jk}} - \Delta_{(j\parallel k)} + \Delta_{(i \parallel j)} \right),
\end{equation}
with 
\begin{equation}
	\Delta_{(i\parallel j)} = \frac{E_i}{E_j} \frac{1}{S_{ik}S_{ij}}, \quad \Delta_{(j\parallel k)} = \frac{E_k}{E_j} \frac{1}{S_{ik}S_{jk}}.
\end{equation}
It fulfils the expectation of giving back the original singular
scaling behaviour in the $(i \parallel j)$ case while being
non-singular when $(j \parallel k)$.

In fact, this construction is algorithmic and can be generalized to
more than one emission; the only complication which arises is in the
fact that several collinear limits might be overlapping, \ie
some of the $\Delta$ factors themselves contain sub-singularities, and
in turn these need to be subtracted out of the possible subtraction
terms to guarantee that one resulting partition summand truly only
reflects one collinearly singular configuration.

In particular we can write the procedure to construct subtraction
terms as
\begin{equation}
	\Delta_{\tau_1;\tau_2,...,\tau_m}[P] =
	{\mathbb F}_{\tau_1}[P]\left({\mathbb S}_{\tau_1}[P] -
	\overline{\sum\limits_{\mathcal{S}/ \tau_1}}
	\Delta_{\tau_{i_1};\tau_{i_2},...,\tau_{i_{m-1}}}\left[{\mathbb S}_{\tau_1}[P]\right] \right) \ ,
\end{equation}
where ${\mathbb F}_{\tau_1}[P]$ indicates to collect those factors
from the propagator factor $P$ which are non-singular in the limit
identified by $\tau_1$ and to replace them by their limiting
expressions in this limit. Conversely, ${\mathbb S}_{\tau_1}[P]$
indicates to take the singular factors.  Moreover, $\mathcal{S}$ is
the set of singular configurations of ${\mathbb S}_{\tau_1}[P]$ and
the summation symbol stands for dividing by the number of terms of the
sum.  The indices after the semicolon in $\Delta$ signify the order of
the limits taken beforehand which can play a role if these limits do
not commute in terms of the partitioning $\Delta$'s.  We then proceed
to subtract the remaining overlapping singular limits, and we define
\begin{equation}
	\Delta_{\tau_1;\tau_2,...,\tau_m}[P] = \Delta_{\tau_2;\tau_3,...,\tau_m}[P] \qquad \text{if }P \text{ not }\tau_1\text{ collinear singular,}
\end{equation}
$\Delta_{\sigma;}(P) = 0$ if $P$ is non-singular in $\sigma$, and
$\Delta_{\sigma;}(P) = {\mathbb F}_{\sigma}[P]{\mathbb S}_{\sigma}[P]$
otherwise.  The resulting partitioned propagator then is
\begin{equation}\label{eq:AO-partitioned-prop}
	{\mathbb P}_{\sigma}[P] = \frac{1}{m} \left( P +(m-1)\Delta_{\sigma;\tau_1,...,\tau_{m-1}}[P] -
	\sum_{i=1}^{m-1} \Delta_{\tau_i;\tau_1,...,\tau_{i-1},\sigma,\tau_{i+1},...,\tau_{m-1}}[P]\right) \ ,
\end{equation}
where $\sigma,\tau_1,...,\tau_{m-1}$ denote the $m$ configurations in
which $P$ can develop collinear singularities.

As an example, we show this version of the partitioning algorithm
again for the amplitude $A^{(1)}$ with propagators factors of
\eqref{eq:propagators-A1}.  The configurations where
$\mathcal{P}(A^{(1)})$ has a leading singularity are
\begin{equation}
	\begin{split}
		\tau_1 &= (i\parallel j \parallel k),\\
		\tau_2 &= (j \parallel k \parallel l), \\
		\tau_3 &= (i\parallel j) (k \parallel l).
	\end{split}
\end{equation}
Therefore, $m=3$ in this case.  First, we collect the limiting
expressions in these configurations, \ie
\begin{equation}
	\begin{split}
		\mathbb{F}_{\tau_1} [\mathcal{P}(A^{(1)})] 
		&= \frac{1}{S_{kl} S_{jkl}}\bigg\rvert_{\tau_1} = \frac{E_i^2}{E_k(E_j + E_k)} \frac{1}{S_{il}^2}, \\
		\mathbb{F}_{\tau_2} [\mathcal{P}(A^{(1)})] 
		&= \frac{1}{S_{ij} S_{ijk}}\bigg\rvert_{\tau_2} = \frac{E_l^2}{E_j(E_j + E_k)} \frac{1}{S_{il}^2}, \\
		\mathbb{F}_{\tau_3} [\mathcal{P}(A^{(1)})] 
		&= \frac{1}{S_{ijk} S_{jkl}}\bigg\rvert_{\tau_3} = \frac{E_i^2 E_l^2}{E_k(E_i + E_j) \, E_j(E_k +E_l)} \frac{1}{S_{il}^2}.	
	\end{split}
\end{equation}
Note that a single index specification is sufficient here, because the
order of taking the limits does not matter in this example.  Then, we
can construct the first subtraction term, \viz
\begin{equation}\label{eq:AO-examples-A1-Delta1}
	\Delta_{\tau_1} [\mathcal{P}(A^{(1)})]  =  \mathbb{F}_{\tau_1} [\mathcal{P}(A^{(1)})] 
	\sum\limits_{\tau \in \mathbf{S}/\tau_1} \left( \frac{1}{S_{ij}S_{ijk}} -  \Delta_{\tau}\left[ \frac{1}{S_{ij}S_{ijk}} \right]\right).
\end{equation}
This term is supposed to only cancel $(i\parallel j \parallel
k)$-singularities.  Therefore, we subtract off the overlapping
sub-singularity in the parentheses via
\begin{equation}
	\Delta_{\tau_3}\left[\frac{1}{S_{ij}S_{ijk}} \right]
	= \mathbb{F}_{\tau_3}\left[\frac{1}{S_{ijk}}\right] \frac{1}{S_{ij}},
\end{equation}
where $\tau_3$ is the only relevant configuration in the sum of
\eqref{eq:AO-examples-A1-Delta1}, because this sub-propagator is
non-singular in $\tau_2$.  The same procedure applies to the other two
subtraction terms.  Eventually, we find
\begin{equation}
	\begin{split}
		\Delta_{\tau_1} [\mathcal{P}(A^{(1)})] 
		&= \frac{E_i^2}{ E_k(E_j+E_k)} \frac{1}{S_{il}^2} \left(\frac{1}{S_{ij}S_{ijk}} - \frac{E_i E_l }{E_k (E_i + E_j)} \frac{1}{S_{il} S_{ij}} \right), \\
		\Delta_{\tau_2} [\mathcal{P}(A^{(1)})] 
		&= \frac{E_l^2}{ E_j(E_j+E_k)} \frac{1}{S_{il}^2} \left(\frac{1}{S_{kl}S_{jkl}} - \frac{E_i E_l }{E_j (E_k + E_l)} \frac{1}{S_{il} S_{kl}} \right), \\
		\Delta_{\tau_3} [\mathcal{P}(A^{(1)})] 
		&= \frac{E_i^2 E_l^2}{ E_k(E_i + E_j) \, E_j(E_k +E_l)} \frac{1}{S_{il}^2} \left(\frac{1}{S_{ij}S_{kl}} - \frac{E_i }{E_k} \frac{1}{S_{il} S_{ij}} - \frac{E_l }{E_j} \frac{1}{S_{il} S_{kl}} \right).
	\end{split}
\end{equation}
Then, using \eqref{eq:AO-partitioned-prop}, the partitioned propagator
of $A^{(1)}$ is given by
\begin{equation}
	\begin{split}
		\mathcal{P}(A^{(1)})  
		&=  \frac{1}{3} \left(\frac{1}{S_{ij} S_{ijk} S_{kl} S_{jkl}} + 2 \Delta_{\tau_1}[\mathcal{P}(A^{(1)})]  - \Delta_{\tau_2}[\mathcal{P}(A^{(1)})]  -\Delta_{\tau_3}[\mathcal{P}(A^{(1)})] \right), \\
		&+ \frac{1}{3} \left(\frac{1}{S_{ij} S_{ijk} S_{kl} S_{jkl}} -  \Delta_{\tau_1}[\mathcal{P}(A^{(1)})]  + 2 \Delta_{\tau_2}[\mathcal{P}(A^{(1)})]  -\Delta_{\tau_3}[\mathcal{P}(A^{(1)})] \right), \\
		&+ \frac{1}{3} \left(\frac{1}{S_{ij} S_{ijk} S_{kl} S_{jkl}} -  \Delta_{\tau_1}[\mathcal{P}(A^{(1)})]  - \Delta_{\tau_2}[\mathcal{P}(A^{(1)})]  +2 \Delta_{\tau_3}[\mathcal{P}(A^{(1)})] \right),
	\end{split}
\end{equation}
where the first, second and third line are ${\mathbb
  P}_{\tau_1}[\mathcal{P}(A^{(1)})]$, ${\mathbb
  P}_{\tau_2}[\mathcal{P}(A^{(1)})]$ and ${\mathbb
  P}_{\tau_3}[\mathcal{P}(A^{(1)})]$, respectively.

\section{Momentum mapping}
\label{sec:Mapping}

In this section, we define two instances of a momentum mapping in
terms of the Sudakov-like decomposition
of~\eqref{eq:Sudakov-decomp-single} where momentum conservation
between emitters and emissions is manifest and a global
Lorentz-transformation is introduced for momentum conservation when
emissions are added to a given process, as already advertised in
Sec.~\ref{sec:recoil-iteration}.  Using the dictionary of
Sec.~\ref{sec:local-backward}, both versions of the mapping are
compatible with the power counting rules introduced in
Sec.~\ref{sec:scaling-of-lines}.

The momentum mapping is set up in a language inspired by parton
showers.  In a parton shower setup, one starts from a set of massless
on-shell momenta of a hard process which are then dressed up with
emissions, which again are massless and on-shell.  Diagrammatically,
this can be represented by 
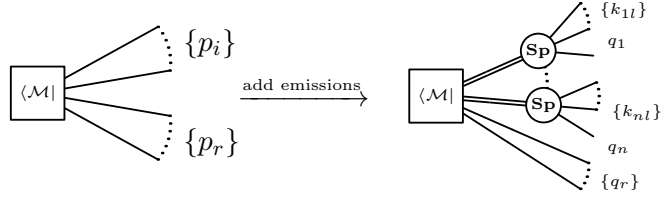
\begin{fmffile}{adding-emissions}
	\fmfset{thin}{.7pt}
	\fmfset{dot_len}{1.2mm}
	\fmfset{dot_size}{3}
	\fmfset{arrow_len}{2.5mm}
	\fmfset{curly_len}{1.5mm}
	\begin{align*}
	\begin{gathered}
	\begin{fmfgraph*}(50,50)
	\fmfleft{l}
	\fmfright{prd,pru,pid,piu}
	\fmf{plain}{l,piu}
	\fmf{plain}{l,pid}
	\fmf{plain}{l,pru}
	\fmf{plain}{l,prd}
	\fmffreeze
	\fmf{dots,left=.2,width=1,label=$\{p_i\}$}{piu,pid}
	\fmf{dots,left=.2,width=1,label=$\{p_r\}$}{pru,prd}
	\fmfv{decor.shape=square,decor.filled=empty,decor.size=20,label=\tiny{$\bra{\mM}$},label.dist=0}{l}
	\end{fmfgraph*}
	\end{gathered}
	\quad\quad \xrightarrow{\makebox[1.5cm]{\tiny\text{add emissions}}}
	\qquad
	\begin{gathered}
	\begin{fmfgraph*}(60,70)
	\fmfleft{l}
	\fmfright{rqr2,rqr1,rn3,rn2,rn1,r13,r12,r11}
	\fmf{dots,left=.2,width=1,tension=0.}{v1,v2}
	\fmf{dbl_plain,tension=2,background=white}{l,v1d,v1}
	\fmf{plain}{v1,r11}
	\fmf{plain,tension=.1}{v1,r12}
	\fmf{plain}{v1,r13}
	\fmf{dbl_plain,tension=2,background=white}{l,v2d,v2}
	\fmf{plain}{v2,rn1}
	\fmf{plain,tension=.1}{v2,rn2}
	\fmf{plain}{v2,rn3}
	\fmf{plain}{l,vqr1,rqr1}
	\fmf{plain}{l,vqr2,rqr2}
	\fmflabel{\tiny{$\{k_{1l}\}$}}{r12}
	\fmflabel{\tiny{$q_{1}$}}{r13}
	\fmflabel{\tiny{$\{k_{nl}\}$}}{rn2}
	\fmflabel{\tiny{$q_{n}$}}{rn3}
	\fmflabel{\tiny{$\{q_{r}\}$}}{rqr1}
	\fmffreeze
	\fmf{dots,left=.2,width=1}{r11,r12}
	\fmf{dots,left=.2,width=1}{rn1,rn2}
	\fmf{dots,left=.2,width=1}{rqr1,rqr2}
	\fmfv{decor.shape=square,decor.filled=empty,decor.size=20,label=\tiny{$\bra{\mM}$},label.dist=0}{l}
	\fmfv{decor.shape=circle,decor.filled=empty,decor.size=13,label=\tiny{$\Sp$},label.dist=0}{v1}
	\fmfv{decor.shape=circle,decor.filled=empty,decor.size=13,label=\tiny{$\Sp$},label.dist=0}{v2}
	\end{fmfgraph*}
	\end{gathered}
	\end{align*}
\end{fmffile}%
where
the $p_i$ are the momenta to which emissions are added with $i\in
\mathbf{S}$ (``splitters'') and the $p_r$ are available for recoil
absorption with $r\in \mathbf{R}$ (``recoilers'').  This leads us to
define the momenta after emissions in terms of the forward ($p_i$),
backward ($n_i$) and transverse components ($n_{\perp,l}^{(i)}$) as
\begin{subequations}\label{mapping}
	\begin{align}
		q_r &\equiv \frac{1}{\hat{\alpha}} \Lambda p_r \ ,
		\\
		\bar{q}_i &\equiv \frac{1}{\hat{\alpha}} \Lambda \left[ (1 - A_i)p_i
		\right] \quad{\text{(unbalanced)}} \ ,
		\\
		q_i &\equiv \frac{1}{\hat{\alpha}} \Lambda \left[ (1 - A_i)p_i + \big(y_i - \tilde{B}_i\big) n_i - \tilde{n}^{(i)}_\perp \right] \quad{\text{(balanced)}}, 
		\\
		\label{eq:kil-mapped}
		k_{il} &\equiv \frac{1}{\hat{\alpha}} \Lambda \left[ \alpha_{il} \, p_i + \tilde{\beta}_{il}\,  n_i + \sqrt{\alpha_{il} \tilde{\beta}_{il}\, } n^{(i)}_{\perp,l}\right] \ ,
	\end{align}
\end{subequations}
where we show one version with balanced ($q_i$) and one with unbalanced ($\bar{q}_i$) transverse components and use the shorthands
\begin{align}
  A_i \equiv \sum\limits_{l \in \mathbf{E}_i}
  \alpha_{il} \ , \quad
  \tilde{B}_i \equiv \sum\limits_{l \in \mathbf{E}_i} \tilde{\beta}_{il} \ , \quad
  \tilde{\beta}_{il} = (1-A_i)\beta_{il} \ , \quad
  \tilde{n}^{(i)}_\perp \equiv \sum\limits_{l \in
    \mathbf{E}_i} \sqrt{\alpha_{il} \tilde{\beta}_{il}}
  \, n^{(i)}_{\perp,l} \ ,
\end{align}
Note the inclusion of a Lorentz transformation $\Lambda$ together with
a scaling $\hat{\alpha}$ which are needed for the non-trivial global
recoil and momentum conservation.  The latter leads to
\begin{equation}\label{mom-conservation-1}
 \sum\limits_{i\in \mathbf{S}} \Big( q_i + \sum\limits_{l \in \mathbf{E}_i} k_{il} \Big) + \sum\limits_{r \in \mathbf{R}} q_r = Q \ ,
\end{equation}
where $Q$ is the original overall momentum transfer
 \begin{equation}\label{mom-transfer-Q}
	Q \equiv \sum\limits_{i\in \mathbf{S}} p_i + \sum\limits_{r\in \mathbf{R}} p_r  \ .
\end{equation}
Inserting \eqref{mapping} into \eqref{mom-conservation-1} gives
\begin{equation}\label{mom-conservation-2}
 Q = \frac{1}{\hat{\alpha}} \Lambda \Big[ \sum\limits_{r\in \mathbf{R}} p_r + \Big(\sum\limits_{i \in \mathbf{S}}  p_i + y_i   n_i \Big) \Big]  \ .
\end{equation}
Squaring this equation fixes the scaling to
\begin{equation}
 \hat{\alpha}^2 = \frac{(Q + N)^2}{Q^2} \ ,
\end{equation}
where
\begin{equation}
 N\equiv \sum\limits_{i \in \mathbf{S}} y_i \, n_i  \ .
\end{equation}
Using Lorentz-invariance of amplitudes, $\hat{\alpha}$ provides the
means to implement a global recoil as shown
in~\eqref{eq:amplitudeglobalrecoil}.

The numbers $\alpha_{il}$ and $\beta_{il}$ take values in $(0,1)$ and
parametrize the soft and collinear behaviour of the emission momenta
$k_{il}$, as well as the emitter momenta $q_i$.
In the balanced version of the mapping, $y_i$
quantifies the off-shellness of the emitter-emission system.  This can
be seen by squaring the momentum sum for emissions and the emitter,
\ie
\begin{align}\label{overall-mom-transfer}
	\Big(q_i + \sum\limits_l k_{il}\Big)^2 &= y_i \, 2 p_i\scdot n_i \ .
\end{align}
Therefore, both the soft and collinear limits are parametrized by the
$y_i\rightarrow 0$ limit.  The $y_i$ are fixed via the on-shell
relations for the $q_i$, \ie
\begin{align}\label{eq:yi-fixed}
	y_i = (1-A_i) B_i - \frac{(\tilde{n}_\perp^{(i)})^2}{ 2 p_i \cdotp n_i (1-A_i)}  \ ,
\end{align}
which shows that a scaling in $y_i$ is fixed by scaling the
$\alpha_{il}$ and $\beta_{il}$.  Instead, for the unbalanced version
of the mapping, we formally have $y_i=\tilde{B_i}$ and $\tilde
n_\perp^{(i)} = 0$ and overall momentum transfer is given by
\begin{equation}
	\Big(\overline{q}_i + \sum\limits_l k_{il} \Big)^2 = \tilde{B}_i \, 2 p_i \cdot n_i + (\tilde{n}_\perp^{(i)})^2  \ .
\end{equation}
The on-shellness of the emission momenta $k_{il}$ fixes the virtuality
of the transverse components as
\begin{equation}\label{eq:n-perp-square}
	\left(n^{(i)}_{\perp,l}\right)^2 = -2 p_i \cdotp n_i  \ .
\end{equation}
These components are further determined by the transversality relations
\begin{equation}\label{eq:orthogonality}
  p_i \cdotp n^{(i)}_{\perp,l} = 0 \quad \text{and} \quad n_i \cdotp n^{(i)}_{\perp,l} = 0,
  \quad \forall i \in \mathbf{S} \text{ and } \forall l \in \mathbf{E}_i  \ .
\end{equation}
We can choose $n_i$ locally w.r.t.\ jet directions while
simultaneously using the gauge vector $n$ to expand in a global basis
of spinors and polarization vectors.  The effect of such a choice is
discussed at the end of Sec.~\ref{sec:recoil-iteration}.  A suitable
choice for the (in principle arbitrary) lightlike backwards components
$n_i$ is
\begin{equation}
 n_i^\mu = Q^\mu - \frac{Q^2}{2 p_i \scdot Q}\, p_i^\mu  \ .
\end{equation}
In App.~\ref{app:phase-space}, we discuss the phase space
factorisation in terms using this type of momentum mapping.

The momentum mapping allows us to parametrise the scaling of the
momentum components in the style of SCET \cite{Becher:2014oda}
according to the following table
\begin{center}
 \begin{tabular}{c|c c}
 $k_{il}$ & $(p_i, n_i , n^{(i)}_{\perp,l})$ & $(\alpha_{il}, y_i, \beta_{il})$ \\ \hline
 (forward) collinear & $Q( 1, \lambda^2, \lambda)$ & $(1,\lambda^2,\lambda^2)$\\ 
 soft & $Q(\lambda, \lambda, \lambda)$ & $(\lambda,\lambda,\lambda)$.
\end{tabular}
\end{center}
The overall scaling in $\lambda$ can subsequently be used to study the
IR singular behaviour of an emission amplitude.  For this purpose, we
present all relevant dot-products which can appear in squared emission
amplitudes in terms of the momentum mapping of \eqref{mapping}, \viz
\begin{subequations}
	\begin{align} \label{eq:S-invariants-new-mapping}
		S(q_i,K_i) &=  y_i \, 2 p_i \cdotp n_i,\\
		S(q_i, k_{il}) &= \frac{2}{\hat{\alpha}^2} \left[ \left( \alpha_{il} (y_i - \tilde{B}_i) + \tilde\beta_{il}(1-A_i)  \right) p_i\cdotp n_i - \sqrt{(1-A_i)\alpha_{il} \tilde\beta_{il}}\, n^{(i)}_{\perp,l} \cdotp \tilde{n}^{(i)}_{\perp} \right] \ , 
		\\
		S(k_{il}, k_{il'}) &= \frac{2}{\hat{\alpha}^2} \left[ (\alpha_{il} \tilde\beta_{il'} + \alpha_{il'} \tilde\beta_{il}) \, p_i \cdotp n_i + \sqrt{\alpha_{il} \tilde\beta_{il} \alpha_{il'} \tilde\beta_{il'}}\, n^{(i)}_{\perp,l} \cdotp n^{(i)}_{\perp,l'} \right] \ ,
		\\
		S(k_{il}, q_r) &=
		\frac{2}{\hat{\alpha}^2} \left[ \alpha_{il}\, p_i \cdotp p_r  + \tilde\beta_{il}\, n_i \cdotp p_r 
		+ \sqrt{\alpha_{il} \tilde\beta_{il}}\, n^{(i)}_{\perp,l} \cdotp p_r
		\right] \ , \\
		S(q_i, q_r) &=
		\frac{2}{\hat{\alpha}^2} \left[ \left(1-A_i\right) p_i \cdotp p_r
		+ \left(y_i - \tilde{B}_i\right) n_i \cdotp p_r 
		- \sqrt{1-A_i}\,\tilde{n}^{(i)}_\perp \cdotp p_r
		\right] \ ,
	\end{align}
\end{subequations}
where we have used the shorthand
\begin{align}
	 K_i \equiv \sum\limits_{l \in \mathbf{E}_i} k_{il} \ .
\end{align}
In the unbalanced version of the mapping, the following $S$-invariants change:
\begin{subequations}
	\begin{align} \label{eq:S-invariants-new-mapping-unbalanced}
		S(\bar{q}_i,K_i) &= \tilde{B}_i \, 2 p_i \cdot n_i + (\tilde{n}_\perp^{(i)})^2 \ ,
		\\
		S(\bar{q}_i, k_{il}) &= \frac{2 p_i\cdotp n_i}{\hat{\alpha}^2} (1-A_i) \tilde{\beta}_{il} \ , 
		\\
		S(\bar{q}_i, q_r) &=
		\frac{2}{\hat{\alpha}^2} \left[ \left(1-A_i\right) p_i \cdotp p_r
		\right] \ ,
	\end{align}
\end{subequations}
The $\lambda$-scaling of these invariants for both variants of the mapping can be summarized as
\begin{center}
 \begin{tabular}{c|c c c c c}
 $k_{il}$ & $S(q_i,K_i)$ & $S(q_i, k_{il})$ & $S(k_{il}, k_{il'})$ & $S(k_{il}, q_r) $ & $S(q_i, q_r) $ \\ \hline
 collinear & $\lambda^2$ & $\lambda^2$ & $\lambda^2$ & $1$ & $1$ \\ 
 soft & $\lambda$ & $\lambda$ & $\lambda^2 $ & $\lambda$ & $1$
\end{tabular}.
\end{center}
This information is helpful in determining the set of diagrams that
contributes to a given soft, collinear or mixed limit and it is, of course,
compatible with the scaling we have used in deriving our effective
Feynman rules.

\section{Splitting kernels}
\label{sec:Kernels}
In this section, we will discuss how we can built up full {\it
  splitting kernels} in combination of our findings on the power
counting, the partitioning and the underlying momentum mapping.  Using
the partitioning factors of Sec.~\ref{sec:partitioning-algorithm},
they can be defined as follows:
\begin{align}\label{kernel-definition}
 \mathbb{U}_{c} \equiv  \sum\limits_d \left[\mathbb{P}^{(d)}_{c} \mathcal{A}^{(d)} \right].
\end{align}
$\mathcal{A}^{(d)} $ represents a certain topology under consideration,
    \ie a configuration with a fixed set of propagators in the
    amplitude and the conjugate amplitude. We find it convenient to
    represent these still as cut diagrams, however it should be clear
    that their meaning is to be taken in the sense of the density
    operator as discussed in Sec.~\ref{sec:Factorization}.
    
We choose the latter language with the intent of making an iterative
procedure in treating emissions more tangible.  The subscript $c$
stands for the collinear configuration that the kernel is addressing.
An example is the triple-collinear configuration $c = (i \parallel j
\parallel k)$ for the two emission case, where $\mathbb{U}_{(i j k)}$ contains
all leading singular contributions for the momenta $q_i$, $q_j$ and
$q_k$ becoming collinear.  The fact that $\mathbb{U}_{(i j k)}$ contains no
other leading singular structures is assured via the partitioning
factors $\mathbb{P}^{(d)}_{ijk}$.  Schematically,
\eqref{kernel-definition} can be paraphrased as
\vspace{5pt}
\begin{fmffile}{splitting-kernel-master}
	\fmfset{thin}{.7pt}
	\fmfset{dot_len}{1.2mm}
	\fmfset{dot_size}{3}
	\fmfset{arrow_len}{2.5mm}
	\fmfset{curly_len}{1.5mm}
	\begin{align}\label{splitting-kernel-master-diagram}
	\mathbb{U}_{c} = 
	\qquad
	&
	\begin{gathered}
	\begin{tikzpicture}
	\node (diagram) {%
	  \begin{fmfgraph*}(50,60)
	    \fmfstraight
	    \fmfleft{l1,l2,l3b,l3,l3t}
	    \fmfright{r1,r2,r3b,r3,r3t}
	    \fmf{plain}{r1,v1,l1}
	    \fmf{plain}{r2,v2,l2}
	    \fmf{phantom}{r3,v,l3}
	    \fmffreeze
	    \fmf{plain}{r3,v}
	    \fmf{plain}{l3b,v3b,v,v3t,l3t}
	    \fmfv{decor.shape=circle,decor.filled=empty,decor.size=13,label=\tiny{$\Sp$},label.dist=0}{v}
	    \fmffreeze
	    \fmf{dots,width=1}{v1,v2}
	    \fmf{dots,right=.2,width=1}{v3t,v3b}
	  \end{fmfgraph*}
	};
	\tikzset{shift={(0,0)}}
	\draw[thick,fill=black!10!white] (.7,-1.2) rectangle (1,.8);
	\end{tikzpicture}
	\;
	\begin{tikzpicture}
	\node (diagram) {%
	  \begin{fmfgraph*}(50,60)
	    \fmfstraight
	    \fmfleft{l1,l2,l3b,l3,l3t}
	    \fmfright{r1,r2,r3b,r3,r3t}
	    \fmf{plain}{r1,v1,l1}
	    \fmf{plain}{r2,v2,l2}
	    \fmf{phantom}{l3,v,r3}
	    \fmffreeze
	    \fmf{plain}{l3,v}
	    \fmf{plain}{r3b,v3b,v,v3t,r3t}
	    \fmfv{decor.shape=circle,decor.filled=empty,decor.size=13,label=\tiny{$\Sp$},label.dist=0}{v}
	    \fmffreeze
	    \fmf{dots,width=1}{v1,v2}
	    \fmf{dots,left=.2,width=1}{v3t,v3b}
	  \end{fmfgraph*}
	};
	\tikzset{shift={(0,0)}}
	\draw[thick,fill=black!10!white] (-1,-1.2) rectangle (-.7,.8);
	\end{tikzpicture}
	\end{gathered}
	\;+\; \mathbb{P}^{(d_M)}_c \times
	\begin{gathered}
	\begin{tikzpicture}
	\node (diagram) {%
	  \begin{fmfgraph*}(50,60)
	    \fmfstraight
	    \fmfleft{l1,l2,l3b,l3,l3t}
	    \fmfright{r1,r2,r3b,r3,r3t}
	    \fmf{plain}{r1,v1,l1}
	    \fmf{plain}{r2,v2,l2}
	    \fmf{phantom}{r3,v,l3}
	    \fmffreeze
	    \fmf{plain}{r3,v}
	    \fmf{plain}{l3b,v3b,v,v3t,l3t}
	    \fmfv{decor.shape=circle,decor.filled=empty,decor.size=13,label=\tiny{$\Sp$},label.dist=0}{v}
	    \fmffreeze
	    \fmf{dots,width=1}{v1,v2}
	    \fmf{dots,right=.2,width=1}{v3t,v3b}
	  \end{fmfgraph*}
	};
	\tikzset{shift={(0,0)}}
	\draw[thick,fill=black!10!white] (.7,-1.2) rectangle (1,.8);
	\end{tikzpicture}
	\;
	\begin{tikzpicture}
	\node (diagram) {%
	  \begin{fmfgraph*}(50,60)
	    \fmfstraight
	    \fmfleft{l1,l1m,l2,l3,l4,l5,l5m,l6,lt}
	    \fmfright{r1,r1m,r2,r3,r4,r5,r5m,r6,rt}
	    \fmf{plain}{l1m,v1}
	    \fmf{plain}{r2,v12,v1,v11,r1}
	    \fmf{plain}{l3,v3,r3}
	    \fmf{plain}{l4,v4,r4}
	    \fmf{plain}{l5m,v5}
	    \fmf{plain}{r6,v52,v5,v51,r5}
	    \fmffreeze
	    \fmfv{decor.shape=circle,decor.filled=empty,decor.size=13,label=\tiny{$\Sp$},label.dist=0}{v1}
	    \fmfv{decor.shape=circle,decor.filled=empty,decor.size=13,label=\tiny{$\Sp$},label.dist=0}{v5}
	    \fmffreeze
	    \fmf{dots,right=.2,width=.8}{v11,v12}
	    \fmf{dots,right=.2,width=.8}{v51,v52}
	    \fmf{dots,width=.8}{v3,v4}
	  \end{fmfgraph*}
	};
	\tikzset{shift={(0,0)}}
	\draw[thick,fill=black!10!white] (-1,-1.2) rectangle (-.7,.8);
	\end{tikzpicture}
	\end{gathered}
	\;+\;
	\nonumber \\[5pt]
	\mathbb{P}^{(d_I)}_c \times
	&\begin{gathered}
	\begin{tikzpicture}
	\node (diagram) {%
	  \begin{fmfgraph*}(50,60)
	    \fmfstraight
	    \fmfleft{l1,l2,l3b,l3,l3t}
	    \fmfright{r1,r2,r3b,r3,r3t}
	    \fmf{plain}{r1,v1,l1}
	    \fmf{plain}{r2,v2,l2}
	    \fmf{phantom}{r3,v,l3}
	    \fmffreeze
	    \fmf{plain}{r3,v}
	    \fmf{plain}{l3b,v3b,v,v3t,l3t}
	    \fmfv{decor.shape=circle,decor.filled=empty,decor.size=13,label=\tiny{$\Sp$},label.dist=0}{v}
	    \fmffreeze
	    \fmf{dots,width=1}{v1,v2}
	    \fmf{dots,right=.2,width=1}{v3t,v3b}
	  \end{fmfgraph*}
	};
	\tikzset{shift={(0,0)}}
	\draw[thick,fill=black!10!white] (.7,-1.2) rectangle (1,.8);
	\end{tikzpicture}
	\;
	\begin{tikzpicture}
	\node (diagram) {%
	  \begin{fmfgraph*}(50,60)
	    \fmfstraight
	    \fmfleft{l3t,l3,l3b,l2,l1,lt}
	    \fmfright{r3t,r3,r3b,r2,r1,rt}
	    \fmf{plain}{r1,v1,l1}
	    \fmf{plain}{r2,v2,l2}
	    \fmf{phantom}{l3,v,r3}
	    \fmffreeze
	    \fmf{plain}{l3,v}
	    \fmf{plain}{r3b,v3b,v,v3t,r3t}
	    \fmfv{decor.shape=circle,decor.filled=empty,decor.size=13,label=\tiny{$\Sp$},label.dist=0}{v}
	    \fmffreeze
	    \fmf{dots,width=1}{v1,v2}
	    \fmf{dots,left=.2,width=1}{v3t,v3b}
	  \end{fmfgraph*}
	};
	\tikzset{shift={(0,0)}}
	\draw[thick,fill=black!10!white] (-1,-1.2) rectangle (-.7,.8);
	\end{tikzpicture}
	\end{gathered}
	\end{align}
\end{fmffile}%
In the language of cut diagrams, the first term can be read as
\emph{self-energy like} contributions, \ie contributions where the
partons on the amplitude and conjugate amplitude side originate from
the same splitter parton.  In lightcone gauge, these contributions
contain the leading collinear singularities.  The second term in
\eqref{splitting-kernel-master-diagram} refers to contributions where
partons on the amplitude side identified with partons from several
different splitting groups on the conjugate side.  These will be
relevant for \emph{mixed} soft-collinear limits.  Lastly, the third
term stands for diagrams where none of the splitter partons on the
conjugate side are identified with the splitter on the amplitude side.
They correspond to \emph{interference} diagrams which in lightcone
gauge are relevant for soft limits only.

In addition to the contributions depicted in
\eqref{splitting-kernel-master-diagram}, one can have multiple
emitters on the amplitude side.  For two emissions, this corresponds
to configurations contributing to double-unresolved limits such as
$c'=(i \parallel j)(k \parallel l)$ where two sets of partons exhibit
independent collinearities.  The construction of the respective
splitting kernels follow the same logic as for the single emitter
case, the main difference being that one can not categorize the
various contributions into self-energy like, mixed and interference
diagrams as easily.

As a next step, we can insert the momentum mapping of
Sec.~\ref{sec:Mapping} for a given configuration $c$ into the
amplitudes entering the splitting kernel $\mathbb{U}_c$.  This step
essentially fixes the notion of which partons are considered as
emitters, recoilers and emissions.  Having made sure that there are no
other singular configurations that need addressing here, we can be
certain that all leading singular contributions can be captured via
the insertion of the respective momentum mapping.

\subsection{Power counting algorithm}
In order to provide a general rule for which contributions to keep in
terms of our power counting in splitting kernels, several remarks are
in order.  First of all, it is easy to show the fact that interference
contributions do not contribute to leading collinear limits using our
formalism.  The most important rule to make use of here is the fact
that amplitudes which only contain forward components of the same
collinear sector vanish exactly (this was already mentioned at the end
of Sec.~\ref{sec:decompositions-intro}), \ie
\begin{fmffile}{boxed-vertex-whites}
	\fmfset{thin}{.7pt}
	\fmfset{arrow_len}{2.5mm}
	\fmfset{curly_len}{1.8mm}
	\fmfset{dash_len}{1.5mm}
	\begin{align}\label{eq:vanishing-vertices-all-white}
		\begin{gathered}
			\vspace{-4pt}
			\begin{fmfgraph*}(60,60)
				\fmftop{r,m,l}
				\fmfbottom{b}
				\fmf{phantom}{m,v,b}
				\fmffreeze
				\fmf{plain}{l,v,r}
				\fmf{plain}{b,v}
				\fmf{phantom}{b,bb,v}
				\fmf{phantom}{l,lb,v}
				\fmf{phantom}{r,rb,v}
				\fmfv{decor.shape=square,decor.filled=empty,decor.size=9,label=\tiny{$p_i$},label.dist=0}{bb}
				\fmfv{decor.shape=square,decor.filled=empty,decor.size=9,label=\tiny{$p_i$},label.dist=0}{lb}
				\fmfv{decor.shape=square,decor.filled=empty,decor.size=9,label=\tiny{$p_i$},label.dist=0}{rb}
				\fmfv{decor.shape=circle,decor.filled=empty,decor.size=12,label=\tiny{$\parallel$},label.dist=0}{v}
			\end{fmfgraph*}
		\end{gathered}
		\; = \; 0 \ ,
	\end{align}
\end{fmffile}%
Now, in order for a diagram to contribute in the leading collinear
limit, it must have a scaling of $1/\lambda^{2k}$, where $k$ is the
number of emissions.  In interference contributions, only the
propagator factors of the splitter side give a collinear scaling as
long as we use either version of our partitioning algorithm to get rid
of collinearities on the interferer legs.  Then, the propagator
factors of the diagram shown in \eqref{eq:k-emmission-interference}
give a scaling of $1/\lambda^{2k}$.  This means that such
contributions can only give rise to leading collinearities when the
numerator does not scale, meaning that one would need only the forward
components of each line, or white boxes respectively.  Knowing that
the respective vertices exactly vanish on the splitter side, we find
\vspace{4pt}
\begin{fmffile}{k-emission-interference}
	\fmfset{thin}{.7pt}
	\fmfset{dot_len}{1.2mm}
	\fmfset{dot_size}{5}
	\fmfset{arrow_len}{2.5mm}
	\fmfset{curly_len}{1.5mm}
	\begin{align}\label{eq:k-emmission-interference}
		\mathbb{P}_{(i\parallel \dots)} \;
		\begin{gathered}
			\begin{fmfgraph*}(100,50)
				\fmfstraight
				\fmfleft{lb,lt}
				\fmfright{rb,rm,rt}
				\fmf{plain}{lb,mb,rb}
				\fmf{phantom}{lb,b1,bv1,b2,b3,b4,b5,bv2,b6,rb}
				\fmf{plain}{lt,t1,tv1,t2,t3}
				\fmf{dots}{t3,t4}
				\fmf{plain}{t4,t5,tv2,t6,rt}
				\fmffreeze
				\fmf{phantom}{tv1,m1,bv1}
				\fmf{phantom}{tv2,m2,bv2}
				\fmf{plain}{tv1,tb1,m1}
				\fmf{plain}{tv2,tb2,m2}
				\fmfv{decor.shape=square,decor.filled=empty,decor.angle=0,decor.size=6}{mb}
				\fmfv{decor.shape=square,decor.filled=empty,decor.angle=0,decor.size=6}{t1}
				\fmfv{decor.shape=square,decor.filled=empty,decor.angle=0,decor.size=6}{t2}
				\fmfv{decor.shape=square,decor.filled=empty,decor.angle=0,decor.size=6}{t5}
				\fmfv{decor.shape=square,decor.filled=empty,decor.angle=0,decor.size=6}{t6}
				\fmfv{decor.shape=square,decor.filled=empty,decor.angle=0,decor.size=6}{tb1}
				\fmfv{decor.shape=square,decor.filled=empty,decor.angle=0,decor.size=6}{tb2}
				\fmfv{label=\tiny{$i$},label.dist=2,label.angle=180}{lt}
				\fmfrectangle{10}{7}{rm}
			\end{fmfgraph*}
		\end{gathered}
		\quad \;
		\begin{gathered}
			\begin{fmfgraph*}(100,50)
				\fmfstraight
				\fmfleft{lb,lm,lt}
				\fmfright{rb,rt}
				\fmf{plain}{lt,mt,rt}
				\fmf{phantom}{lt,t1,tv1,t2,t3,t4,t5,tv2,t6,rt}
				\fmf{plain}{lb,b1,bv1,b2,b3}
				\fmf{dots}{b3,b4}
				\fmf{plain}{b4,b5,bv2,b6,rb}
				\fmffreeze
				\fmf{phantom}{bv1,m1,tv1}
				\fmf{phantom}{bv2,m2,tv2}
				\fmf{plain}{bv1,bb1,m1}
				\fmf{plain}{bv2,bb2,m2}
				\fmfv{decor.shape=square,decor.filled=empty,decor.angle=0,decor.size=6}{mt}
				\fmfv{decor.shape=square,decor.filled=empty,decor.angle=0,decor.size=6}{b1}
				\fmfv{decor.shape=square,decor.filled=empty,decor.angle=0,decor.size=6}{b2}
				\fmfv{decor.shape=square,decor.filled=empty,decor.angle=0,decor.size=6}{b5}
				\fmfv{decor.shape=square,decor.filled=empty,decor.angle=0,decor.size=6}{b6}
				\fmfv{decor.shape=square,decor.filled=empty,decor.angle=0,decor.size=6}{bb1}
				\fmfv{decor.shape=square,decor.filled=empty,decor.angle=0,decor.size=6}{bb2}
				\fmfrectangle{10}{7}{lm}
			\end{fmfgraph*}
		\end{gathered}
		\quad = 0 \ .
	\end{align}
\end{fmffile}%
The above holds generically also for appearances of the collinear
three-gluon vertex of \eqref{eq:three-gluon-col-vertex}.  This shows
that interferences do not contribute to the leading collinear limits.

For contributions that are relevant for mixed soft-collinear
behaviour, a general discussion is more involved.  We start here by
discussing of two emissions, but expect this pattern to hold also for
$k>2$.  The leading singular amplitudes will have at least two
$(\perp)$-boxes on the splitter and one on the conjugate side.  An
example is
\vspace{4pt}
\begin{fmffile}{mixed-amp-2-emissions}
	\fmfset{thin}{.7pt}
	\fmfset{arrow_len}{1.8mm}
	\fmfset{curly_len}{1.8mm}
	\fmfset{dash_len}{1.5mm}
	\begin{align}
		\begin{gathered}
			\begin{fmfgraph*}(80,50)
				\fmfstraight
				\fmfleft{lb,lm,lt}
				\fmfright{rb,rm,rt}
				\fmf{plain}{rt,b1,v1,b2,v2,b3,lt}
				\fmf{phantom}{rb,vb1,vb2,lb}
				\fmffreeze
				\fmf{phantom}{v1,bm1,e1,vb1}
				\fmf{phantom}{v2,bm2,e2,vb2}
				\fmf{plain}{v1,e1}
				\fmf{plain}{v2,e2}
				\fmfrectangle{10}{7}{rm}
				\fmfv{decor.shape=square,decor.filled=empty,decor.angle=0,decor.size=6}{b1}
				\fmfv{decor.shape=square,decor.filled=empty,decor.angle=0,decor.size=6}{b2}
				\fmfv{decor.shape=square,decor.filled=empty,decor.angle=0,decor.size=6}{b3}
				\fmfv{decor.shape=square,decor.filled=empty,decor.angle=0,decor.size=6,label=\tiny{$\perp$},label.dist=0}{bm1}
				\fmfv{decor.shape=square,decor.filled=empty,decor.angle=0,decor.size=6,label=\tiny{$\perp$},label.dist=0}{bm2}
			\end{fmfgraph*}
		\end{gathered}
		\quad \;
		\begin{gathered}
			\begin{fmfgraph*}(80,50)
				\fmfstraight
				\fmfleft{lb,lm,lt}
				\fmfright{rb,rm,rt}
				\fmf{plain}{lt,b1,v1,b2,v2,b3,rt}
				\fmf{plain}{lb,bb1,vb1,bb2,vb2,bb3,rb}
				\fmffreeze
				\fmf{phantom}{v1,bm1,e1,vb1}
				\fmf{phantom}{v2,e2,bm2,vb2}
				\fmf{plain}{v1,e1}
				\fmf{plain}{vb2,e2}
				\fmfrectangle{10}{7}{lm}
				\fmfv{decor.shape=square,decor.filled=empty,decor.angle=0,decor.size=6,label=\tiny{$\perp$},label.dist=0}{b1}
				\fmfv{decor.shape=square,decor.filled=empty,decor.angle=0,decor.size=6}{b2}
				\fmfv{decor.shape=square,decor.filled=empty,decor.angle=0,decor.size=6}{bb2}
				\fmfv{decor.shape=square,decor.filled=empty,decor.angle=0,decor.size=6}{bb3}
				\fmfv{decor.shape=square,decor.filled=empty,decor.angle=0,decor.size=6}{bm1}
				\fmfv{decor.shape=square,decor.filled=empty,decor.angle=0,decor.size=6}{bm2}
			\end{fmfgraph*}
		\end{gathered}
	\end{align}
\end{fmffile}%

The numerator of these leading amplitudes goes as $\lambda_c^3$ while
the denominator goes as $\lambda_c^6$ in the purely collinear limit.
Therefore, soft-collinear mixed contributions do not contribute
leadingly here.

Naively, the collinearly leading power amplitude on the splitter side
is the one with just one internal $(\perp)$-component, but an explicit
check of all possible gluon and quark insertions shows that it
vanishes exactly, \ie
\smallskip
\begin{fmffile}{internal-perp}
	\fmfset{thin}{.7pt}
	\fmfset{arrow_len}{1.8mm}
	\fmfset{curly_len}{1.8mm}
	\fmfset{dash_len}{1.5mm}
	\begin{align}\label{eq:vanishing-internal-perp}
		\begin{gathered}
			\vspace{4pt}
			\begin{fmfgraph*}(75,25)
				\fmfstraight
				\fmfleft{l2,lm,l1}
				\fmfright{r2,rm,r1}
				\fmf{phantom}{l2,k1,k2,r2}
				\fmffreeze
				\fmf{plain}{l1,bl1,m1,bm1,m2,br1,r1}
				\fmffreeze
				\fmf{plain}{m1,bk1,k1}
				\fmf{plain}{m2,bk2,k2}
				\fmffreeze
				\fmfrectangle{4}{4}{r1}
				\fmfv{decor.shape=square,decor.filled=empty,decor.angle=0,decor.size=6,label=\tiny{$\perp$},label.dist=0}{bm1}
				\fmfv{decor.shape=square,decor.filled=empty,decor.angle=0,decor.size=6}{br1}
				\fmfv{decor.shape=square,decor.filled=empty,decor.angle=0,decor.size=6}{bl1}
				\fmfv{decor.shape=square,decor.filled=empty,decor.angle=0,decor.size=6}{bk1}
				\fmfv{decor.shape=square,decor.filled=empty,decor.angle=0,decor.size=6}{bk2}
			\end{fmfgraph*}
		\end{gathered}
		\quad  =  \quad
		\begin{gathered}
			\vspace{4pt}
			\begin{fmfgraph*}(75,50)
				\fmfstraight
				\fmfleft{l2,lm,l1}
				\fmfright{r2,rm,r1}
				\fmf{phantom}{l2,k1,m2,k2,r2}
				\fmf{phantom,tension=0.5}{lm,mm,rm}
				\fmffreeze
				\fmf{plain}{l1,bl1,m1,br1,r1}
				\fmffreeze
				\fmf{plain}{m1,bm1,mm}
				\fmf{plain}{mm,bk1,k1}
				\fmf{plain}{mm,bk2,k2}
				\fmffreeze
				\fmfrectangle{4}{4}{r1}
				\fmfv{decor.shape=square,decor.filled=empty,decor.angle=0,decor.size=6,label=\tiny{$\perp$},label.dist=0}{bm1}
				\fmfv{decor.shape=square,decor.filled=empty,decor.angle=0,decor.size=6}{br1}
				\fmfv{decor.shape=square,decor.filled=empty,decor.angle=0,decor.size=6}{bl1}
				\fmfv{decor.shape=square,decor.filled=empty,decor.angle=0,decor.size=6}{bk1}
				\fmfv{decor.shape=square,decor.filled=empty,decor.angle=0,decor.size=6}{bk2}
			\end{fmfgraph*}
		\end{gathered}
		\quad = 0 \ .
	\end{align} 
\end{fmffile}%
This holds when both vertices are of the same type, \ie either both
collinear or both with $(\perp)$-only components.

With neither the interference, nor the mixed topologies contributing
to the purely collinear limits, we can now use the fact that any
interferer leg connecting to an emission will contribute only if this
emission becomes soft. Thus we can additionally carry out a soft
decomposition for the interferer legs here using
\eqref{eqs:linearity}, neglecting the terms which scale as ${\cal
  O}(\lambda)$ in this limit.

Finally, we come to the self-energy type amplitudes.  For $k$
emissions, the respective propagator factors will induce a scaling of
$1/\lambda_c^{4k}$ while from phase space considerations, we again
expect a leading scaling of $1/\lambda_c^{2k}$.  This shows that the
numerators will give a scaling of $\lambda_c^{2k}$ to arrive at the
expected leading scaling.  For two emissions, we have shown this fact
already due to vanishing of \eqref{eq:vanishing-internal-perp}.  This
means that the numerator for self-energy like contributions for two
emissions starts at $\lambda_c^4$.  We expect this pattern to hold for
any number of emissions.  An argument in favour of this is the
vanishing of the amplitude with the naively lowest scaling, \ie
\begin{fmffile}{internal-perp-general}
	\fmfset{thin}{.7pt}
	\fmfset{arrow_len}{1.8mm}
	\fmfset{curly_len}{1.8mm}
	\fmfset{dot_len}{1.2mm}
	\begin{align}\label{eq:vanishing-internal-perp-general}
		\begin{gathered}
			\begin{fmfgraph*}(100,80)
				\fmfstraight
				\fmfleft{lb,lm,lt}
				\fmfright{rb,rm,rt}
				\fmf{phantom}{lb,b1,bv1,b2,b3,b4,b5,bv2,b6,rb}
				\fmf{plain}{lm,t1,tv1,t2,t3}
				\fmf{dots}{t3,t4}
				\fmf{plain}{t4,t5,tv2,t6,rm}
				\fmffreeze
				\fmf{phantom}{tv1,m1,bv1}
				\fmf{phantom}{tv2,m2,bv2}
				\fmf{plain}{tv1,tb1,m1}
				\fmf{plain}{tv2,tb2,m2}
				\fmfv{decor.shape=square,decor.filled=empty,decor.angle=0,decor.size=6}{t1}
				\fmfv{decor.shape=square,decor.filled=empty,label=\tiny{$\perp$},label.dist=0,decor.size=6}{t2}
				\fmfv{decor.shape=square,decor.filled=empty,label=\tiny{$\perp$},label.dist=0,decor.size=6}{t5}
				\fmfv{decor.shape=square,decor.filled=empty,decor.angle=0,decor.size=6}{t6}
				\fmfv{decor.shape=square,decor.filled=empty,decor.angle=0,decor.size=6}{tb1}
				\fmfv{decor.shape=square,decor.filled=empty,decor.angle=0,decor.size=6}{tb2}
				\fmfrectangle{4}{4}{rm}
			\end{fmfgraph*}
		\end{gathered}
		\; = 0 \ .
	\end{align} 
\end{fmffile}%

In conclusion, we can formulate the power counting algorithm as
follows: we begin by using the collinear power counting for the
numerator scaling from Tab.~\ref{tab:power-counting1} and
\ref{tab:power-counting2} to find the a leading collinear
contribution.  This includes finding and using rules such as
\eqref{eq:vanishing-vertices-all-white} or
\eqref{eq:vanishing-internal-perp}.  Next, knowing that interferer
lines do not contribute in leading collinear limits, we can carry out
an additional soft decomposition using \eqref{eqs:linearity}.  The
result is a splitting kernel which yields the correct leading power
soft and collinear behaviour.  We will discuss this in more detail for
one and two emissions in
Sec.~\ref{sec:single-emission-splitting-kernel} and
\ref{sec:two-emission-splitting-kernel}.

\subsection{Factorisation to hard amplitude}
In this section, we want to discuss the factorisation properties of
the splitting kernels.  For this purpose, we discuss the following
splitting kernel where we omit non-splitter lines for clarity, \ie
\begin{fmffile}{splitting-kernel-factorisation}
	\fmfset{thin}{.7pt}
	\fmfset{dot_len}{1.2mm}
	\fmfset{dot_size}{3}
	\fmfset{arrow_len}{2.5mm}
	\fmfset{curly_len}{1.5mm}
	\begin{align}\label{splitting-kernel-factorisation}
		\mathbb{P}^{(d)}_{c \parallel h_1} 
		\begin{gathered}
			\begin{fmfgraph*}(50,40)
				\fmfstraight
				\fmfleft{l1,l2,l3b,l3,l3t}
				\fmfright{r1,r2,r3b,r3,r3t}
				\fmf{plain}{r3t,v}
				\fmf{plain}{v,l3t}
				\fmffreeze
				\fmf{plain}{l3,v3b,v,v3t,l3b}
				\fmfv{decor.shape=circle,decor.filled=empty,decor.size=13,label=\tiny{$\Sp$},label.dist=0}{v}
				\fmfv{label=\footnotesize{$h_1$},l.d=0.8,l.a=180}{l3t}
				\fmffreeze
				\fmf{dots,width=.8}{l3,l3b}
				\fmfrectangle{8}{7}{r3b}
			\end{fmfgraph*}
			\quad \;
			\begin{fmfgraph*}(50,40)
				\fmfstraight
				\fmfleft{l1,lm,l2}
				\fmfright{r1,r2,r3,r4,r5,r6}
				\fmf{plain}{l1,v1}
				\fmf{plain}{v1,r1}
				\fmf{plain}{l2,v2}
				\fmf{plain}{v2,r6}
				\fmffreeze
				\fmf{plain}{r3,v12,v1,v11,r2}
				\fmf{plain}{r5,v22,v2,v21,r4}
				\fmfv{decor.shape=circle,decor.filled=empty,decor.size=13,label=\tiny{$\Sp$},label.dist=0}{v1}
				\fmfv{decor.shape=circle,decor.filled=empty,decor.size=13,label=\tiny{$\Sp$},label.dist=0}{v2}
				\fmfv{label=\footnotesize{$h_1$},l.d=0.8,l.a=0}{r6}
				\fmfv{label=\footnotesize{$h_2$},l.d=0.8,l.a=0}{r1}
				\fmffreeze
				\fmf{dots,width=.8}{r2,r3}
				\fmf{dots,width=.8}{r4,r5}
				\fmfrectangle{8}{7}{lm}
			\end{fmfgraph*}
		\end{gathered}
	\end{align}
\end{fmffile}
Here, we denote hard partons providing different forward directions by
$h_1$ and $h_2$.  The existence of these hard partons can be
guaranteed via the respective choice of an observable, such as a
two-jet observable in this example.  In other words, phase space
configurations where the latter would become soft or collinear are cut
off by the observable.  All other partons in the process, \ie the
\emph{emissions}, can become soft or collinear w.r.t.\ the hard ones.
Now, the partitioning factor $\mathbb{P}^{(d)}_{c \parallel h_1}$
works such that only configurations with collinearities to parton
$h_1$ give rise to a leading singular behaviour of the kernel.
Therefore, parton $h_2$ can be treated without loss of generality as a
\emph{recoiler} which will only be of importance for soft limits.  Let
us denote the set of partons on the amplitude side by $I$, the the
ones connected to $h_1$ on the conjugate side by $\bar{I}$ and the
ones connected to $h_2$ by $\bar{J}$.  The observable guarantees that
$I$ contains a hard parton.

We can use the projector rules of \eqref{eq:line-rules} to establish
factorisation in the following sense:
\begin{fmffile}{splitting-kernel-factorisation-3}
	\fmfset{thin}{.7pt}
	\fmfset{dot_len}{1.2mm}
	\fmfset{dot_size}{3}
	\fmfset{arrow_len}{2.5mm}
	\fmfset{curly_len}{1.5mm}
	\begin{align}\label{splitting-kernel-factorisation-3}
		\mathbb{P}^{(d)}_{c \parallel h_1}  \quad
		\begin{gathered}
			\begin{fmfgraph*}(90,40)
				\fmfstraight
				\fmfleft{l1,l2,l3b,l3,l3t}
				\fmfright{r1,r2,r3b,r3,r3t}
				\fmf{phantom}{r3t,vboxr,vr,vP,vl,vboxl,v}
				\fmf{phantom,tension=0.8}{vr,vl}
				\fmf{plain}{v,l3t}
				\fmffreeze
				\fmf{plain}{r3t,vboxr,vr}
				\fmf{plain}{vl,vboxl,v}
				\fmf{plain}{l3,v3b,v,v3t,l3b}
				\fmfv{decor.shape=square,decor.filled=empty,decor.size=10,label=\tiny{$\phantom{a} $},label.dist=0}{vboxr}
				\fmfv{label=\small{$\mathbf{P}$},label.dist=0}{vP}
				\fmfv{decor.shape=square,decor.filled=empty,decor.size=10,label=\tiny{$\phantom{a} $},label.dist=0}{vboxl}
				\fmfv{decor.shape=circle,decor.filled=empty,decor.size=13,label=\tiny{$\Sp$},label.dist=0}{v}
				\fmfv{label=\footnotesize{$h_1$},l.d=0.8,l.a=180}{l3t}
				\fmffreeze
				\fmf{dots,width=.8}{l3,l3b}
				\fmfrectangle{8}{7}{r3b}
			\end{fmfgraph*}
			\quad \;
			\begin{fmfgraph*}(90,40)
				\fmfstraight
				\fmfleft{l1,lm,l2}
				\fmfright{r1,r2,r3,r4,r5,r6}
				\fmf{phantom}{l1,vbox1l,v1l,v1P,v1r,vbox1r,v1}
				\fmf{phantom,tension=0.8}{v1r,v1l}
				\fmfv{decor.shape=square,decor.filled=empty,decor.size=10,label=\tiny{$\phantom{a} $},label.dist=0}{vbox1l}
				\fmfv{decor.shape=square,decor.filled=empty,decor.size=10,label=\tiny{$\phantom{a} $},label.dist=0}{vbox1r}
				\fmfv{label=\small{$\mathbf{P}$},label.dist=0}{v1P}
				\fmf{plain}{v1,r1}
				\fmf{phantom}{l2,vbox2l,v2l,v2P,v2r,vbox2r,v2}
				\fmf{phantom,tension=0.8}{v2r,v2l}
				\fmfv{decor.shape=square,decor.filled=empty,decor.size=10,label=\tiny{$\phantom{a}$},label.dist=0}{vbox2l}
				\fmfv{decor.shape=square,decor.filled=empty,decor.size=11,label=\tiny{$ (\perp) $},label.dist=0}{vbox2r}
				\fmfv{label=\small{$\mathbf{P}$},label.dist=0}{v2P}
				\fmf{plain}{v2,r6}
				\fmffreeze
				\fmf{plain}{l1,v1l}
				\fmf{plain}{v1r,v1}
				\fmf{plain}{l2,v2l}
				\fmf{plain}{v2r,v2}
				\fmf{plain}{r3,v12,v1,v11,r2}
				\fmf{plain}{r5,v22,v2,v21,r4}
				\fmfv{decor.shape=circle,decor.filled=empty,decor.size=13,label=\tiny{$\Sp$},label.dist=0}{v1}
				\fmfv{decor.shape=circle,decor.filled=empty,decor.size=13,label=\tiny{$\Sp$},label.dist=0}{v2}
				\fmfv{label=\footnotesize{$h_1$},l.d=0.8,l.a=0}{r6}
				\fmfv{label=\footnotesize{$h_2$},l.d=0.8,l.a=0}{r1}
				\fmffreeze
				\fmf{dots,width=.8}{r2,r3}
				\fmf{dots,width=.8}{r4,r5}
				\fmfrectangle{8}{7}{lm}
			\end{fmfgraph*}
		\end{gathered}
	\end{align}
\end{fmffile}
For the splitter line on the amplitude side, there are no
$(\perp)$-components coming from the hard amplitude when transverse
momenta are balanced between the emitter and emissions.  Then, the
leading contribution from this side comes just from the forward
momentum component.\footnote{Note that a few of the two emission
amplitudes have a black box appearing on the hard amplitude
line. Nevertheless, Tab.~\ref{tab:two-emissions-sp1} shows that these
could only contribute to the leading soft-collinear limits. Moreover,
one has to check whether such contributions vanish identically on a case by case basis.}  The
leading contributions on recoiler lines come from its forward
components which are from different collinear sectors and therefore,
\eqref{eq:vanishing-vertices-all-white} does not apply here.  The
situation is more complicated for the conjugate side splitter line.
Here, the $(\perp)$-components entering the hard amplitude are in
general unbalanced and could give contributions to the leading
singular limits.  Nevertheless, we suspect that only the combination
shown in \eqref{splitting-kernel-factorisation-3} could give such a
contribution.  Then, by including the projection operators in the
splitting amplitude, one can neglect all but the forward momenta
components from the hard amplitude.

The reason for our suspicion comes from the one and two emission
examples.  For one emission, there are no $(\perp)$-components coming
from the hard amplitude at all by choice of a respective recoil
scheme.  For two emissions, there are contributions from such
components, but we can use \eqref{eq:line-rules} to decompose them,
\ie
\begin{fmffile}{perp-components-hard-amp}
	\fmfset{thin}{.7pt}
	\fmfset{dot_len}{1.2mm}
	\fmfset{dot_size}{3}
	\fmfset{arrow_len}{2.5mm}
	\fmfset{curly_len}{1.5mm}
	\begin{align}\label{eq:perp-components-hard-amp}
		\begin{gathered}
			\begin{fmfgraph*}(60,50)
				\fmfstraight
				\fmfleft{lb,l,lt}
				\fmfright{rb,r,rt}
				\fmf{plain}{l,b1,v,b2,r}
				\fmf{phantom}{lb,b1b,vb,b2b,rb}
				\fmffreeze
				\fmf{plain}{v,bb,vb}
				\fmfv{decor.shape=square,decor.filled=empty,decor.size=6,label=\tiny{$\perp$},label.dist=0}{b1}
				\fmfv{decor.shape=square,decor.filled=empty,decor.size=6}{b2}
				\fmfv{decor.shape=square,decor.filled=empty,decor.size=6}{bb}
				\fmfrectangle{5}{7}{l}
			\end{fmfgraph*}
		\end{gathered}
		\; = \;
		\begin{gathered}
			\begin{fmfgraph*}(60,50)
				\fmfstraight
				\fmfleft{lb,l,lt}
				\fmfright{rb,r,rt}
				\fmf{phantom}{l,b1,p,bp,v,b2,r}
				\fmffreeze
				\fmf{plain}{l,b1}
				\fmf{plain}{bp,r}
				\fmf{phantom}{lb,b1b,pb,bpb,vb,b2b,rb}
				\fmf{plain}{v,bb,vb}
				\fmfv{decor.shape=square,decor.filled=empty,decor.size=6,label=\tiny{$\perp$},label.dist=0}{bp}
				\fmfv{decor.shape=square,decor.filled=empty,decor.size=6}{b2}
				\fmfv{decor.shape=square,decor.filled=empty,decor.size=6}{b1}
				\fmfv{decor.shape=square,decor.filled=empty,decor.size=6}{bb}
				\fmfv{label=\small{$\mathbf{P}$},label.dist=0}{p}
				\fmfrectangle{5}{7}{l}
			\end{fmfgraph*}
		\end{gathered}
		\; + \;
		\begin{gathered}
			\begin{fmfgraph*}(60,50)
				\fmfstraight
				\fmfleft{lb,l,lt}
				\fmfright{rb,r,rt}
				\fmf{phantom}{l,b1,p,bp,v,b2,r}
				\fmffreeze
				\fmf{plain}{l,b1}
				\fmf{plain}{bp,r}
				\fmf{phantom}{lb,b1b,pb,bpb,vb,b2b,rb}
				\fmf{plain}{v,bb,vb}
				\fmfv{decor.shape=square,decor.filled=empty,decor.size=6,label=\tiny{$\perp$},label.dist=0}{b1}
				\fmfv{decor.shape=square,decor.filled=empty,decor.size=6}{b2}
				\fmfv{decor.shape=square,decor.filled=empty,decor.size=6}{bp}
				\fmfv{decor.shape=square,decor.filled=empty,decor.size=6}{bb}
				\fmfv{label=\small{$\mathbf{P}$},label.dist=0}{p}
				\fmfrectangle{5}{7}{l}
			\end{fmfgraph*}
		\end{gathered}
	\end{align}
\end{fmffile}
The last diagram vanishes due to
\eqref{eq:vanishing-vertices-all-white}, making all momenta components
from the hard amplitude other than the forward one obsolete.

Therefore, in the sense of \eqref{splitting-kernel-factorisation-3},
we can establish the factorisation of the leading soft and collinear
singularities to the hard amplitude with only its forward components.
Note that the projector insertions for the splitter and recoiler line
are redundant here, but essential for the conjugate splitter line.

The discussion above is also relevant for recoil schemes that include
unbalanced $(\perp)$-components coming from the hard amplitude.  Even
in this case, we find that the leading $\lambda_c$-contributions come
solely from diagrams of the type of the middle one in
\eqref{eq:perp-components-hard-amp}.  The reason is that the same
logic as in the balanced $(\perp)$-momentum case applies when the
splitting amplitudes are factored using the projector, as in
\begin{fmffile}{perp-components-hard-amp-2}
	\fmfset{thin}{.7pt}
	\fmfset{arrow_len}{1.8mm}
	\fmfset{curly_len}{1.8mm}
	\fmfset{dot_len}{1.2mm}
	\begin{align}
		\begin{gathered}
			\begin{fmfgraph*}(100,50)
				\fmfstraight
				\fmfleft{lb,lm,lt}
				\fmfright{rb,rm,rt}
				\fmf{phantom}{lm,b1,v1,b2,d1,d2,b3,v2,b4,p,b5,rm}
				\fmf{phantom}{lb,b1b,v1b,b2b,d1b,d2b,b3b,v2b,b4b,pb,b5b,rb}
				\fmffreeze
				\fmf{plain}{lm,d1}
				\fmf{dots}{d1,d2}
				\fmf{plain}{d2,b4}
				\fmf{plain}{b5,rm}
				\fmf{plain}{v1,bb1,v1b}
				\fmf{plain}{v2,bb2,v2b}
				\fmfv{decor.shape=square,decor.filled=30,decor.angle=0,decor.size=6}{b1}
				\fmfv{decor.shape=square,decor.filled=30,decor.angle=0,decor.size=6}{b2}
				\fmfv{decor.shape=square,decor.filled=30,decor.angle=0,decor.size=6}{b3}
				\fmfv{decor.shape=square,decor.filled=empty,decor.angle=0,decor.size=6}{b4}
				\fmfv{decor.shape=square,decor.filled=empty,decor.angle=0,decor.size=6,label=\tiny{$\perp$},label.dist=0}{b5}
				\fmfv{decor.shape=square,decor.filled=30,decor.angle=0,decor.size=6}{bb1}
				\fmfv{decor.shape=square,decor.filled=30,decor.angle=0,decor.size=6}{bb2}
				\fmfv{label=\small{$\mathbf{P}$},label.dist=0}{p}
				\fmfrectangle{5}{7}{rm}
			\end{fmfgraph*}
		\end{gathered}
	\end{align} 
\end{fmffile}%
Here, the grey boxes represent any component of the momentum
decomposition.  The left part of the diagram follows the same rules as
in the balanced case.  Then, the $(\perp)$-box on the hard amplitudes
side will add a power of $\lambda_c$ to any contribution and will
therefore be subleading.

\subsection{Single emission case}
\label{sec:single-emission-splitting-kernel}

In this section, we discuss explicit results for the one-emission
quark and gluon splitting kernels.  The relevant subamplitudes are
listed in Tab.~\ref{tab:one-emission-pc}.\footnote{ Note that those
with a backward component on an external leg, \ie a black box, do not
contribute because they identically vanish when contracted with the
projectors of \eqref{eq:projector-gluon} and
\eqref{eq:projector-fermion}.}
\begin{table}[h]
	\begin{center}
		\begin{tabular}{c|cc|c|cc}
			 & $\mathrm{C}$ & $\mathrm{S}$ & & $\mathrm{C}$ & $\mathrm{S}$  \\
			\hline
			\begin{minipage}{2cm}\begin{fmffile}{one-emission1}
	\fmfset{thin}{.7pt}
	\fmfset{arrow_len}{1.8mm}
	\fmfset{curly_len}{1.8mm}
	\fmfset{dash_len}{1.5mm}
	\begin{align*}
	\begin{gathered}
			\begin{fmfgraph*}(50,25)
				\fmfstraight
				\fmfleft{l2,lm,l1}
				\fmfright{r2,rm,r1}
				\fmf{phantom}{l2,m2,r2}
				\fmffreeze
				\fmf{plain}{l1,bl,mm,br,r1}
				\fmffreeze
				\fmf{plain}{mm,bm,m2}
				\fmffreeze
				\fmfrectangle{4}{4}{r1}
				\fmfv{decor.shape=square,decor.filled=empty,decor.angle=0,decor.size=6}{bm}
				\fmfv{decor.shape=square,decor.filled=empty,decor.angle=0,decor.size=6}{br}
				\fmfv{decor.shape=square,decor.filled=empty,decor.angle=0,decor.size=6,label=\tiny{$\perp$},label.dist=0}{bl}
			\end{fmfgraph*}
		\end{gathered}
	\end{align*} 
	\end{fmffile}\end{minipage} 
	& $\lambda$ & $\lambda$ &
	\begin{minipage}{2cm}\begin{fmffile}{one-emission12}
	\fmfset{thin}{.7pt}
	\fmfset{arrow_len}{1.8mm}
	\fmfset{curly_len}{1.8mm}
	\fmfset{dash_len}{1.5mm}
	\begin{align*}
	\begin{gathered}
			\begin{fmfgraph*}(50,25)
				\fmfstraight
				\fmfleft{l2,lm,l1}
				\fmfright{r2,rm,r1}
				\fmf{phantom}{l2,m2,r2}
				\fmffreeze
				\fmf{plain}{l1,bl,mm,br,r1}
				\fmffreeze
				\fmf{plain}{mm,bm,m2}
				\fmffreeze
				\fmfrectangle{4}{4}{r1}
				\fmfv{decor.shape=square,decor.filled=empty,decor.angle=0,decor.size=6}{bm}
				\fmfv{decor.shape=square,decor.filled=empty,decor.angle=0,decor.size=6}{br}
				\fmfv{decor.shape=square,decor.filled=empty,decor.angle=0,decor.size=6}{bl}
				\fmfv{decor.shape=circle,decor.filled=empty,decor.size=10,label=\tiny{$\perp$},label.dist=0}{mm}
			\end{fmfgraph*}
		\end{gathered}
	\end{align*} 
	\end{fmffile}\end{minipage} 
	&  $\lambda$	&  $\lambda$	\\
	\hline
	\begin{minipage}{2cm}\begin{fmffile}{one-emission2}
	\fmfset{thin}{.7pt}
	\fmfset{arrow_len}{1.8mm}
	\fmfset{curly_len}{1.8mm}
	\fmfset{dash_len}{1.5mm}
	\begin{align*}
	\begin{gathered}
			\begin{fmfgraph*}(50,25)
				\fmfstraight
				\fmfleft{l2,lm,l1}
				\fmfright{r2,rm,r1}
				\fmf{phantom}{l2,m2,r2}
				\fmffreeze
				\fmf{plain}{l1,bl,mm,br,r1}
				\fmffreeze
				\fmf{plain}{mm,bm,m2}
				\fmffreeze
				\fmfrectangle{4}{4}{r1}
				\fmfv{decor.shape=square,decor.filled=empty,decor.angle=0,decor.size=6,label=\tiny{$\perp$},label.dist=0}{bm}
				\fmfv{decor.shape=square,decor.filled=empty,decor.angle=0,decor.size=6}{br}
				\fmfv{decor.shape=square,decor.filled=empty,decor.angle=0,decor.size=6}{bl}
			\end{fmfgraph*}
		\end{gathered}
	\end{align*} 
	\end{fmffile}\end{minipage}
	& $\lambda$	& $1$ &
	\begin{minipage}{2cm}\begin{fmffile}{one-emission22}
	\fmfset{thin}{.7pt}
	\fmfset{arrow_len}{1.8mm}
	\fmfset{curly_len}{1.8mm}
	\fmfset{dash_len}{1.5mm}
	\begin{align*}
	\begin{gathered}
			\begin{fmfgraph*}(50,25)
				\fmfstraight
				\fmfleft{l2,lm,l1}
				\fmfright{r2,rm,r1}
				\fmf{phantom}{l2,m2,r2}
				\fmffreeze
				\fmf{plain}{l1,bl,mm,br,r1}
				\fmffreeze
				\fmf{plain}{mm,bm,m2}
				\fmffreeze
				\fmfrectangle{4}{4}{r1}
				\fmfv{decor.shape=square,decor.filled=empty,decor.angle=0,decor.size=6}{bm}
				\fmfv{decor.shape=square,decor.filled=empty,decor.angle=0,decor.size=6}{br}
				\fmfv{decor.shape=square,decor.filled=empty,decor.angle=0,decor.size=6,label=\tiny{$\perp$},label.dist=0}{bl}
				\fmfv{decor.shape=circle,decor.filled=empty,decor.size=10,label=\tiny{$\perp$},label.dist=0}{mm}
			\end{fmfgraph*}
		\end{gathered}
	\end{align*} 
	\end{fmffile}\end{minipage} 
	&  $\lambda^2$	&  $\lambda^2$	\\
	\hline
	\begin{minipage}{2cm}\begin{fmffile}{one-emission3}
	\fmfset{thin}{.7pt}
	\fmfset{arrow_len}{1.8mm}
	\fmfset{curly_len}{1.8mm}
	\fmfset{dash_len}{1.5mm}
	\begin{align*}
	\begin{gathered}
			\begin{fmfgraph*}(50,25)
				\fmfstraight
				\fmfleft{l2,lm,l1}
				\fmfright{r2,rm,r1}
				\fmf{phantom}{l2,m2,r2}
				\fmffreeze
				\fmf{plain}{l1,bl,mm,br,r1}
				\fmffreeze
				\fmf{plain}{mm,bm,m2}
				\fmffreeze
				\fmfrectangle{4}{4}{r1}
				\fmfv{decor.shape=square,decor.filled=empty,decor.angle=0,decor.size=6}{bm}
				\fmfv{decor.shape=square,decor.filled=empty,decor.angle=0,decor.size=6,label=\tiny{$\perp$},label.dist=0}{br}
				\fmfv{decor.shape=square,decor.filled=empty,decor.angle=0,decor.size=6}{bl}
			\end{fmfgraph*}
		\end{gathered}
	\end{align*} 
	\end{fmffile}
	\end{minipage}
	& $\lambda$	& $\lambda$  &
	\begin{minipage}{2cm}\begin{fmffile}{one-emission32}
	\fmfset{thin}{.7pt}
	\fmfset{arrow_len}{1.8mm}
	\fmfset{curly_len}{1.8mm}
	\fmfset{dash_len}{1.5mm}
	\begin{align*}
	\begin{gathered}
			\begin{fmfgraph*}(50,25)
				\fmfstraight
				\fmfleft{l2,lm,l1}
				\fmfright{r2,rm,r1}
				\fmf{phantom}{l2,m2,r2}
				\fmffreeze
				\fmf{plain}{l1,bl,mm,br,r1}
				\fmffreeze
				\fmf{plain}{mm,bm,m2}
				\fmffreeze
				\fmfrectangle{4}{4}{r1}
				\fmfv{decor.shape=square,decor.filled=empty,decor.angle=0,decor.size=6}{bm}
				\fmfv{decor.shape=square,decor.filled=empty,decor.angle=0,decor.size=6,label=\tiny{$\perp$},label.dist=0}{br}
				\fmfv{decor.shape=square,decor.filled=empty,decor.angle=0,decor.size=6}{bl}
				\fmfv{decor.shape=circle,decor.filled=empty,decor.size=10,label=\tiny{$\perp$},label.dist=0}{mm}
			\end{fmfgraph*}
		\end{gathered}
	\end{align*} 
	\end{fmffile}	\end{minipage} 
	&  $\lambda^2$	&  $\lambda^2$	\\
	\hline
	\begin{minipage}{2cm}\begin{fmffile}{one-emission4}
	\fmfset{thin}{.7pt}
	\fmfset{arrow_len}{1.8mm}
	\fmfset{curly_len}{1.8mm}
	\fmfset{dash_len}{1.5mm}
	\begin{align*}
	\begin{gathered}
			\begin{fmfgraph*}(50,25)
				\fmfstraight
				\fmfleft{l2,lm,l1}
				\fmfright{r2,rm,r1}
				\fmf{phantom}{l2,m2,r2}
				\fmffreeze
				\fmf{plain}{l1,bl,mm,br,r1}
				\fmffreeze
				\fmf{plain}{mm,bm,m2}
				\fmffreeze
				\fmfrectangle{4}{4}{r1}
				\fmfv{decor.shape=square,decor.filled=empty,decor.angle=0,decor.size=6,label=\tiny{$\perp$},label.dist=0}{bm}
				\fmfv{decor.shape=square,decor.filled=empty,decor.angle=0,decor.size=6}{br}
				\fmfv{decor.shape=square,decor.filled=empty,decor.angle=0,decor.size=6,label=\tiny{$\perp$},label.dist=0}{bl}
			\end{fmfgraph*}
		\end{gathered}
	\end{align*} 
	\end{fmffile}\end{minipage}
	& $\lambda^2$	& $\lambda$&
	\begin{minipage}{2cm}\begin{fmffile}{one-emission42}
	\fmfset{thin}{.7pt}
	\fmfset{arrow_len}{1.8mm}
	\fmfset{curly_len}{1.8mm}
	\fmfset{dash_len}{1.5mm}
	\begin{align*}
	\begin{gathered}
			\begin{fmfgraph*}(50,25)
				\fmfstraight
				\fmfleft{l2,lm,l1}
				\fmfright{r2,rm,r1}
				\fmf{phantom}{l2,m2,r2}
				\fmffreeze
				\fmf{plain}{l1,bl,mm,br,r1}
				\fmffreeze
				\fmf{plain}{mm,bm,m2}
				\fmffreeze
				\fmfrectangle{4}{4}{r1}
				\fmfv{decor.shape=square,decor.filled=empty,decor.angle=0,decor.size=6,label=\tiny{$\perp$},label.dist=0}{bm}
				\fmfv{decor.shape=square,decor.filled=empty,decor.angle=0,decor.size=6}{br}
				\fmfv{decor.shape=square,decor.filled=empty,decor.angle=0,decor.size=6}{bl}
				\fmfv{decor.shape=circle,decor.filled=empty,decor.size=10,label=\tiny{$\perp$},label.dist=0}{mm}
			\end{fmfgraph*}
		\end{gathered}
	\end{align*} 
	\end{fmffile}\end{minipage} 
	&  $\lambda^2$	&  $\lambda$ \\
	\hline
	\begin{minipage}{2cm}\begin{fmffile}{one-emission5}
	\fmfset{thin}{.7pt}
	\fmfset{arrow_len}{1.8mm}
	\fmfset{curly_len}{1.8mm}
	\fmfset{dash_len}{1.5mm}
	\begin{align*}
	\begin{gathered}
			\begin{fmfgraph*}(50,25)
				\fmfstraight
				\fmfleft{l2,lm,l1}
				\fmfright{r2,rm,r1}
				\fmf{phantom}{l2,m2,r2}
				\fmffreeze
				\fmf{plain}{l1,bl,mm,br,r1}
				\fmffreeze
				\fmf{plain}{mm,bm,m2}
				\fmffreeze
				\fmfrectangle{4}{4}{r1}
				\fmfv{decor.shape=square,decor.filled=empty,decor.angle=0,decor.size=6}{bm}
				\fmfv{decor.shape=square,decor.filled=empty,decor.angle=0,decor.size=6,label=\tiny{$\perp$},label.dist=0}{br}
				\fmfv{decor.shape=square,decor.filled=empty,decor.angle=0,decor.size=6,label=\tiny{$\perp$},label.dist=0}{bl}
			\end{fmfgraph*}
	\end{gathered}
	\end{align*} 
	\end{fmffile}\end{minipage}
	& $\lambda^2$	& $\lambda^2$ \\
	\hline
	\begin{minipage}{2cm}\begin{fmffile}{one-emission6}
	\fmfset{thin}{.7pt}
	\fmfset{arrow_len}{1.8mm}
	\fmfset{curly_len}{1.8mm}
	\fmfset{dash_len}{1.5mm}
	\begin{align*}
\begin{gathered}
			\begin{fmfgraph*}(50,25)
				\fmfstraight
				\fmfleft{l2,lm,l1}
				\fmfright{r2,rm,r1}
				\fmf{phantom}{l2,m2,r2}
				\fmffreeze
				\fmf{plain}{l1,bl,mm,br,r1}
				\fmffreeze
				\fmf{plain}{mm,bm,m2}
				\fmffreeze
				\fmfrectangle{4}{4}{r1}
				\fmfv{decor.shape=square,decor.filled=empty,decor.angle=0,decor.size=6,label=\tiny{$\perp$},label.dist=0}{bm}
				\fmfv{decor.shape=square,decor.filled=empty,decor.angle=0,decor.size=6,label=\tiny{$\perp$},label.dist=0}{br}
				\fmfv{decor.shape=square,decor.filled=empty,decor.angle=0,decor.size=6}{bl}
			\end{fmfgraph*}
		\end{gathered}
	\end{align*} 
	\end{fmffile}\end{minipage}
	& $\lambda^2$	& $\lambda$  \\
		\hline
	\begin{minipage}{2cm}\begin{fmffile}{one-emission7}
			\fmfset{thin}{.7pt}
			\fmfset{arrow_len}{1.8mm}
			\fmfset{curly_len}{1.8mm}
			\fmfset{dash_len}{1.5mm}
			\begin{align*}
				\begin{gathered}
					\begin{fmfgraph*}(50,25)
						\fmfstraight
						\fmfright{l1,lm,l2}
						\fmfleft{r1,rm,r2}
						\fmf{phantom}{l2,m2,r2}
						\fmffreeze
						\fmf{plain}{l1,bl,mm,br,r1}
						\fmffreeze
						\fmf{plain}{mm,bm,m2}
						\fmffreeze
						\fmfrectangle{4}{4}{r1}
						\fmfv{decor.shape=square,decor.filled=empty,decor.angle=0,decor.size=6}{bm}
						\fmfv{decor.shape=square,decor.filled=empty,decor.angle=0,decor.size=6}{br}
						\fmfv{decor.shape=square,decor.filled=empty,decor.angle=0,decor.size=6}{bl}
					\end{fmfgraph*}
				\end{gathered}
			\end{align*} 
	\end{fmffile}\end{minipage}
	& $1$	& $1$  \\
			\hline
	\begin{minipage}{2cm}\begin{fmffile}{one-emission8}
			\fmfset{thin}{.7pt}
			\fmfset{arrow_len}{1.8mm}
			\fmfset{curly_len}{1.8mm}
			\fmfset{dash_len}{1.5mm}
			\begin{align*}
				\begin{gathered}
					\begin{fmfgraph*}(50,25)
						\fmfstraight
						\fmfright{l1,lm,l2}
						\fmfleft{r1,rm,r2}
						\fmf{phantom}{l2,m2,r2}
						\fmffreeze
						\fmf{plain}{l1,bl,mm,br,r1}
						\fmffreeze
						\fmf{plain}{mm,bm,m2}
						\fmffreeze
						\fmfrectangle{4}{4}{r1}
						\fmfv{decor.shape=square,decor.filled=empty,decor.angle=0,decor.size=6,label=\tiny{$\perp$},label.dist=0}{bm}
						\fmfv{decor.shape=square,decor.filled=empty,decor.angle=0,decor.size=6}{br}
						\fmfv{decor.shape=square,decor.filled=empty,decor.angle=0,decor.size=6}{bl}
					\end{fmfgraph*}
				\end{gathered}
			\end{align*} 
	\end{fmffile}\end{minipage}
	& $\lambda$	& $1$  \\
		\end{tabular}
	\end{center}
	\caption{Single emission subamplitudes up to $\mathcal{O}(\lambda^2)$ collinear (C) or soft (S) scaling for both longitudinal and transverse vertices. Emission lines going upwards (downwards) signify interferer (splitter) subamplitudes.}
	\label{tab:one-emission-pc}
\end{table} From these, we construct the splitting
kernel $\mathbb{U}_{(ij)}$ with the lowest possible overall collinear
power scaling, \ie the leading singular contributions when an emission
$j$ becomes collinear to a hard parton $i$.  Inserting splitter lines
into the amplitude and conjugate amplitude side, we generate
self-energy like contributions.  Their propagator factors give a power
scaling of $\mathcal{O}(1/\lambda_c^4)$ in the collinear and
$\mathcal{O}(1/\lambda_s^2)$ in the soft limit.  The lowest numerator
scaling is of $\mathcal{O}(\lambda_c^2)$, giving an overall scaling of
$\mathcal{O}(1/\lambda_c^2)$ in this case.  Simultaneously, the
amplitudes with the lowest soft scaling contained in here, namely the
ones with a $(\perp)$-momentum component on the emission lines.
Combining a splitter and interferer line with hard partons $i$ and
$k$, respectively, gives interference-like contributions.  These have
propagator factors from two different collinear sectors.  We extract
the $(ij)$-collinear behaviour by applying the partitioning
$\mathbb{P}_{(ij)}$ which allows us to use the collinear power
counting of Tab.~\ref{tab:one-emission-pc} in the first place.
Otherwise, additional subamplitudes with the correct collinear scaling
w.r.t. to parton $k$ would have to be taken into account.  In this
setting, both the leading collinear and soft scaling is of
$\mathcal{O}(1/\lambda_{c/s}^2)$.  The lowest collinear numerator
scaling if of $\mathcal{O}(\lambda_{c})$ which shows again that the
interferences do not contribute in the leading collinear limit.

Eventually, we find that the full one emission $(ij)$-splitting kernel
consists of
\vspace{4pt}
	\begin{fmffile}{one-emission-kernel}
	\fmfset{thin}{.7pt}
	\fmfset{dot_len}{1.2mm}
	\fmfset{dot_size}{5}
	\fmfset{arrow_len}{2.5mm}
	\fmfset{curly_len}{1.5mm}
	\begin{align}\label{eq:one-emission-kernel}
		\mathbb{U}_{(ij)} =	\mathbb{P}_{(ij)} 
		&\Bigg(\quad
		\begin{gathered}
			\begin{fmfgraph*}(50,30)
				\fmfstraight
				\fmfleft{lb,lm,lt}
				\fmfright{rb,rm,rt}
				\fmf{plain}{rt,bt1,v,bt2,lt}
				\fmf{plain}{rb,bb,lb}
				\fmffreeze
				\fmf{curly}{v,bm,lm}
				\fmfrectangle{8}{6}{rm}
				\fmfv{decor.shape=square,decor.filled=empty,label=\tiny{},label.dist=0,decor.size=6}{bt1}
				\fmfv{decor.shape=square,decor.filled=empty,label=\tiny{},label.dist=0,decor.size=6}{bt2}
				\fmfv{decor.shape=square,decor.filled=empty,label=\tiny{$\perp$},label.dist=0,decor.size=6}{bm}
				\fmfv{decor.shape=square,decor.filled=empty,label=\tiny{},label.dist=0,decor.size=6}{bb}
			\end{fmfgraph*}
			\quad \;
			\begin{fmfgraph*}(50,30)
				\fmfstraight
				\fmfleft{lb,lm,lt}
				\fmfright{rb,rm,rt}
				\fmf{plain}{lt,bt,rt}
				\fmf{plain}{lb,bb1,v,bb2,rb}
				\fmffreeze
				\fmf{curly}{v,bm,rm}
				\fmfrectangle{8}{6}{lm}
				\fmfv{decor.shape=square,decor.filled=empty,label=\tiny{},label.dist=0,decor.size=6}{bb1}
				\fmfv{decor.shape=square,decor.filled=empty,label=\tiny{},label.dist=0,decor.size=6}{bb2}
				\fmfv{decor.shape=square,decor.filled=empty,label=\tiny{},label.dist=0,decor.size=6}{bm}
				\fmfv{decor.shape=square,decor.filled=empty,label=\tiny{},label.dist=0,decor.size=6}{bt}
			\end{fmfgraph*}
		\end{gathered}
		\quad + \quad
		\begin{gathered}
			\begin{fmfgraph*}(50,30)
				\fmfstraight
				\fmfleft{lb,lm,lt}
				\fmfright{rb,rm,rt}
				\fmf{plain}{rt,bt1,v,bt2,lt}
				\fmf{plain}{rb,bb,lb}
				\fmffreeze
				\fmf{curly}{v,bm,lm}
				\fmfrectangle{8}{6}{rm}
				\fmfv{decor.shape=square,decor.filled=empty,label=\tiny{},label.dist=0,decor.size=6}{bt1}
				\fmfv{decor.shape=square,decor.filled=empty,label=\tiny{},label.dist=0,decor.size=6}{bt2}
				\fmfv{decor.shape=square,decor.filled=empty,label=\tiny{$\perp$},label.dist=0,decor.size=6}{bm}
				\fmfv{decor.shape=square,decor.filled=empty,label=\tiny{},label.dist=0,decor.size=6}{bb}
			\end{fmfgraph*}
			\quad \;
			\begin{fmfgraph*}(50,30)
				\fmfstraight
				\fmfleft{lb,lm,lt}
				\fmfright{rb,rm,rt}
				\fmf{plain}{lt,bt,rt}
				\fmf{plain}{lb,bb1,v,bb2,rb}
				\fmffreeze
				\fmf{curly}{v,bm,rm}
				\fmfrectangle{8}{6}{lm}
				\fmfv{decor.shape=square,decor.filled=empty,label=\tiny{},label.dist=0,decor.size=6}{bb1}
				\fmfv{decor.shape=square,decor.filled=empty,label=\tiny{},label.dist=0,decor.size=6}{bb2}
				\fmfv{decor.shape=square,decor.filled=empty,label=\tiny{$\perp$},label.dist=0,decor.size=6}{bm}
				\fmfv{decor.shape=square,decor.filled=empty,label=\tiny{},label.dist=0,decor.size=6}{bt}
			\end{fmfgraph*}
		\end{gathered}
		\quad\Bigg)
		\\[20pt] \nonumber
		&
		+ \quad
		\begin{gathered}
			\begin{fmfgraph*}(90,50)
				\fmfstraight
				\fmfleft{l1,lm,l2}
				\fmfright{r1,rm,r2}
				\fmf{phantom}{l1,vl,l2}
				\fmf{phantom}{r1,vr,r2}
				\fmf{phantom,tension=8}{vl,vr}
				\fmffreeze
				\fmf{plain}{vl,bl1,v1l,bl2,lm}
				\fmf{plain}{vr,br1,v1r,br2,rm}
				\fmf{plain,tension=0.25}{lm,bl1}
				\fmf{plain,tension=0.25}{rm,br1}
				\fmffreeze
				\fmf{curly}{v1l,bgl,l1}
				\fmf{curly}{r1,bgr,v1r}
				\fmfrectangle{6}{5}{vl}
				\fmfrectangle{6}{5}{vr}
				\fmfv{decor.shape=square,decor.filled=empty,decor.angle=0,decor.size=6}{bl1}
				\fmfv{decor.shape=square,decor.filled=empty,decor.angle=0,decor.size=6,label=\tiny{$\perp$},label.dist=0}{bl2}
				\fmfv{decor.shape=square,decor.filled=empty,decor.angle=0,decor.size=6}{br1}
				\fmfv{decor.shape=square,decor.filled=empty,decor.angle=0,decor.size=6,label=\tiny{$\perp$},label.dist=0}{br2}
				\fmfv{decor.shape=square,decor.filled=empty,decor.angle=0,decor.size=6}{bgl}
				\fmfv{decor.shape=square,decor.filled=empty,decor.angle=0,decor.size=6}{bgr}
			\end{fmfgraph*}
		\end{gathered}
		\quad + \quad
		\begin{gathered}
			\begin{fmfgraph*}(90,50)
				\fmfstraight
				\fmfleft{l1,lm,l2}
				\fmfright{r1,rm,r2}
				\fmf{phantom}{l1,vl,l2}
				\fmf{phantom}{r1,vr,r2}
				\fmf{phantom,tension=8}{vl,vr}
				\fmffreeze
				\fmf{plain}{vl,bl1,v1l,bl2,lm}
				\fmf{plain}{vr,br1,v1r,br2,rm}
				\fmf{plain,tension=0.25}{lm,bl1}
				\fmf{plain,tension=0.25}{rm,br1}
				\fmffreeze
				\fmf{curly}{v1l,bgl,l1}
				\fmf{curly}{r1,bgr,v1r}
				\fmfrectangle{6}{5}{vr}
				\fmfrectangle{6}{5}{vl}
				\fmfv{decor.shape=square,decor.filled=empty,decor.angle=0,decor.size=6}{bl1}
				\fmfv{decor.shape=square,decor.filled=empty,decor.angle=0,decor.size=6,label=\tiny{$\perp$},label.dist=0}{bgl}
				\fmfv{decor.shape=square,decor.filled=empty,decor.angle=0,decor.size=6}{br1}
				\fmfv{decor.shape=square,decor.filled=empty,decor.angle=0,decor.size=6,label=\tiny{$\perp$},label.dist=0}{bgr}
				\fmfv{decor.shape=square,decor.filled=empty,decor.angle=0,decor.size=6}{bl2}
				\fmfv{decor.shape=square,decor.filled=empty,decor.angle=0,decor.size=6}{br2}
			\end{fmfgraph*}
		\end{gathered}
		\\ \nonumber
		& + \quad 
		\begin{gathered}
			\begin{fmfgraph*}(90,50)
				\fmfstraight
				\fmfleft{l1,lm,l2}
				\fmfright{r1,rm,r2}
				\fmf{phantom}{l1,vl,l2}
				\fmf{phantom}{r1,vr,r2}
				\fmf{phantom,tension=8}{vl,vr}
				\fmffreeze
				\fmf{plain}{vl,bl1,v1l,bl2,lm}
				\fmf{plain}{vr,br1,v1r,br2,rm}
				\fmf{plain,tension=0.25}{lm,bl1}
				\fmf{plain,tension=0.25}{rm,br1}
				\fmffreeze
				\fmf{curly}{v1l,bgl,l1}
				\fmf{curly}{r1,bgr,v1r}
				\fmfrectangle{6}{5}{vl}
				\fmfrectangle{6}{5}{vr}
				\fmfv{decor.shape=square,decor.filled=empty,decor.angle=0,decor.size=6}{bl1}
				\fmfv{decor.shape=square,decor.filled=empty,decor.angle=0,decor.size=6,label=\tiny{$\perp$},label.dist=0}{bl2}
				\fmfv{decor.shape=square,decor.filled=empty,decor.angle=0,decor.size=6}{br1}
				\fmfv{decor.shape=square,decor.filled=empty,decor.angle=0,decor.size=6,label=\tiny{$\perp$},label.dist=0}{bgr}
				\fmfv{decor.shape=square,decor.filled=empty,decor.angle=0,decor.size=6}{bgl}
				\fmfv{decor.shape=square,decor.filled=empty,decor.angle=0,decor.size=6}{br2}
			\end{fmfgraph*}
		\end{gathered}
		\quad + \quad 
		\begin{gathered}
			\begin{fmfgraph*}(90,50)
				\fmfstraight
				\fmfleft{l1,lm,l2}
				\fmfright{r1,rm,r2}
				\fmf{phantom}{l1,vl,l2}
				\fmf{phantom}{r1,vr,r2}
				\fmf{phantom,tension=8}{vl,vr}
				\fmffreeze
				\fmf{plain}{vl,bl1,v1l,bl2,lm}
				\fmf{plain}{vr,br1,v1r,br2,rm}
				\fmf{plain,tension=0.25}{lm,bl1}
				\fmf{plain,tension=0.25}{rm,br1}
				\fmffreeze
				\fmf{curly}{v1l,bgl,l1}
				\fmf{curly}{r1,bgr,v1r}
				\fmfrectangle{6}{5}{vr}
				\fmfrectangle{6}{5}{vl}
				\fmfv{decor.shape=square,decor.filled=empty,decor.angle=0,decor.size=6}{bl1}
				\fmfv{decor.shape=square,decor.filled=empty,decor.angle=0,decor.size=6,label=\tiny{$\perp$},label.dist=0}{bgl}
				\fmfv{decor.shape=square,decor.filled=empty,decor.angle=0,decor.size=6}{br1}
				\fmfv{decor.shape=square,decor.filled=empty,decor.angle=0,decor.size=6,label=\tiny{$\perp$},label.dist=0}{br2}
				\fmfv{decor.shape=square,decor.filled=empty,decor.angle=0,decor.size=6}{bl2}
				\fmfv{decor.shape=square,decor.filled=empty,decor.angle=0,decor.size=6}{bgr}
			\end{fmfgraph*}
		\end{gathered}
	\end{align}
\end{fmffile}
We now want to exhibit the interplay between soft and collinear
contributions in this splitting kernel and base this on a general
discussion of Eikonal currents.  First, we note that the emission of a
soft gluon with momentum $q_j$ off of a hard quark line $i$ leads to
the Eikonal vertex rule \cite{BASSETTO1983201}
\begin{fmffile}{soft-qqg-vertex}
	\fmfset{thin}{.7pt}
	\fmfset{arrow_len}{2.5mm}
	\fmfset{curly_len}{1.5mm}
	\begin{align}\label{eq:soft-qqg-vertex}
		\begin{gathered}
			\begin{fmfgraph*}(50,40)
				\fmfleft{l}
				\fmfright{rb,rm,rt}
				\fmf{fermion}{l,v,rm}
				\fmffreeze
				\fmf{phantom,label=\tiny{$q_i$},label.side=left}{v,rm}
				\fmf{curly,label=\tiny{$q_j$},label.side=right}{v,rb}
				\fmfv{label=\tiny{$\mu$},label.dist=5}{rb}
				\fmfv{decor.shape=circle,decor.filled=full,decor.size=2}{v}
			\end{fmfgraph*}
		\end{gathered}
	\;
		\propto \frac{q_i^\mu}{q_i \scdot q_j} +\mathcal{O}\left(1\right),
	\end{align}
\end{fmffile}%
where $Q$ is the hard scale of the full process.
Using this vertex rule in a self-energy like contribution, we find
\begin{fmffile}{one-emission-self-energy-eikonal}
	\fmfset{thin}{.7pt}
	\fmfset{dash_len}{1.5mm}
	\fmfset{arrow_len}{2.5mm}
	\fmfset{curly_len}{1.5mm}
	\begin{align}\label{eq:eikonal-self-energy}
		\begin{gathered}
			\begin{tikzpicture}
				\node (diagram) {%
					\begin{fmfgraph*}(70,50)
						\fmfleft{bl}
						\fmfright{br}
						\fmf{phantom}{br,v1,vf,v2,bl}
						\fmf{phantom,tension=1.1}{v1,vf,v2}
						\fmffreeze
						\fmf{fermion}{v2,v1}
						\fmf{plain,background=white,rubout}{bl,v2}
						\fmf{plain,background=white,rubout}{v1,br}
						\fmffreeze
						\fmf{curly,right}{v1,v2}
						\fmfv{label=\tiny{$q_j$},label.angle=130,label.dist=6mm}{vf}
						\fmfv{label=\tiny{$q_i$},label.angle=-40,label.dist=2mm}{v2}
						\fmfv{decor.shape=circle,decor.filled=full,decor.size=2.5}{v1}
						\fmfv{decor.shape=circle,decor.filled=full,decor.size=2.5}{v2}
						\fmfrectangle{4}{4}{bl}
						\fmfrectangle{4}{4}{br}
					\end{fmfgraph*}
				};
				\tikzset{shift={(0,0)}}
				\draw[thick, dashed] (-0.3,-0.6) arc(-90:0:0.3) (0,-0.3) -- (0,0.8) arc(180:90:0.3);
			\end{tikzpicture}
		\end{gathered}
		\;  \propto 4 \pi \alpha_s \mathbf{T}_i^2 \, \frac{q_i^\mu}{q_i \scdot q_j} d_{\mu\nu} (q_j) \frac{q_i^\nu}{q_i \scdot q_j} = \frac{8 \pi \alpha_s \mathbf{T}_i^2}{q_i \scdot q_j} \frac{q_i \scdot n}{q_j\scdot n} \ ,
	\end{align}
\end{fmffile}%
%
where $\mathbf{T}_i$ is the colour charge operator associated with
parton $i$.  We will shortly see that this is just the soft-divergent
part of the splitting function $\hat{P}_{qg}$ (\ie the second term of
\eqref{eq:one-emission-coll}).  Now we compare this to the
contribution from the corresponding exchange diagram.  Inserting the
Eikonal couplings here, we find
\begin{fmffile}{one-emission-interference-eikonal}
	\fmfset{thin}{.7pt}
	\fmfset{dash_len}{1.5mm}
	\fmfset{arrow_len}{2.5mm}
	\fmfset{curly_len}{1.5mm}
	\begin{align} \label{eq:eikonal-calc-interference}
		2\mathrm{Re} \; \;
		\begin{gathered}
			\begin{tikzpicture}
				\node (diagram) {%
					\begin{fmfgraph*}(70,30)
						\fmfstraight
						\fmfleft{ld,lm,lu}
						\fmfright{rd,rm,ru}
						\fmf{plain}{lu,vul,vur,ru}
						\fmf{plain}{ld,vdl,vdr,rd}
						\fmf{fermion}{vur,ru}
						\fmf{fermion}{vdl,ld}
						\fmf{phantom,tension=1.6}{lu,vul,vur}
						\fmf{phantom,tension=1.6}{rd,vdr,vdl}
						\fmffreeze
						\fmfv{label=\tiny{$q_i$},label.angle=160,label.dist=6mm}{ru}
						\fmfv{label=\tiny{$q_k$},label.angle=-20,label.dist=6mm}{ld}
						\fmfv{label=\tiny{$q_j$},label.angle=-90,label.dist=5mm}{vul}
						\fmf{curly}{vul,vdr}
						\fmf{plain}{lu,vul}
						\fmf{plain}{vdr,rd}
						\fmf{phantom}{vur,vdl}
						\fmfv{decor.shape=circle,decor.filled=full,decor.size=2.5}{vul}
						\fmfv{decor.shape=circle,decor.filled=full,decor.size=2.5}{vdr}
						\fmfrectangle{10}{4}{lm}
						\fmfrectangle{10}{4}{rm}
					\end{fmfgraph*}
				};
				\tikzset{shift={(0,0)}}
				\draw[thick, dashed] (-0.3,-1.1) arc(-90:0:0.3) (0,-0.8) -- (0,0.8) arc(180:90:0.3);
			\end{tikzpicture}
		\end{gathered}
		\;
		&\propto 2\times 4 \pi \alpha_s \mathbf{T}_i \scdot \mathbf{T}_k \, \frac{q_i^\mu}{q_i \scdot q_j} d_{\mu\nu} (q_j) \frac{q_k^\nu}{q_k \scdot q_j} 
		\nonumber \\
		&
		= 8 \pi \alpha_s (-\mathbf{T}_i \scdot \mathbf{T}_k) \left[ \frac{q_i \scdot q_k}{(q_i \scdot q_j)(q_k \scdot q_j)} - \frac{1}{q_i \scdot q_j} \frac{q_i \scdot n}{q_j \scdot n} - \frac{1}{q_k \scdot q_j} \frac{q_k \scdot n}{q_j \scdot n}\right].
	\end{align}
\end{fmffile}
The first term of this contribution is the squared Eikonal current
representing the leading soft singular behaviour.  The second (third)
term is independent of the parton momentum $q_k$ ($q_i$).  This allows
the use of colour conservation when summing over all partons, \ie
\begin{equation}
	\sum\limits_{k\neq i} \mathbf{T}_k = -\mathbf{T}_i \ .
\end{equation}
Therefore, the last two terms in \eqref{eq:eikonal-calc-interference}
cancel exactly against the soft singular contribution from
\eqref{eq:eikonal-self-energy} for parton $i$ and $k$ and one is left
only with the (gauge-invariant) squared Eikonal current from the first
term of \eqref{eq:eikonal-calc-interference} in the soft limit.  This
analysis in the one emission case highlights a fundamental difference
of our approach as compared to constructing fixed-order subtraction
terms as in \cite{Catani:1996vz}: Our formalism would keep the
structure of the interference diagrams, which are collinear finite due
to the presence of the gauge-vector dependent terms as in
Eq.~(\ref{eq:eikonal-calc-interference}). At the same time,
Eq.~(\ref{eq:eikonal-self-energy}) will contain the entire collinear
splitting function. Approaches like the dipole formalism start from
the known soft and collinear behaviours, and explicitly remove the
overlap in between them through partitioning the soft behaviour,
including the soft-collinear singularity, in between different kernels
for collinear sectors. Especially in view of the more complicated
colour structures pertaining to the interference diagrams, our
formalism might be beneficial as it explicitly appreciates the fact
that the colour correlations are collinear finite.

Our partitioning algorithms allow us, however, to apply a similar
logic to the interplay of soft and collinear limits at the level of
cross section factorization: we further discuss this idea in
Sec.~\ref{sec:soft-collinear-functions}, which can serve as an
additional starting point for parton branching algorithms beyond the
leading order\footnote{In fact, while we were finalizing the present
work, Ref.~\cite{Gellersen:2021eci} which discusses a similar
question}.

\subsubsection{Quark-gluon splitting}

In order to discuss these results in terms of our power counting
algorithm, we employ the Sudakov decomposition of
\eqref{eq:Sudakov-decomp-single}, \ie
\begin{align}\label{eq:mapping-alpha-1E}
	q_i^\mu &= z_i\, p_i^\mu + \frac{p_{\perp,i}^2}{z_i \, 2p_i \scdot n}n^\mu + k_{\perp,i}^\mu,  \nonumber\\
	q_j^\mu &= z_j\, p_i^\mu + \frac{p_{\perp,j}^2}{z_j \, 2p_i \scdot n}n^\mu + k_{\perp,j}^\mu,\nonumber \\
	q_k^\mu &= z_k \, p_k^\mu.
\end{align}
Using this mapping, we find vertex rules for splitter lines, namely
\begin{fmffile}{one-emission-amps}
	\fmfset{thin}{.7pt}
	\fmfset{dot_len}{1.2mm}
	\fmfset{dot_size}{5}
	\fmfset{arrow_len}{2.5mm}
	\fmfset{curly_len}{1.5mm}
	\begin{subequations}
		\begin{align}
			\begin{gathered}
				\begin{fmfgraph*}(50,25)
					\fmfstraight
					\fmfleft{l2,lm,l1}
					\fmfright{r2,rm,r1}
					\fmf{phantom}{l2,m2,r2}
					\fmffreeze
					\fmf{plain}{l1,bl,mm,br,r1}
					\fmffreeze
					\fmf{curly}{mm,bm,m2}
					\fmffreeze
					\fmfrectangle{4}{4}{r1}
					\fmfv{decor.shape=square,decor.filled=empty,decor.angle=0,decor.size=6}{bm}
					\fmfv{decor.shape=square,decor.filled=empty,decor.angle=0,decor.size=6}{br}
					\fmfv{decor.shape=square,decor.filled=empty,decor.angle=0,decor.size=6,label=\tiny{$\perp$},label.dist=0}{bl}
				\end{fmfgraph*}
			\end{gathered}
			\quad &= \frac{g_s}{S_{ij}}\, \sqrt{\frac{z_i + z_j}{z_i}} \left(\frac{\slashed{k}_{\perp,i} \slashed{n}\, \slashed{p_i}}{n\scdot p_i} p_i^\nu - \slashed{k}_{\perp,i} \gamma^\nu \slashed{p}_i\right) \ ,
			\\[5pt]
			\begin{gathered}
				\begin{fmfgraph*}(50,25)
					\fmfstraight
					\fmfleft{l2,lm,l1}
					\fmfright{r2,rm,r1}
					\fmf{phantom}{l2,m2,r2}
					\fmffreeze
					\fmf{plain}{l1,bl,mm,br,r1}
					\fmffreeze
					\fmf{curly}{mm,bm,m2}
					\fmffreeze
					\fmfrectangle{4}{4}{r1}
					\fmfv{decor.shape=square,decor.filled=empty,decor.angle=0,decor.size=6}{bl}
					\fmfv{decor.shape=square,decor.filled=empty,decor.angle=0,decor.size=6}{br}
					\fmfv{decor.shape=square,decor.filled=empty,decor.angle=0,decor.size=6,label=\tiny{$\perp$},label.dist=0}{bm}
				\end{fmfgraph*}
			\end{gathered}
			\quad &=  2 \frac{g_s}{S_{ij}}\, \frac{\sqrt{z_i(z_i+z_j)}}{z_j} \slashed{p}_i  k_{\perp,j}^{\nu} \ ,
			\\[5pt]
			\begin{gathered}
				\begin{fmfgraph*}(50,25)
					\fmfstraight
					\fmfleft{r1,lm,r2}
					\fmfright{l1,rm,l2}
					\fmf{phantom}{l2,m2,r2}
					\fmffreeze
					\fmf{plain}{l1,bl,mm,br,r1}
					\fmffreeze
					\fmf{curly}{mm,bm,m2}
					\fmffreeze
					\fmfrectangle{4}{4}{r1}
					\fmfv{decor.shape=square,decor.filled=empty,decor.angle=0,decor.size=6}{bl}
					\fmfv{decor.shape=square,decor.filled=empty,decor.angle=0,decor.size=6}{br}
					\fmfv{decor.shape=square,decor.filled=empty,decor.angle=0,decor.size=6}{bm}
				\end{fmfgraph*}
			\end{gathered}
			\quad &=  2\frac{g_s}{S_{jk}}\,  z_k \frac{p_i\scdot p_k \,n^\sigma -n \scdot p_i \, p_k^\sigma + n\scdot p_k \, p_i^\sigma}{n\scdot p_i} \slashed{p}_k \  ,
			\\[5pt]
			\begin{gathered}
				\begin{fmfgraph*}(50,25)
					\fmfstraight
					\fmfleft{r1,lm,r2}
					\fmfright{l1,rm,l2}
					\fmf{phantom}{l2,m2,r2}
					\fmffreeze
					\fmf{plain}{l1,bl,mm,br,r1}
					\fmffreeze
					\fmf{curly}{mm,bm,m2}
					\fmffreeze
					\fmfrectangle{4}{4}{r1}
					\fmfv{decor.shape=square,decor.filled=empty,decor.angle=0,decor.size=6}{bl}
					\fmfv{decor.shape=square,decor.filled=empty,decor.angle=0,decor.size=6}{br}
					\fmfv{decor.shape=square,decor.filled=empty,decor.angle=0,decor.size=6,label=\tiny{$\perp$},label.dist=0}{bm}
				\end{fmfgraph*}
			\end{gathered}
			\quad &=  2 \frac{g_s}{S_{jk}}\, \frac{ z_k}{z_j} \frac{n\scdot p_k \, k_{\perp,j}^\sigma + k_{\perp,j} \scdot p_k \, n^\sigma}{n\scdot p_i} \slashed{p}_k  \ .
		\end{align}
	\end{subequations}
\end{fmffile}
The second diagram represents the Eikonal coupling in terms of our
mapping. In order to reproduce the splitting function and soft limits,
we instate one specific version of the momentum mapping,\ie
\begin{align}\label{eq:to-z-mapping}
	z_i &= z, \quad  z_j = 1-z,\quad  z_k=1, \nonumber\\
	k_{\perp,i} &= - k_{\perp,k} = k_{\perp}, \nonumber \\
	p_{\perp,i}^2 &= p_{\perp,j}^2 = p_{\perp}^2  \ .
\end{align}
Using these rules together with the projectors of
\eqref{eq:projectors-general} for connecting the amplitude and
conjugate amplitude, the self-energy like contributions of
\eqref{eq:one-emission-kernel} give
\begin{fmffile}{one-emission-self-energy-decomp-density}
	\fmfset{thin}{.7pt}
	\fmfset{dot_len}{.8mm}
	\fmfset{dot_size}{5}
	\fmfset{arrow_len}{2.5mm}
	\fmfset{curly_len}{1.5mm}
	\begin{align}
		\;\;\;
		&\begin{gathered}
			\begin{fmfgraph*}(80,50)
				\fmfstraight
				\fmfleft{l1,lm,l2}
				\fmfright{r1,rm,r2}
				\fmf{phantom}{l1,vl,l2}
				\fmf{phantom}{r1,vr,r2}
				\fmf{phantom,tension=8}{vl,vr}
				\fmffreeze
				\fmf{plain}{vl,bl1,v1l,bl2,lm}
				\fmf{plain}{vr,br1,v1r,br2,rm}
				\fmffreeze
				\fmf{curly}{v1l,bgl,l1}
				\fmf{curly}{r1,bgr,v1r}
				\fmfrectangle{6}{5}{vl}
				\fmfrectangle{6}{5}{vr}
				\fmfv{decor.shape=square,decor.filled=empty,decor.angle=0,decor.size=6}{bl1}
				\fmfv{decor.shape=square,decor.filled=empty,decor.angle=0,decor.size=6,label=\tiny{$\perp$},label.dist=0}{bl2}
				\fmfv{decor.shape=square,decor.filled=empty,decor.angle=0,decor.size=6}{br1}
				\fmfv{decor.shape=square,decor.filled=empty,decor.angle=0,decor.size=6,label=\tiny{$\perp$},label.dist=0}{br2}
				\fmfv{decor.shape=square,decor.filled=empty,decor.angle=0,decor.size=6}{bgl}
				\fmfv{decor.shape=square,decor.filled=empty,decor.angle=0,decor.size=6}{bgr}
			\end{fmfgraph*}
		\end{gathered}
		\; + \;
		\begin{gathered}
			\begin{fmfgraph*}(80,50)
				\fmfstraight
				\fmfleft{l1,lm,l2}
				\fmfright{r1,rm,r2}
				\fmf{phantom}{l1,vl,l2}
				\fmf{phantom}{r1,vr,r2}
				\fmf{phantom,tension=8}{vl,vr}
				\fmffreeze
				\fmf{plain}{vl,bl1,v1l,bl2,lm}
				\fmf{plain}{vr,br1,v1r,br2,rm}
				\fmffreeze
				\fmf{curly}{v1l,bgl,l1}
				\fmf{curly}{r1,bgr,v1r}
				\fmfrectangle{6}{5}{vl}
				\fmfrectangle{6}{5}{vr}
				\fmfv{decor.shape=square,decor.filled=empty,decor.angle=0,decor.size=6}{bl1}
				\fmfv{decor.shape=square,decor.filled=empty,decor.angle=0,decor.size=6}{bl2}
				\fmfv{decor.shape=square,decor.filled=empty,decor.angle=0,decor.size=6}{br1}
				\fmfv{decor.shape=square,decor.filled=empty,decor.angle=0,decor.size=6}{br2}
				\fmfv{decor.shape=square,decor.filled=empty,decor.angle=0,decor.size=6,label=\tiny{$\perp$},label.dist=0}{bgl}
				\fmfv{decor.shape=square,decor.filled=empty,decor.angle=0,decor.size=6,label=\tiny{$\perp$},label.dist=0}{bgr}
			\end{fmfgraph*}
		\end{gathered}
		\;+\;
		\begin{gathered}
			\begin{fmfgraph*}(80,50)
				\fmfstraight
				\fmfleft{l1,lm,l2}
				\fmfright{r1,rm,r2}
				\fmf{phantom}{l1,vl,l2}
				\fmf{phantom}{r1,vr,r2}
				\fmf{phantom,tension=8}{vl,vr}
				\fmffreeze
				\fmf{plain}{vl,bl1,v1l,bl2,lm}
				\fmf{plain}{vr,br1,v1r,br2,rm}
				\fmffreeze
				\fmf{curly}{v1l,bgl,l1}
				\fmf{curly}{r1,bgr,v1r}
				\fmfrectangle{6}{5}{vl}
				\fmfrectangle{6}{5}{vr}
				\fmfv{decor.shape=square,decor.filled=empty,decor.angle=0,decor.size=6}{bl1}
				\fmfv{decor.shape=square,decor.filled=empty,decor.angle=0,decor.size=6,label=\tiny{$\perp$},label.dist=0}{bl2}
				\fmfv{decor.shape=square,decor.filled=empty,decor.angle=0,decor.size=6}{br1}
				\fmfv{decor.shape=square,decor.filled=empty,decor.angle=0,decor.size=6}{br2}
				\fmfv{decor.shape=square,decor.filled=empty,decor.angle=0,decor.size=6}{bgl}
				\fmfv{decor.shape=square,decor.filled=empty,decor.angle=0,decor.size=6,label=\tiny{$\perp$},label.dist=0}{bgr}
			\end{fmfgraph*}
		\end{gathered}
		\;+\;
		\begin{gathered}
			\begin{fmfgraph*}(80,50)
				\fmfstraight
				\fmfleft{l1,lm,l2}
				\fmfright{r1,rm,r2}
				\fmf{phantom}{l1,vl,l2}
				\fmf{phantom}{r1,vr,r2}
				\fmf{phantom,tension=8}{vl,vr}
				\fmffreeze
				\fmf{plain}{vl,bl1,v1l,bl2,lm}
				\fmf{plain}{vr,br1,v1r,br2,rm}
				\fmffreeze
				\fmf{curly}{v1l,bgl,l1}
				\fmf{curly}{r1,bgr,v1r}
				\fmfrectangle{6}{5}{vl}
				\fmfrectangle{6}{5}{vr}
				\fmfv{decor.shape=square,decor.filled=empty,decor.angle=0,decor.size=6}{bl1}
				\fmfv{decor.shape=square,decor.filled=empty,decor.angle=0,decor.size=6}{bl2}
				\fmfv{decor.shape=square,decor.filled=empty,decor.angle=0,decor.size=6}{br1}
				\fmfv{decor.shape=square,decor.filled=empty,decor.angle=0,decor.size=6,label=\tiny{$\perp$},label.dist=0}{br2}
				\fmfv{decor.shape=square,decor.filled=empty,decor.angle=0,decor.size=6,label=\tiny{$\perp$},label.dist=0}{bgl}
				\fmfv{decor.shape=square,decor.filled=empty,decor.angle=0,decor.size=6}{bgr}
			\end{fmfgraph*}
		\end{gathered} 
		\nonumber \\
		&\overset{\mathrm{Tr}}{\rightarrow} \frac{4 \pi \alpha_s C_F}{S_{ij}} \Big[(d-2)(1-z) +  \frac{4 z^2}{1-z} + 4 z	\Big]  \slashed{p}_i \ ,
		\label{eq:one-emission-coll}
	\end{align}
\end{fmffile}
where the soft-singular term in square brackets solely comes about via
the second diagram above.  In total, we reproduce the well-known spin
averaged quark gluon splitting function in $d=4-2\epsilon$ dimensions
\cite{Catani:1996vz}
\begin{equation}\label{eq:Pqg-CS}
	\langle \hat{P}_{qg} (z)\rangle = C_F \left[ \frac{1+z^2}{1-z} - \epsilon (1-z)\right].
\end{equation}

For the interference contributions of \eqref{eq:one-emission-kernel},
we find
\vspace{4pt}
\begin{fmffile}{qg-exchange-diagrams}
	\fmfset{thin}{.7pt}
	\fmfset{dot_len}{1.2mm}
	\fmfset{dot_size}{5}
	\fmfset{arrow_len}{2.5mm}
	\fmfset{curly_len}{1.5mm}
	\begin{subequations}\label{eq:interference-qg-1E}
		\begin{align}\label{eq:interference-one-perp}
			\mathbb{P}_{(ij)} \;
			\begin{gathered}
				\begin{fmfgraph*}(50,30)
					\fmfstraight
					\fmfleft{lb,lm,lt}
					\fmfright{rb,rm,rt}
					\fmf{plain}{rt,bt1,v,bt2,lt}
					\fmf{plain}{rb,bb,lb}
					\fmffreeze
					\fmf{curly}{v,bm,lm}
					\fmfrectangle{8}{6}{rm}
					\fmfv{decor.shape=square,decor.filled=empty,label=\tiny{},label.dist=0,decor.size=6}{bt1}
					\fmfv{decor.shape=square,decor.filled=empty,label=\tiny{},label.dist=0,decor.size=6}{bt2}
					\fmfv{decor.shape=square,decor.filled=empty,label=\tiny{$\perp$},label.dist=0,decor.size=6}{bm}
					\fmfv{decor.shape=square,decor.filled=empty,label=\tiny{},label.dist=0,decor.size=6}{bb}
				\end{fmfgraph*}
				\quad \;
				\begin{fmfgraph*}(50,30)
					\fmfstraight
					\fmfleft{lb,lm,lt}
					\fmfright{rb,rm,rt}
					\fmf{plain}{lt,bt,rt}
					\fmf{plain}{lb,bb1,v,bb2,rb}
					\fmffreeze
					\fmf{curly}{v,bm,rm}
					\fmfrectangle{8}{6}{lm}
					\fmfv{decor.shape=square,decor.filled=empty,label=\tiny{},label.dist=0,decor.size=6}{bb1}
					\fmfv{decor.shape=square,decor.filled=empty,label=\tiny{},label.dist=0,decor.size=6}{bb2}
					\fmfv{decor.shape=square,decor.filled=empty,label=\tiny{},label.dist=0,decor.size=6}{bm}
					\fmfv{decor.shape=square,decor.filled=empty,label=\tiny{},label.dist=0,decor.size=6}{bt}
				\end{fmfgraph*}
			\end{gathered}
			\quad &= \mathbb{P}_{(ij)}  \frac{- 4\pi \alpha_s \mathbf{T}_i \scdot \mathbf{T}_k}{S_{ij}S_{jk}} 
			\frac{4 z}{1-z} k_\perp \cdot p_k
			[\slashed{p}_i] [\slashed{p}_k] \ , \\[20pt]
			\mathbb{P}_{(ij)} \;
			\begin{gathered}\label{eq:interference-two-perp}
				\begin{fmfgraph*}(50,30)
					\fmfstraight
					\fmfleft{lb,lm,lt}
					\fmfright{rb,rm,rt}
					\fmf{plain}{rt,bt1,v,bt2,lt}
					\fmf{plain}{rb,bb,lb}
					\fmffreeze
					\fmf{curly}{v,bm,lm}
					\fmfrectangle{8}{6}{rm}
					\fmfv{decor.shape=square,decor.filled=empty,label=\tiny{},label.dist=0,decor.size=6}{bt1}
					\fmfv{decor.shape=square,decor.filled=empty,label=\tiny{},label.dist=0,decor.size=6}{bt2}
					\fmfv{decor.shape=square,decor.filled=empty,label=\tiny{$\perp$},label.dist=0,decor.size=6}{bm}
					\fmfv{decor.shape=square,decor.filled=empty,label=\tiny{},label.dist=0,decor.size=6}{bb}
				\end{fmfgraph*}
				\quad \;
				\begin{fmfgraph*}(50,30)
					\fmfstraight
					\fmfleft{lb,lm,lt}
					\fmfright{rb,rm,rt}
					\fmf{plain}{lt,bt,rt}
					\fmf{plain}{lb,bb1,v,bb2,rb}
					\fmffreeze
					\fmf{curly}{v,bm,rm}
					\fmfrectangle{8}{6}{lm}
					\fmfv{decor.shape=square,decor.filled=empty,label=\tiny{},label.dist=0,decor.size=6}{bb1}
					\fmfv{decor.shape=square,decor.filled=empty,label=\tiny{},label.dist=0,decor.size=6}{bb2}
					\fmfv{decor.shape=square,decor.filled=empty,label=\tiny{$\perp$},label.dist=0,decor.size=6}{bm}
					\fmfv{decor.shape=square,decor.filled=empty,label=\tiny{},label.dist=0,decor.size=6}{bt}
				\end{fmfgraph*}
			\end{gathered}
			\quad &=  \mathbb{P}_{(ij)} \frac{4\pi \alpha_s \mathbf{T}_i \scdot \mathbf{T}_k}{S_{ij}S_{jk}}
			\frac{4 z p_\perp^2}{(1-z)^2}  \frac{n \cdot p_k}{n \cdot p_i}	
			[\slashed{p}_i] [\slashed{p}_k] \ .
		\end{align}
	\end{subequations}
\end{fmffile}
Note that one has to include a factor of $\sqrt{z}$ coming from parton
$i$ on the conjugate side.  Inserting the mapping of
\eqref{eq:mapping-alpha-1E} in the partitioned version of
\eqref{eq:eikonal-calc-interference}, \ie
\begin{align*}
	 &\mathbb{P}_{(ij)} \, 8 \pi \alpha_s (-\mathbf{T}_i \scdot \mathbf{T}_k) \frac{4}{S_{ij} S_{jk}}\left[ q_i \scdot q_k - q_k \scdot q_j \frac{q_i \scdot n}{q_j \scdot n} - q_i \scdot q_j \frac{q_k \scdot n}{q_j \scdot n}\right].
\end{align*}
This has the same soft limit as \eqref{eq:interference-qg-1E} (modulo
a factor of two from using two times the real part in
\eqref{eq:eikonal-calc-interference}) This represents a non-trivial
check of our power counting rules.  Also notice the similarity between
\eqref{eq:interference-two-perp} and the soft singular term in
\eqref{eq:one-emission-coll}.  We do not have an immediate
cancellation between both contributions as was the case for
\eqref{eq:eikonal-self-energy}
vs.\ \eqref{eq:eikonal-calc-interference}.  This is due to the fact
that we implement the partitioning before carrying out these
cancellations and because with each of the three terms in the
partitioned Eikonal above containing various powers of soft
contributions, these terms mix and we can not isolate each term via a
power counting.  Nevertheless, when adding up the kernels
$\mathbb{U}_{(ij)}$ and $\mathbb{U}_{(jk)}$,
\eqref{eq:eikonal-calc-interference} is recovered in the soft limit
and the soft divergent parts of the splitting functions are cancelled
against the respective contributions from the interferences (with the
sum over partitioning factors collapsing to 1), leaving only soft
interference contributions of the kind in
\eqref{eq:interference-one-perp}. 

Notice that we can, of course, acquire the same behaviour from a
different momentum parametrization.  In order to show this, it is
interesting to look at a momentum mapping with unbalanced
($\perp$)-component of the emission (in \eqref{eq:mapping-alpha-1E},
these components are balanced between emittee end emission), because
it shows how different contributions in terms of our power counting
algorithm lead to the same results.  This version of the mapping reads
\begin{align}
	q_i^\mu &= z p_i^\mu \ , \nonumber \\
	q_j^\mu &= (1-z)p_i^\mu + \frac{p_\perp^2}{(1-z) 2 p_i \scdot n} n^\mu - k_\perp^\mu \ , \nonumber \\
	q_k^\mu &= p_k^\mu \ .
\end{align}
With this choice, momentum conservation implies the shift of the ($\perp$)-component to the momentum of the splitter line, \ie
\begin{equation}
	q_{I}^\mu = q_i^\mu + q_j^\mu = p_i^\mu + \frac{p_\perp^2}{(1-z) 2 p_i \scdot n} n^\mu - k_\perp^\mu \ .
\end{equation}
This is consistent with the Sudakov decomposition of \eqref{eq:Sudakov-decomp-combined} when 
\begin{equation}
	S_{I} = \frac{z}{1-z} p_\perp^2 
\end{equation}
is inserted.
Now, we need to take into account the ($\perp$)-components in the internal emitter line instead of the external one as compared to \eqref{eq:one-emission-coll} when carrying out the power counting.
Therefore, the splitting function now comes about via
\begin{fmffile}{one-emission-self-energy-decomp-density-kT-emitter}
	\fmfset{thin}{.7pt}
	\fmfset{dot_len}{.8mm}
	\fmfset{dot_size}{5}
	\fmfset{arrow_len}{2.5mm}
	\fmfset{curly_len}{1.5mm}
	\begin{align}
		\;\;\;
		&\begin{gathered}
			\begin{fmfgraph*}(80,50)
				\fmfstraight
				\fmfleft{l1,lm,l2}
				\fmfright{r1,rm,r2}
				\fmf{phantom}{l1,vl,l2}
				\fmf{phantom}{r1,vr,r2}
				\fmf{phantom,tension=8}{vl,vr}
				\fmffreeze
				\fmf{plain}{vl,bl1,v1l,bl2,lm}
				\fmf{plain}{vr,br1,v1r,br2,rm}
				\fmffreeze
				\fmf{curly}{v1l,bgl,l1}
				\fmf{curly}{r1,bgr,v1r}
				\fmfrectangle{6}{5}{vl}
				\fmfrectangle{6}{5}{vr}
				\fmfv{decor.shape=square,decor.filled=empty,decor.angle=0,decor.size=6}{bl2}
				\fmfv{decor.shape=square,decor.filled=empty,decor.angle=0,decor.size=6,label=\tiny{$\perp$},label.dist=0}{bl1}
				\fmfv{decor.shape=square,decor.filled=empty,decor.angle=0,decor.size=6}{br2}
				\fmfv{decor.shape=square,decor.filled=empty,decor.angle=0,decor.size=6,label=\tiny{$\perp$},label.dist=0}{br1}
				\fmfv{decor.shape=square,decor.filled=empty,decor.angle=0,decor.size=6}{bgl}
				\fmfv{decor.shape=square,decor.filled=empty,decor.angle=0,decor.size=6}{bgr}
			\end{fmfgraph*}
		\end{gathered}
		\; + \;
		\begin{gathered}
			\begin{fmfgraph*}(80,50)
				\fmfstraight
				\fmfleft{l1,lm,l2}
				\fmfright{r1,rm,r2}
				\fmf{phantom}{l1,vl,l2}
				\fmf{phantom}{r1,vr,r2}
				\fmf{phantom,tension=8}{vl,vr}
				\fmffreeze
				\fmf{plain}{vl,bl1,v1l,bl2,lm}
				\fmf{plain}{vr,br1,v1r,br2,rm}
				\fmffreeze
				\fmf{curly}{v1l,bgl,l1}
				\fmf{curly}{r1,bgr,v1r}
				\fmfrectangle{6}{5}{vl}
				\fmfrectangle{6}{5}{vr}
				\fmfv{decor.shape=square,decor.filled=empty,decor.angle=0,decor.size=6}{bl1}
				\fmfv{decor.shape=square,decor.filled=empty,decor.angle=0,decor.size=6}{bl2}
				\fmfv{decor.shape=square,decor.filled=empty,decor.angle=0,decor.size=6}{br1}
				\fmfv{decor.shape=square,decor.filled=empty,decor.angle=0,decor.size=6}{br2}
				\fmfv{decor.shape=square,decor.filled=empty,decor.angle=0,decor.size=6,label=\tiny{$\perp$},label.dist=0}{bgl}
				\fmfv{decor.shape=square,decor.filled=empty,decor.angle=0,decor.size=6,label=\tiny{$\perp$},label.dist=0}{bgr}
			\end{fmfgraph*}
		\end{gathered}
		\;+\;
		\begin{gathered}
			\begin{fmfgraph*}(80,50)
				\fmfstraight
				\fmfleft{l1,lm,l2}
				\fmfright{r1,rm,r2}
				\fmf{phantom}{l1,vl,l2}
				\fmf{phantom}{r1,vr,r2}
				\fmf{phantom,tension=8}{vl,vr}
				\fmffreeze
				\fmf{plain}{vl,bl1,v1l,bl2,lm}
				\fmf{plain}{vr,br1,v1r,br2,rm}
				\fmffreeze
				\fmf{curly}{v1l,bgl,l1}
				\fmf{curly}{r1,bgr,v1r}
				\fmfrectangle{6}{5}{vl}
				\fmfrectangle{6}{5}{vr}
				\fmfv{decor.shape=square,decor.filled=empty,decor.angle=0,decor.size=6}{bl2}
				\fmfv{decor.shape=square,decor.filled=empty,decor.angle=0,decor.size=6,label=\tiny{$\perp$},label.dist=0}{bl1}
				\fmfv{decor.shape=square,decor.filled=empty,decor.angle=0,decor.size=6}{br1}
				\fmfv{decor.shape=square,decor.filled=empty,decor.angle=0,decor.size=6}{br2}
				\fmfv{decor.shape=square,decor.filled=empty,decor.angle=0,decor.size=6}{bgl}
				\fmfv{decor.shape=square,decor.filled=empty,decor.angle=0,decor.size=6,label=\tiny{$\perp$},label.dist=0}{bgr}
			\end{fmfgraph*}
		\end{gathered}
		\;+\;
		\begin{gathered}
			\begin{fmfgraph*}(80,50)
				\fmfstraight
				\fmfleft{l1,lm,l2}
				\fmfright{r1,rm,r2}
				\fmf{phantom}{l1,vl,l2}
				\fmf{phantom}{r1,vr,r2}
				\fmf{phantom,tension=8}{vl,vr}
				\fmffreeze
				\fmf{plain}{vl,bl1,v1l,bl2,lm}
				\fmf{plain}{vr,br1,v1r,br2,rm}
				\fmffreeze
				\fmf{curly}{v1l,bgl,l1}
				\fmf{curly}{r1,bgr,v1r}
				\fmfrectangle{6}{5}{vl}
				\fmfrectangle{6}{5}{vr}
				\fmfv{decor.shape=square,decor.filled=empty,decor.angle=0,decor.size=6}{bl1}
				\fmfv{decor.shape=square,decor.filled=empty,decor.angle=0,decor.size=6}{bl2}
				\fmfv{decor.shape=square,decor.filled=empty,decor.angle=0,decor.size=6}{br2}
				\fmfv{decor.shape=square,decor.filled=empty,decor.angle=0,decor.size=6,label=\tiny{$\perp$},label.dist=0}{br1}
				\fmfv{decor.shape=square,decor.filled=empty,decor.angle=0,decor.size=6,label=\tiny{$\perp$},label.dist=0}{bgl}
				\fmfv{decor.shape=square,decor.filled=empty,decor.angle=0,decor.size=6}{bgr}
			\end{fmfgraph*}
		\end{gathered} 
		\nonumber \\
		&\rightarrow \frac{4 \pi \alpha_s \mathbf{T}_i^2}{S_{ij}} \Big[(d-2)(1-z) + \frac{4}{1-z} - 4 \Big]  \slashed{p}_i  \ .
		\label{eq:one-emission-coll-kT-emitter}
	\end{align}
\end{fmffile}
Note how the individual contributions differ from
\eqref{eq:one-emission-coll}, yet the splitting function of
\eqref{eq:Pqg-CS} is reproduced again when combining the terms.  The
interference contributions become
\vspace{5pt}
\begin{fmffile}{qg-exchange-diagrams-2}
	\fmfset{thin}{.7pt}
	\fmfset{dot_len}{1.2mm}
	\fmfset{dot_size}{5}
	\fmfset{arrow_len}{2.5mm}
	\fmfset{curly_len}{1.5mm}
	\begin{subequations}\label{eq:interference-qg-1E-2}
		\begin{align}\label{eq:interference-one-perp-2}
			\mathbb{P}_{(ij)} \;
			\begin{gathered}
				\begin{fmfgraph*}(50,30)
					\fmfstraight
					\fmfleft{lb,lm,lt}
					\fmfright{rb,rm,rt}
					\fmf{plain}{rt,bt1,v,bt2,lt}
					\fmf{plain}{rb,bb,lb}
					\fmffreeze
					\fmf{curly}{v,bm,lm}
					\fmfrectangle{8}{6}{rm}
					\fmfv{decor.shape=square,decor.filled=empty,label=\tiny{},label.dist=0,decor.size=6}{bt1}
					\fmfv{decor.shape=square,decor.filled=empty,label=\tiny{},label.dist=0,decor.size=6}{bt2}
					\fmfv{decor.shape=square,decor.filled=empty,label=\tiny{$\perp$},label.dist=0,decor.size=6}{bm}
					\fmfv{decor.shape=square,decor.filled=empty,label=\tiny{},label.dist=0,decor.size=6}{bb}
				\end{fmfgraph*}
				\quad \;
				\begin{fmfgraph*}(50,30)
					\fmfstraight
					\fmfleft{lb,lm,lt}
					\fmfright{rb,rm,rt}
					\fmf{plain}{lt,bt,rt}
					\fmf{plain}{lb,bb1,v,bb2,rb}
					\fmffreeze
					\fmf{curly}{v,bm,rm}
					\fmfrectangle{8}{6}{lm}
					\fmfv{decor.shape=square,decor.filled=empty,label=\tiny{},label.dist=0,decor.size=6}{bb1}
					\fmfv{decor.shape=square,decor.filled=empty,label=\tiny{},label.dist=0,decor.size=6}{bb2}
					\fmfv{decor.shape=square,decor.filled=empty,label=\tiny{},label.dist=0,decor.size=6}{bm}
					\fmfv{decor.shape=square,decor.filled=empty,label=\tiny{},label.dist=0,decor.size=6}{bt}
				\end{fmfgraph*}
			\end{gathered}
			\quad &=  - \mathbb{P}_{(ij)} \frac{4\pi \alpha_s \mathbf{T}_i \scdot \mathbf{T}_k}{S_{ij} S_{jk}} 
			\frac{4 z}{1-z} k_\perp \cdot p_k
			[\slashed{p}_i] [\slashed{p}_k] \ , \\[20pt]
			\mathbb{P}_{(ij)} \;
			\begin{gathered}\label{eq:interference-two-perp-2}
				\begin{fmfgraph*}(50,30)
					\fmfstraight
					\fmfleft{lb,lm,lt}
					\fmfright{rb,rm,rt}
					\fmf{plain}{rt,bt1,v,bt2,lt}
					\fmf{plain}{rb,bb,lb}
					\fmffreeze
					\fmf{curly}{v,bm,lm}
					\fmfrectangle{8}{6}{rm}
					\fmfv{decor.shape=square,decor.filled=empty,label=\tiny{},label.dist=0,decor.size=6}{bt1}
					\fmfv{decor.shape=square,decor.filled=empty,label=\tiny{},label.dist=0,decor.size=6}{bt2}
					\fmfv{decor.shape=square,decor.filled=empty,label=\tiny{$\perp$},label.dist=0,decor.size=6}{bm}
					\fmfv{decor.shape=square,decor.filled=empty,label=\tiny{},label.dist=0,decor.size=6}{bb}
				\end{fmfgraph*}
				\quad \;
				\begin{fmfgraph*}(50,30)
					\fmfstraight
					\fmfleft{lb,lm,lt}
					\fmfright{rb,rm,rt}
					\fmf{plain}{lt,bt,rt}
					\fmf{plain}{lb,bb1,v,bb2,rb}
					\fmffreeze
					\fmf{curly}{v,bm,rm}
					\fmfrectangle{8}{6}{lm}
					\fmfv{decor.shape=square,decor.filled=empty,label=\tiny{},label.dist=0,decor.size=6}{bb1}
					\fmfv{decor.shape=square,decor.filled=empty,label=\tiny{},label.dist=0,decor.size=6}{bb2}
					\fmfv{decor.shape=square,decor.filled=empty,label=\tiny{$\perp$},label.dist=0,decor.size=6}{bm}
					\fmfv{decor.shape=square,decor.filled=empty,label=\tiny{},label.dist=0,decor.size=6}{bt}
				\end{fmfgraph*}
			\end{gathered}
			\quad &=  \mathbb{P}_{(ij)}  \frac{4\pi \alpha_s \mathbf{T}_i \scdot \mathbf{T}_k}{S_{ij}S_{jk}}
			\frac{4 p_\perp^2}{z(1-z)^2}  \frac{n \cdot p_k}{n \cdot p_i}	
			[\slashed{p}_i] [\slashed{p}_k] \ .
			\vspace{5pt}
		\end{align}
	\end{subequations}
\end{fmffile}
The Eikonal contribution of \eqref{eq:interference-one-perp-2}
coincides with the one from before while the soft-singular one
differs.  Nevertheless, we find the same correspondence of
soft-singular terms between interference- and self-energy like
contributions when comparing \eqref{eq:one-emission-coll-kT-emitter}
to \eqref{eq:interference-two-perp-2}.

\subsubsection{Gluon-gluon splitting}
The same argumentation holds for the emission from hard gluon.  In
this case, one reproduces \eqref{eq:eikonal-self-energy} and
\eqref{eq:eikonal-calc-interference} by using the vertex rule
\begin{fmffile}{soft-ggg-vertex}
	\fmfset{thin}{.7pt}
	\fmfset{arrow_len}{2.5mm}
	\fmfset{curly_len}{1.5mm}
	\begin{align}\label{eq:soft-ggg-vertex}
		\begin{gathered}
			\begin{fmfgraph*}(50,40)
				\fmfleft{l}
				\fmfright{rb,rm,rt}
				\fmfbottom{b}
				\fmf{curly}{l,v,rm}
				\fmffreeze
				\fmf{phantom,label=\tiny{$q_i$},label.side=left}{v,rm}
				\fmf{curly,label=\tiny{$q_j$},label.side=right}{v,b}
				\fmfv{label=\tiny{$\mu$},label.dist=5}{rm}
				\fmfv{label=\tiny{$\nu$},label.dist=5}{l}
				\fmfv{label=\tiny{$\rho$},label.dist=5}{b}
				\fmfv{decor.shape=circle,decor.filled=full,decor.size=2}{v}
			\end{fmfgraph*}
		\end{gathered}
	\quad
	\propto \frac{q_i^\rho}{q_i \scdot q_j} \eta^{ \mu \nu} +\mathcal{O}\left(1\right),
	\end{align}
\end{fmffile}%
and by realizing that longitudinal contributions proportional to
$q_i^\mu$ vanish due to gauge invariance of the underlying hard
amplitude \cite{PhysRevD.55.6819}.  Then, we again need to show that
our power-counting rules reproduce this soft behaviour leading to the
same cancellations between interference and self-energy like
contributions as in the quark case.  One complication arises here due
to the three-gluon vertex carrying a momentum dependence which we need
to account for using \eqref{eq:vertex-rules}.  The respective rules
for a splitter line are
\vspace{5pt}
\begin{fmffile}{one-emission-amps-ggg}
	\fmfset{thin}{.7pt}
	\fmfset{dot_len}{1.2mm}
	\fmfset{dot_size}{5}
	\fmfset{arrow_len}{2.5mm}
	\fmfset{curly_len}{1.5mm}
	\begin{subequations}
		\begin{align}
			\begin{gathered}
				\begin{fmfgraph*}(60,30)
					\fmfstraight
					\fmfleft{l2,lm,l1}
					\fmfright{r2,rm,r1}
					\fmf{phantom}{l2,m2,r2}
					\fmffreeze
					\fmf{curly}{l1,bl,mm,br,r1}
					\fmffreeze
					\fmf{curly}{mm,bm,m2}
					\fmffreeze
					\fmfrectangle{4}{4}{r1}
					\fmfv{decor.shape=circle,decor.filled=empty,decor.size=12,label=\tiny{$\parallel$},label.dist=0}{mm}
					\fmfv{decor.shape=square,decor.filled=empty,decor.angle=0,decor.size=6}{bm}
					\fmfv{decor.shape=square,decor.filled=empty,decor.angle=0,decor.size=6}{br}
					\fmfv{decor.shape=square,decor.filled=empty,decor.angle=0,decor.size=6,label=\tiny{$\perp$},label.dist=0}{bl}
					\fmfv{label=\tiny{$\nu$},label.dist=5,label.angle=180}{l1}
					\fmfv{label=\tiny{$\rho$},label.dist=5,label.angle=-90}{m2}
					\fmfv{label=\tiny{$\mu$},label.dist=5,label.angle=90}{br}
				\end{fmfgraph*}
			\end{gathered}
			\quad &= 
			\frac{g_s \mathbf{T}_i}{S_{ij}} \, \frac{z_i+2z_j}{z_i} \left[ - \eta^{\mu \rho} + \frac{n^\mu p_i^\rho}{p_i \scdot n} 	 
			\right] k_\perp^\nu \ ,
			\\[20pt]
			\begin{gathered}
				\begin{fmfgraph*}(60,30)
					\fmfstraight
					\fmfleft{l2,lm,l1}
					\fmfright{r2,rm,r1}
					\fmf{phantom}{l2,m2,r2}
					\fmffreeze
					\fmf{curly}{l1,bl,mm,br,r1}
					\fmffreeze
					\fmf{curly}{mm,bm,m2}
					\fmffreeze
					\fmfrectangle{4}{4}{r1}
					\fmfv{decor.shape=circle,decor.filled=empty,decor.size=12,label=\tiny{$\parallel$},label.dist=0}{mm}
					\fmfv{decor.shape=square,decor.filled=empty,decor.angle=0,decor.size=6}{bl}
					\fmfv{decor.shape=square,decor.filled=empty,decor.angle=0,decor.size=6}{br}
					\fmfv{decor.shape=square,decor.filled=empty,decor.angle=0,decor.size=6,label=\tiny{$\perp$},label.dist=0}{bm}
					\fmfv{label=\tiny{$\nu$},label.dist=5,label.angle=180}{l1}
					\fmfv{label=\tiny{$\rho$},label.dist=5,label.angle=-90}{m2}
					\fmfv{label=\tiny{$\mu$},label.dist=5,label.angle=90}{br}
				\end{fmfgraph*}
			\end{gathered}
			\quad &=
			\frac{g_s \mathbf{T}_i}{S_{ij}} \, \frac{2z_i + z_j}{z_j} 
			\left[
			\eta^{\mu\nu} - \frac{n^\mu p_i^\nu}{p_i \scdot n} 
			\right] k_\perp^\rho \ ,
			\\[20pt]
			\begin{gathered}
				\begin{fmfgraph*}(60,30)
					\fmfstraight
					\fmfleft{l2,lm,l1}
					\fmfright{r2,rm,r1}
					\fmf{phantom}{l2,m2,r2}
					\fmffreeze
					\fmf{curly}{l1,bl,mm,br,r1}
					\fmffreeze
					\fmf{curly}{mm,bm,m2}
					\fmffreeze
					\fmfrectangle{4}{4}{r1}
					\fmfv{decor.shape=circle,decor.filled=empty,decor.size=12,label=\tiny{$\perp$},label.dist=0}{mm}
					\fmfv{decor.shape=square,decor.filled=empty,decor.angle=0,decor.size=6}{bl}
					\fmfv{decor.shape=square,decor.filled=empty,decor.angle=0,decor.size=6}{br}
					\fmfv{decor.shape=square,decor.filled=empty,decor.angle=0,decor.size=6,label=\tiny{},label.dist=0}{bm}
					\fmfv{label=\tiny{$\nu$},label.dist=5,label.angle=180}{l1}
					\fmfv{label=\tiny{$\rho$},label.dist=5,label.angle=-90}{m2}
					\fmfv{label=\tiny{$\mu$},label.dist=5,label.angle=90}{br}
				\end{fmfgraph*}
			\end{gathered}
			\quad& = 
			\frac{g \mathbf{T}_i }{S_{ij}} \Bigg\{ 
			\eta^{\nu \rho} (k_{\perp,i}^\mu -k_{\perp,j}^\mu) 
			+ \eta^{\mu \rho} (k_{\perp,i}^\nu + 2 k_{\perp,j}^\nu)
			- \eta^{\mu \nu} (2 k_{\perp,i}^\rho - k_{\perp,j}^\rho) \\
			& \hphantom{ = \frac{g \mathbf{T}_i }{S_{ij}} \Bigg\{ }
			\begin{aligned}
				- \frac{1}{n \scdot p_i}\Big[
				&(2k_{\perp,i}^\mu - k_{\perp,j}^\mu) (n^\nu p_i^\rho + n^\rho p_i^\nu)
				\\
				- &( k_{\perp,i}^\nu -2k_{\perp,j}^\nu) (n^\mu p_i^\rho + n^\rho p_i^\mu)
				\\
				- &(2k_{\perp,i}^\rho- k_{\perp,j}^\rho) (n^\mu p_i^\nu + n^\nu p_i^\mu)
				\Big]
				\Bigg\} \ .
			\end{aligned}
		\end{align}
	\end{subequations}
\end{fmffile}
The rules for interferer vertices are
\vspace{5pt}
\begin{fmffile}{one-emission-amps-ggg-interferer}
	\fmfset{thin}{.7pt}
	\fmfset{dot_len}{1.2mm}
	\fmfset{dot_size}{5}
	\fmfset{arrow_len}{2.5mm}
	\fmfset{curly_len}{1.5mm}
	\begin{subequations}
		\begin{align}
			\begin{gathered}
				\begin{fmfgraph*}(60,30)
					\fmfstraight
					\fmfleft{r1,lm,r2}
					\fmfright{l1,rm,l2}
					\fmf{phantom}{l2,m2,r2}
					\fmffreeze
					\fmf{curly}{l1,bl,mm,br,r1}
					\fmffreeze
					\fmf{curly}{mm,bm,m2}
					\fmffreeze
					\fmfrectangle{4}{4}{r1}
					\fmfv{decor.shape=circle,decor.filled=empty,decor.size=12,label=\tiny{$\parallel$},label.dist=0}{mm}
					\fmfv{decor.shape=square,decor.filled=empty,decor.angle=0,decor.size=6}{bm}
					\fmfv{decor.shape=square,decor.filled=empty,decor.angle=0,decor.size=6}{br}
					\fmfv{decor.shape=square,decor.filled=empty,decor.angle=0,decor.size=6}{bl}
					\fmfv{label=\tiny{$\mu$},label.dist=5,label.angle=90}{br}
					\fmfv{label=\tiny{$\nu$},label.dist=5,label.angle=0}{l1}
					\fmfv{label=\tiny{$\rho$},label.dist=5,label.angle=0}{m2}
				\end{fmfgraph*}
			\end{gathered}
			\quad &= 
			\frac{2 g_s \mathbf{T}_k}{S_{jk}} \eta^{\mu \nu} \frac{z_k n\scdot p_k}{z_j n\scdot p_i + z_k n\scdot p_k}\left(p_k^\rho - \frac{p_i \scdot p_k \, n^\rho  + n\scdot p_k \, p_i^\rho}{n \scdot p_i} \right) \ ,
			\\[20pt]
			\begin{gathered}
				\begin{fmfgraph*}(60,30)
					\fmfstraight
					\fmfleft{r1,lm,r2}
					\fmfright{l1,rm,l2}
					\fmf{phantom}{l2,m2,r2}
					\fmffreeze
					\fmf{curly}{l1,bl,mm,br,r1}
					\fmffreeze
					\fmf{curly}{mm,bm,m2}
					\fmffreeze
					\fmfrectangle{4}{4}{r1}
					\fmfv{decor.shape=circle,decor.filled=empty,decor.size=12,label=\tiny{$\parallel$},label.dist=0}{mm}
					\fmfv{decor.shape=square,decor.filled=empty,decor.angle=0,decor.size=6,label=\tiny{$\perp$},label.dist=0}{bm}
					\fmfv{decor.shape=square,decor.filled=empty,decor.angle=0,decor.size=6}{br}
					\fmfv{decor.shape=square,decor.filled=empty,decor.angle=0,decor.size=6}{bl}
					\fmfv{label=\tiny{$\mu$},label.dist=5,label.angle=90}{br}
					\fmfv{label=\tiny{$\nu$},label.dist=5,label.angle=0}{l1}
					\fmfv{label=\tiny{$\rho$},label.dist=5,label.angle=0}{m2}
				\end{fmfgraph*}
			\end{gathered}
			\quad &= 
			\frac{2 g_s \mathbf{T}_k}{S_{jk}}\frac{z_k n\scdot p_k}{z_j (z_j n\scdot p_i + z_k n\scdot p_k)}\frac{1}{n\scdot p_i} 
			\left(
			-\eta^{\mu\nu} n \scdot p_k + p_k^\mu n^\nu + p_k^\nu n^\mu
			\right)k_{\perp,j}^\rho \ .
			\vspace{5pt}
		\end{align}
	\end{subequations}
\end{fmffile}
Note that these vertices are only relevant for interferer-like
amplitudes and therefore only contribute in soft limits.  This is why
we include the soft partitioning factors and vertex decomposition of
\eqref{eqs:linearity} and \eqref{eq:linearity-three-gluon-vertex} here
which greatly simplifies the vertex rule.  Note also that in the
strict soft limit, $z_j \sim \mathcal{O}(\lambda)$ while $z_k = 1$ and
the soft partitioning factor tends to unity.

Then, after applying \eqref{eq:to-z-mapping} again, the splitting
function $P_{gg}$ comes about via
\begin{fmffile}{one-emission-gluon-self-energy-decomp-density}
	\fmfset{thin}{.7pt}
	\fmfset{dot_len}{.8mm}
	\fmfset{dot_size}{5}
	\fmfset{arrow_len}{2.5mm}
	\fmfset{curly_len}{1.5mm}
	\begin{align}
		\;\;\;
		&\begin{gathered}
			\begin{fmfgraph*}(120,50)
				\fmfstraight
				\fmfleft{l1,lm,l2}
				\fmfright{r1,rm,r2}
				\fmf{phantom}{l1,vl,l2}
				\fmf{phantom}{r1,vr,r2}
				\fmf{phantom,tension=12}{vl,vr}
				\fmffreeze
				\fmf{curly}{lm,bl2,v1l,bl1,vl}
				\fmf{curly}{vr,br1,v1r,br2,rm}
				\fmffreeze
				\fmf{curly}{v1l,bgl,l1}
				\fmf{curly}{r1,bgr,v1r}
				\fmfrectangle{6}{5}{vl}
				\fmfrectangle{6}{5}{vr}
				\fmfv{decor.shape=square,decor.filled=empty,decor.angle=0,decor.size=6}{bl1}
				\fmfv{decor.shape=square,decor.filled=empty,decor.angle=0,decor.size=6,label=\tiny{},label.dist=0}{bl2}
				\fmfv{decor.shape=square,decor.filled=empty,decor.angle=0,decor.size=6}{br1}
				\fmfv{decor.shape=square,decor.filled=empty,decor.angle=0,decor.size=6,label=\tiny{},label.dist=0}{br2}
				\fmfv{decor.shape=square,decor.filled=empty,decor.angle=0,decor.size=6}{bgl}
				\fmfv{decor.shape=square,decor.filled=empty,decor.angle=0,decor.size=6}{bgr}
				\fmfv{decor.shape=circle,decor.filled=empty,decor.size=12,label=\tiny{$\perp$},label.dist=0}{v1l}
				\fmfv{decor.shape=circle,decor.filled=empty,decor.size=12,label=\tiny{$\perp$},label.dist=0}{v1r}
			\end{fmfgraph*}
		\end{gathered}
		\nonumber \\
		\; + \;
		&\begin{gathered}
			\begin{fmfgraph*}(120,50)
				\fmfstraight
				\fmfleft{l1,lm,l2}
				\fmfright{r1,rm,r2}
				\fmf{phantom}{l1,vl,l2}
				\fmf{phantom}{r1,vr,r2}
				\fmf{phantom,tension=12}{vl,vr}
				\fmffreeze
				\fmf{curly}{lm,bl2,v1l,bl1,vl}
				\fmf{curly}{vr,br1,v1r,br2,rm}
				\fmffreeze
				\fmf{curly}{v1l,bgl,l1}
				\fmf{curly}{r1,bgr,v1r}
				\fmfrectangle{6}{5}{vl}
				\fmfrectangle{6}{5}{vr}
				\fmfv{decor.shape=square,decor.filled=empty,decor.angle=0,decor.size=6}{bl1}
				\fmfv{decor.shape=square,decor.filled=empty,decor.angle=0,decor.size=6,label=\tiny{$\perp$},label.dist=0}{bl2}
				\fmfv{decor.shape=square,decor.filled=empty,decor.angle=0,decor.size=6}{br1}
				\fmfv{decor.shape=square,decor.filled=empty,decor.angle=0,decor.size=6,label=\tiny{},label.dist=0}{br2}
				\fmfv{decor.shape=square,decor.filled=empty,decor.angle=0,decor.size=6}{bgl}
				\fmfv{decor.shape=square,decor.filled=empty,decor.angle=0,decor.size=6,label=\tiny{$\perp$},label.dist=0}{bgr}
				\fmfv{decor.shape=circle,decor.filled=empty,decor.size=12,label=\tiny{$\parallel$},label.dist=0}{v1l}
				\fmfv{decor.shape=circle,decor.filled=empty,decor.size=12,label=\tiny{$\parallel$},label.dist=0}{v1r}
			\end{fmfgraph*}
		\end{gathered}
		\; + \;
		\begin{gathered}
			\begin{fmfgraph*}(120,50)
				\fmfstraight
				\fmfleft{l1,lm,l2}
				\fmfright{r1,rm,r2}
				\fmf{phantom}{l1,vl,l2}
				\fmf{phantom}{r1,vr,r2}
				\fmf{phantom,tension=12}{vl,vr}
				\fmffreeze
				\fmf{curly}{lm,bl2,v1l,bl1,vl}
				\fmf{curly}{vr,br1,v1r,br2,rm}
				\fmffreeze
				\fmf{curly}{v1l,bgl,l1}
				\fmf{curly}{r1,bgr,v1r}
				\fmfrectangle{6}{5}{vl}
				\fmfrectangle{6}{5}{vr}
				\fmfv{decor.shape=square,decor.filled=empty,decor.angle=0,decor.size=6}{bl1}
				\fmfv{decor.shape=square,decor.filled=empty,decor.angle=0,decor.size=6,label=\tiny{},label.dist=0}{bl2}
				\fmfv{decor.shape=square,decor.filled=empty,decor.angle=0,decor.size=6}{br1}
				\fmfv{decor.shape=square,decor.filled=empty,decor.angle=0,decor.size=6,label=\tiny{$\perp$},label.dist=0}{br2}
				\fmfv{decor.shape=square,decor.filled=empty,decor.angle=0,decor.size=6,label=\tiny{$\perp$},label.dist=0}{bgl}
				\fmfv{decor.shape=square,decor.filled=empty,decor.angle=0,decor.size=6}{bgr}
				\fmfv{decor.shape=circle,decor.filled=empty,decor.size=12,label=\tiny{$\parallel$},label.dist=0}{v1l}
				\fmfv{decor.shape=circle,decor.filled=empty,decor.size=12,label=\tiny{$\parallel$},label.dist=0}{v1r}
			\end{fmfgraph*}
		\end{gathered}
		\nonumber \\
		+ \; 
		&\begin{gathered}
			\begin{fmfgraph*}(120,50)
				\fmfstraight
				\fmfleft{l1,lm,l2}
				\fmfright{r1,rm,r2}
				\fmf{phantom}{l1,vl,l2}
				\fmf{phantom}{r1,vr,r2}
				\fmf{phantom,tension=12}{vl,vr}
				\fmffreeze
				\fmf{curly}{lm,bl2,v1l,bl1,vl}
				\fmf{curly}{vr,br1,v1r,br2,rm}
				\fmffreeze
				\fmf{curly}{v1l,bgl,l1}
				\fmf{curly}{r1,bgr,v1r}
				\fmfrectangle{6}{5}{vl}
				\fmfrectangle{6}{5}{vr}
				\fmfv{decor.shape=square,decor.filled=empty,decor.angle=0,decor.size=6}{bl1}
				\fmfv{decor.shape=square,decor.filled=empty,decor.angle=0,decor.size=6,label=\tiny{$\perp$},label.dist=0}{bl2}
				\fmfv{decor.shape=square,decor.filled=empty,decor.angle=0,decor.size=6}{br1}
				\fmfv{decor.shape=square,decor.filled=empty,decor.angle=0,decor.size=6,label=\tiny{$\perp$},label.dist=0}{br2}
				\fmfv{decor.shape=square,decor.filled=empty,decor.angle=0,decor.size=6}{bgl}
				\fmfv{decor.shape=square,decor.filled=empty,decor.angle=0,decor.size=6,label=\tiny{},label.dist=0}{bgr}
				\fmfv{decor.shape=circle,decor.filled=empty,decor.size=12,label=\tiny{$\parallel$},label.dist=0}{v1l}
				\fmfv{decor.shape=circle,decor.filled=empty,decor.size=12,label=\tiny{$\parallel$},label.dist=0}{v1r}
			\end{fmfgraph*}
		\end{gathered}
		\; + \;
		\begin{gathered}
			\begin{fmfgraph*}(120,50)
				\fmfstraight
				\fmfleft{l1,lm,l2}
				\fmfright{r1,rm,r2}
				\fmf{phantom}{l1,vl,l2}
				\fmf{phantom}{r1,vr,r2}
				\fmf{phantom,tension=12}{vl,vr}
				\fmffreeze
				\fmf{curly}{lm,bl2,v1l,bl1,vl}
				\fmf{curly}{vr,br1,v1r,br2,rm}
				\fmffreeze
				\fmf{curly}{v1l,bgl,l1}
				\fmf{curly}{r1,bgr,v1r}
				\fmfrectangle{6}{5}{vl}
				\fmfrectangle{6}{5}{vr}
				\fmfv{decor.shape=square,decor.filled=empty,decor.angle=0,decor.size=6}{bl1}
				\fmfv{decor.shape=square,decor.filled=empty,decor.angle=0,decor.size=6,label=\tiny{},label.dist=0}{bl2}
				\fmfv{decor.shape=square,decor.filled=empty,decor.angle=0,decor.size=6}{br1}
				\fmfv{decor.shape=square,decor.filled=empty,decor.angle=0,decor.size=6,label=\tiny{},label.dist=0}{br2}
				\fmfv{decor.shape=square,decor.filled=empty,decor.angle=0,decor.size=6,label=\tiny{$\perp$},label.dist=0}{bgl}
				\fmfv{decor.shape=square,decor.filled=empty,decor.angle=0,decor.size=6,label=\tiny{$\perp$},label.dist=0}{bgr}
				\fmfv{decor.shape=circle,decor.filled=empty,decor.size=12,label=\tiny{$\parallel$},label.dist=0}{v1l}
				\fmfv{decor.shape=circle,decor.filled=empty,decor.size=12,label=\tiny{$\parallel$},label.dist=0}{v1r}
			\end{fmfgraph*}
		\end{gathered}
		\nonumber \\
		\; + \; \Bigg( \;
		&\begin{gathered}
			\begin{fmfgraph*}(120,50)
				\fmfstraight
				\fmfleft{l1,lm,l2}
				\fmfright{r1,rm,r2}
				\fmf{phantom}{l1,vl,l2}
				\fmf{phantom}{r1,vr,r2}
				\fmf{phantom,tension=12}{vl,vr}
				\fmffreeze
				\fmf{curly}{lm,bl2,v1l,bl1,vl}
				\fmf{curly}{vr,br1,v1r,br2,rm}
				\fmffreeze
				\fmf{curly}{v1l,bgl,l1}
				\fmf{curly}{r1,bgr,v1r}
				\fmfrectangle{6}{5}{vl}
				\fmfrectangle{6}{5}{vr}
				\fmfv{decor.shape=square,decor.filled=empty,decor.angle=0,decor.size=6}{bl1}
				\fmfv{decor.shape=square,decor.filled=empty,decor.angle=0,decor.size=6,label=\tiny{$\perp$},label.dist=0}{bl2}
				\fmfv{decor.shape=square,decor.filled=empty,decor.angle=0,decor.size=6}{br1}
				\fmfv{decor.shape=square,decor.filled=empty,decor.angle=0,decor.size=6,label=\tiny{},label.dist=0}{br2}
				\fmfv{decor.shape=square,decor.filled=empty,decor.angle=0,decor.size=6,label=\tiny{},label.dist=0}{bgl}
				\fmfv{decor.shape=square,decor.filled=empty,decor.angle=0,decor.size=6,label=\tiny{},label.dist=0}{bgr}
				\fmfv{decor.shape=circle,decor.filled=empty,decor.size=12,label=\tiny{$\parallel$},label.dist=0}{v1l}
				\fmfv{decor.shape=circle,decor.filled=empty,decor.size=12,label=\tiny{$\perp$},label.dist=0}{v1r}
			\end{fmfgraph*}
		\end{gathered}
		\; + \;
		\begin{gathered}
			\begin{fmfgraph*}(120,50)
				\fmfstraight
				\fmfleft{l1,lm,l2}
				\fmfright{r1,rm,r2}
				\fmf{phantom}{l1,vl,l2}
				\fmf{phantom}{r1,vr,r2}
				\fmf{phantom,tension=12}{vl,vr}
				\fmffreeze
				\fmf{curly}{lm,bl2,v1l,bl1,vl}
				\fmf{curly}{vr,br1,v1r,br2,rm}
				\fmffreeze
				\fmf{curly}{v1l,bgl,l1}
				\fmf{curly}{r1,bgr,v1r}
				\fmfrectangle{6}{5}{vl}
				\fmfrectangle{6}{5}{vr}
				\fmfv{decor.shape=square,decor.filled=empty,decor.angle=0,decor.size=6}{bl1}
				\fmfv{decor.shape=square,decor.filled=empty,decor.angle=0,decor.size=6,label=\tiny{},label.dist=0}{bl2}
				\fmfv{decor.shape=square,decor.filled=empty,decor.angle=0,decor.size=6}{br1}
				\fmfv{decor.shape=square,decor.filled=empty,decor.angle=0,decor.size=6,label=\tiny{$\perp$},label.dist=0}{br2}
				\fmfv{decor.shape=square,decor.filled=empty,decor.angle=0,decor.size=6,label=\tiny{},label.dist=0}{bgl}
				\fmfv{decor.shape=square,decor.filled=empty,decor.angle=0,decor.size=6,label=\tiny{},label.dist=0}{bgr}
				\fmfv{decor.shape=circle,decor.filled=empty,decor.size=12,label=\tiny{$\parallel$},label.dist=0}{v1l}
				\fmfv{decor.shape=circle,decor.filled=empty,decor.size=12,label=\tiny{$\perp$},label.dist=0}{v1r}
			\end{fmfgraph*}
		\end{gathered}
		\; +\mathrm{c.c.} \Bigg)
		\nonumber \\
		\; + \; \Bigg( \;
		&\begin{gathered}
			\begin{fmfgraph*}(120,50)
				\fmfstraight
				\fmfleft{l1,lm,l2}
				\fmfright{r1,rm,r2}
				\fmf{phantom}{l1,vl,l2}
				\fmf{phantom}{r1,vr,r2}
				\fmf{phantom,tension=12}{vl,vr}
				\fmffreeze
				\fmf{curly}{lm,bl2,v1l,bl1,vl}
				\fmf{curly}{vr,br1,v1r,br2,rm}
				\fmffreeze
				\fmf{curly}{v1l,bgl,l1}
				\fmf{curly}{r1,bgr,v1r}
				\fmfrectangle{6}{5}{vl}
				\fmfrectangle{6}{5}{vr}
				\fmfv{decor.shape=square,decor.filled=empty,decor.angle=0,decor.size=6}{bl1}
				\fmfv{decor.shape=square,decor.filled=empty,decor.angle=0,decor.size=6,label=\tiny{},label.dist=0}{bl2}
				\fmfv{decor.shape=square,decor.filled=empty,decor.angle=0,decor.size=6}{br1}
				\fmfv{decor.shape=square,decor.filled=empty,decor.angle=0,decor.size=6,label=\tiny{},label.dist=0}{br2}
				\fmfv{decor.shape=square,decor.filled=empty,decor.angle=0,decor.size=6,label=\tiny{$\perp$},label.dist=0}{bgl}
				\fmfv{decor.shape=square,decor.filled=empty,decor.angle=0,decor.size=6,label=\tiny{},label.dist=0}{bgr}
				\fmfv{decor.shape=circle,decor.filled=empty,decor.size=12,label=\tiny{$\parallel$},label.dist=0}{v1l}
				\fmfv{decor.shape=circle,decor.filled=empty,decor.size=12,label=\tiny{$\perp$},label.dist=0}{v1r}
			\end{fmfgraph*}
		\end{gathered}
		\; + \;
		\begin{gathered}
			\begin{fmfgraph*}(120,50)
				\fmfstraight
				\fmfleft{l1,lm,l2}
				\fmfright{r1,rm,r2}
				\fmf{phantom}{l1,vl,l2}
				\fmf{phantom}{r1,vr,r2}
				\fmf{phantom,tension=12}{vl,vr}
				\fmffreeze
				\fmf{curly}{lm,bl2,v1l,bl1,vl}
				\fmf{curly}{vr,br1,v1r,br2,rm}
				\fmffreeze
				\fmf{curly}{v1l,bgl,l1}
				\fmf{curly}{r1,bgr,v1r}
				\fmfrectangle{6}{5}{vl}
				\fmfrectangle{6}{5}{vr}
				\fmfv{decor.shape=square,decor.filled=empty,decor.angle=0,decor.size=6}{bl1}
				\fmfv{decor.shape=square,decor.filled=empty,decor.angle=0,decor.size=6,label=\tiny{},label.dist=0}{bl2}
				\fmfv{decor.shape=square,decor.filled=empty,decor.angle=0,decor.size=6}{br1}
				\fmfv{decor.shape=square,decor.filled=empty,decor.angle=0,decor.size=6,label=\tiny{},label.dist=0}{br2}
				\fmfv{decor.shape=square,decor.filled=empty,decor.angle=0,decor.size=6,label=\tiny{},label.dist=0}{bgl}
				\fmfv{decor.shape=square,decor.filled=empty,decor.angle=0,decor.size=6,label=\tiny{$\perp$},label.dist=0}{bgr}
				\fmfv{decor.shape=circle,decor.filled=empty,decor.size=12,label=\tiny{$\parallel$},label.dist=0}{v1l}
				\fmfv{decor.shape=circle,decor.filled=empty,decor.size=12,label=\tiny{$\perp$},label.dist=0}{v1r}
			\end{fmfgraph*}
		\end{gathered} +\mathrm{c.c.} \Bigg)
		\nonumber \\
		\label{eq:one-emission-coll-gluon}
		&= \frac{8 \pi \alpha_s \mathbf{T}_i^2}{S_{ij}} \Big[ - 2 \,\eta^{\mu \nu}  \left(\frac{z}{1-z} + \frac{1-z}{z}\right)
		+ (d-2) \, z (1-z) \frac{k_\perp^\mu k_\perp^\nu}{p_\perp^2}	\Big].
	\end{align}
\end{fmffile}
The first term of this expression shows the soft-divergent part of
$P_{gg}$.  Turning to the interferences, we have fewer diagrams to
take care of because the $(\perp)$-vertices give soft-subleading
contributions.  The Eikonal contribution is then given by
\vspace{10pt}
\begin{fmffile}{gg-exchange-diagrams-eikonal}
	\fmfset{thin}{.7pt}
	\fmfset{dot_len}{1.2mm}
	\fmfset{dot_size}{5}
	\fmfset{arrow_len}{2.5mm}
	\fmfset{curly_len}{1.5mm}
	\begin{align}
		& \mathbb{P}_{(ij)} \;
		\begin{gathered}
			\begin{fmfgraph*}(60,40)
				\fmfstraight
				\fmfleft{lb,lm,lt}
				\fmfright{rb,rm,rt}
				\fmf{curly}{rt,bt1,v,bt2,lt}
				\fmf{curly}{rb,bb,lb}
				\fmffreeze
				\fmf{curly}{v,bm,lm}
				\fmfrectangle{10}{6}{rm}
				\fmfv{decor.shape=circle,decor.filled=empty,decor.size=10,label=\tiny{$\parallel$},label.dist=0}{v}
				\fmfv{decor.shape=square,decor.filled=empty,label=\tiny{},label.dist=0,decor.size=6}{bt1}
				\fmfv{decor.shape=square,decor.filled=empty,label=\tiny{},label.dist=0,decor.size=6}{bt2}
				\fmfv{decor.shape=square,decor.filled=empty,label=\tiny{$\perp$},label.dist=0,decor.size=6}{bm}
				\fmfv{decor.shape=square,decor.filled=empty,label=\tiny{},label.dist=0,decor.size=6}{bb}
			\end{fmfgraph*}
			\quad \;
			\begin{fmfgraph*}(60,40)
				\fmfstraight
				\fmfleft{lb,lm,lt}
				\fmfright{rb,rm,rt}
				\fmf{curly}{lt,bt,rt}
				\fmf{curly}{lb,bb1,v,bb2,rb}
				\fmffreeze
				\fmf{curly}{v,bm,rm}
				\fmfrectangle{10}{6}{lm}
				\fmfv{decor.shape=circle,decor.filled=empty,decor.size=10,label=\tiny{$\parallel$},label.dist=0}{v}
				\fmfv{decor.shape=square,decor.filled=empty,label=\tiny{},label.dist=0,decor.size=6}{bb1}
				\fmfv{decor.shape=square,decor.filled=empty,label=\tiny{},label.dist=0,decor.size=6}{bb2}
				\fmfv{decor.shape=square,decor.filled=empty,label=\tiny{},label.dist=0,decor.size=6}{bm}
				\fmfv{decor.shape=square,decor.filled=empty,label=\tiny{},label.dist=0,decor.size=6}{bt}
			\end{fmfgraph*}
		\end{gathered}
		\nonumber \\[20pt]
		& \qquad = \mathbb{P}_{(ij)}  \frac{8 \pi \alpha_s \mathbf{T}_i\scdot\mathbf{T}_k}{S_{ij} S_{jk}} \,
		\frac{1+z}{1-z}
		\frac{ n\scdot p_k}{n\scdot p_k + (1-z) n\scdot p_i } \, p_k \scdot k_\perp  \, \eta^{\mu_i \nu_i} \eta^{\mu_k \nu_k} \ .
		\label{eq:1E-eikonal}
	\end{align}
\end{fmffile}
Here, we have used the fact that the hard amplitudes are transverse
meaning that terms with a $p^{\mu_i/\nu_i}$ and $p^{\mu_k/\nu_k}$
vanish.  By the same logic, the leading-soft divergent part is given
by
\vspace{10pt}
\begin{fmffile}{gg-exchange-diagrams}
	\fmfset{thin}{.7pt}
	\fmfset{dot_len}{1.2mm}
	\fmfset{dot_size}{5}
	\fmfset{arrow_len}{2.5mm}
	\fmfset{curly_len}{1.5mm}
	\begin{align}\label{eq:interference-gg-1E}
		& \mathbb{P}_{(ij)} \;
		\begin{gathered}
			\begin{fmfgraph*}(60,40)
				\fmfstraight
				\fmfleft{lb,lm,lt}
				\fmfright{rb,rm,rt}
				\fmf{curly}{rt,bt1,v,bt2,lt}
				\fmf{curly}{rb,bb,lb}
				\fmffreeze
				\fmf{curly}{v,bm,lm}
				\fmfrectangle{10}{6}{rm}
				\fmfv{decor.shape=circle,decor.filled=empty,decor.size=10,label=\tiny{$\parallel$},label.dist=0}{v}
				\fmfv{decor.shape=square,decor.filled=empty,label=\tiny{},label.dist=0,decor.size=6}{bt1}
				\fmfv{decor.shape=square,decor.filled=empty,label=\tiny{},label.dist=0,decor.size=6}{bt2}
				\fmfv{decor.shape=square,decor.filled=empty,label=\tiny{$\perp$},label.dist=0,decor.size=6}{bm}
				\fmfv{decor.shape=square,decor.filled=empty,label=\tiny{},label.dist=0,decor.size=6}{bb}
			\end{fmfgraph*}
			\quad \;
			\begin{fmfgraph*}(60,40)
				\fmfstraight
				\fmfleft{lb,lm,lt}
				\fmfright{rb,rm,rt}
				\fmf{curly}{lt,bt,rt}
				\fmf{curly}{lb,bb1,v,bb2,rb}
				\fmffreeze
				\fmf{curly}{v,bm,rm}
				\fmfrectangle{10}{6}{lm}
				\fmfv{decor.shape=circle,decor.filled=empty,decor.size=10,label=\tiny{$\parallel$},label.dist=0}{v}
				\fmfv{decor.shape=square,decor.filled=empty,label=\tiny{},label.dist=0,decor.size=6}{bb1}
				\fmfv{decor.shape=square,decor.filled=empty,label=\tiny{},label.dist=0,decor.size=6}{bb2}
				\fmfv{decor.shape=square,decor.filled=empty,label=\tiny{$\perp$},label.dist=0,decor.size=6}{bm}
				\fmfv{decor.shape=square,decor.filled=empty,label=\tiny{},label.dist=0,decor.size=6}{bt}
			\end{fmfgraph*}
		\end{gathered}
		\nonumber \\[20pt]
		& \qquad = \mathbb{P}_{(ij)} \frac{8 \pi \alpha_s \mathbf{T}_i\scdot\mathbf{T}_k}{S_{ij} S_{jk}}
		\, \frac{p_\perp^2(1+z)}{(1-z)^2} \,
		\frac{(n \scdot p_k)^2}{n \scdot p_i(n\scdot p_k + (1-z) n\scdot p_i)} \,  \eta^{\mu_i \nu_i}  \eta^{\mu_k \nu_k} \ .
	\end{align}
\end{fmffile}

\subsection{Two emission case}
\label{sec:two-emission-splitting-kernel}

\begin{table}[ht]
	\begin{center}
		\begin{tabular}{c|cccc|c|ccc}
			 & $\mathrm{C}_1\mathrm{C}_2$ & $\mathrm{C}_1\mathrm{S}_2$ & $\mathrm{S}_1\mathrm{C}_2$ & $\mathrm{S}_1\mathrm{S}_2$ & & $\mathrm{C}_1\mathrm{C}_2$ & $\mathrm{S}_1\mathrm{C}_2$ & $\mathrm{S}_1\mathrm{S}_2$  \\
			\hline
			\begin{minipage}{2cm}\begin{fmffile}{two-emission-sp1}
	\fmfset{thin}{.7pt}
	\fmfset{arrow_len}{1.8mm}
	\fmfset{curly_len}{1.8mm}
	\fmfset{dash_len}{1.5mm}
	\begin{align*}
	\begin{gathered}
	\vspace{4pt}
	\begin{fmfgraph*}(75,25)
				\fmfstraight
				\fmfleft{l2,lm,l1}
				\fmfright{r2,rm,r1}
				\fmf{phantom}{l2,k1,k2,r2}
				\fmffreeze
				\fmf{plain}{l1,bl1,m1,bm1,m2,br1,r1}
				\fmffreeze
				\fmf{plain}{m1,bk1,k1}
				\fmf{plain}{m2,bk2,k2}
				\fmffreeze
				\fmfrectangle{4}{4}{r1}
				\fmfv{decor.shape=square,decor.filled=full,decor.angle=0,decor.size=6}{bm1}
				\fmfv{decor.shape=square,decor.filled=empty,decor.angle=0,decor.size=6}{br1}
				\fmfv{decor.shape=square,decor.filled=empty,decor.angle=0,decor.size=6}{bl1}
				\fmfv{decor.shape=square,decor.filled=empty,decor.angle=0,decor.size=6}{bk1}
				\fmfv{decor.shape=square,decor.filled=empty,decor.angle=0,decor.size=6}{bk2}
				\fmfv{label=\tiny{1}, l.a=180}{k1}
				\fmfv{label=\tiny{2}, l.a=180}{k2}
			\end{fmfgraph*}
	\end{gathered}
	\end{align*} 
	\end{fmffile}\end{minipage} 
	& $\lambda^2$ & $\lambda^2$ & $\lambda$ & $\lambda$
	& \begin{minipage}{2cm}\begin{fmffile}{two-emission-sp21}
	\fmfset{thin}{.7pt}
	\fmfset{arrow_len}{1.8mm}
	\fmfset{curly_len}{1.8mm}
	\fmfset{dash_len}{1.5mm}
	\begin{align*}
	\begin{gathered}
	\vspace{4pt}
	\begin{fmfgraph*}(60,30)
				\fmfstraight
				\fmfleft{l2,lm,l1}
				\fmfright{r2,rm,r1}
				\fmf{phantom}{l2,k1,m2,k2,r2}
				\fmf{phantom}{lm,mm,rm}
				\fmffreeze
				\fmf{plain}{l1,bl1,m1,br1,r1}
				\fmffreeze
				\fmf{plain}{m1,bm1,mm}
				\fmf{plain}{mm,bk1,k1}
				\fmf{plain}{mm,bk2,k2}
				\fmffreeze
				\fmfrectangle{4}{4}{r1}
				\fmfv{decor.shape=square,decor.filled=full,decor.angle=0,decor.size=6}{bm1}
				\fmfv{decor.shape=square,decor.filled=empty,decor.angle=0,decor.size=6}{br1}
				\fmfv{decor.shape=square,decor.filled=empty,decor.angle=0,decor.size=6}{bl1}
				\fmfv{decor.shape=square,decor.filled=empty,decor.angle=0,decor.size=6}{bk1}
				\fmfv{decor.shape=square,decor.filled=empty,decor.angle=0,decor.size=6}{bk2}
				\fmfv{label=\tiny{1}, l.a=180}{k1}
				\fmfv{label=\tiny{2}, l.a=180}{k2}
			\end{fmfgraph*}
	\end{gathered}
	\end{align*} 
	\end{fmffile}\end{minipage} &$\lambda^2$ & $\lambda$ & $\lambda$
		\\
	\hline
			\begin{minipage}{2cm}\begin{fmffile}{two-emission-sp2}
	\fmfset{thin}{.7pt}
	\fmfset{arrow_len}{1.8mm}
	\fmfset{curly_len}{1.8mm}
	\fmfset{dash_len}{1.5mm}
	\begin{align*}
	\begin{gathered}
	\vspace{4pt}
	\begin{fmfgraph*}(75,25)
				\fmfstraight
				\fmfleft{l2,lm,l1}
				\fmfright{r2,rm,r1}
				\fmf{phantom}{l2,k1,k2,r2}
				\fmffreeze
				\fmf{plain}{l1,bl1,m1,bm1,m2,br1,r1}
				\fmffreeze
				\fmf{plain}{m1,bk1,k1}
				\fmf{plain}{m2,bk2,k2}
				\fmffreeze
				\fmfrectangle{4}{4}{r1}
				\fmfv{decor.shape=square,decor.filled=empty,decor.angle=0,decor.size=6,label=\tiny{$\perp$},label.dist=0}{bm1}
				\fmfv{decor.shape=square,decor.filled=empty,decor.angle=0,decor.size=6}{br1}
				\fmfv{decor.shape=square,decor.filled=empty,decor.angle=0,decor.size=6}{bl1}
				\fmfv{decor.shape=square,decor.filled=empty,decor.angle=0,decor.size=6,label=\tiny{$\perp$},label.dist=0}{bk1}
				\fmfv{decor.shape=square,decor.filled=empty,decor.angle=0,decor.size=6}{bk2}
			\end{fmfgraph*}
	\end{gathered}
	\end{align*} 
	\end{fmffile}\end{minipage}
	& $\lambda^2$	& $\lambda^2$ & $\lambda$ & $\lambda$ & 
	\begin{minipage}{2cm}\begin{fmffile}{two-emission-sp22}
	\fmfset{thin}{.7pt}
	\fmfset{arrow_len}{1.8mm}
	\fmfset{curly_len}{1.8mm}
	\fmfset{dash_len}{1.5mm}
	\begin{align*}
	\begin{gathered}
	\vspace{4pt}
	\begin{fmfgraph*}(60,30)
				\fmfstraight
				\fmfleft{l2,lm,l1}
				\fmfright{r2,rm,r1}
				\fmf{phantom}{l2,k1,m2,k2,r2}
				\fmf{phantom}{lm,mm,rm}
				\fmffreeze
				\fmf{plain}{l1,bl1,m1,br1,r1}
				\fmffreeze
				\fmf{plain}{m1,bm1,mm}
				\fmf{plain}{mm,bk1,k1}
				\fmf{plain}{mm,bk2,k2}
				\fmffreeze
				\fmfrectangle{4}{4}{r1}
				\fmfv{decor.shape=square,decor.filled=empty,decor.angle=0,decor.size=6,label=\tiny{$\perp$},label.dist=0}{bm1}
				\fmfv{decor.shape=square,decor.filled=empty,decor.angle=0,decor.size=6}{br1}
				\fmfv{decor.shape=square,decor.filled=empty,decor.angle=0,decor.size=6}{bl1}
				\fmfv{decor.shape=square,decor.filled=empty,decor.angle=0,decor.size=6,label=\tiny{$\perp$},label.dist=0}{bk1}
				\fmfv{decor.shape=square,decor.filled=empty,decor.angle=0,decor.size=6}{bk2}
			\end{fmfgraph*}
	\end{gathered}
	\end{align*} 
	\end{fmffile}\end{minipage} &$\lambda^2$ & $\lambda$ & $\lambda$
	\\
	\hline
	\begin{minipage}{2cm}\begin{fmffile}{two-emission-sp3}
	\fmfset{thin}{.7pt}
	\fmfset{arrow_len}{1.8mm}
	\fmfset{curly_len}{1.8mm}
	\fmfset{dash_len}{1.5mm}
	\begin{align*}
	\begin{gathered}
	\vspace{4pt}
	\begin{fmfgraph*}(75,25)
				\fmfstraight
				\fmfleft{l2,lm,l1}
				\fmfright{r2,rm,r1}
				\fmf{phantom}{l2,k1,k2,r2}
				\fmffreeze
				\fmf{plain}{l1,bl1,m1,bm1,m2,br1,r1}
				\fmffreeze
				\fmf{plain}{m1,bk1,k1}
				\fmf{plain}{m2,bk2,k2}
				\fmffreeze
				\fmfrectangle{4}{4}{r1}
				\fmfv{decor.shape=square,decor.filled=empty,decor.angle=0,decor.size=6}{bm1}
				\fmfv{decor.shape=square,decor.filled=empty,decor.angle=0,decor.size=6}{br1}
				\fmfv{decor.shape=square,decor.filled=empty,decor.angle=0,decor.size=6,label=\tiny{$\perp$},label.dist=0}{bl1}
				\fmfv{decor.shape=square,decor.filled=empty,decor.angle=0,decor.size=6}{bk1}
				\fmfv{decor.shape=square,decor.filled=empty,decor.angle=0,decor.size=6,label=\tiny{$\perp$},label.dist=0}{bk2}
			\end{fmfgraph*}
	\end{gathered}
	\end{align*} 
	\end{fmffile}\end{minipage}
	& $\lambda^2$	& $\lambda$ & $\lambda^2$ & $\lambda$ & 
	\begin{minipage}{2cm}\begin{fmffile}{two-emission-sp23}
	\fmfset{thin}{.7pt}
	\fmfset{arrow_len}{1.8mm}
	\fmfset{curly_len}{1.8mm}
	\fmfset{dash_len}{1.5mm}
	\begin{align*}
	\begin{gathered}
	\vspace{4pt}
	\begin{fmfgraph*}(60,30)
				\fmfstraight
				\fmfleft{l2,lm,l1}
				\fmfright{r2,rm,r1}
				\fmf{phantom}{l2,k1,m2,k2,r2}
				\fmf{phantom}{lm,mm,rm}
				\fmffreeze
				\fmf{plain}{l1,bl1,m1,br1,r1}
				\fmffreeze
				\fmf{plain}{m1,bm1,mm}
				\fmf{plain}{mm,bk1,k1}
				\fmf{plain}{mm,bk2,k2}
				\fmffreeze
				\fmfrectangle{4}{4}{r1}
				\fmfv{decor.shape=square,decor.filled=empty,decor.angle=0,decor.size=6,label=\tiny{$\perp$},label.dist=0}{bm1}
				\fmfv{decor.shape=square,decor.filled=empty,decor.angle=0,decor.size=6}{br1}
				\fmfv{decor.shape=square,decor.filled=empty,decor.angle=0,decor.size=6}{bl1}
				\fmfv{decor.shape=square,decor.filled=empty,decor.angle=0,decor.size=6}{bk1}
				\fmfv{decor.shape=square,decor.filled=empty,decor.angle=0,decor.size=6,label=\tiny{$\perp$},label.dist=0}{bk2}
			\end{fmfgraph*}
	\end{gathered}
	\end{align*} 
	\end{fmffile}\end{minipage} &$\lambda^2$ & $\lambda^2$ & $\lambda$
	 \\
	\hline
	\begin{minipage}{2cm}\begin{fmffile}{two-emission-sp4}
	\fmfset{thin}{.7pt}
	\fmfset{arrow_len}{1.8mm}
	\fmfset{curly_len}{1.8mm}
	\fmfset{dash_len}{1.5mm}
	\begin{align*}
	\begin{gathered}
	\vspace{4pt}
	\begin{fmfgraph*}(75,25)
				\fmfstraight
				\fmfleft{l2,lm,l1}
				\fmfright{r2,rm,r1}
				\fmf{phantom}{l2,k1,k2,r2}
				\fmffreeze
				\fmf{plain}{l1,bl1,m1,bm1,m2,br1,r1}
				\fmffreeze
				\fmf{plain}{m1,bk1,k1}
				\fmf{plain}{m2,bk2,k2}
				\fmffreeze
				\fmfrectangle{4}{4}{r1}
				\fmfv{decor.shape=square,decor.filled=empty,decor.angle=0,decor.size=6,label=\tiny{$\perp$},label.dist=0}{bm1}
				\fmfv{decor.shape=square,decor.filled=empty,decor.angle=0,decor.size=6}{br1}
				\fmfv{decor.shape=square,decor.filled=empty,decor.angle=0,decor.size=6}{bl1}
				\fmfv{decor.shape=square,decor.filled=empty,decor.angle=0,decor.size=6}{bk1}
				\fmfv{decor.shape=square,decor.filled=empty,decor.angle=0,decor.size=6,label=\tiny{$\perp$},label.dist=0}{bk2}
			\end{fmfgraph*}
	\end{gathered}
	\end{align*} 
	\end{fmffile}\end{minipage}
	& $\lambda^2$	& $\lambda$ & $\lambda^2$ & $\lambda$ &
	\begin{minipage}{2cm}\begin{fmffile}{two-emission-sp24}
	\fmfset{thin}{.7pt}
	\fmfset{arrow_len}{1.8mm}
	\fmfset{curly_len}{1.8mm}
	\fmfset{dash_len}{1.5mm}
	\begin{align*}
	\begin{gathered}
	\vspace{4pt}
	\begin{fmfgraph*}(60,30)
				\fmfstraight
				\fmfleft{l2,lm,l1}
				\fmfright{r2,rm,r1}
				\fmf{phantom}{l2,k1,m2,k2,r2}
				\fmf{phantom}{lm,mm,rm}
				\fmffreeze
				\fmf{plain}{l1,bl1,m1,br1,r1}
				\fmffreeze
				\fmf{plain}{m1,bm1,mm}
				\fmf{plain}{mm,bk1,k1}
				\fmf{plain}{mm,bk2,k2}
				\fmffreeze
				\fmfrectangle{4}{4}{r1}
				\fmfv{decor.shape=square,decor.filled=empty,decor.angle=0,decor.size=6,label=\tiny{$\perp$},label.dist=0}{bm1}
				\fmfv{decor.shape=square,decor.filled=empty,decor.angle=0,decor.size=6}{br1}
				\fmfv{decor.shape=square,decor.filled=empty,decor.angle=0,decor.size=6,label=\tiny{$\perp$},label.dist=0}{bl1}
				\fmfv{decor.shape=square,decor.filled=empty,decor.angle=0,decor.size=6}{bk1}
				\fmfv{decor.shape=square,decor.filled=empty,decor.angle=0,decor.size=6}{bk2}
			\end{fmfgraph*}
	\end{gathered}
	\end{align*} 
	\end{fmffile}\end{minipage} &$\lambda^2$ & $\lambda^2$ & $\lambda^2$
	 \\
	\hline
	\begin{minipage}{2cm}\begin{fmffile}{two-emission-sp5}
	\fmfset{thin}{.7pt}
	\fmfset{arrow_len}{1.8mm}
	\fmfset{curly_len}{1.8mm}
	\fmfset{dash_len}{1.5mm}
	\begin{align*}
	\begin{gathered}
	\vspace{4pt}
	\begin{fmfgraph*}(75,25)
				\fmfstraight
				\fmfleft{l2,lm,l1}
				\fmfright{r2,rm,r1}
				\fmf{phantom}{l2,k1,k2,r2}
				\fmffreeze
				\fmf{plain}{l1,bl1,m1,bm1,m2,br1,r1}
				\fmffreeze
				\fmf{plain}{m1,bk1,k1}
				\fmf{plain}{m2,bk2,k2}
				\fmffreeze
				\fmfrectangle{4}{4}{r1}
			         \fmfv{decor.shape=square,decor.filled=empty,decor.angle=0,decor.size=6}{bm1}
				\fmfv{decor.shape=square,decor.filled=empty,decor.angle=0,decor.size=6}{br1}
				\fmfv{decor.shape=square,decor.filled=empty,decor.angle=0,decor.size=6}{bl1}
				\fmfv{decor.shape=square,decor.filled=empty,decor.angle=0,decor.size=6,label=\tiny{$\perp$},label.dist=0}{bk1}
				\fmfv{decor.shape=square,decor.filled=empty,decor.angle=0,decor.size=6,label=\tiny{$\perp$},label.dist=0}{bk2}
			\end{fmfgraph*}
	\end{gathered}
	\end{align*} 
	\end{fmffile}\end{minipage}
	& $\lambda^2$	& $\lambda$ & $\lambda$ & $1$ 	& 
	\begin{minipage}{2cm}\begin{fmffile}{two-emission-sp25}
	\fmfset{thin}{.7pt}
	\fmfset{arrow_len}{1.8mm}
	\fmfset{curly_len}{1.8mm}
	\fmfset{dash_len}{1.5mm}
	\begin{align*}
	\begin{gathered}
	\vspace{4pt}
	\begin{fmfgraph*}(60,30)
				\fmfstraight
				\fmfleft{l2,lm,l1}
				\fmfright{r2,rm,r1}
				\fmf{phantom}{l2,k1,m2,k2,r2}
				\fmf{phantom}{lm,mm,rm}
				\fmffreeze
				\fmf{plain}{l1,bl1,m1,br1,r1}
				\fmffreeze
				\fmf{plain}{m1,bm1,mm}
				\fmf{plain}{mm,bk1,k1}
				\fmf{plain}{mm,bk2,k2}
				\fmffreeze
				\fmfrectangle{4}{4}{r1}
				\fmfv{decor.shape=square,decor.filled=empty,decor.angle=0,decor.size=6}{bm1}
				\fmfv{decor.shape=square,decor.filled=empty,decor.angle=0,decor.size=6}{br1}
				\fmfv{decor.shape=square,decor.filled=empty,decor.angle=0,decor.size=6,label=\tiny{$\perp$},label.dist=0}{bl1}
				\fmfv{decor.shape=square,decor.filled=empty,decor.angle=0,decor.size=6,label=\tiny{$\perp$},label.dist=0}{bk1}
				\fmfv{decor.shape=square,decor.filled=empty,decor.angle=0,decor.size=6}{bk2}
			\end{fmfgraph*}
	\end{gathered}
	\end{align*} 
	\end{fmffile}\end{minipage} &$\lambda^2$ & $\lambda$ & $\lambda^2$
	\\
	\hline
	\begin{minipage}{2cm}\begin{fmffile}{two-emission-sp6}
	\fmfset{thin}{.7pt}
	\fmfset{arrow_len}{1.8mm}
	\fmfset{curly_len}{1.8mm}
	\fmfset{dash_len}{1.5mm}
	\begin{align*}
	\begin{gathered}
	\vspace{4pt}
	\begin{fmfgraph*}(75,25)
				\fmfstraight
				\fmfleft{l2,lm,l1}
				\fmfright{r2,rm,r1}
				\fmf{phantom}{l2,k1,k2,r2}
				\fmffreeze
				\fmf{plain}{l1,bl1,m1,bm1,m2,br1,r1}
				\fmffreeze
				\fmf{plain}{m1,bk1,k1}
				\fmf{plain}{m2,bk2,k2}
				\fmffreeze
				\fmfrectangle{4}{4}{r1}
				\fmfv{decor.shape=square,decor.filled=empty,decor.angle=0,decor.size=6,label=\tiny{$\perp$},label.dist=0}{bm1}
				\fmfv{decor.shape=square,decor.filled=empty,decor.angle=0,decor.size=6}{br1}
				\fmfv{decor.shape=square,decor.filled=empty,decor.angle=0,decor.size=6,label=\tiny{$\perp$},label.dist=0}{bl1}
				\fmfv{decor.shape=square,decor.filled=empty,decor.angle=0,decor.size=6}{bk1}
				\fmfv{decor.shape=square,decor.filled=empty,decor.angle=0,decor.size=6}{bk2}
			\end{fmfgraph*}
	\end{gathered}
	\end{align*} 
	\end{fmffile}\end{minipage}
	& $\lambda^2$	& $\lambda^2$ & $\lambda^2$ & $\lambda^2$ & 
	\begin{minipage}{2cm}\begin{fmffile}{two-emission-sp26}
	\fmfset{thin}{.7pt}
	\fmfset{arrow_len}{1.8mm}
	\fmfset{curly_len}{1.8mm}
	\fmfset{dash_len}{1.5mm}
	\begin{align*}
	\begin{gathered}
	\vspace{4pt}
	\begin{fmfgraph*}(60,30)
				\fmfstraight
				\fmfleft{l2,lm,l1}
				\fmfright{r2,rm,r1}
				\fmf{phantom}{l2,k1,m2,k2,r2}
				\fmf{phantom}{lm,mm,rm}
				\fmffreeze
				\fmf{plain}{l1,bl1,m1,br1,r1}
				\fmffreeze
				\fmf{plain}{m1,bm1,mm}
				\fmf{plain}{mm,bk1,k1}
				\fmf{plain}{mm,bk2,k2}
				\fmffreeze
				\fmfrectangle{4}{4}{r1}
				\fmfv{decor.shape=square,decor.filled=empty,decor.angle=0,decor.size=6}{bm1}
				\fmfv{decor.shape=square,decor.filled=empty,decor.angle=0,decor.size=6}{br1}
				\fmfv{decor.shape=square,decor.filled=empty,decor.angle=0,decor.size=6,label=\tiny{$\perp$},label.dist=0}{bl1}
				\fmfv{decor.shape=square,decor.filled=empty,decor.angle=0,decor.size=6}{bk1}
				\fmfv{decor.shape=square,decor.filled=empty,decor.angle=0,decor.size=6,label=\tiny{$\perp$},label.dist=0}{bk2}
			\end{fmfgraph*}
	\end{gathered}
	\end{align*} 
	\end{fmffile}\end{minipage} &$\lambda^2$ & $\lambda^2$ & $\lambda^2$
	 \\
	 \hline
	 \begin{minipage}{2cm}\begin{fmffile}{two-emission-sp7}
	 		\fmfset{thin}{.7pt}
	 		\fmfset{arrow_len}{1.8mm}
	 		\fmfset{curly_len}{1.8mm}
	 		\fmfset{dash_len}{1.5mm}
	 		\begin{align*}
	 			\begin{gathered}
	 				\vspace{4pt}
	 				\begin{fmfgraph*}(75,25)
	 					\fmfstraight
	 					\fmfleft{l2,lm,l1}
	 					\fmfright{r2,rm,r1}
	 					\fmf{phantom}{l2,k1,k2,r2}
	 					\fmffreeze
	 					\fmf{plain}{l1,bl1,m1,bm1,m2,br1,r1}
	 					\fmffreeze
	 					\fmf{plain}{m1,bk1,k1}
	 					\fmf{plain}{m2,bk2,k2}
	 					\fmffreeze
	 					\fmfrectangle{4}{4}{r1}
	 					\fmfv{decor.shape=square,decor.filled=empty,decor.angle=0,decor.size=6}{bm1}
	 					\fmfv{decor.shape=square,decor.filled=full,decor.angle=0,decor.size=6}{br1}
	 					\fmfv{decor.shape=square,decor.filled=empty,decor.angle=0,decor.size=6}{bl1}
	 					\fmfv{decor.shape=square,decor.filled=empty,decor.angle=0,decor.size=6,label=\tiny{$\perp$},label.dist=0}{bk1}
	 					\fmfv{decor.shape=square,decor.filled=empty,decor.angle=0,decor.size=6}{bk2}
	 				\end{fmfgraph*}
	 			\end{gathered}
	 		\end{align*} 
	 \end{fmffile}\end{minipage}
	 & $\lambda^3$	& $\lambda^2$ & $\lambda$ & $\lambda$ & 
	 \begin{minipage}{2cm}\begin{fmffile}{two-emission-sp27}
	\fmfset{thin}{.7pt}
	\fmfset{arrow_len}{1.8mm}
	\fmfset{curly_len}{1.8mm}
	\fmfset{dash_len}{1.5mm}
	\begin{align*}
	\begin{gathered}
	\vspace{4pt}
	\begin{fmfgraph*}(60,30)
				\fmfstraight
				\fmfleft{l2,lm,l1}
				\fmfright{r2,rm,r1}
				\fmf{phantom}{l2,k1,m2,k2,r2}
				\fmf{phantom}{lm,mm,rm}
				\fmffreeze
				\fmf{plain}{l1,bl1,m1,br1,r1}
				\fmffreeze
				\fmf{plain}{m1,bm1,mm}
				\fmf{plain}{mm,bk1,k1}
				\fmf{plain}{mm,bk2,k2}
				\fmffreeze
				\fmfrectangle{4}{4}{r1}
				\fmfv{decor.shape=square,decor.filled=empty,decor.angle=0,decor.size=6}{bm1}
				\fmfv{decor.shape=square,decor.filled=full,decor.angle=0,decor.size=6}{br1}
				\fmfv{decor.shape=square,decor.filled=empty,decor.angle=0,decor.size=6}{bl1}
				\fmfv{decor.shape=square,decor.filled=empty,decor.angle=0,decor.size=6,label=\tiny{$\perp$},label.dist=0}{bk1}
				\fmfv{decor.shape=square,decor.filled=empty,decor.angle=0,decor.size=6}{bk2}
			\end{fmfgraph*}
	\end{gathered}
	\end{align*} 
	\end{fmffile}\end{minipage} &$\lambda^3$ & $\lambda$ & $\lambda^2$
	 	 \\
	 \hline
	 \begin{minipage}{2cm}\begin{fmffile}{two-emission-sp8}
	 		\fmfset{thin}{.7pt}
	 		\fmfset{arrow_len}{1.8mm}
	 		\fmfset{curly_len}{1.8mm}
	 		\fmfset{dash_len}{1.5mm}
	 		\begin{align*}
	 			\begin{gathered}
	 				\vspace{4pt}
	 				\begin{fmfgraph*}(75,25)
	 					\fmfstraight
	 					\fmfleft{l2,lm,l1}
	 					\fmfright{r2,rm,r1}
	 					\fmf{phantom}{l2,k1,k2,r2}
	 					\fmffreeze
	 					\fmf{plain}{l1,bl1,m1,bm1,m2,br1,r1}
	 					\fmffreeze
	 					\fmf{plain}{m1,bk1,k1}
	 					\fmf{plain}{m2,bk2,k2}
	 					\fmffreeze
	 					\fmfrectangle{4}{4}{r1}
	 					\fmfv{decor.shape=square,decor.filled=full,decor.angle=0,decor.size=6}{bm1}
	 					\fmfv{decor.shape=square,decor.filled=empty,decor.angle=0,decor.size=6}{br1}
	 					\fmfv{decor.shape=square,decor.filled=empty,decor.angle=0,decor.size=6}{bl1}
	 					\fmfv{decor.shape=square,decor.filled=empty,decor.angle=0,decor.size=6,label=\tiny{$\perp$},label.dist=0}{bk1}
	 					\fmfv{decor.shape=square,decor.filled=empty,decor.angle=0,decor.size=6}{bk2}
	 				\end{fmfgraph*}
	 			\end{gathered}
	 		\end{align*} 
	 \end{fmffile}\end{minipage}
	 & $\lambda^3$	& $\lambda^3$ & $\lambda$ & $\lambda$ & 
	 \begin{minipage}{2cm}\begin{fmffile}{two-emission-sp28}
	\fmfset{thin}{.7pt}
	\fmfset{arrow_len}{1.8mm}
	\fmfset{curly_len}{1.8mm}
	\fmfset{dash_len}{1.5mm}
	\begin{align*}
	\begin{gathered}
	\vspace{4pt}
	\begin{fmfgraph*}(60,30)
				\fmfstraight
				\fmfleft{l2,lm,l1}
				\fmfright{r2,rm,r1}
				\fmf{phantom}{l2,k1,m2,k2,r2}
				\fmf{phantom}{lm,mm,rm}
				\fmffreeze
				\fmf{plain}{l1,bl1,m1,br1,r1}
				\fmffreeze
				\fmf{plain}{m1,bm1,mm}
				\fmf{plain}{mm,bk1,k1}
				\fmf{plain}{mm,bk2,k2}
				\fmffreeze
				\fmfrectangle{4}{4}{r1}
				\fmfv{decor.shape=square,decor.filled=full,decor.angle=0,decor.size=6}{bm1}
				\fmfv{decor.shape=square,decor.filled=empty,decor.angle=0,decor.size=6}{br1}
				\fmfv{decor.shape=square,decor.filled=empty,decor.angle=0,decor.size=6}{bl1}
				\fmfv{decor.shape=square,decor.filled=empty,decor.angle=0,decor.size=6,label=\tiny{$\perp$},label.dist=0}{bk1}
				\fmfv{decor.shape=square,decor.filled=empty,decor.angle=0,decor.size=6}{bk2}
			\end{fmfgraph*}
	\end{gathered}
	\end{align*} 
	\end{fmffile}\end{minipage} &$\lambda^3$ & $\lambda$ & $\lambda$
		\end{tabular}
	\end{center}
	\caption{Leading collinear contributions for splitter line with collinear vertices.}
	\label{tab:two-emissions-sp1}
\end{table}

\begin{table}[h]
	\begin{center}
		\begin{tabular}{c|cccc|c|ccc}
			 & $\mathrm{C}_1\mathrm{C}_2$ & $\mathrm{C}_1\mathrm{S}_2$ & $\mathrm{S}_1\mathrm{C}_2$ & $\mathrm{S}_1\mathrm{S}_2$ & & $\mathrm{C}_1\mathrm{C}_2$ & $\mathrm{S}_1\mathrm{C}_2$ & $\mathrm{S}_1\mathrm{S}_2$ \\
			\hline
			\begin{minipage}{3cm}\begin{fmffile}{two-emission-spt1}
	\fmfset{thin}{.7pt}
	\fmfset{arrow_len}{1.8mm}
	\fmfset{curly_len}{1.8mm}
	\fmfset{dash_len}{1.5mm}
	\begin{align*}
	\begin{gathered}
	\vspace{4pt}
	\begin{fmfgraph*}(75,25)
				\fmfstraight
				\fmfleft{l2,lm,l1}
				\fmfright{r2,rm,r1}
				\fmf{phantom}{l2,k1,k2,r2}
				\fmffreeze
				\fmf{plain}{l1,bl1,m1,bm1,m2,br1,r1}
				\fmffreeze
				\fmf{plain}{m1,bk1,k1}
				\fmf{plain}{m2,bk2,k2}
				\fmffreeze
				\fmfrectangle{4}{4}{r1}
				\fmfv{decor.shape=square,decor.filled=empty,decor.angle=0,decor.size=6,label=\tiny{$\perp$},label.dist=0}{bm1}
				\fmfv{decor.shape=square,decor.filled=empty,decor.angle=0,decor.size=6}{br1}
				\fmfv{decor.shape=square,decor.filled=empty,decor.angle=0,decor.size=6}{bl1}
				\fmfv{decor.shape=square,decor.filled=empty,decor.angle=0,decor.size=6}{bk1}
				\fmfv{decor.shape=square,decor.filled=empty,decor.angle=0,decor.size=6}{bk2}
				\fmfv{label=\tiny{1}, l.a=180}{k1}
				\fmfv{label=\tiny{2}, l.a=180}{k2}
				\fmfv{decor.shape=circle,decor.filled=empty,decor.size=10,label=\tiny{$\parallel$},label.dist=0}{m2}
				\fmfv{decor.shape=circle,decor.filled=empty,decor.size=10,label=\tiny{$\perp$},label.dist=0}{m1}
			\end{fmfgraph*}
	\end{gathered}
	\end{align*} 
	\end{fmffile}\end{minipage} 
	& $\lambda^2$ & $\lambda^2$ & $\lambda^2$ & $\lambda^2$ & 
	\begin{minipage}{2cm}\begin{fmffile}{two-emission-spt21}
	\fmfset{thin}{.7pt}
	\fmfset{arrow_len}{1.8mm}
	\fmfset{curly_len}{1.8mm}
	\fmfset{dash_len}{1.5mm}
	\begin{align*}
	\begin{gathered}
	\vspace{4pt}
	\begin{fmfgraph*}(60,40)
				\fmfstraight
				\fmfleft{l2,lm,l1}
				\fmfright{r2,rm,r1}
				\fmf{phantom}{l2,k1,m2,k2,r2}
				\fmf{phantom}{lm,mm,rm}
				\fmffreeze
				\fmf{plain}{l1,bl1,m1,br1,r1}
				\fmffreeze
				\fmf{plain}{m1,bm1,mm}
				\fmf{plain}{mm,bk1,k1}
				\fmf{plain}{mm,bk2,k2}
				\fmffreeze
				\fmfrectangle{4}{4}{r1}
				\fmfv{decor.shape=square,decor.filled=empty,decor.angle=0,decor.size=6}{bm1}
				\fmfv{decor.shape=square,decor.filled=empty,decor.angle=0,decor.size=6}{br1}
				\fmfv{decor.shape=square,decor.filled=empty,decor.angle=0,decor.size=6,label=\tiny{$\perp$},label.dist=0}{bl1}
				\fmfv{decor.shape=square,decor.filled=empty,decor.angle=0,decor.size=6}{bk1}
				\fmfv{decor.shape=square,decor.filled=empty,decor.angle=0,decor.size=6}{bk2}
				\fmfv{label=\tiny{1}, l.a=180}{k1}
				\fmfv{label=\tiny{2}, l.a=180}{k2}
				\fmfv{decor.shape=circle,decor.filled=empty,decor.size=10,label=\tiny{$\parallel$},label.dist=0}{m1}
				\fmfv{decor.shape=circle,decor.filled=empty,decor.size=10,label=\tiny{$\perp$},label.dist=0}{mm}
			\end{fmfgraph*}
	\end{gathered}
	\end{align*} 
	\end{fmffile}\end{minipage} &$\lambda^2$ & $\lambda^2$ & $\lambda^2$
	 \\
	\hline
	\begin{minipage}{3cm}\begin{fmffile}{two-emission-spt2}
	\fmfset{thin}{.7pt}
	\fmfset{arrow_len}{1.8mm}
	\fmfset{curly_len}{1.8mm}
	\fmfset{dash_len}{1.5mm}
	\begin{align*}
	\begin{gathered}
	\vspace{4pt}
	\begin{fmfgraph*}(75,25)
				\fmfstraight
				\fmfleft{l2,lm,l1}
				\fmfright{r2,rm,r1}
				\fmf{phantom}{l2,k1,k2,r2}
				\fmffreeze
				\fmf{plain}{l1,bl1,m1,bm1,m2,br1,r1}
				\fmffreeze
				\fmf{plain}{m1,bk1,k1}
				\fmf{plain}{m2,bk2,k2}
				\fmffreeze
				\fmfrectangle{4}{4}{r1}
				\fmfv{decor.shape=square,decor.filled=empty,decor.angle=0,decor.size=6,label=\tiny{$\perp$},label.dist=0}{bm1}
				\fmfv{decor.shape=square,decor.filled=empty,decor.angle=0,decor.size=6}{br1}
				\fmfv{decor.shape=square,decor.filled=empty,decor.angle=0,decor.size=6}{bl1}
				\fmfv{decor.shape=square,decor.filled=empty,decor.angle=0,decor.size=6}{bk1}
				\fmfv{decor.shape=square,decor.filled=empty,decor.angle=0,decor.size=6}{bk2}
				\fmfv{decor.shape=circle,decor.filled=empty,decor.size=10,label=\tiny{$\perp$},label.dist=0}{m2}
				\fmfv{decor.shape=circle,decor.filled=empty,decor.size=10,label=\tiny{$\parallel$},label.dist=0}{m1}
	\end{fmfgraph*}
	\end{gathered}
	\end{align*} 
	\end{fmffile}\end{minipage} 
	& $\lambda^2$ & $\lambda^2$ & $\lambda^2$ & $\lambda^2$ & 
	\begin{minipage}{2cm}\begin{fmffile}{two-emission-spt22}
	\fmfset{thin}{.7pt}
	\fmfset{arrow_len}{1.8mm}
	\fmfset{curly_len}{1.8mm}
	\fmfset{dash_len}{1.5mm}
	\begin{align*}
	\begin{gathered}
	\vspace{4pt}
	\begin{fmfgraph*}(60,40)
				\fmfstraight
				\fmfleft{l2,lm,l1}
				\fmfright{r2,rm,r1}
				\fmf{phantom}{l2,k1,m2,k2,r2}
				\fmf{phantom}{lm,mm,rm}
				\fmffreeze
				\fmf{plain}{l1,bl1,m1,br1,r1}
				\fmffreeze
				\fmf{plain}{m1,bm1,mm}
				\fmf{plain}{mm,bk1,k1}
				\fmf{plain}{mm,bk2,k2}
				\fmffreeze
				\fmfrectangle{4}{4}{r1}
				\fmfv{decor.shape=square,decor.filled=empty,decor.angle=0,decor.size=6,label=\tiny{$\perp$},label.dist=0}{bm1}
				\fmfv{decor.shape=square,decor.filled=empty,decor.angle=0,decor.size=6}{br1}
				\fmfv{decor.shape=square,decor.filled=empty,decor.angle=0,decor.size=6}{bl1}
				\fmfv{decor.shape=square,decor.filled=empty,decor.angle=0,decor.size=6}{bk1}
				\fmfv{decor.shape=square,decor.filled=empty,decor.angle=0,decor.size=6}{bk2}
				\fmfv{decor.shape=circle,decor.filled=empty,decor.size=10,label=\tiny{$\parallel$},label.dist=0}{m1}
				\fmfv{decor.shape=circle,decor.filled=empty,decor.size=10,label=\tiny{$\perp$},label.dist=0}{mm}
			\end{fmfgraph*}
	\end{gathered}
	\end{align*} 
	\end{fmffile}\end{minipage} &$\lambda^2$ & $\lambda^2$ & $\lambda$
	\\
	\hline
	\begin{minipage}{3cm}\begin{fmffile}{two-emission-spt3}
	\fmfset{thin}{.7pt}
	\fmfset{arrow_len}{1.8mm}
	\fmfset{curly_len}{1.8mm}
	\fmfset{dash_len}{1.5mm}
	\begin{align*}
	\begin{gathered}
	\vspace{4pt}
	\begin{fmfgraph*}(75,25)
				\fmfstraight
				\fmfleft{l2,lm,l1}
				\fmfright{r2,rm,r1}
				\fmf{phantom}{l2,k1,k2,r2}
				\fmffreeze
				\fmf{plain}{l1,bl1,m1,bm1,m2,br1,r1}
				\fmffreeze
				\fmf{plain}{m1,bk1,k1}
				\fmf{plain}{m2,bk2,k2}
				\fmffreeze
				\fmfrectangle{4}{4}{r1}
				\fmfv{decor.shape=square,decor.filled=empty,decor.angle=0,decor.size=6}{bm1}
				\fmfv{decor.shape=square,decor.filled=empty,decor.angle=0,decor.size=6}{br1}
				\fmfv{decor.shape=square,decor.filled=empty,decor.angle=0,decor.size=6,label=\tiny{$\perp$},label.dist=0}{bl1}
				\fmfv{decor.shape=square,decor.filled=empty,decor.angle=0,decor.size=6}{bk1}
				\fmfv{decor.shape=square,decor.filled=empty,decor.angle=0,decor.size=6}{bk2}
				\fmfv{decor.shape=circle,decor.filled=empty,decor.size=10,label=\tiny{$\perp$},label.dist=0}{m2}
				\fmfv{decor.shape=circle,decor.filled=empty,decor.size=10,label=\tiny{$\parallel$},label.dist=0}{m1}
	\end{fmfgraph*}
	\end{gathered}
	\end{align*} 
	\end{fmffile}\end{minipage} 
	& $\lambda^2$ & $\lambda^2$ & $\lambda^2$  & $\lambda^2$  & 
	\begin{minipage}{2cm}\begin{fmffile}{two-emission-spt23}
	\fmfset{thin}{.7pt}
	\fmfset{arrow_len}{1.8mm}
	\fmfset{curly_len}{1.8mm}
	\fmfset{dash_len}{1.5mm}
	\begin{align*}
	\begin{gathered}
	\vspace{4pt}
	\begin{fmfgraph*}(60,40)
				\fmfstraight
				\fmfleft{l2,lm,l1}
				\fmfright{r2,rm,r1}
				\fmf{phantom}{l2,k1,m2,k2,r2}
				\fmf{phantom}{lm,mm,rm}
				\fmffreeze
				\fmf{plain}{l1,bl1,m1,br1,r1}
				\fmffreeze
				\fmf{plain}{m1,bm1,mm}
				\fmf{plain}{mm,bk1,k1}
				\fmf{plain}{mm,bk2,k2}
				\fmffreeze
				\fmfrectangle{4}{4}{r1}
				\fmfv{decor.shape=square,decor.filled=empty,decor.angle=0,decor.size=6,label=\tiny{$\perp$},label.dist=0}{bm1}
				\fmfv{decor.shape=square,decor.filled=empty,decor.angle=0,decor.size=6}{br1}
				\fmfv{decor.shape=square,decor.filled=empty,decor.angle=0,decor.size=6}{bl1}
				\fmfv{decor.shape=square,decor.filled=empty,decor.angle=0,decor.size=6}{bk1}
				\fmfv{decor.shape=square,decor.filled=empty,decor.angle=0,decor.size=6}{bk2}
				\fmfv{decor.shape=circle,decor.filled=empty,decor.size=10,label=\tiny{$\perp$},label.dist=0}{m1}
				\fmfv{decor.shape=circle,decor.filled=empty,decor.size=10,label=\tiny{$\parallel$},label.dist=0}{mm}
			\end{fmfgraph*}
	\end{gathered}
	\end{align*} 
	\end{fmffile}\end{minipage} &$\lambda^2$ & $\lambda^2$ & $\lambda^2$
	 \\
	\hline
	\begin{minipage}{3cm}\begin{fmffile}{two-emission-spt4}
	\fmfset{thin}{.7pt}
	\fmfset{arrow_len}{1.8mm}
	\fmfset{curly_len}{1.8mm}
	\fmfset{dash_len}{1.5mm}
	\begin{align*}
	\begin{gathered}
	\vspace{4pt}
	\begin{fmfgraph*}(75,25)
				\fmfstraight
				\fmfleft{l2,lm,l1}
				\fmfright{r2,rm,r1}
				\fmf{phantom}{l2,k1,k2,r2}
				\fmffreeze
				\fmf{plain}{l1,bl1,m1,bm1,m2,br1,r1}
				\fmffreeze
				\fmf{plain}{m1,bk1,k1}
				\fmf{plain}{m2,bk2,k2}
				\fmffreeze
				\fmfrectangle{4}{4}{r1}
				\fmfv{decor.shape=square,decor.filled=empty,decor.angle=0,decor.size=6}{bm1}
				\fmfv{decor.shape=square,decor.filled=empty,decor.angle=0,decor.size=6}{br1}
				\fmfv{decor.shape=square,decor.filled=empty,decor.angle=0,decor.size=6}{bl1}
				\fmfv{decor.shape=square,decor.filled=empty,decor.angle=0,decor.size=6,label=\tiny{$\perp$},label.dist=0}{bk1}
				\fmfv{decor.shape=square,decor.filled=empty,decor.angle=0,decor.size=6}{bk2}
				\fmfv{decor.shape=circle,decor.filled=empty,decor.size=10,label=\tiny{$\perp$},label.dist=0}{m2}
				\fmfv{decor.shape=circle,decor.filled=empty,decor.size=10,label=\tiny{$\parallel$},label.dist=0}{m1}
	\end{fmfgraph*}
	\end{gathered}
	\end{align*} 
	\end{fmffile}\end{minipage} 
	& $\lambda^2$ & $\lambda^2$ & $\lambda$ & $\lambda$  & 
	\begin{minipage}{2cm}\begin{fmffile}{two-emission-spt24}
	\fmfset{thin}{.7pt}
	\fmfset{arrow_len}{1.8mm}
	\fmfset{curly_len}{1.8mm}
	\fmfset{dash_len}{1.5mm}
	\begin{align*}
	\begin{gathered}
	\vspace{4pt}
	\begin{fmfgraph*}(60,40)
				\fmfstraight
				\fmfleft{l2,lm,l1}
				\fmfright{r2,rm,r1}
				\fmf{phantom}{l2,k1,m2,k2,r2}
				\fmf{phantom}{lm,mm,rm}
				\fmffreeze
				\fmf{plain}{l1,bl1,m1,br1,r1}
				\fmffreeze
				\fmf{plain}{m1,bm1,mm}
				\fmf{plain}{mm,bk1,k1}
				\fmf{plain}{mm,bk2,k2}
				\fmffreeze
				\fmfrectangle{4}{4}{r1}
				\fmfv{decor.shape=square,decor.filled=empty,decor.angle=0,decor.size=6}{bm1}
				\fmfv{decor.shape=square,decor.filled=empty,decor.angle=0,decor.size=6}{br1}
				\fmfv{decor.shape=square,decor.filled=empty,decor.angle=0,decor.size=6}{bl1}
				\fmfv{decor.shape=square,decor.filled=empty,decor.angle=0,decor.size=6}{bk1}
				\fmfv{decor.shape=square,decor.filled=empty,decor.angle=0,decor.size=6,label=\tiny{$\perp$},label.dist=0}{bk2}
				\fmfv{decor.shape=circle,decor.filled=empty,decor.size=10,label=\tiny{$\perp$},label.dist=0}{m1}
				\fmfv{decor.shape=circle,decor.filled=empty,decor.size=10,label=\tiny{$\parallel$},label.dist=0}{mm}
			\end{fmfgraph*}
	\end{gathered}
	\end{align*} 
	\end{fmffile}\end{minipage} &$\lambda^2$ & $\lambda^2$ & $\lambda^2$
	\\
	\hline
		\begin{minipage}{3cm}\begin{fmffile}{two-emission-spt5}
	\fmfset{thin}{.7pt}
	\fmfset{arrow_len}{1.8mm}
	\fmfset{curly_len}{1.8mm}
	\fmfset{dash_len}{1.5mm}
	\begin{align*}
	\begin{gathered}
	\vspace{4pt}
	\begin{fmfgraph*}(75,25)
				\fmfstraight
				\fmfleft{l2,lm,l1}
				\fmfright{r2,rm,r1}
				\fmf{phantom}{l2,k1,k2,r2}
				\fmffreeze
				\fmf{plain}{l1,bl1,m1,bm1,m2,br1,r1}
				\fmffreeze
				\fmf{plain}{m1,bk1,k1}
				\fmf{plain}{m2,bk2,k2}
				\fmffreeze
				\fmfrectangle{4}{4}{r1}
				\fmfv{decor.shape=square,decor.filled=empty,decor.angle=0,decor.size=6}{bm1}
				\fmfv{decor.shape=square,decor.filled=empty,decor.angle=0,decor.size=6}{br1}
				\fmfv{decor.shape=square,decor.filled=empty,decor.angle=0,decor.size=6}{bl1}
				\fmfv{decor.shape=square,decor.filled=empty,decor.angle=0,decor.size=6}{bk1}
				\fmfv{decor.shape=square,decor.filled=empty,decor.angle=0,decor.size=6,label=\tiny{$\perp$},label.dist=0}{bk2}
				\fmfv{decor.shape=circle,decor.filled=empty,decor.size=10,label=\tiny{$\parallel$},label.dist=0}{m2}
				\fmfv{decor.shape=circle,decor.filled=empty,decor.size=10,label=\tiny{$\perp$},label.dist=0}{m1}
	\end{fmfgraph*}
	\end{gathered}
	\end{align*} 
	\end{fmffile}\end{minipage} 
	& $\lambda^2$ & $\lambda$ & $\lambda^2$ & $\lambda$ & 
	\begin{minipage}{2cm}\begin{fmffile}{two-emission-spt25}
	\fmfset{thin}{.7pt}
	\fmfset{arrow_len}{1.8mm}
	\fmfset{curly_len}{1.8mm}
	\fmfset{dash_len}{1.5mm}
	\begin{align*}
	\begin{gathered}
	\vspace{4pt}
	\begin{fmfgraph*}(60,40)
				\fmfstraight
				\fmfleft{l2,lm,l1}
				\fmfright{r2,rm,r1}
				\fmf{phantom}{l2,k1,m2,k2,r2}
				\fmf{phantom}{lm,mm,rm}
				\fmffreeze
				\fmf{plain}{l1,bl1,m1,br1,r1}
				\fmffreeze
				\fmf{plain}{m1,bm1,mm}
				\fmf{plain}{mm,bk1,k1}
				\fmf{plain}{mm,bk2,k2}
				\fmffreeze
				\fmfrectangle{4}{4}{r1}
				\fmfv{decor.shape=square,decor.filled=empty,decor.angle=0,decor.size=6}{bm1}
				\fmfv{decor.shape=square,decor.filled=empty,decor.angle=0,decor.size=6}{br1}
				\fmfv{decor.shape=square,decor.filled=empty,decor.angle=0,decor.size=6}{bl1}
				\fmfv{decor.shape=square,decor.filled=empty,decor.angle=0,decor.size=6,label=\tiny{$\perp$},label.dist=0}{bk1}
				\fmfv{decor.shape=square,decor.filled=empty,decor.angle=0,decor.size=6}{bk2}
				\fmfv{decor.shape=circle,decor.filled=empty,decor.size=10,label=\tiny{$\perp$},label.dist=0}{m1}
				\fmfv{decor.shape=circle,decor.filled=empty,decor.size=10,label=\tiny{$\parallel$},label.dist=0}{mm}
			\end{fmfgraph*}
	\end{gathered}
	\end{align*} 
	\end{fmffile}\end{minipage} &$\lambda^2$ & $\lambda$ & $\lambda^2$
	\\
	\hline
		\begin{minipage}{3cm}\begin{fmffile}{two-emission-spt6}
	\fmfset{thin}{.7pt}
	\fmfset{arrow_len}{1.8mm}
	\fmfset{curly_len}{1.8mm}
	\fmfset{dash_len}{1.5mm}
	\begin{align*}
	\begin{gathered}
	\vspace{4pt}
	\begin{fmfgraph*}(75,25)
				\fmfstraight
				\fmfleft{l2,lm,l1}
				\fmfright{r2,rm,r1}
				\fmf{phantom}{l2,k1,k2,r2}
				\fmffreeze
				\fmf{plain}{l1,bl1,m1,bm1,m2,br1,r1}
				\fmffreeze
				\fmf{plain}{m1,bk1,k1}
				\fmf{plain}{m2,bk2,k2}
				\fmffreeze
				\fmfrectangle{4}{4}{r1}
				\fmfv{decor.shape=square,decor.filled=empty,decor.angle=0,decor.size=6}{bm1}
				\fmfv{decor.shape=square,decor.filled=empty,decor.angle=0,decor.size=6}{br1}
				\fmfv{decor.shape=square,decor.filled=empty,decor.angle=0,decor.size=6}{bl1}
				\fmfv{decor.shape=square,decor.filled=empty,decor.angle=0,decor.size=6}{bk1}
				\fmfv{decor.shape=square,decor.filled=empty,decor.angle=0,decor.size=6}{bk2}
				\fmfv{decor.shape=circle,decor.filled=empty,decor.size=10,label=\tiny{$\perp$},label.dist=0}{m2}
				\fmfv{decor.shape=circle,decor.filled=empty,decor.size=10,label=\tiny{$\perp$},label.dist=0}{m1}
	\end{fmfgraph*}
	\end{gathered}
	\end{align*} 
	\end{fmffile}\end{minipage} 
	& $\lambda^2$ & $\lambda^2$ & $\lambda^2$ &  $\lambda^2$ & 
	\begin{minipage}{2cm}\begin{fmffile}{two-emission-spt26}
	\fmfset{thin}{.7pt}
	\fmfset{arrow_len}{1.8mm}
	\fmfset{curly_len}{1.8mm}
	\fmfset{dash_len}{1.5mm}
	\begin{align*}
	\begin{gathered}
	\vspace{4pt}
	\begin{fmfgraph*}(60,40)
				\fmfstraight
				\fmfleft{l2,lm,l1}
				\fmfright{r2,rm,r1}
				\fmf{phantom}{l2,k1,m2,k2,r2}
				\fmf{phantom}{lm,mm,rm}
				\fmffreeze
				\fmf{plain}{l1,bl1,m1,br1,r1}
				\fmffreeze
				\fmf{plain}{m1,bm1,mm}
				\fmf{plain}{mm,bk1,k1}
				\fmf{plain}{mm,bk2,k2}
				\fmffreeze
				\fmfrectangle{4}{4}{r1}
				\fmfv{decor.shape=square,decor.filled=empty,decor.angle=0,decor.size=6}{bm1}
				\fmfv{decor.shape=square,decor.filled=empty,decor.angle=0,decor.size=6}{br1}
				\fmfv{decor.shape=square,decor.filled=empty,decor.angle=0,decor.size=6}{bl1}
				\fmfv{decor.shape=square,decor.filled=empty,decor.angle=0,decor.size=6}{bk1}
				\fmfv{decor.shape=square,decor.filled=empty,decor.angle=0,decor.size=6}{bk2}
				\fmfv{decor.shape=circle,decor.filled=empty,decor.size=10,label=\tiny{$\perp$},label.dist=0}{m1}
				\fmfv{decor.shape=circle,decor.filled=empty,decor.size=10,label=\tiny{$\perp$},label.dist=0}{mm}
			\end{fmfgraph*}
	\end{gathered}
	\end{align*} 
	\end{fmffile}\end{minipage} &$\lambda^2$ & $\lambda^2$ & $\lambda^2$
	\end{tabular}
	\end{center}
	\caption{Leading collinear contributions for splitter line with transverse vertices.}
	\label{tab:two-emissions-sp2}
\end{table} 
\begin{table}[ht]
	\begin{center}
		\begin{tabular}{c|cccc|c|ccc}
			 & $\mathrm{C}_1\mathrm{C}_2$ & $\mathrm{C}_1\mathrm{S}_2$ & $\mathrm{S}_1\mathrm{C}_2$ & $\mathrm{S}_1\mathrm{S}_2$ & & $\mathrm{C}_1\mathrm{C}_2$ & $\mathrm{S}_1\mathrm{C}_2$ & $\mathrm{S}_1\mathrm{S}_2$  \\
			\hline
			\begin{minipage}{2cm}\begin{fmffile}{two-emission-int1}
	\fmfset{thin}{.7pt}
	\fmfset{arrow_len}{1.8mm}
	\fmfset{curly_len}{1.8mm}
	\fmfset{dash_len}{1.5mm}
	\begin{align*}
	\begin{gathered}
	\vspace{4pt}
	\begin{fmfgraph*}(75,25)
				\fmfstraight
				\fmfright{l1,lm,l2}
				\fmfleft{r1,rm,r2}
				\fmf{phantom}{l2,k1,k2,r2}
				\fmffreeze
				\fmf{plain}{l1,bl1,m1,bm1,m2,br1,r1}
				\fmffreeze
				\fmf{plain}{m1,bk1,k1}
				\fmf{plain}{m2,bk2,k2}
				\fmffreeze
				\fmfrectangle{4}{4}{r1}
				\fmfv{decor.shape=square,decor.filled=empty,decor.angle=0,decor.size=6}{bm1}
				\fmfv{decor.shape=square,decor.filled=empty,decor.angle=0,decor.size=6}{br1}
				\fmfv{decor.shape=square,decor.filled=empty,decor.angle=0,decor.size=6}{bl1}
				\fmfv{decor.shape=square,decor.filled=empty,decor.angle=0,decor.size=6}{bk1}
				\fmfv{decor.shape=square,decor.filled=empty,decor.angle=0,decor.size=6}{bk2}
				\fmfv{label=\tiny{1}, l.a=0}{k1}
				\fmfv{label=\tiny{2}, l.a=0}{k2}
			\end{fmfgraph*}
	\end{gathered}
	\end{align*} 
	\end{fmffile}\end{minipage} 
	& $1$ & $1$ & $1$ & $1$
	& \begin{minipage}{2cm}\begin{fmffile}{two-emission-int21}
	\fmfset{thin}{.7pt}
	\fmfset{arrow_len}{1.8mm}
	\fmfset{curly_len}{1.8mm}
	\fmfset{dash_len}{1.5mm}
	\begin{align*}
	\begin{gathered}
	\vspace{4pt}
	\begin{fmfgraph*}(60,30)
				\fmfstraight
				\fmfright{l1,lm,l2}
				\fmfleft{r1,rm,r2}
				\fmf{phantom}{l2,k1,m2,k2,r2}
				\fmf{phantom}{lm,mm,rm}
				\fmffreeze
				\fmf{plain}{l1,bl1,m1,br1,r1}
				\fmffreeze
				\fmf{plain}{m1,bm1,mm}
				\fmf{plain}{mm,bk1,k1}
				\fmf{plain}{mm,bk2,k2}
				\fmffreeze
				\fmfrectangle{4}{4}{r1}
				\fmfv{decor.shape=square,decor.filled=empty,decor.angle=0,decor.size=6}{bm1}
				\fmfv{decor.shape=square,decor.filled=empty,decor.angle=0,decor.size=6}{br1}
				\fmfv{decor.shape=square,decor.filled=empty,decor.angle=0,decor.size=6}{bl1}
				\fmfv{decor.shape=square,decor.filled=empty,decor.angle=0,decor.size=6,label=\tiny{$\perp$},label.dist=0}{bk1}
				\fmfv{decor.shape=square,decor.filled=empty,decor.angle=0,decor.size=6}{bk2}
				\fmfv{label=\tiny{1}, l.a=0}{k1}
				\fmfv{label=\tiny{2}, l.a=0}{k2}
			\end{fmfgraph*}
	\end{gathered}
	\end{align*} 
	\end{fmffile}\end{minipage} &$\lambda $ & $1$ & $ \lambda $
		\\
	\hline
	\begin{minipage}{2cm}\begin{fmffile}{two-emission-int2}
			\fmfset{thin}{.7pt}
			\fmfset{arrow_len}{1.8mm}
			\fmfset{curly_len}{1.8mm}
			\fmfset{dash_len}{1.5mm}
			\begin{align*}
				\begin{gathered}
					\vspace{4pt}
					\begin{fmfgraph*}(75,25)
						\fmfstraight
						\fmfright{l1,lm,l2}
						\fmfleft{r1,rm,r2}
						\fmf{phantom}{l2,k1,k2,r2}
						\fmffreeze
						\fmf{plain}{l1,bl1,m1,bm1,m2,br1,r1}
						\fmffreeze
						\fmf{plain}{m1,bk1,k1}
						\fmf{plain}{m2,bk2,k2}
						\fmffreeze
						\fmfrectangle{4}{4}{r1}
						\fmfv{decor.shape=square,decor.filled=empty,decor.angle=0,decor.size=6}{bm1}
						\fmfv{decor.shape=square,decor.filled=empty,decor.angle=0,decor.size=6}{br1}
						\fmfv{decor.shape=square,decor.filled=empty,decor.angle=0,decor.size=6}{bl1}
						\fmfv{decor.shape=square,decor.filled=empty,decor.angle=0,decor.size=6,label=\tiny{$\perp$},label.dist=0}{bk1}
						\fmfv{decor.shape=square,decor.filled=empty,decor.angle=0,decor.size=6}{bk2}
					\end{fmfgraph*}
				\end{gathered}
			\end{align*} 
	\end{fmffile}\end{minipage} 
	& $\lambda$ & $\lambda$ & $1$ & $1$
	\\
	\hline
				\begin{minipage}{2cm}\begin{fmffile}{two-emission-int3}
			\fmfset{thin}{.7pt}
			\fmfset{arrow_len}{1.8mm}
			\fmfset{curly_len}{1.8mm}
			\fmfset{dash_len}{1.5mm}
			\begin{align*}
				\begin{gathered}
					\vspace{4pt}
					\begin{fmfgraph*}(75,25)
						\fmfstraight
						\fmfright{l1,lm,l2}
						\fmfleft{r1,rm,r2}
						\fmf{phantom}{l2,k1,k2,r2}
						\fmffreeze
						\fmf{plain}{l1,bl1,m1,bm1,m2,br1,r1}
						\fmffreeze
						\fmf{plain}{m1,bk1,k1}
						\fmf{plain}{m2,bk2,k2}
						\fmffreeze
						\fmfrectangle{4}{4}{r1}
						\fmfv{decor.shape=square,decor.filled=empty,decor.angle=0,decor.size=6}{bm1}
						\fmfv{decor.shape=square,decor.filled=empty,decor.angle=0,decor.size=6}{br1}
						\fmfv{decor.shape=square,decor.filled=empty,decor.angle=0,decor.size=6}{bl1}
						\fmfv{decor.shape=square,decor.filled=empty,decor.angle=0,decor.size=6}{bk1}
						\fmfv{decor.shape=square,decor.filled=empty,decor.angle=0,decor.size=6,label=\tiny{$\perp$},label.dist=0}{bk2}
					\end{fmfgraph*}
				\end{gathered}
			\end{align*} 
	\end{fmffile}\end{minipage} 
	& $\lambda$ & $1$ & $\lambda$ & $1$
	\\
	\hline
					\begin{minipage}{2cm}\begin{fmffile}{two-emission-int4}
			\fmfset{thin}{.7pt}
			\fmfset{arrow_len}{1.8mm}
			\fmfset{curly_len}{1.8mm}
			\fmfset{dash_len}{1.5mm}
			\begin{align*}
				\begin{gathered}
					\vspace{4pt}
					\begin{fmfgraph*}(75,25)
						\fmfstraight
						\fmfright{l1,lm,l2}
						\fmfleft{r1,rm,r2}
						\fmf{phantom}{l2,k1,k2,r2}
						\fmffreeze
						\fmf{plain}{l1,bl1,m1,bm1,m2,br1,r1}
						\fmffreeze
						\fmf{plain}{m1,bk1,k1}
						\fmf{plain}{m2,bk2,k2}
						\fmffreeze
						\fmfrectangle{4}{4}{r1}
						\fmfv{decor.shape=square,decor.filled=empty,decor.angle=0,decor.size=6}{bm1}
						\fmfv{decor.shape=square,decor.filled=empty,decor.angle=0,decor.size=6}{br1}
						\fmfv{decor.shape=square,decor.filled=empty,decor.angle=0,decor.size=6}{bl1}
						\fmfv{decor.shape=square,decor.filled=empty,decor.angle=0,decor.size=6,label=\tiny{$\perp$},label.dist=0}{bk1}
						\fmfv{decor.shape=square,decor.filled=empty,decor.angle=0,decor.size=6,label=\tiny{$\perp$},label.dist=0}{bk2}
					\end{fmfgraph*}
				\end{gathered}
			\end{align*} 
	\end{fmffile}\end{minipage} 
	& $\lambda^2$ & $\lambda$ & $\lambda$ & $1$
	\\
		\end{tabular}
	\end{center}
	\caption{Leading numerator scaling for interferer lines with collinear vertices.}
	\label{tab:two-emissions-interferer-table}
\end{table}

 In
\tabref{tab:two-emissions-sp1}, we show the leading splitter
amplitudes with collinear vertex insertions for all relevant two
emission singular configurations, while \tabref{tab:two-emissions-sp2}
shows $(\perp)$-vertex insertions.  In
\tabref{tab:two-emissions-interferer-table} we show the respective
interferer amplitudes.  From here, one can deduce the numerator
scaling for all relevant two-emission topologies.

These tables feature several noteworthy aspects we want to mention
here.  Concerning the triple collinear limit $(i \parallel j \parallel
k)$, we see that many of the splitter amplitudes shown minimally scale
as $\lambda^2$.  All of these need to be taken into account for this
limit while for interferer lines, there is only one relevant
amplitude, namely the first one shown in
\tabref{tab:two-emissions-interferer-table}.  For self-energy
topologies, where one finds a propagator scaling of $1/\lambda^8$ for
$E^{(1)}$, $E^{(2)}$ and $E^{(3)}$ this means that the overall scaling
drops to $1/\lambda^4$ as one would expect for a contribution to a
splitting function.  Moreover, we find that these are the \emph{only}
relevant topologies for two emission splitting functions as can be
seen in \eqref{splitting-abelian} and \eqref{splitting-non-abelian}.

Instead, for the double soft limit, there is only one amplitude whose
numerator does not scale, namely the one with $(\perp)$-momentum
components on the emission lines.  This will be the only relevant
numerator structure in said limit on the amplitude side.  On the
interferer side, shown in
Tab.~\ref{tab:two-emissions-interferer-table} all amplitudes shown are
relevant, because none of them scale in the double soft-limit.
Nevertheless, none of the amplitudes with a $(\perp)$-vertex shown in
\tabref{tab:two-emissions-sp2} can contribute on the interferer line
in a double soft limit, because the inclusion of such a vertex always
adds a power of $\lambda$ to the scaling.  This can already be seen in
the one emission example, where there are only two relevant
interference contributions of \eqref{eq:interference-gg-1E} and
\eqref{eq:1E-eikonal} for the same reasons.

Another interesting finding is the appearance of amplitudes with a
black box connecting to the hard amplitude.  If such an amplitude were
to contribute in a leading singular limit, it would be a sign of a
factorisation breakdown with the reason being that the backwards
components of momenta from the hard amplitude can not be neglected.
Nevertheless, a closer inspection of said amplitude shows that in
conjunction with the scaling of the relevant partitioning and
propagator factors, they do not contribute in any leading singular
limit for two emissions.

In general, it is algorithmically an easy task to collect all
contributions relevant for a specific leading singularity.  One can
first check the scaling of partitioning times propagator factors to
find potentially relevant topologies (\ie the ones which scale as
$1/\lambda^{4}$ or worse) and then pick the corresponding amplitudes
from our lists that feature the lowest scaling in the respective
limit.  As mentioned in the one emission example, one can finally
apply the decomposition of \eqref{eqs:linearity} for interferer lines
to single out leading soft-singular contributions.
We show the scaling of the partitioned propagator factors together with their respective numerator scaling in Tab.~\ref{tab:scaling-two-emission-with-num}.
All contributions with a scaling of $1/\lambda^4$ should enter the kernel for the collinear configuration in question, in this case the $\mathbb{U}_{(ijk)}$-kernel.
Several remarks for this collection of data are in order.
Firstly, we see that only the self-energy like topologies $E$ contribute in the triple collinear limit.
For soft-collinear configurations, only a handful of topologies give rise to leading contributions, while almost all of them contribute in the double-soft limit.
An interesting difference between the partitioning types is that topology $A^{(3)}$ does not contribute in the double-soft limit for fractional partitioning while it does in the angular ordered version.
Both partitionings give rise to the same leading contributions otherwise.
\begin{table}[h]
	\begin{center}
		\begin{tabular}{c|cccc}
			& CC & CS & SC & SS  \\
			\hline
			$ A^{(1)} $ & $1/\lambda^2$ & $1/\lambda^3$ & $1/\lambda$ & \red{$1/\lambda^4$}\\
			$ A^{(2)} $ & $1/\lambda^2$ & $1/\lambda^2$ & $1/\lambda^2$ & \red{$1/\lambda^4$} \\
			$ A^{(3)} $ & $1/\lambda^2$ & $1/\lambda^2$ & $1/\lambda$ & $1/\lambda^3$ \\
			$ A^{(4)} $ & $1/\lambda^3$ & $1/\lambda^2$& $1/\lambda$ & \red{$1/\lambda^4$} \\
			$ A^{(5)} $ & $1/\lambda^3$ & $1/\lambda$ & $1/\lambda$  & \red{$1/\lambda^4$} \\
			$ B^{(1)} $ & $1/\lambda^3$ & \red{$1/\lambda^4$}& $1/\lambda^2$ & \red{$1/\lambda^4$} \\
			$ B^{(2)} $ & $1/\lambda^3$ & $1/\lambda^3$ & $1/\lambda^3$ & \red{$1/\lambda^4$} \\
			$ B^{(3)} $ & $1/\lambda^3$ & $1/\lambda^3$ & $1/\lambda^2$ & \red{$1/\lambda^4$} \\
			$ B^{(4)} $ & $1/\lambda^3$ & \red{$1/\lambda^4$}& $1/\lambda^2$ & \red{$1/\lambda^4$} \\
			$ B^{(5)} $ & $1/\lambda^3$ & $1/\lambda^3$ & $1/\lambda^3$ & \red{$1/\lambda^4$} \\
			$ B^{(6)} $ & $1/\lambda^3$ & $1/\lambda^3$ & $1/\lambda^2$ & \red{$1/\lambda^4$} \\
			$ X^{(1)} $ & $1/\lambda^2$ & $1/\lambda^2$ & \red{$1/\lambda^4$} & \red{$1/\lambda^4$} \\
			$ X^{(2)} $ & $1/\lambda^2$ & $1/\lambda^2$ & $1/\lambda^2$ & $1/\lambda^2$ \\
			$ E^{(1)} $ & \red{$1/\lambda^4$}& \red{$1/\lambda^4$} & $1/\lambda^2$ & \red{$1/\lambda^4$} \\
			$ E^{(2)} $ & \red{$1/\lambda^4$} & $1/\lambda^3$ &$1/\lambda^3$ & \red{$1/\lambda^4$}  \\
			$ E^{(3)} $ & \red{$1/\lambda^4$}& $1/\lambda^3$ & $1/\lambda^2$& \red{$1/\lambda^4$} \\
			$ E^{(4)} $ & \red{$1/\lambda^4$}& $1/\lambda^3$ & $1/\lambda^2$& \red{$1/\lambda^4$} \\
			$ E^{(5)} $ & \red{$1/\lambda^4$}& $1/\lambda^2$ & $1/\lambda^2$& \red{$1/\lambda^4$} \\			
		\end{tabular}
	\end{center}
	\caption{Scaling for propagator times partitioning factor of two emission single emitter topologies when partitioned to $(i\parallel j \parallel k)$ together with the respective numerator scaling. Here, `CC' refers to the triple collinear limit where $i||j||k$, `CS' refers to $(i \parallel j)$ with soft $k$, `SC' to $(i \parallel k)$ and $j$ soft and `SS' is the double soft limit where both $j$ and $k$ are soft. The leading terms are marked in red.}
	\label{tab:scaling-two-emission-with-num}
\end{table}

As an illustrative example for a single diagram, we discuss the power counting for the
$B^{(1)}$-topology for the emission of two gluons, \ie
\begin{fmffile}{B1-density-operator-like}
	\fmfset{thin}{.7pt}
	\fmfset{arrow_len}{1.8mm}
	\fmfset{curly_len}{1.5mm}
	\fmfset{dash_len}{1.5mm}
	\begin{align}
		B^{(1)}_{ijkl} \simeq
		\begin{gathered}
			\begin{fmfgraph*}(0,0)
			\end{fmfgraph*}
		\end{gathered}
		\begin{gathered}
			\begin{fmfgraph*}(80,50)
				\fmfstraight
				\fmfleft{lb,lm,lt}
				\fmfright{rb,rm,rt}
				\fmf{plain}{rt,b1,v1,b2,v2,b3,lt}
				\fmf{phantom}{rb,bb1,vb1,bb2,vb2,bb3,lb}
				\fmffreeze
				\fmf{phantom}{v1,bm1,e1,vb1}
				\fmf{phantom}{v2,bm2,e2,vb2}
				\fmf{curly}{v1,bm1,e1}
				\fmf{curly}{v2,bm2,e2}
				\fmfrectangle{10}{7}{rm}
				\fmfv{decor.shape=square,decor.filled=30,decor.angle=0,decor.size=6}{b1}
				\fmfv{decor.shape=square,decor.filled=30,decor.angle=0,decor.size=6}{b2}
				\fmfv{decor.shape=square,decor.filled=30,decor.angle=0,decor.size=6}{b3}
				\fmfv{decor.shape=square,decor.filled=30,decor.angle=0,decor.size=6}{bm1}
				\fmfv{decor.shape=square,decor.filled=30,decor.angle=0,decor.size=6}{bm2}
				\fmffreeze
				\fmfv{label=\tiny{$i$},label.dist=2,label.angle=180}{lt}
				\fmfv{label=\tiny{$k$},label.dist=2,label.angle=-90}{e1}
				\fmfv{label=\tiny{$j$},label.dist=2,label.angle=-90}{e2}
			\end{fmfgraph*}
		\end{gathered}
		\quad \;
		\begin{gathered}
			\begin{fmfgraph*}(80,50)
				\fmfstraight
				\fmfleft{lb,lm,lt}
				\fmfright{rb,rm,rt}
				\fmf{plain}{rt,b1,v1,b2,v2,b3,lt}
				\fmf{plain}{rb,bb1,vb1,bb2,vb2,bb3,lb}
				\fmffreeze
				\fmf{phantom}{v2,bm2,e2,vb2}
				\fmf{phantom}{v1,bm1,e1,vb1}
				\fmf{curly}{v2,bm2,e2}
				\fmf{curly}{vb1,e1,bm1}
				\fmfrectangle{10}{7}{lm}
				\fmffreeze
				\fmfv{decor.shape=square,decor.filled=30,decor.angle=0,decor.size=6}{b3}
				\fmfv{decor.shape=square,decor.filled=30,decor.angle=0,decor.size=6}{b2}
				\fmfv{decor.shape=square,decor.filled=30,decor.angle=0,decor.size=6}{bb2}
				\fmfv{decor.shape=square,decor.filled=30,decor.angle=0,decor.size=6}{bb1}
				\fmfv{decor.shape=square,decor.filled=30,decor.angle=0,decor.size=6}{e1}
				\fmfv{decor.shape=square,decor.filled=30,decor.angle=0,decor.size=6}{bm2}
				\fmffreeze
				\fmfv{label=\tiny{$j$},label.dist=2,label.angle=-90}{e2}
				\fmfv{label=\tiny{$k$},label.dist=2,label.angle=90}{bm1}
				\fmfv{label=\tiny{$l$},label.dist=2,label.angle=0}{rb}
			\end{fmfgraph*}
		\end{gathered}
		\; \propto \frac{1}{S_{ij}^2 S_{ijk} S_{kl}} \ .
	\end{align}
\end{fmffile}\\%
First, we check the scaling of the diagram's propagator times the
partitioning factors for all limits where the partons $j,k$ can become
unresolved explicitly for the partial fractioning version of the
partitioning.  An example is the triple collinear $(i\parallel j
\parallel k)$-limit.  The respective partitioning factor which
eliminates the collinear singularity w.r.t.\ parton $l$ is
\begin{align}
	\mathbb{P}_{(ijk)}^{B^{(1)}} =
	\frac{\varsigma^4 S_{k l}}{\varsigma^2 S_{k l} S_{i j k}+\varsigma^4 S_{i j k}+2 S_{i j}^2 S_{i j k}+\varsigma^4 S_{k l}} \ .
\end{align}
The configurations $c$ where $\mathbb{P}_{c}^{B^{(1)}} \times B^{(1)}$
could give rise to a leading singular contribution are shown in
\tabref{tab:PtimesB1-scaling}.
\bgroup \def\arraystretch{1.5}%
\begin{table}[h]
	\begin{center}
		\begin{tabular}{l|cccc}
			& $(i\parallel j \parallel k)$ & $(i\parallel j) (k \parallel l)$ & $(i \parallel j), k$ soft & $j,k$ soft
			\\[5pt] \hline
			$\mathbb{P}_{(ijk)}^{B^{(1)}}$ & $\frac{1}{\lambda^6}$ &  $\frac{1}{\lambda^4}$ &  $\frac{1}{\lambda^6}$ &  $\frac{1}{\lambda^4}$
			\\
			$\mathbb{P}_{(ij)(kl)}^{B^{(1)}}$ &  $\frac{1}{\lambda^4}$ &  $\frac{1}{\lambda^6}$ &  $\frac{1}{\lambda^6}$ &  $\frac{1}{\lambda^4}$
		\end{tabular}
		\caption{Leading propagator scalings of $\mathbb{P}_{c}^{B^{(1)}} \times B^{(1)}$ for four different configurations where the partons $j,k$ can become unresolved.}
		\label{tab:PtimesB1-scaling}
	\end{center}
\end{table}
\egroup Next, we check the numerator scaling, \ie the possible
insertions of subamplitudes from \tabref{tab:two-emissions-sp1} for
the amplitude side and \tabref{tab:one-emission-pc} for the conjugate
side.  In each case, the $(ij)$-collinearity leads to a scaling of at
least $\lambda^2$ for the numerator.  Additionally, the $(ik)$- and
$(kl)$-collinearities give an additional power of $\lambda$ in the
configurations where $k$ is not soft.  Then, non of the purely
collinear settings give rise to a leading singular contribution (\ie
the partitioned amplitude scales as $1/\lambda^n$ with $n<4$).  The
only relevant contributions come in the cases where $k$ is soft and
$j$ is either collinear to $i$ or also soft from a partitioning into
either $(i \parallel j \parallel k)$ or $(i\parallel j) (k\parallel
l)$ (meaning the soft limits in \tabref{tab:PtimesB1-scaling}).  When
applying $\mathbb{P}_{(ijk)}^{B^{(1)}}$ in the double soft limit,
there are only two numerator structures which have no
$\lambda$-scaling and will therefore give rise to a leading
contribution, \viz
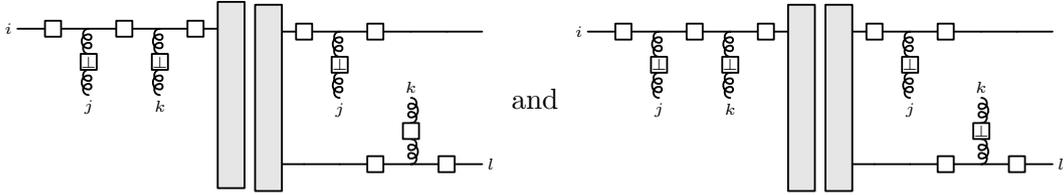
\vspace{8pt}
\begin{fmffile}{B2-density-operator-like-leading-soft}
	\fmfset{thin}{.7pt}
	\fmfset{arrow_len}{1.8mm}
	\fmfset{curly_len}{1.5mm}
	\fmfset{dash_len}{1.5mm}
	\begin{align*}
		\begin{gathered}
			\begin{fmfgraph*}(0,0)
			\end{fmfgraph*}
		\end{gathered}
		\begin{gathered}
			\begin{fmfgraph*}(80,50)
				\fmfstraight
				\fmfleft{lb,lm,lt}
				\fmfright{rb,rm,rt}
				\fmf{plain}{rt,b1,v1,b2,v2,b3,lt}
				\fmf{phantom}{rb,vb1,vb2,lb}
				\fmffreeze
				\fmf{phantom}{v1,bm1,e1,vb1}
				\fmf{phantom}{v2,bm2,e2,vb2}
				\fmf{curly}{v1,bm1,e1}
				\fmf{curly}{v2,bm2,e2}
				\fmfrectangle{10}{7}{rm}
				\fmfv{decor.shape=square,decor.filled=empty,decor.angle=0,decor.size=6}{b1}
				\fmfv{decor.shape=square,decor.filled=empty,decor.angle=0,decor.size=6}{b2}
				\fmfv{decor.shape=square,decor.filled=empty,decor.angle=0,decor.size=6}{b3}
				\fmfv{decor.shape=square,decor.filled=empty,decor.angle=0,decor.size=6,label=\tiny{$\perp$},label.dist=0}{bm1}
				\fmfv{decor.shape=square,decor.filled=empty,decor.angle=0,decor.size=6,label=\tiny{$\perp$},label.dist=0}{bm2}
				\fmfv{label=\tiny{$i$},label.dist=2,label.angle=180}{lt}
				\fmfv{label=\tiny{$k$},label.dist=2,label.angle=-90}{e1}
				\fmfv{label=\tiny{$j$},label.dist=2,label.angle=-90}{e2}
			\end{fmfgraph*}
		\end{gathered}
		\quad \;
		\begin{gathered}
			\begin{fmfgraph*}(80,50)
				\fmfstraight
				\fmfleft{lb,lm,lt}
				\fmfright{rb,rm,rt}
				\fmf{plain}{lt,b1,v1,b2,v2,b3,rt}
				\fmf{plain}{lb,bb1,vb1,bb2,vb2,bb3,rb}
				\fmffreeze
				\fmf{phantom}{v1,bm1,e1,vb1}
				\fmf{phantom}{v2,e2,bm2,vb2}
				\fmf{curly}{v1,bm1,e1}
				\fmf{curly}{vb2,bm2,e2}
				\fmfrectangle{10}{7}{lm}
				\fmfv{decor.shape=square,decor.filled=empty,decor.angle=0,decor.size=6}{b1}
				\fmfv{decor.shape=square,decor.filled=empty,decor.angle=0,decor.size=6}{b2}
				\fmfv{decor.shape=square,decor.filled=empty,decor.angle=0,decor.size=6}{bb2}
				\fmfv{decor.shape=square,decor.filled=empty,decor.angle=0,decor.size=6}{bb3}
				\fmfv{decor.shape=square,decor.filled=empty,decor.angle=0,decor.size=6,label=\tiny{$\perp$},label.dist=0}{bm1}
				\fmfv{decor.shape=square,decor.filled=empty,decor.angle=0,decor.size=6}{bm2}
				\fmfv{label=\tiny{$j$},label.dist=2,label.angle=-90}{e1}
				\fmfv{label=\tiny{$k$},label.dist=2,label.angle=90}{e2}
				\fmfv{label=\tiny{$l$},label.dist=2,label.angle=0}{rb}
			\end{fmfgraph*}
		\end{gathered}
		\quad \text{and} \quad 
		\begin{gathered}
			\begin{fmfgraph*}(80,50)
				\fmfstraight
				\fmfleft{lb,lm,lt}
				\fmfright{rb,rm,rt}
				\fmf{plain}{rt,b1,v1,b2,v2,b3,lt}
				\fmf{phantom}{rb,vb1,vb2,lb}
				\fmffreeze
				\fmf{phantom}{v1,bm1,e1,vb1}
				\fmf{phantom}{v2,bm2,e2,vb2}
				\fmf{curly}{v1,bm1,e1}
				\fmf{curly}{v2,bm2,e2}
				\fmfrectangle{10}{7}{rm}
				\fmfv{decor.shape=square,decor.filled=empty,decor.angle=0,decor.size=6}{b1}
				\fmfv{decor.shape=square,decor.filled=empty,decor.angle=0,decor.size=6}{b2}
				\fmfv{decor.shape=square,decor.filled=empty,decor.angle=0,decor.size=6}{b3}
				\fmfv{decor.shape=square,decor.filled=empty,decor.angle=0,decor.size=6,label=\tiny{$\perp$},label.dist=0}{bm1}
				\fmfv{decor.shape=square,decor.filled=empty,decor.angle=0,decor.size=6,label=\tiny{$\perp$},label.dist=0}{bm2}
				\fmfv{label=\tiny{$i$},label.dist=2,label.angle=180}{lt}
				\fmfv{label=\tiny{$k$},label.dist=2,label.angle=-90}{e1}
				\fmfv{label=\tiny{$j$},label.dist=2,label.angle=-90}{e2}
			\end{fmfgraph*}
		\end{gathered}
		\quad \;
		\begin{gathered}
			\begin{fmfgraph*}(80,50)
				\fmfstraight
				\fmfleft{lb,lm,lt}
				\fmfright{rb,rm,rt}
				\fmf{plain}{lt,b1,v1,b2,v2,b3,rt}
				\fmf{plain}{lb,bb1,vb1,bb2,vb2,bb3,rb}
				\fmffreeze
				\fmf{phantom}{v1,bm1,e1,vb1}
				\fmf{phantom}{v2,e2,bm2,vb2}
				\fmf{curly}{v1,bm1,e1}
				\fmf{curly}{vb2,bm2,e2}
				\fmfrectangle{10}{7}{lm}
				\fmfv{decor.shape=square,decor.filled=empty,decor.angle=0,decor.size=6}{b1}
				\fmfv{decor.shape=square,decor.filled=empty,decor.angle=0,decor.size=6}{b2}
				\fmfv{decor.shape=square,decor.filled=empty,decor.angle=0,decor.size=6}{bb2}
				\fmfv{decor.shape=square,decor.filled=empty,decor.angle=0,decor.size=6}{bb3}
				\fmfv{decor.shape=square,decor.filled=empty,decor.angle=0,decor.size=6,label=\tiny{$\perp$},label.dist=0}{bm1}
				\fmfv{decor.shape=square,decor.filled=empty,decor.angle=0,decor.size=6,label=\tiny{$\perp$},label.dist=0}{bm2}
				\fmfv{label=\tiny{$j$},label.dist=2,label.angle=-90}{e1}
				\fmfv{label=\tiny{$k$},label.dist=2,label.angle=90}{e2}
				\fmfv{label=\tiny{$l$},label.dist=2,label.angle=0}{rb}
			\end{fmfgraph*}
		\end{gathered}
	\end{align*}
\end{fmffile}\\
The only other leading contributions appear in the
$(i\parallel j)$ and $k$ soft configuration.  The relevant numerator
structures here are the ones from above and the combinations where the
$(\perp)$-boxes of emission $j$ are moved to one of the nearest
neighbour boxes.

In conclusion, we find that, as apparent already from its topology,
$B^{(1)}$ contributes leadingly only in the soft-collinear and double
soft limits with its only relevant partitions being
$\mathbb{P}^{B^{(1)}}_{(ijk)}$ and $\mathbb{P}^{B^{(1)}}_{(ij)(kl)}$.

Two other interesting examples are the $B^{(2)}_{ijkl}$- and
$X^{(1)}_{ijkl}$-topologies, \ie
\begin{fmffile}{B2-density-operator-like}
	\fmfset{thin}{.7pt}
	\fmfset{arrow_len}{1.8mm}
	\fmfset{curly_len}{1.5mm}
	\fmfset{dash_len}{1.5mm}
	\begin{align}
		B^{(2)}_{ijkl} \simeq \;
		\begin{gathered}
			\begin{fmfgraph*}(0,0)
			\end{fmfgraph*}
		\end{gathered}
		\begin{gathered}
			\begin{fmfgraph*}(80,50)
				\fmfstraight
				\fmfleft{lb,lm,lt}
				\fmfright{rb,rm,rt}
				\fmf{plain}{rt,b1,v1,b2,v2,b3,lt}
				\fmf{phantom}{rb,bb1,vb1,bb2,vb2,bb3,lb}
				\fmffreeze
				\fmf{phantom}{v1,bm1,e1,vb1}
				\fmf{phantom}{v2,bm2,e2,vb2}
				\fmf{curly}{v1,bm1,e1}
				\fmf{curly}{v2,bm2,e2}
				\fmfrectangle{10}{7}{rm}
				\fmfv{decor.shape=square,decor.filled=30,decor.angle=0,decor.size=6}{b1}
				\fmfv{decor.shape=square,decor.filled=30,decor.angle=0,decor.size=6}{b2}
				\fmfv{decor.shape=square,decor.filled=30,decor.angle=0,decor.size=6}{b3}
				\fmfv{decor.shape=square,decor.filled=30,decor.angle=0,decor.size=6}{bm1}
				\fmfv{decor.shape=square,decor.filled=30,decor.angle=0,decor.size=6}{bm2}
				\fmffreeze
				\fmfv{label=\tiny{$i$},label.dist=2,label.angle=180}{lt}
				\fmfv{label=\tiny{$k$},label.dist=2,label.angle=-90}{e1}
				\fmfv{label=\tiny{$j$},label.dist=2,label.angle=-90}{e2}
			\end{fmfgraph*}
		\end{gathered}
		\quad \;
		\begin{gathered}
			\begin{fmfgraph*}(80,50)
				\fmfstraight
				\fmfleft{lb,lm,lt}
				\fmfright{rb,rm,rt}
				\fmf{plain}{rt,b1,v1,b2,v2,b3,lt}
				\fmf{plain}{rb,bb1,vb1,bb2,vb2,bb3,lb}
				\fmffreeze
				\fmf{phantom}{v2,bm2,e2,vb2}
				\fmf{phantom}{v1,bm1,e1,vb1}
				\fmf{curly}{v2,bm2,e2}
				\fmf{curly}{vb1,e1,bm1}
				\fmfrectangle{10}{7}{lm}
				\fmffreeze
				\fmfv{decor.shape=square,decor.filled=30,decor.angle=0,decor.size=6}{b3}
				\fmfv{decor.shape=square,decor.filled=30,decor.angle=0,decor.size=6}{b2}
				\fmfv{decor.shape=square,decor.filled=30,decor.angle=0,decor.size=6}{bb2}
				\fmfv{decor.shape=square,decor.filled=30,decor.angle=0,decor.size=6}{bb1}
				\fmfv{decor.shape=square,decor.filled=30,decor.angle=0,decor.size=6}{e1}
				\fmfv{decor.shape=square,decor.filled=30,decor.angle=0,decor.size=6}{bm2}
				\fmffreeze
				\fmfv{label=\tiny{$k$},label.dist=2,label.angle=-90}{e2}
				\fmfv{label=\tiny{$j$},label.dist=2,label.angle=90}{bm1}
				\fmfv{label=\tiny{$l$},label.dist=2,label.angle=0}{rb}
			\end{fmfgraph*}
		\end{gathered}
		\; \propto \frac{1}{S_{ij} S_{ik} S_{ijk} S_{jl}} \ ,
	\end{align}
\end{fmffile}
\vspace{5pt}
and
\begin{fmffile}{X-density-operator-like}
	\fmfset{thin}{.7pt}
	\fmfset{arrow_len}{1.8mm}
	\fmfset{curly_len}{1.5mm}
	\fmfset{dash_len}{1.5mm}
	\begin{align}
		X_{ijkl} \simeq \;
		\begin{gathered}
			\begin{fmfgraph*}(0,0)
			\end{fmfgraph*}
		\end{gathered}
		\begin{gathered}
			\begin{fmfgraph*}(80,50)
				\fmfstraight
				\fmfright{lb,lm,lt}
				\fmfleft{rb,rm,rt}
				\fmf{plain}{rt,b1,v1,b2,v2,b3,lt}
				\fmf{plain}{rb,bb1,vb1,bb2,vb2,bb3,lb}
				\fmffreeze
				\fmf{phantom}{v2,bm2,e2,vb2}
				\fmf{phantom}{v1,bm1,e1,vb1}
				\fmf{curly}{v2,bm2,e2}
				\fmf{curly}{vb1,e1,bm1}
				\fmfrectangle{10}{7}{lm}
				\fmffreeze
				\fmfv{decor.shape=square,decor.filled=30,decor.angle=0,decor.size=6}{b3}
				\fmfv{decor.shape=square,decor.filled=30,decor.angle=0,decor.size=6}{b2}
				\fmfv{decor.shape=square,decor.filled=30,decor.angle=0,decor.size=6}{bb2}
				\fmfv{decor.shape=square,decor.filled=30,decor.angle=0,decor.size=6}{bb1}
				\fmfv{decor.shape=square,decor.filled=30,decor.angle=0,decor.size=6}{e1}
				\fmfv{decor.shape=square,decor.filled=30,decor.angle=0,decor.size=6}{bm2}
				\fmffreeze
				\fmfv{label=\tiny{$j$},label.dist=2,label.angle=-90}{e2}
				\fmfv{label=\tiny{$k$},label.dist=2,label.angle=90}{bm1}
				\fmfv{label=\tiny{$l$},label.dist=2,label.angle=180}{rb}
			\end{fmfgraph*}
		\end{gathered}
		\quad \;
		\begin{gathered}
			\begin{fmfgraph*}(80,50)
				\fmfstraight
				\fmfleft{lb,lm,lt}
				\fmfright{rb,rm,rt}
				\fmf{plain}{rt,b1,v1,b2,v2,b3,lt}
				\fmf{plain}{rb,bb1,vb1,bb2,vb2,bb3,lb}
				\fmffreeze
				\fmf{phantom}{v2,bm2,e2,vb2}
				\fmf{phantom}{v1,bm1,e1,vb1}
				\fmf{curly}{v2,bm2,e2}
				\fmf{curly}{vb1,e1,bm1}
				\fmfrectangle{10}{7}{lm}
				\fmffreeze
				\fmfv{decor.shape=square,decor.filled=30,decor.angle=0,decor.size=6}{b3}
				\fmfv{decor.shape=square,decor.filled=30,decor.angle=0,decor.size=6}{b2}
				\fmfv{decor.shape=square,decor.filled=30,decor.angle=0,decor.size=6}{bb2}
				\fmfv{decor.shape=square,decor.filled=30,decor.angle=0,decor.size=6}{bb1}
				\fmfv{decor.shape=square,decor.filled=30,decor.angle=0,decor.size=6}{e1}
				\fmfv{decor.shape=square,decor.filled=30,decor.angle=0,decor.size=6}{bm2}
				\fmffreeze
				\fmfv{label=\tiny{$k$},label.dist=2,label.angle=-90}{e2}
				\fmfv{label=\tiny{$j$},label.dist=2,label.angle=90}{bm1}
				\fmfv{label=\tiny{$l$},label.dist=2,label.angle=0}{rb}
			\end{fmfgraph*}
		\end{gathered}
		\; \propto \frac{1}{S_{ij} S_{ik} S_{jl} S_{kl}} \ .
	\end{align}
\end{fmffile}\\
For $B^{(2)}_{ijkl}$, the same numerator power counting
applies as for $B^{(1)}_{ijkl}$.  By checking the scaling of its
propagator factors times the partitioning, we find that \emph{only}
leading singular contribution appears in the double soft limit in
$\mathbb{P}^{B^{(2)}}_{(ijk)} B^{(2)}_{ijkl}$.  By the same logic, we
find that $X^{(1)}_{ijkl}$ exclusively contributes in the double soft limit.
The reason in this case is that each partition of $X^{(1)}_{ijkl}$ is at
most $1/\lambda^4$-singular.  In any collinear or soft-collinear
configuration, the numerator will contribute at least one power of
$\lambda$, making these contributions subleading.  In the double-soft
limit though, each partition (except for the one into $(il)(jk)$) will
contribute equally with the respective leading numerator structures
being of $\mathcal{O}(1)$.

\section{Applications and Outlook}
\label{sec:Applications}

In the previous section we have been outlining how one can, given a
partitioning of the soft radiation into different collinear sectors,
systematically factor the leading behaviour of multi-leg
amplitudes. Taking this as a starting point, several applications are
in reach: At a fixed order in perturbation theory, one can simply
square the amplitude thus obtained, and derive an interpolating
formula for the singular behaviour of a fixed number of emissions.
While it was not our primary goal to use this as a subtraction
term (and we have thus possibly not created functions which are most
easily integrated analytically), one can still use the results to
migrate certain contributions in between real emission and virtual
corrections if there is no immediate need for an analytic integration
\eg within the context of the loop-tree duality local subtractions
\cite{Bierenbaum:2010xg,Driencourt-Mangin:2019yhu} and similar
approaches \cite{Platzer:2020lbr}.

Within the context of a specific (class of) observables, which exhibit
a definite perturbation around a set of hard jets, our findings are a
vital input to a resummation programme. In particular, when using
azimuthal averaging, and our partitioning which generalizes the
subtractions behind angular ordering, we expect that we can obtain a
generalization of the coherent branching formalism beyond the usual
next-to-leading logarithmic accurate algorithm for global event
shapes. Colour correlations can directly be addressed within our
analysis and more recent approaches to simplify the structure of
colour in the soft limit might be vital to supplement such a
formalism, which we will address in future work. An immediate
difference of our approach is that we have deliberately not chosen to
analytically move soft-collinear contributions in between different
classes of diagrams. This implies that the colour-diagonal
contributions which involve the same colour charge acting in the
amplitude and its conjugate, will still deliver the full soft- and
hard-collinear singularity. They do not need to rely on the soft,
colour-correlated contribution to collapse into the collinear
singularity upon relying on colour conservation.  This might be
advantageous in a numerical implementation, but also implies that our
non-trivially colour correlated contributions truly describe the
effect of large-angle soft radiation, which could thus be separated in
an analytic way; this should be confronted with the approaches of
collinear subtractions outlined in \cite{Forshaw:2019ver}. Within the
same parton-branching at the amplitude level approach, our analysis
can be exploited to derive splitting kernels beyond the limit of
iterated singly-unresolved emissions in order to build up the leading
behaviour of an amplitude with many legs. In this case, we will be
able to extend the approaches of doubly-unresolved soft radiation
\cite{Platzer:2020lbr} to include soft- and hard-collinear limits and
thus arrive at a more general algorithm.  This is the primary scope of
our work, together with the generalized angular-ordered partitioning,
as well as the flexible recoil schemes.  These allow us to analyse the
effect in comparison to previous work in this direction at the level
of existing parton branching algorithms \cite{Forshaw:2020wrq} for
which we are convinced that there will be a deeper link established
between recoil, partitioning, the form of the evolution kernel and
the accuracy of an overall resulting parton branching algorithm.

To be definite, we here give expressions for the factoring vertices in
terms of cutting apart the internal lines using the definition of our
projector operators in \eqref{eq:projectors-general}, \ie we do
contract the vertices which are dressed up with the respective hard,
transverse or backward propagator components with polarization vectors
and spinors in the respective collinear sectors.  From this we obtain
(complex) weights for each colour structure, which we can then iterate
in a amplitude-level Monte Carlo algorithm.  This would in turn,
besides the colour quantum numbers, also sample spin quantum
numbers. The procedure is sketched below in the case of a gluon
emission attaching to a quark splitter and quark spectator line,
where the $\lambda$ and $\bar{\lambda}$ refer to the helicities
on the amplitude and conjugate amplitude side, respectively:
\vspace{5pt}
\begin{fmffile}{projector-insertion-example}
	\fmfset{thin}{.7pt}
	\fmfset{dot_len}{1.2mm}
	\fmfset{dot_size}{3}
	\fmfset{arrow_len}{2.5mm}
	\fmfset{curly_len}{1.5mm}
	\begin{align}
		\begin{gathered}
			\begin{fmfgraph*}(60,50)
				\fmfstraight
				\fmfright{lb,l,lt}
				\fmfleft{rb,r,rt}
				\fmf{phantom}{lt,b1,p,bp,v,b2,rt}
				\fmffreeze
				\fmf{plain}{lt,b1}
				\fmf{plain}{bp,rt}
				\fmf{phantom}{lb,b1b,pb,bpb,vb,b2b,rb}
				\fmf{curly}{v,bb,vb}
				\fmfv{decor.shape=square,decor.filled=empty,decor.size=6}{b1}
				\fmfv{decor.shape=square,decor.filled=empty,decor.size=6,label=\tiny{$\perp$},label.dist=0}{b2}
				\fmfv{decor.shape=square,decor.filled=empty,decor.size=6}{bp}
				\fmfv{decor.shape=square,decor.filled=empty,decor.size=6}{bb}
				\fmfv{label=\small{$\mathbf{P}$},label.dist=0}{p}
				\fmfv{label=\small{$\mathbf{P}$},label.dist=5,label.angle=180}{rt}
				\fmfv{label=\small{$\mathbf{P}$},label.dist=5,label.angle=-90}{vb}
				\fmfrectangle{8}{8}{l}
			\end{fmfgraph*}
		\end{gathered}
		\quad \;
		\begin{gathered}
			\begin{fmfgraph*}(60,50)
				\fmfstraight
				\fmfleft{lb,l,lt}
				\fmfright{rb,r,rt}
				\fmf{phantom}{lb,b1,p,bp,v,b2,rb}
				\fmffreeze
				\fmf{plain}{lb,b1}
				\fmf{plain}{bp,rb}
				\fmf{phantom}{lt,b1b,pb,bpb,vb,b2b,rt}
				\fmf{curly}{v,bb,vb}
				\fmfv{decor.shape=square,decor.filled=empty,decor.size=6,label.dist=0}{b1}
				\fmfv{decor.shape=square,decor.filled=empty,decor.size=6}{b2}
				\fmfv{decor.shape=square,decor.filled=empty,decor.size=6}{bp}
				\fmfv{decor.shape=square,decor.filled=empty,decor.size=6}{bb}
				\fmfv{label=\small{$\mathbf{P}$},label.dist=0}{p}
				\fmfv{label=\small{$\mathbf{P}$},label.dist=5,label.angle=0}{rb}
				\fmfv{label=\small{$\mathbf{P}$},label.dist=5,label.angle=90}{vb}
				\fmfrectangle{8}{8}{l}
			\end{fmfgraph*}
		\end{gathered}
		\quad \rightarrow \sum \limits_{\lambda_i, \bar{\lambda}_i} &\frac{u_{\lambda_1}}{\sqrt{2 n\scdot p_i}} 
		\Big[ \frac{\bar{u}_{\lambda_1}}{\sqrt{2 n\scdot p_i}} \slashed{k}_\perp \slashed{\epsilon}_{\lambda_3} \slashed{p}_i \frac{u_{\lambda_2}}{\sqrt{2 n\scdot p_i}}\Big]
		\frac{\bar{u}_{\lambda_2}}{\sqrt{2 n\scdot p_i}} \epsilon^\sigma_{\lambda_3} \nonumber \\
		\times &\frac{u_{\bar\lambda_1}}{\sqrt{2 n\scdot p_i}} 
		\Big[ \frac{\bar{u}_{\bar\lambda_1}}{\sqrt{2 n\scdot p_k}} \slashed{p}_k  \frac{u_{\bar\lambda_2}}{\sqrt{2 n\scdot p_k}} \, p_k\cdot \epsilon_{\bar{\lambda}_3} \Big] 
		\frac{\bar{u}_{\bar\lambda_2}}{\sqrt{2 n\scdot p_k}} \epsilon_{\bar{\sigma},{\bar\lambda_3}}.
	\end{align}
\end{fmffile}%

At the end of an evolution built up this way, the amplitude would then
be squared by evaluating the matrix element of the final projector
corresponding to the most external lines we have been starting with in
the first place, and our decomposition and normalization guarantees
that there is no need to take into account additional factors. The
other vertices we encounter in the case of a single and double
emission are outlined below.  Here and in the following, the symbol
``$\simeq$'' stands for only showing the scalar emission quantity that
factorizes to the hard amplitude while leaving the rest of the
polarization sum and the hard amplitude itself implicit.
The relevant expressions for a quark-gluon splitting are
\vspace{5pt}
\begin{fmffile}{quark-vertex-rules-projectors}
	\fmfset{thin}{.7pt}
	\fmfset{dot_len}{1.2mm}
	\fmfset{dot_size}{5}
	\fmfset{arrow_len}{2.5mm}
	\fmfset{curly_len}{1.5mm}
	\begin{subequations}
		\begin{align}
			\begin{gathered}
				\begin{fmfgraph*}(60,30)
					\fmfstraight
					\fmfright{lb,l,lt}
					\fmfleft{rb,r,rt}
					\fmf{phantom}{lt,b1,p,bp,v,b2,rt}
					\fmffreeze
					\fmf{plain}{lt,b1}
					\fmf{plain}{bp,rt}
					\fmf{phantom}{lb,b1b,pb,bpb,vb,b2b,rb}
					\fmf{curly}{v,bb,vb}
					\fmfv{decor.shape=square,decor.filled=empty,decor.size=6}{b1}
					\fmfv{decor.shape=square,decor.filled=empty,decor.size=6,label=\tiny{$\perp$},label.dist=0}{b2}
					\fmfv{decor.shape=square,decor.filled=empty,decor.size=6}{bp}
					\fmfv{decor.shape=square,decor.filled=empty,decor.size=6}{bb}
					\fmfv{label=\small{$\mathbf{P}$},label.dist=0}{p}
					\fmfv{label=\small{$\mathbf{P}$},label.dist=5,label.angle=180}{rt}
					\fmfv{label=\small{$\mathbf{P}$},label.dist=5,label.angle=-90}{vb}
					\fmfrectangle{4}{4}{lt}
				\end{fmfgraph*}
			\end{gathered}
			\quad & \simeq - \frac{g_s}{S_{ij}}\, \sqrt{\frac{z_i + z_j}{z_i}}
			\frac{\bar{u}_{\lambda_1}}{\sqrt{2 n\scdot p_i}}
			\left(\slashed{k}_{\perp,i} \slashed{\epsilon}_{\lambda_3} \slashed{p}_i\right)
			\frac{u_{\lambda_2}}{\sqrt{2 n\scdot p_i}},
			\\[10pt]
			\begin{gathered}
				\begin{fmfgraph*}(60,30)
					\fmfstraight
					\fmfright{lb,l,lt}
					\fmfleft{rb,r,rt}
					\fmf{phantom}{lt,b1,p,bp,v,b2,rt}
					\fmffreeze
					\fmf{plain}{lt,b1}
					\fmf{plain}{bp,rt}
					\fmf{phantom}{lb,b1b,pb,bpb,vb,b2b,rb}
					\fmf{curly}{v,bb,vb}
					\fmfv{decor.shape=square,decor.filled=empty,decor.size=6}{b1}
					\fmfv{decor.shape=square,decor.filled=empty,decor.size=6}{b2}
					\fmfv{decor.shape=square,decor.filled=empty,decor.size=6}{bp}
					\fmfv{decor.shape=square,decor.filled=empty,decor.size=6,label=\tiny{$\perp$},label.dist=0}{bb}
					\fmfv{label=\small{$\mathbf{P}$},label.dist=0}{p}
					\fmfv{label=\small{$\mathbf{P}$},label.dist=5,label.angle=180}{rt}
					\fmfv{label=\small{$\mathbf{P}$},label.dist=5,label.angle=-90}{vb}
					\fmfrectangle{4}{4}{lt}
				\end{fmfgraph*}
			\end{gathered}
			\quad & \simeq 2\frac{g_s}{S_{ij}}\, \frac{\sqrt{z_i(z_i + z_j)}}{z_j}
			\frac{\bar{u}_{\lambda_1}}{\sqrt{2 n\scdot p_i}}
			\,\slashed{p}_i\,
			\frac{u_{\lambda_2}}{\sqrt{2 n\scdot p_i}} \, k_{\perp,j}\cdot \epsilon_{\lambda_3},
			\\[10pt]
			\begin{gathered}
				\begin{fmfgraph*}(60,30)
					\fmfstraight
					\fmfright{rt,l,rb}
					\fmfleft{lt,r,lb}
					\fmf{phantom}{lt,b1,p,bp,v,b2,rt}
					\fmffreeze
					\fmf{plain}{lt,b1}
					\fmf{plain}{bp,rt}
					\fmf{phantom}{lb,b1b,pb,bpb,vb,b2b,rb}
					\fmf{curly}{v,bb,vb}
					\fmfv{decor.shape=square,decor.filled=empty,decor.size=6}{b1}
					\fmfv{decor.shape=square,decor.filled=empty,decor.size=6}{b2}
					\fmfv{decor.shape=square,decor.filled=empty,decor.size=6}{bp}
					\fmfv{decor.shape=square,decor.filled=empty,decor.size=6}{bb}
					\fmfv{label=\small{$\mathbf{P}$},label.dist=0}{p}
					\fmfv{label=\small{$\mathbf{P}$},label.dist=5,label.angle=0}{rt}
					\fmfv{label=\small{$\mathbf{P}$},label.dist=5,label.angle=90}{vb}
					\fmfrectangle{4}{4}{lt}
				\end{fmfgraph*}
			\end{gathered}
			\quad & \simeq -2\frac{g_s}{S_{jk}}\, z_k \,
			\frac{\bar{u}_{\bar{\lambda}_1}}{\sqrt{2 n\scdot p_k}}
			\left(\slashed{p}_k\, p_k\cdot \epsilon_{\bar{\lambda}_3} \right)
			\frac{u_{\bar{\lambda}_2}}{\sqrt{2 n\scdot p_k}} ,
			\\[10pt]
			\begin{gathered}
				\begin{fmfgraph*}(60,30)
					\fmfstraight
					\fmfright{rt,l,rb}
					\fmfleft{lt,r,lb}
					\fmf{phantom}{lt,b1,p,bp,v,b2,rt}
					\fmffreeze
					\fmf{plain}{lt,b1}
					\fmf{plain}{bp,rt}
					\fmf{phantom}{lb,b1b,pb,bpb,vb,b2b,rb}
					\fmf{curly}{v,bb,vb}
					\fmfv{decor.shape=square,decor.filled=empty,decor.size=6}{b1}
					\fmfv{decor.shape=square,decor.filled=empty,decor.size=6}{b2}
					\fmfv{decor.shape=square,decor.filled=empty,decor.size=6}{bp}
					\fmfv{decor.shape=square,decor.filled=empty,decor.size=6,label=\tiny{$\perp$},label.dist=0}{bb}
					\fmfv{label=\small{$\mathbf{P}$},label.dist=0}{p}
					\fmfv{label=\small{$\mathbf{P}$},label.dist=5,label.angle=0}{rt}
					\fmfv{label=\small{$\mathbf{P}$},label.dist=5,label.angle=90}{vb}
					\fmfrectangle{4}{4}{lt}
				\end{fmfgraph*}
			\end{gathered}
			\quad & \simeq 2\frac{g_s}{S_{jk}}\, \frac{z_k}{z_j} \frac{n\scdot p_k}{n \scdot p_i}\,
			\frac{\bar{u}_{\bar{\lambda}_1}}{\sqrt{2 n\scdot p_k}}
			\left(\slashed{p}_k k_{\perp,j} \cdot \epsilon_{\bar{\lambda}_3}\right)
			\frac{u_{\bar{\lambda}_2}}{\sqrt{2 n\scdot p_k}} \,. \\[10pt] \nonumber
		\end{align}
	\end{subequations}
\end{fmffile}%

The amplitudes for a gluon-gluon splitting read
\vspace{5pt}
\begin{fmffile}{gluon-vertex-rules-projectors}
	\fmfset{thin}{.7pt}
	\fmfset{dot_len}{1.2mm}
	\fmfset{dot_size}{5}
	\fmfset{arrow_len}{2.5mm}
	\fmfset{curly_len}{1.5mm}
	\begin{subequations}
		\begin{align}
			\begin{gathered}
				\begin{fmfgraph*}(80,40)
					\fmfstraight
					\fmfright{lb,l,lt}
					\fmfleft{rb,r,rt}
					\fmf{phantom}{lt,b1,p,bp,v,b2,rt}
					\fmffreeze
					\fmf{curly}{b1,lt}
					\fmf{curly}{rt,bp}
					\fmf{phantom}{lb,b1b,pb,bpb,vb,b2b,rb}
					\fmf{curly}{v,bb,vb}
					\fmfv{decor.shape=square,decor.filled=empty,decor.size=6}{b1}
					\fmfv{decor.shape=square,decor.filled=empty,decor.size=6,label=\tiny{$\perp$},label.dist=0}{b2}
					\fmfv{decor.shape=square,decor.filled=empty,decor.size=6}{bp}
					\fmfv{decor.shape=square,decor.filled=empty,decor.size=6}{bb}
					\fmfv{label=\small{$\mathbf{P}$},label.dist=0}{p}
					\fmfv{label=\small{$\mathbf{P}$},label.dist=5,label.angle=180}{rt}
					\fmfv{label=\small{$\mathbf{P}$},label.dist=5,label.angle=-90}{vb}
					\fmfv{decor.shape=circle,decor.filled=empty,decor.size=12,label=\tiny{$\parallel$},label.dist=0}{v}
					\fmfrectangle{4}{4}{lt}
				\end{fmfgraph*}
			\end{gathered}
		 	\quad & \simeq  \frac{g_s \mathbf{T}_i}{S_{ij}}\, \frac{z_i+ 2z_j}{z_i} \left(k_{\perp,i}\cdot \epsilon^{(i)}_{\lambda_1}\right)  \delta_{\lambda_2 \lambda_3} ,
			\\[10pt]
			\begin{gathered}
				\begin{fmfgraph*}(80,40)
					\fmfstraight
					\fmfright{lb,l,lt}
					\fmfleft{rb,r,rt}
					\fmf{phantom}{lt,b1,p,bp,v,b2,rt}
					\fmffreeze
					\fmf{curly}{b1,lt}
					\fmf{curly}{rt,bp}
					\fmf{phantom}{lb,b1b,pb,bpb,vb,b2b,rb}
					\fmf{curly}{v,bb,vb}
					\fmfv{decor.shape=square,decor.filled=empty,decor.size=6}{b1}
					\fmfv{decor.shape=square,decor.filled=empty,decor.size=6}{b2}
					\fmfv{decor.shape=square,decor.filled=empty,decor.size=6}{bp}
					\fmfv{decor.shape=square,decor.filled=empty,decor.size=6,label=\tiny{$\perp$},label.dist=0}{bb}
					\fmfv{label=\small{$\mathbf{P}$},label.dist=0}{p}
					\fmfv{label=\small{$\mathbf{P}$},label.dist=5,label.angle=180}{rt}
					\fmfv{label=\small{$\mathbf{P}$},label.dist=5,label.angle=-90}{vb}
					\fmfv{decor.shape=circle,decor.filled=empty,decor.size=12,label=\tiny{$\parallel$},label.dist=0}{v}
					\fmfrectangle{4}{4}{lt}
				\end{fmfgraph*}
			\end{gathered}
			\quad & \simeq - \frac{g_s \mathbf{T}_i}{S_{ij}}\, \frac{2z_i+ z_j}{z_j} \left(k_{\perp,j}\cdot \epsilon^{(i)}_{\lambda_2}\right) \delta_{\lambda_1 \lambda_3},
			\\[10pt]
			\begin{gathered}
				\begin{fmfgraph*}(80,40)
					\fmfstraight
					\fmfright{lb,l,lt}
					\fmfleft{rb,r,rt}
					\fmf{phantom}{lt,b1,p,bp,v,b2,rt}
					\fmffreeze
					\fmf{curly}{b1,lt}
					\fmf{curly}{rt,bp}
					\fmf{phantom}{lb,b1b,pb,bpb,vb,b2b,rb}
					\fmf{curly}{v,bb,vb}
					\fmfv{decor.shape=square,decor.filled=empty,decor.size=6}{b1}
					\fmfv{decor.shape=square,decor.filled=empty,decor.size=6}{b2}
					\fmfv{decor.shape=square,decor.filled=empty,decor.size=6}{bp}
					\fmfv{decor.shape=square,decor.filled=empty,decor.size=6}{bb}
					\fmfv{label=\small{$\mathbf{P}$},label.dist=0}{p}
					\fmfv{label=\small{$\mathbf{P}$},label.dist=5,label.angle=180}{rt}
					\fmfv{label=\small{$\mathbf{P}$},label.dist=5,label.angle=-90}{vb}
					\fmfv{decor.shape=circle,decor.filled=empty,decor.size=12,label=\tiny{$\perp$},label.dist=0}{v}
					\fmfrectangle{4}{4}{lt}
				\end{fmfgraph*}
			\end{gathered}
			\quad & \simeq \frac{g_s \mathbf{T}_i}{S_{ij}}\, \left[
			(k_{\perp,j} - k_{\perp,i})\cdot \epsilon^{(i)}_{\lambda_3} \delta_{\lambda_1 \lambda_2} \right.
			\nonumber \\
			& \left. \hphantom{\simeq \frac{g_s \mathbf{T}_i}{S_{ij}}\,}
			+ (k_{\perp,i} \cdot \epsilon^{(i)}_{\lambda_2}) \delta_{\lambda_1 \lambda_3}
			- (k_{\perp,j} \cdot \epsilon^{(i)}_{\lambda_1}) \delta_{\lambda_2 \lambda_3}  \right]  ,
			\\[10pt]
			\begin{gathered}
				\begin{fmfgraph*}(80,40)
					\fmfstraight
					\fmfright{rt,l,rb}
					\fmfleft{lt,r,lb}
					\fmf{phantom}{lt,b1,p,bp,v,b2,rt}
					\fmffreeze
					\fmf{curly}{b1,lt}
					\fmf{curly}{rt,bp}
					\fmf{phantom}{lb,b1b,pb,bpb,vb,b2b,rb}
					\fmf{curly}{v,bb,vb}
					\fmfv{decor.shape=square,decor.filled=empty,decor.size=6}{b1}
					\fmfv{decor.shape=square,decor.filled=empty,decor.size=6}{b2}
					\fmfv{decor.shape=square,decor.filled=empty,decor.size=6}{bp}
					\fmfv{decor.shape=square,decor.filled=empty,decor.size=6}{bb}
					\fmfv{label=\small{$\mathbf{P}$},label.dist=0}{p}
					\fmfv{label=\small{$\mathbf{P}$},label.dist=5,label.angle=0}{rt}
					\fmfv{label=\small{$\mathbf{P}$},label.dist=5,label.angle=90}{vb}
					\fmfv{decor.shape=circle,decor.filled=empty,decor.size=12,label=\tiny{$\parallel$},label.dist=0}{v}
					\fmfrectangle{4}{4}{lt}
				\end{fmfgraph*}
			\end{gathered}
			\quad & \simeq 2 \frac{g_s \mathbf{T}_k}{S_{jk}} 
			\frac{z_k n\scdot p_k}{(z_j n\scdot p_i + z_k n\scdot p_k) n\scdot p_i} 
			\nonumber \\
			&\hphantom{\simeq}\times
			\left[(n\scdot p_k) (p_i\cdot \epsilon^{(i)}_{\bar{\lambda}_3})  - (n\scdot p_i)( p_k \cdot \epsilon^{(i)}_{\bar{\lambda}_3})\right] \, \delta_{\bar{\lambda}_1 \bar{\lambda}_2},
			\\[10pt]
			\begin{gathered}
				\begin{fmfgraph*}(80,40)
					\fmfstraight
					\fmfright{rt,l,rb}
					\fmfleft{lt,r,lb}
					\fmf{phantom}{lt,b1,p,bp,v,b2,rt}
					\fmffreeze
					\fmf{curly}{b1,lt}
					\fmf{curly}{rt,bp}
					\fmf{phantom}{lb,b1b,pb,bpb,vb,b2b,rb}
					\fmf{curly}{v,bb,vb}
					\fmfv{decor.shape=square,decor.filled=empty,decor.size=6}{b1}
					\fmfv{decor.shape=square,decor.filled=empty,decor.size=6}{b2}
					\fmfv{decor.shape=square,decor.filled=empty,decor.size=6}{bp}
					\fmfv{decor.shape=square,decor.filled=empty,decor.size=6,label=\tiny{$\perp$},label.dist=0}{bb}
					\fmfv{label=\small{$\mathbf{P}$},label.dist=0}{p}
					\fmfv{label=\small{$\mathbf{P}$},label.dist=5,label.angle=0}{rt}
					\fmfv{label=\small{$\mathbf{P}$},label.dist=5,label.angle=90}{vb}
					\fmfv{decor.shape=circle,decor.filled=empty,decor.size=12,label=\tiny{$\parallel$},label.dist=0}{v}
					\fmfrectangle{4}{4}{lt}
				\end{fmfgraph*}
			\end{gathered}
			\quad & \simeq 2 \frac{g_s \mathbf{T}_k}{S_{jk}} 
			\frac{z_k n\scdot p_k}{z_j(z_j n\scdot p_i + z_k n\scdot p_k)} \frac{n \scdot p_k}{n \scdot p_i}
			\left( k_{\perp,j} \cdot \epsilon^{(i)}_{\bar{\lambda}_3}\right) \, \delta_{\bar{\lambda}_1 \bar{\lambda}_2}.
		\end{align}
	\end{subequations}
\end{fmffile}%

The combination of these vertices together with one choice of
partitioning then constitutes a complete set of splitting kernels in
the single emission case.  Also note that one will encounter iterations
of these when building up the double emission case.  Differences to
solely iterating single emissions mostly arise because of the
partitioning and additional two-emission diagrams which can not be
acquired by iterations.  The only other changes to the kernels are
then encoded by the difference of the full double emission mapping to
an iteration of the single emission one and how these relate to the
global parametrization we have chosen to derive the splitting
amplitudes. Details of this will be subject to a follow-up
publication.

\section{Summary and Conclusions}
\label{sec:Summary}

The construction of parton branching algorithms requires a command of
the singular limits of QCD amplitudes which give rise to
logarithmically enhanced contributions. While the extraction of a
definitive singular limit is an established task, parton branching
algorithms, much like the development of subtraction terms, require a
removal of overlapping singularities in between the various soft and
collinear limits. The most general parton branching algorithms will
need to address the iteration of emissions at the amplitude level in
order to reflect the full set of correlations or to set the framework
to derive improved algorithms. In the present work we have set out a
framework which starts from the structure of QCD amplitudes using a
physical gauge in order to systematically factorize emissions at the
level of the density operator, organizing emission kernels through
various ``key'' topologies of collinear splittings. This enabled us to
use a power counting and effective set of Feynman rules to determine
the singularity structures without resorting to analysing the overlap
in between different limits: our algorithm will directly provide an
interpolating formula to extract the respective splitting kernels. We
have found that the way the kinematics is parametrized and how recoil
is handled, has a significant impact on the form of the final kernels
and more notably on the possibility to iterate them. This fact is
crucial, since one needs to remove iterated strongly ordered emissions
from a full kernel addressing the doubly unresolved limits. We have
also developed more general partitioning algorithms, which are able to
distribute the various interference contributions among leading
collinear limits, generalizing both the Catani-Seymour partial
fractioning idea as well as a differential version of angular
ordering, which effectively subtracts out collinear divergences. We
have demonstrated how our partitioning algorithm can be used to
distribute the known double-soft behaviour between different classes
of contributions, however we stress that this is not our primary
strategy to attack the problem of constructing splitting kernels for
parton branching algorithms.

Our formalism leads us to branching amplitudes, which carry full spin
and colour information and can be used within existing Monte Carlo
efforts such as the CVolver \cite{Platzer:2013fha,DeAngelis:2020rvq}
framework. One particular fact, which is worth highlighting, is that
we keep the hard-collinear behaviour separate in terms of the full
splitting function and find that our interference contributions are
manifestly suppressed in the collinear limit. We are able to perform
this distinction at the expense of an explicit dependence of the gauge
vector, for which we have explicitly shown how we can translate this
dependence into a backward direction local to each collinear
sector. In an upcoming publication we will use the present formalism
to outline the full set of double-emission kernels \cite{double:22}. We also anticipate
that a similar formalism, extended to virtual corrections and combined
with the techniques presented in \cite{Platzer:2020lbr} will allow us
to construct a full second-order evolution at the amplitude level which
can be used as a rigorous starting point for improved parton branching
algorithms.

\section*{Acknowledgements}

The work of ML is supported partially by the DFG Collaborative
Research Center TRR 257 ``Particle Physics Phenomenology after the
Higgs Discovery''. ESD has been supported by the Marie
Skłodowska-Curie Innovative Training Network MCnetITN3 (grant
agreement no. 722104). This work has also been supported in part by
the COST actions CA16201 ``PARTICLEFACE'' and CA16108 ``VBSCAN''. We
are grateful to the Erwin Schr\"odinger Institute Vienna for
hospitality and support while significant parts of this work have been
achieved within the Research in Teams programmes ``Higher-order
Corrections to Parton Branching at the Amplitude Level'' (RIT2020) and
``Amplitude Level Evolution I: Initial State Evolution.'' (RIT0421).
We are deeply obliged to Stefan Gieseke for his trust and support, and
we would like to thank Jeffrey Forshaw, Jack Holguin, Kirill Melnikov,
Ines Ruffa and Malin Sj\"odahl for fruitful discussions.

\newpage

\appendix

\section{Phase space factorization}\label{app:phase-space}
In this section, we discuss how phase space factorisation is achieved
with a generic version of the momentum mapping of
Sec.~\ref{sec:Mapping}.  Compared to the discussion of various
momentum mappings in~\cite{Del_Duca_2019}, we face the difficulty of
treating collinear and soft configurations simultaneously, together
with a global Lorentz-transformation.  The latter can have a
non-trivial dependence on emission and progenitor momenta, which needs
to be carefully traced when switching integration variables.  In the
following, we will use the notation
\begin{equation}
	[\dd q] \equiv \frac{\dd^{d} q}{(2\pi)^{d-1}} \delta (q^2) \Theta(q^0).
\end{equation}

We are interested in the phase space
\begin{align}\label{eq:phase-space-start}
	\mathrm{PS} = \delta^{(d)}\Big(\sum\limits_{i}q_i + K - Q\Big)\, \prod\limits_i
	\, [\dd q_i] \, \prod\limits_l [\dd k_{i,l}],
\end{align}
where $K$ is the sum over all emission momenta $k_{i,l}$, $Q$ is the
overall momentum transfer and we think of the $q_i$ as emitter and
recoiler momenta.  The goal is to express this phase space in terms of
the progenitor momenta $p_i$ of an emission process.

\subsection{Emission phase space}
We begin the discussion with a generic version of the emission momenta, \ie
\begin{equation}\label{eq:khat-def}
	k_{il}^\mu = \frac{1}{\hat{\alpha}} \Lambda^\mu_{\ \nu} \hat{k}_{il}^\nu, \qquad
	\hat{k}_{il}^\nu = \alpha_{il} p_i^\nu + \beta_{il} n^\nu + k_{\perp,il}^\nu.
\end{equation}
The Lorentz-transformation $\Lambda$ and scaling factor $\hat{\alpha}$
will be functions of $p_i$, $n$, $\alpha_{il}$, $\beta_{il}$ and
$k_{\perp,il}$.  The use of a local backwards direction $n_i$ will be
discussed in the next section.  We fix this mapping by inserting
\begin{align}
	[\dd k_{il}] = [\dd k_{il}] &\times \delta\left( \alpha_{il} - \frac{1}{\hat{\alpha}} \frac{(\Lambda n)\cdot k_{il}}{n \cdot p_i} \right) \dd \alpha_{il} \nonumber \\
	&\times \delta\left( \beta_{il} - \frac{1}{\hat{\alpha}} \frac{(\Lambda p_i)\cdot k_{il}}{n \cdot p_i}  - \frac{p\cdot k_{\perp,il}}{n \cdot p_i} \right) \dd \beta_{il} \nonumber \\
	& \times \delta \left(k_{\perp, il} - \alpha \Lambda^{(-1)} k_{il} +\alpha_{il} p_i + \beta_{il} n\right) \dd^d k_{\perp,il}.
\end{align}
Now, we switch the integration from $k_{il} \to \hat{k}_{il}$, use the
fact that $\Lambda \in SO^+(1,3)$ and assume $n^2=0$ and $n\cdot p_i >
0$.  This yields
\begin{align}\label{eq:emission-PS-1}
	[\dd k_{il}] = \hat{\alpha}^{2-d} [\dd \hat{k}_{il}] &\times  \delta\left( \alpha_{il} - \frac{n\cdot \hat{k}_{il}}{n \cdot p_i} \right) \dd \alpha_{il} \nonumber \\
	&\times \delta\left( \beta_{il} - \frac{ p_i \cdot \hat{k}_{il}}{n \cdot p_i} - \frac{p_i\cdot k_{\perp,il}}{n \cdot p_i}  \right) \dd \beta_{il} \nonumber \\
	& \times \delta \left(\hat{k}_{il} -\alpha_{il} p_i - \beta_{il} n -k_{\perp, il}   \right) \dd^d k_{\perp,il}
\end{align}
and we can remove the integration $\dd^d \hat{k}_{il}$ via the last
Dirac-delta.  This also fixes the projections of $n$ and $p_i$ on
$\hat{k}_{il}$ and we can replace
\begin{align}
	\frac{n\cdot \hat{k}_{il}}{n \cdot p_i} &= \alpha_{il} + \frac{n \cdot k_{\perp,il}}{n \cdot p_i} \nonumber \\
	\frac{p_i\cdot \hat{k}_{il}}{n \cdot p_i}  &= \beta_{il} + \frac{p_i \cdot k_{\perp,il}}{n\cdot p_i} + \alpha_{il}\frac{ p_i^2}{n \cdot p_i} .
\end{align}
in \eqref{eq:emission-PS-1}.
Moreover, we have 
\begin{align}
	\delta (\hat{k}_{il}^2) = \delta\big(\alpha_{il}(\alpha_{il} p_i^2 + 2p_i\cdot k_{\perp,il})+ 2\beta_{il} n \cdot k_{\perp,il} + k_{\perp,il}^2 + \beta_{il} 2\alpha_{il} p_i \cdot n\big)
\end{align}
Using the remaining delta-functions in \eqref{eq:emission-PS-1} and
carrying out the $\beta_{il}$-integration then fixes
\begin{align}
	\beta_{il} = \frac{-k_{\perp,il}^2}{\alpha_{il} 2 n \cdot p_i}.
\end{align}
This leads to
\begin{align}\label{eq:emission-ps-final}
	[\dd k_{il}] = &\frac{\hat{\alpha}^{2-d}}{(2\pi)^{d-1}}\frac{n\cdot p_i}{2 \alpha_{il}} \, 
	\delta(n \cdot k_{\perp,il}) \, \delta (\alpha_{il}p_i^2 + 2 p_i \cdot k_{\perp,il})
	\nonumber \\
	&\times  \Theta\!\!\left( \alpha_{il} p_i^0 + \frac{ -k_{\perp,il}^2}{\alpha_{il} 2 n\cdot p_i} \, n^0 + k_{\perp,il}^0 \right) \dd \alpha_{il} \,\dd^d k_{\perp,il}.
\end{align}

One can now consider a frame where $p_i$ and $n$ are back to back and
rewrite the $k_{\perp,il}$-integration in terms of polar coordinates.
What we then find is
\begin{align}\label{eq:emission-ps-in-frame}
	[\dd k_{il}] = \frac{\hat{\alpha}^{2-d}}{2 \alpha_{il}} \, \dd \alpha_{il} \, p_{\perp,il}^{d-3}  \dd p_{\perp,il} \, \dd \Omega^{(d-3)} \Theta (p_{\perp,il}) \Theta (\alpha_{il}),
\end{align}
with the properties
\begin{align}
	&k_{\perp,il} \cdot n = 0, \quad 
	k_{\perp,il} \cdot p_i = -\frac{\alpha_{il} p_i^2}{2} \nonumber \\
	&k_{\perp,il}^2 = -p_{\perp,il}^2, \quad 
	\beta_{il} = \frac{p_{\perp,il}^2}{\alpha_{il} 2 p_i \cdot n}.
\end{align}

\subsection{Emitter and recoiler phase space}
For the discussion of the emitter and recoiler phase space, we use
\begin{equation}\label{eq:qhat}
	q_i^\mu = \frac{1}{\hat{\alpha}} \Lambda^\mu_{\ \nu} \hat{q}_i^\nu,
\end{equation}
and discuss some details related to the introduction of the
Lorentz-transformation.  First, we consider the momentum conserving
delta-function of \eqref{eq:phase-space-start}.  Inserting the
transformed momenta, we demand
\begin{align}
	\delta^{(d)}\Bigg(\frac{\Lambda}{\hat{\alpha}} \sum_i \Big(\hat{q}_i + \sum_l \hat{k}_{il}\Big) - Q\Bigg) = \delta^{(d)}\left(\frac{\Lambda}{\hat{\alpha}}(P-Q)\right),
\end{align} 
where $P=\sum_i p_i$.
This is achieved by choosing the Lorentz transformation such that
\begin{equation}
	\Lambda^\mu_{\ \nu} \frac{Q^\nu+N^\nu}{\hat{\alpha}} = Q^\mu, 
\end{equation}
with
\begin{equation}
	N = \sum_i\left(\hat{q}_i+\sum_l \hat{k}_{il} - p_i \right),
\end{equation}
and 
\begin{equation}
	\hat{\alpha} = \sqrt{\frac{(Q+N)^2}{Q^2}}.
\end{equation}
Explicitly, this Lorentz-transformation reads~\cite{Catani:1996vz}
\begin{equation}\label{eq:Lorentz-CS}
	\Lambda^\mu_{\ \nu} (p_1 \to p_2) = \eta^\mu_{\ \nu} - \frac{2(p_1+p_2)^\mu (p_1+p_2)_\nu}{(p_1 + p_2)^2} + \frac{2 \, p_2^\mu \, p_{1 \nu}}{p_1^2}.
\end{equation}
Due to $\Lambda$ being a function of the $p_i$, \ie via $N =
N(\{p_i\})$, one could in principle find derivative terms of the
transformation when applying the usual transformation laws for Dirac
delta-functions.  As it turns out though, we find
\begin{equation}
	\frac{\partial}{\partial p_j^\rho} \left. \left(\frac{\Lambda^\mu_{\ \nu}}{\hat{\alpha}}(P_\mu-Q_\mu)\right)\right\vert_{P=Q} = \left.\frac{\Lambda^\mu_{\ \rho}}{\hat{\alpha}}\right\vert_{P=Q}.
\end{equation}
This means that with $\Lambda$ having a unit determinant, we find
\begin{equation}
	\delta^{(d)} \! \left(\frac{\Lambda}{\hat{\alpha}}(P-Q)\right) = \hat{\alpha}^d \delta(P-Q). 
\end{equation}
Next, we discuss the variable change in $[\dd q_i]\to [\dd p_i]$.  In
order to study the $p_i$-dependence in the emitter and recoiler
momenta, we define
\begin{equation}
	\hat{q}_i^\nu = (1-A_i) p_i^\nu -R_i ,
\end{equation}
where $R_i$ depends on the choice of a recoil scheme:
\begin{align}
	R_{i} =
	\begin{cases}
		B_i n_i + K_{\perp,i}  &(\text{balanced})\\
		0 &(\text{unbalanced})
	\end{cases}
\end{align}
Demanding
\begin{equation}
	\hat{q}_i^2 = (1-A_i)p_i^2
\end{equation}
leads to
\begin{equation}
	B_i = -\frac{1}{2 n_i\cdot p_i} \left(\frac{K_{\perp,i}^2}{1-A_i} + 2K_{\perp,i}\cdot p_i \right) .
\end{equation}
We note that the $R_i$, as well as $\hat{\alpha}$ and $\Lambda$ can be
functions of the $\alpha_{il}$, $p_{\perp,il}$, $\Omega_{il}$ and
$p_i$.  Therefore, we again need to study this dependence when
switching integration variables.
\begin{equation}
	\dd^d q_i = \mathrm{det} \mathcal{J}(\{q_i\},\{p_i\})\, \dd^d p_i.
\end{equation}
For the Jacobian, we find
\begin{align}
	\mathcal{J}(\{q_i\},\{p_i\}) = \frac{\partial q_i^\mu}{\partial p_j^\rho}
	= \frac{\Lambda^\lambda_{\ \xi}}{\hat{\alpha}} 
	\bigg[
	\underbrace{
		\hat{\alpha} \Lambda^\xi_{\ \mu} \frac{\partial}{\partial N^\sigma } \left(\frac{\Lambda^\mu_{\ \nu}}{\hat{\alpha}}\right) \frac{\partial N^\sigma}{\partial p_j^\rho} \hat{q}_i^\nu + \frac{\partial \hat{q}_i^\xi}{\partial p_j^\rho}}_{=\hat{\mathcal{J}}}
	\bigg].
\end{align}
We can now use 
\begin{align}
	N = \begin{cases}
		\sum_i \left(\sum_l \beta_{il} - B_i\right)n_i & \text{(balanced)} \\
		\sum_i \left( \sum_l \beta_{il} n_i + K_{\perp,i}\right) & \text{(unbalanced)}.
	\end{cases}
\end{align}
In both cases, $N$ has a uniform soft and collinear scaling. For a
global $n$, this shows that the derivative terms in $\hat{J}$ can be
neglected and we have
\begin{equation}
	\frac{\partial \hat{q}_i^\mu}{\partial p_j^\nu} = (1-A_i) \delta_{ij} \delta^\mu_{\ \nu} + \mathcal{O}(\lambda). 
\end{equation}
This means we have
\begin{equation}
	\dd^d q_i = \hat{\alpha}^{-d} (1-A_i)^d  \dd^d p_i + \mathcal{O}(\lambda).
\end{equation}
The $p_i$-dependence for the case of a local $n_i$ is more involved,
but we expect the general features ot this discussion to hold in that
case as well.  Lastly, the delta-function for the on-shell condition
of the $q_i$ is
\begin{equation}
	\delta(q_i^2) = \frac{\hat{\alpha}^2}{(1-A_i)^2} \delta(p_i^2).
\end{equation}

Putting everything together, we have
\begin{align}\label{eq:emitter-ps-final}
	[\dd q_i] = \left(\frac{\hat{\alpha}}{1-A_i}\right)^{2-d} \delta(p_i^2) \, \Theta\big[(1-A_i)p_i^0 - R_i^0\big] \frac{\dd^d p_i}{(2\pi)^{d-1}}\, + \mathcal{O}(\lambda).
\end{align}
In the unbalanced case, where $R_i=0$, the Heaviside function
immediately leads to an upper bound for $A_i$. The same if true for
the balanced case, but in a less obvious way due to the more involved
additional dependence on $K_{\perp,i}$.

Note that \eqref{eq:emitter-ps-final} shows a powerful result:
switching from a momentum mapping with a global gauge vector $n$ to
one with a local $n_i$ only amounts to a redefinition of
$k_{\perp,il}$ via \eqref{eq:kperp-redef-local-ni} in
\eqref{eq:emission-ps-final} because additional modifications only
amount to subleading power.

\section{Two emissions}\label{app:two-emissions}
\subsection{Quark-gluon-gluon splitting function}
In this section, we want to demonstrate how the two-emission splitting
function $ \langle \hat{P}_{g_1 g_2 q}\rangle$ for the emission of two
gluons off of a quark comes about in our formalism.  For brevity, we
show the results in terms of cut diagrams, which can be acquired via
tracing over the amplitude insertions from
Tab.~\ref{tab:two-emissions-sp1} with the lowest numerator scaling.

First, we note that in lightcone gauge all interference contributions
are subleading in the collinear limit.  Therefore, the splitting
function is fully contained in the self energy-like diagrams.  Note
that using our power counting rules, one immediately sees that only
the forward component of the off-shell lines connecting to the hard
amplitude give rise to leading contributions.

The Abelian part of the splitting function  is given by
\begin{fmffile}{abelian-SE}
	\fmfset{thin}{.7pt}
	\fmfset{dash_len}{1.5mm}
	\fmfset{arrow_len}{2.5mm}
	\fmfset{curly_len}{1.5mm}
	\begin{align}\label{splitting-abelian}
	\frac{\mu^{4\varepsilon}}{\hat{\alpha}^2} &\Bigg\{\quad
	\begin{gathered}
	\begin{tikzpicture}
	\node (diagram) {%
		\begin{fmfgraph*}(90,60)
		\fmfleft{bl}
		\fmfright{br}
		\fmf{plain}{bl,vboxl,vcutl,v1,v2,v3,v4,vcutr,vboxr,br}
		\fmf{phantom,tension=0.5}{v1,v2}
		\fmf{phantom,tension=0.5}{v3,v4}
		\fmf{phantom,tension=-0.4}{v2,v3}
		\fmf{phantom,tension=-0.5}{bl,vboxl}
		\fmf{phantom,tension=-0.5}{br,vboxr}
		\fmffreeze
		\fmf{curly,right}{v2,v3}
		\fmf{curly,right}{v4,v1}
		\fmfv{label=\tiny{$i$},label.angle=70,label.dist=.9mm}{v2}
		\fmfv{label=\tiny{$1$},label.angle=-45,label.dist=2mm}{v3}
		\fmfv{label=\tiny{$2$},label.angle=90,label.dist=5mm}{v1}
		\fmfv{decor.shape=square,decor.filled=empty,decor.size=8}{vboxl}
		\fmfv{decor.shape=square,decor.filled=empty,decor.size=8}{vboxr}
		\fmfrectangle{6}{5}{bl}
		\fmfrectangle{6}{5}{br}
		\end{fmfgraph*}
	};
	\tikzset{shift={(0,0)}}
	\draw[thick, dashed] (-0.3,-1.1) arc(-90:0:0.3) (0,-0.8) -- (0,0.8) arc(180:90:0.3);
	\end{tikzpicture}
	\end{gathered}
	\; + \;
	\Bigg[ \;
	\begin{gathered}
	\begin{tikzpicture}
	\node (diagram) {%
		\begin{fmfgraph*}(90,60)\fmfkeep{diagram-E2-qcd}
		\fmfleft{bl}
		\fmfright{br}
		\fmf{plain}{bl,vboxl,vcutl,v1,v2,v3,v4,vcutr,vboxr,br}
		\fmf{phantom,tension=-0.2}{v1,v2}
		\fmf{phantom,tension=-0.2}{v3,v4}
		\fmf{phantom,tension=-0.4}{v2,v3}
		\fmf{phantom,tension=-0.5}{bl,vboxl}
		\fmf{phantom,tension=-0.5}{br,vboxr}
		\fmffreeze
		\fmf{curly,right}{v2,v4}
		\fmf{curly,right}{v3,v1}
		\fmfv{label=\tiny{$i$},label.angle=60,label.dist=1mm}{v2}
		\fmfv{label=\tiny{$1$},label.angle=-90,label.dist=5mm}{v4}
		\fmfv{label=\tiny{$2$},label.angle=90,label.dist=5mm}{v1}
		\fmfv{decor.shape=square,decor.filled=empty,decor.size=8}{vboxl}
		\fmfv{decor.shape=square,decor.filled=empty,decor.size=8}{vboxr}
		\fmfrectangle{6}{5}{bl}
		\fmfrectangle{6}{5}{br}
		\end{fmfgraph*}
	};
	\tikzset{shift={(0,0)}}
	\draw[thick, dashed] (-0.3,-1.1) arc(-90:0:0.3) (0,-0.8) -- (0,0.8) arc(180:90:0.3);
	\end{tikzpicture}
	\end{gathered}
	\; + (1 \leftrightarrow 2)
	\Bigg]
	\;
	\Bigg\}_{C_F^2}
	\nonumber \\
	&= \left( \frac{8\pi \alpha_S}{ \hat{\alpha} S_{i12}}  \mu^{2\varepsilon} \right)^2 C_F^2 \, \langle \hat{P}^{\text{(Ab)}}_{ggq}\rangle \hat{\slashed{p}}_i \, + \mathcal{O}\left(\beta_{il}^{-3/2}\right).
	\end{align}
	\end{fmffile}%
The subscript on the curly brackets stands for an extraction of terms
proportional to $C_F^2$.  This is necessary because the second diagram
has the colour structure
\begin{equation}
T^a T^b T^a T^b = \left(C_F^2 -\frac{1}{2}C_A C_F \right)\mathds{1}_N.
\end{equation}
Therefore, it also contributes to the non-Abelian part of the
splitting function.  The splitting function in terms of our momentum
mapping then reads
\begin{align}	
	\frac{\langle \hat{P}^{\text{(Ab)} }_{ggq}\rangle }{4 \, S_{i12}^2} = \, &\frac{1-\epsilon }{2 S_{i12}^2} \left\{
	2 - (1-\varepsilon)\frac{(S_{i1}+ S_{i2})^2}{S_{i1} S_{i2}}
	\right\}  
	\nonumber\\
	&+ \frac{1}{2 S_{i1} S_{i_2}} \bigg\{
	-(1-\varepsilon)^2 (1- A_i) 
	+\frac{(1-\varepsilon) (1-A_i) (A_i^2+\alpha_{i1}\alpha_{i2}) + 2 (1-A_i)^2}{\alpha_{i1}\alpha_{i2}}	
	\bigg\}
	\nonumber\\
	&
	\begin{aligned}
		{} + \frac{1}{S_{i1} S_{i12}} \bigg\{
		&(1-\varepsilon)^2 (2- A_i) + \frac{(1-\varepsilon)\big[2(A_i-2)\alpha_{i1} \alpha_{i2} +(1-\alpha_{i2})(A_i^2-\alpha_{i1}\alpha_{i2})\big]}{\alpha_{i1} \alpha_{i2}}
		\\
		& +\frac{(1-A_i)(1-\alpha_{i2})}{\alpha_{i1} \alpha_{i2}}
		\bigg\} +\big(1\leftrightarrow 2\big)
	\end{aligned}
\end{align}
The non-Abelian part comes about via
\begin{fmffile}{non-abelian-SE}
	\fmfset{thin}{.7pt}
	\fmfset{dash_len}{1.5mm}
	\fmfset{arrow_len}{2.5mm}
	\fmfset{curly_len}{1.5mm}
	\begin{align}\label{splitting-non-abelian}
	\frac{\mu^{2\varepsilon}}{\hat{\alpha}^2} &\Bigg\{\quad
	\begin{gathered}
	\begin{tikzpicture}
	\node (diagram) {%
		\begin{fmfgraph*}(90,60)
		\fmfleft{bl}
		\fmfright{br}
		\fmftop{t}
		\fmf{plain}{bl,vboxl,vcutl,v1,v2,vcutr,vboxr,br}
		\fmf{phantom,tension=-0.8}{v1,v2}
		\fmf{phantom,tension=-0.5}{bl,vboxl}
		\fmf{phantom,tension=-0.5}{br,vboxr}
		\fmffreeze
		\fmf{phantom,left}{v1,vt1,t,vt2,v2}
		\fmffreeze
		\fmf{curly,right=.2}{vt1,v1}
		\fmf{curly,right=.2}{v2,vt2}
		\fmf{curly,right=.6}{vt1,vt2,vt1}
		\fmfv{label=\tiny{$i$},label.angle=-140,label.dist=2mm}{v2}
		\fmfv{decor.shape=square,decor.filled=empty,decor.size=8}{vboxl}
		\fmfv{decor.shape=square,decor.filled=empty,decor.size=8}{vboxr}
		\fmfrectangle{6}{5}{bl}
		\fmfrectangle{6}{5}{br}
		\end{fmfgraph*}
	};
	\tikzset{shift={(0,0)}}
	\draw[thick, dashed] (-0.3,-0.8) arc(-90:0:0.3) (0,-0.5) -- (0,0.8) arc(180:90:0.3);
	\end{tikzpicture}
	\end{gathered}
	\; + \; \nonumber \\
	&\Bigg[ \;
	\begin{gathered}
	\begin{tikzpicture}
	\node (diagram) {%
		\fmfreuse{diagram-E2-qcd}
	};
	\tikzset{shift={(0,0)}}
	\draw[thick, dashed] (-0.3,-1.1) arc(-90:0:0.3) (0,-0.8) -- (0,0.8) arc(180:90:0.3);
	\end{tikzpicture}
	\end{gathered}
	\; + \;
	\begin{gathered}
	\begin{tikzpicture}
	\node (diagram) {%
		\begin{fmfgraph*}(90,60)
		\fmfleft{bl}
		\fmfright{br}
		\fmftop{t}
		\fmf{plain}{bl,vboxl,vcutl,v1,vm1,vm2,v2,vcutr,vboxr,br}
		\fmf{phantom,tension=-0.7}{vm1,vm2}
		\fmf{phantom,tension=-0.5}{bl,vboxl}
		\fmf{phantom,tension=-0.5}{br,vboxr}
		\fmffreeze
		\fmf{phantom,left}{v1,vt1,t,vt2,v2}
		\fmffreeze
		\fmf{curly}{vm1,vt2}
		\fmf{curly,left=1.}{v1,v2}
		\fmfv{label=\tiny{$i$},label.angle=-140,label.dist=2mm}{v2}
		\fmfv{label=\tiny{$1$},label.angle=-100,label.dist=3mm}{vt2}
		\fmfv{label=\tiny{$2$},label.angle=90,label.dist=5mm}{v1}
		\fmfv{decor.shape=square,decor.filled=empty,decor.size=8}{vboxl}
		\fmfv{decor.shape=square,decor.filled=empty,decor.size=8}{vboxr}
		\fmfrectangle{6}{5}{bl}
		\fmfrectangle{6}{5}{br}
		\end{fmfgraph*}
	};
	\tikzset{shift={(0,0)}}
	\draw[thick, dashed] (-0.3,-0.8) arc(-90:0:0.3) (0,-0.5) -- (0,0.8) arc(180:90:0.3);
	\end{tikzpicture}
	\end{gathered}
	\; + \;
	\begin{gathered}
	\begin{tikzpicture}
	\node (diagram) {%
		\begin{fmfgraph*}(90,60)
		\fmfleft{bl}
		\fmfright{br}
		\fmftop{t}
		\fmf{plain}{bl,vboxl,vcutl,v1,vm1,vm2,v2,vcutr,vboxr,br}
		\fmf{phantom,tension=-0.7}{vm1,vm2}
		\fmf{phantom,tension=-0.5}{bl,vboxl}
		\fmf{phantom,tension=-0.5}{br,vboxr}
		\fmffreeze
		\fmf{phantom,left}{v1,vt1,t,vt2,v2}
		\fmffreeze
		\fmf{curly}{vt1,vm2}
		\fmf{curly,left=1.}{v1,v2}
		\fmfv{label=\tiny{$i$},label.angle=-40,label.dist=2mm}{v1}
		\fmfv{label=\tiny{$1$},label.angle=-80,label.dist=3mm}{vt1}
		\fmfv{label=\tiny{$2$},label.angle=90,label.dist=5mm}{v2}
		\fmfv{decor.shape=square,decor.filled=empty,decor.size=8}{vboxl}
		\fmfv{decor.shape=square,decor.filled=empty,decor.size=8}{vboxr}
		\fmfrectangle{6}{5}{bl}
		\fmfrectangle{6}{5}{br}
		\end{fmfgraph*}
	};
	\tikzset{shift={(0,0)}}
	\draw[thick, dashed] (-0.3,-0.8) arc(-90:0:0.3) (0,-0.5) -- (0,0.8) arc(180:90:0.3);
	\end{tikzpicture}
	\end{gathered}
	\; + 
	(1 \leftrightarrow 2)
	\Bigg]
	\;
	\Bigg\}_{C_A C_F}
	\nonumber \\
	&= \left( \frac{8\pi \alpha_S}{ \hat{\alpha} S_{i12}}  \mu^{2\varepsilon} \right)^2 C_A C_F \, \langle \hat{P}^{\text{(nAb)}}_{ggq}\rangle \hat{\slashed{p}}_i \, + \mathcal{O}\left(\beta_{il}^{-3/2}\right),
	\end{align}
\end{fmffile}%
with
\begin{align}
	\frac{\langle \hat{P}^{\text{(nAb)}}_{ggq}\rangle}{4 S_{i12}^2} = 
	&\hphantom{+}\frac{1}{S_{i12}^2}\Bigg\{\frac{(1-\varepsilon)^2}{2}
	-\frac{(1-\varepsilon)
		\left[\alpha_{i1} \left(S_{i2}+S_{12}\right)-\alpha_{i2}
		S_{i1}\right] \left[\alpha_{i2} \left(S_{i1}+S_{12}\right)-\alpha_{i1} S_{i2}\right]}{S_{12}^2 \, A_i^2 }\Bigg\} \nonumber\\
	&+\frac{1}{ S_{12} S_{i2} \, A_i \alpha_{i1}} \Bigg\{\left(1-\alpha_{i1}\right) \alpha_{i1}+\left(1-A_i\right)
	A_i+\frac{(1-\varepsilon) }{2}\left(\alpha
	_{i1}^3+A_i^3\right)\Bigg\} \nonumber\\
	&+\frac{1}{2 S_{i1} S_{i2}} \Bigg\{\frac{(1-\varepsilon)^2}{2} \left(1-A_i\right)
	- \frac{(1-\varepsilon)}{2} \frac{\left(1- A_i\right) \left(\alpha_{i1}
		\alpha_{i2} + A_i^2\right) }{\alpha_{i1} \alpha_{i2}} - \frac{\left(1-A_i\right)^2}{\alpha_{i1} \alpha_{i2}}\Bigg\} \nonumber\\
	&\begin{aligned}
		{}+\frac{1}{S_{12}S_{i12} \, A_i \alpha_{i1}} \Bigg\{
		&\frac{(1-\varepsilon)}{2} \left[ \alpha _{i,1}^3 - \alpha _{i,2}^3 - 2\alpha _{i,1}\left(\alpha _{i,1}- \alpha _{i,2} \right)\right]\\
		&+\left(2 A_i-3\right) \alpha_{i1}-A_i \alpha_{i2}+\alpha_{i2} \Bigg\}
	\end{aligned}\nonumber\\
	&
	\begin{aligned}
		{}+	\frac{1}{2 S_{i2} S_{i12}} \Bigg\{
		&\frac{(1-\varepsilon)^2}{2 } \left(\alpha _{i1}-1\right)
		+\frac{(1-\varepsilon)}{2} \frac{\left(\alpha_{i1}-1\right) \left(\alpha _{i1}^2-\alpha _{i2} \alpha_{i1}-\alpha_{i2}^2\right)}{ \alpha_{i2} A_i} 
		\\
		&-\frac{\alpha _{i1}^2-2 \alpha
			_{i1}+1}{\alpha _{i2} A_i}
		\Bigg\} 		+\big(1\leftrightarrow 2\big).
	\end{aligned}
\end{align}
Note that the first diagram has the topology of $E_1$, see Fig.~\ref{fig:self-energies}.
Using the identifications
\begin{eqnarray}
	 z_2 = \alpha_{i1}, & z_1 = \alpha_{i2}, & z_3 = 1-A_i = 1 - \alpha_{i1} - \alpha_{i2}, \\
	 s_{123} = S_{i12}, & s_{13} = S_{i2} & \text{etc.}, 
\end{eqnarray}
one can verify that the splitting function from
\eqref{splitting-abelian} and \eqref{splitting-non-abelian} coincide
with the well knowns ones as presented in \cite{Catani:1999ss}.

\subsection{Relation to soft and soft-collinear functions}
\label{sec:soft-collinear-functions}

In this section, we want to study the interplay of soft divergent
terms for two emissions as an extension of what we discussed in
Sec.~\ref{sec:single-emission-splitting-kernel}.  For this purpose, we
investigate the composition of the two-emission splitting function and
check its relation to the double soft and soft-collinear results
of~\cite{Catani:1999ss} by taking its soft-soft and soft-collinear
limits.  For comparability, we use a notation in terms of dot-products
to stay independent of a momentum mapping, \eg $\alpha_{i1} = n \scdot
q_1$.  Also note that in the following, partons $1$ and $2$ are
switched as compared to \eqref{splitting-abelian} for an easier
translation of our results to the ones of~\cite{Catani:1999ss}.

\subsubsection*{Double soft limit}
Of specific interest is the behaviour of the splitting function in the
double soft limit where the Abelian part should reproduce the double
Eikonal function and the non-Abelian part the two-gluon soft function
$\mathcal{S}_{ij}(q_1, q_2 )$.

First, we make use of the following scaling behaviour in the soft-soft limit:
\begin{equation}
	\begin{split}
		S_{12} \to& \lambda^2 S_{12}, \qquad S_{i(1/2)} \to \lambda \, S_{i(1/2)}, \qquad n\scdot q_{1/2} \to \lambda \, n\scdot q_{1/2}
	\end{split}
\end{equation}
Then, the splitting function has the following leading soft terms:
\begin{equation}
	\begin{split}\label{eq:splitting-fct-double-soft-limit}
		\left. \frac{\langle  \hat{P}^{\text{(Ab)}}_{q_i g_1 g_2 } \rangle}{S_{i12}^2}\right\rvert_{SS} =& \frac{ S_{i2}+3 S_{i1}}{\, S_{i2} S_{i1} (S_{i1}+S_{i2})} \, \frac{(n\scdot q_i)^2}{(n\scdot q_1) (n\scdot q_2)} + (1\leftrightarrow 2) = \frac{4}{S_{i2} S_{i1}} \,\frac{ (n\scdot q_i)^2}{(n\scdot q_1) (n\scdot q_2)},
		\\
		\left. \frac{\langle \hat{P}^{\text{(nAb)}}_{q_i g_1 g_2} \rangle}{S_{i12}^2}\right\rvert_{SS} =&  \frac{1}{S_{12}S_{i1}}\frac{n\scdot q_i (n\scdot q_1+2 n\scdot q_2)}{n\scdot q_2  (n\scdot q_1+n\scdot q_2)} -\frac{1}{S_{i1} S_{i2}}\frac{(n\scdot q_i)^2}{2 (n\scdot q_1) (n\scdot q_2) }
		\\
		&\frac{1}{(S_{i1}+S_{i2})(n\scdot q_1 + n\scdot q_2)}\bigg[ \frac{1}{S_{12}} \frac{n\scdot q_i(n \scdot q_1 - 3 n\scdot q_2)}{n \scdot q_2 } -\frac{1}{S_{i1}} \frac{(n\scdot q_i)^2}{n \scdot q_1 } \bigg]
		\\
		&+\frac{1-\varepsilon}{S_{12}^2 } \, \frac{ (n\scdot q_1 S_{i2}-n\scdot q_2 S_{i1})^2}{(S_{i1}+S_{i2})^2 (n\scdot q_1+n\scdot q_2)^2} + (1\leftrightarrow 2).
	\end{split}
\end{equation}
The amplitude factorises in the soft limit with a factor of the two
gluon soft current, $J_{\mu_1 \mu_2}^{a_1 a_2}(q_1,q_2)$, which
contains the double Eikonal and the two-gluon soft function. For the
squared amplitude the factor can be simplified into the following
form~\cite{Catani:1999ss}:
\begin{equation}\label{eq:double-soft-squared}
	\begin{split}
		[J_{\mu \rho}^{a_1 a_2}(q_1,q_2)]^\dagger &d^{\mu\nu}(q_1) d^{\rho\sigma}(q_2) J_{\nu \sigma}^{a_1 a_2}(q_1,q_2)
		\\
		&= \frac{1}{2}\{\mathbf{J}^2(q_1),\mathbf{J}^2(q_2)\}
		-C_{A}\sum_{i,j=3}^n\mathbf{T}_i\scdot \mathbf{T}_j \ \mathcal{S}_{ij}(q_1,q_2)+...
	\end{split}
\end{equation}
where additional terms vanish in combination with the factorised
amplitude as they are proportional to the total colour charge.  The
first term is the anti-commutator of the single emission squared gluon
current of \eqref{eq:eikonal-calc-interference}.  It can be written as
\begin{align}\label{eq:double-soft-Ab}
	\frac{1}{2}\{\mathbf{J}^2(q_1),\mathbf{J}^2(q_2)\} = \frac{1}{2}\sum\limits_{i,j} \sum\limits_{k,l} \big[(\mathbf{T}_i\scdot\mathbf{T}_j) (\mathbf{T}_k \scdot \mathbf{T}_l) + (\mathbf{T}_k \scdot \mathbf{T}_l) (\mathbf{T}_i\scdot\mathbf{T}_j)\big] \frac{4 S_{ij}^2}{S_{i1}S_{j1} S_{k2} S_{l,2}}.
\end{align}
The second term contains the two-gluon soft function which can be written as
\begin{equation}\label{eq:double-soft-non-Ab}
	\begin{split}
		\mathcal{S}_{ij}(q_1,q_2) = 2\bigg\{&\frac{1}{S_{12}S_{i1}} \, \frac{p_j\scdot q_i (p_j\scdot q_1 + 2p_j\scdot q_2)}{p_j\scdot q_2(p_j\scdot q_1 + p_j\scdot q_2)} - \frac{1}{S_{i1}S_{i2}}\frac{(p_j\scdot q_i)^2}{2 (p_j\scdot q_1) (p_j\scdot q_2)}\\
		&+\frac{1}{(S_{i1} + S_{i2}) \, (p_j\scdot q_1 + p_j\scdot q_2) } \bigg[\frac{1}{S_{12}}\frac{ p_j\scdot q_i(p_j\scdot q_1-3p_j\scdot q_2)}{p_j\scdot q_2}-\frac{1}{S_{i1}}\frac{(p_j\scdot q_i)^2}{p_j\scdot q_1}\bigg] \\
		&+\frac{1-\varepsilon}{S_{12}^2} \, 
		\bigg[
		\frac{(p_j\scdot q_1 S_{i2}-p_j\scdot q_2 S_{i1})^2}{(S_{i1}+S_{i2})^2 (p_j\scdot q_1+p_j\scdot q_2)^2}
		+\frac{S_{i1}S_{i2}}{(S_{i1}+S_{i2})^2}
		+\frac{(p_j\scdot q_1)(p_j\scdot q_2)}{ (p_j\scdot q_1+p_j\scdot q_2)^2}
		\bigg] \bigg\}\\
		& + (1\leftrightarrow 2).
	\end{split}
\end{equation}
Note that the last two terms vanish when inserted into
\eqref{eq:double-soft-squared} and colour-conservation is applied,
because they do not depend on either the parton labels $i$ or $j$.
Using the gauge-independence of the splitting function and choosing $n
= p_j$ shows an exact correspondence of the functional dependencies
between \eqref{eq:double-soft-non-Ab} and two times the non-Abelian
part of \eqref{eq:splitting-fct-double-soft-limit}.  It therefore
exhibits the overlap between the soft singular bits of the splitting
function and the two gluon soft current squared.

We can now go one step further and apply our partitioning algorithm to
\eqref{eq:double-soft-Ab} and \eqref{eq:double-soft-non-Ab}.  For the
Abelian part of \eqref{eq:double-soft-Ab}, we realize that for partons
$k=i$ and $l=j$, the propagator factors are given exactly by the ones
of the $X^{(1)}$-topology shown in App.~\ref{app:two-emission-diagrams}
which leads to
\begin{align}
	\frac{1}{2}\{\mathbf{J}^2(q_1),\mathbf{J}^2(q_2)\} = \sum\limits_{i,j} \big[ (\mathbf{T}_i \scdot \mathbf{T}_j)^2\big] \mathcal{P}\left(X^{(1)}_{i12j}\right) \times 4 S_{ij}^2,
\end{align}
where $\mathcal{P}(X^{(1)}_{i12j})$ are the respective propagator factors.
This means that we can partition this contribution into different
collinear sectors with any of the two partitioning variants of
Sec.~\ref{sec:Paritioning}.  Note that the terms where $k\neq i$ and
$l\neq j$ can be viewed as as single emission squared contributions.
The same option presents itself for the non-Abelian part,
\eqref{eq:double-soft-non-Ab}, when rewriting the scalar products in
terms of $S$-invariants and using the fact that we can introduce the
triple invariants $S_{i12}$ and $S_{j12}$ via
\begin{equation}
	S_{i1} + S_{i2} = S_{i12} N_{i12} = S_{i12}(1+\mathcal{O}(\lambda^2)), \qquad N_{i12}=\frac{S_{i1} + S_{i2}}{S_{i12}}.
\end{equation}
The result is
\begin{align}
	\mathcal{S}_{ij}(q_1,q_2) =	\hphantom{+}&\mathcal{P}\left(B^{(4)}_{i1j2}\right) \times 2 S_{ij}(S_{j1}+ 2 S_{j2}) N_{j12} 
	- \mathcal{P}\left(X^{(1)}_{i12j}\right) \times S_{ij}^2 \nonumber \\
	+\ &\mathcal{P}\left(A^{(3)}_{i12j}\right) \times 2S_{ij}(S_{j1} - 3 S_{j2}) N_{i12} N_{j12}
	-\mathcal{P}\left(A^{(2)}_{i12j}\right) \times 2 S_{ij}^2 N_{i12} N_{j12} \nonumber \\
	+\ &\mathcal{P}\left(A^{(5)}_{i12j}\right) \times 2 (1-\varepsilon) (S_{i1}S_{j2} + S_{i2} S_{j1}) N_{i12} N_{j12}
	+ (1 \leftrightarrow 2).
\end{align}
Having extracted these propagator factors and realizing that each of
them corresponds to a known topology, we again can apply partitionings
into different collinear sectors for each term.

What we have just shown presents an alternative way of building up a
splitting operator compared to what we discuss in the main part of
this work.  The procedure of constructing a splitting operator for
some specific collinear setting, \eg $(i\parallel 1\parallel 2)$,
would be as follows:
\begin{enumerate}
	\item Extract double soft and soft-collinear (shown in the
          next subsection) bits of the splitting function by using its
          gauge-independence, \ie by choosing a reference (recoiler)
          momentum and comparing to the known soft-limits.  The
          leftover parts can be viewed as purely collinear-singular
          ones for some specific collinear setting.
	
	\item Recombine the partitioned soft and soft-collinear bits
          for said collinear setting with the purely collinear bits.
	
	\item The resulting object reproduces the original collinear
          behaviour of the splitting function (because the
          partitioning does not spoil the scaling), but do not
          contribute in other collinear sectors due to the
          partitioning.  The correct soft and soft-collinear behaviour
          is reproduced when summing up the splitting operators of the
          different collinear sectors, because the partitioning is
          build up simply as a decomposition of unity.
\end{enumerate}

\subsubsection*{Soft-collinear limit}
The two emission splitting function shows very similar relations in
the soft-collinear limit.  The soft-collinear scaling where $q_1$ is
collinear to $q_i$ and $q_2$ is soft (which we denote as $\dots
\vert_{S_2C_{i1}}$) is as follows:
\begin{equation}
	\begin{split}
		S_{12} \to& \lambda \, S_{12}, \qquad S_{i2} \to \lambda \, S_{i2}, \qquad S_{i1} \to \lambda^2 S_{i1}, \qquad n\scdot q_2 \to \lambda \, n\scdot q_2.
	\end{split}
\end{equation}
The splitting function in this limit has the form
\begin{equation}
	\label{eq:splitting-soft-collinear-2E}
	\begin{split}
		\left.\frac{\langle \hat{P}_{q_i g_1 g_2} \rangle}{S_{i12}^2}\right\rvert_{S_2C_{i1}} 
		= {} C_F &\,
		\frac{(n\scdot q_{1})^2 (1-\varepsilon)+2 (n\scdot q_i)^2+2(n\scdot q_1) (n\scdot q_i)}{ S_{i1} (n\scdot q_{1}) (n\scdot q_1 + n\scdot q_i) } 
		\\
		&\times \bigg\{
		C_F \bigg[
		\frac{n \scdot q_1 + n \scdot q_i}{(S_{12}+S_{i2}) n \scdot q_2} + \frac{n \scdot q_i}{S_{i2} \, n \scdot q_2}
		\bigg]
		- \frac{C_A }{2}
		\bigg[
		\frac{n \scdot q_i}{S_{i2} \, n \scdot q_2} - \frac{n \scdot q_1}{S_{12} \, n \scdot q_2}
		\bigg] \bigg\}.
	\end{split}
\end{equation}
At the same time, we know that the factorised squared amplitude in the soft-collinear limit can be written as~\cite{Catani:1999ss}:
\begin{equation}
	\label{eq:soft-col-factorisation}
	\begin{split}
		|\mathcal{M}&_{g,a_1,a_2,...,a_n}(q,p_1,p_2,...,p_n)|^2\\
		& \simeq -\frac{2}{S_{12}}(4\pi \mu^{2\varepsilon}\alpha_s)^2\bra{\mathcal{M}_{a,...,a_n}(p,...,p_n)} \hat{\mathbf{P}}_{a_1,a_2}[\mathbf{J}^\dagger_{(12)\mu}(q)\mathbf{J}^\mu_{(12)}(q)]\ket{\mathcal{M}_{a,...,a_n}(p,...,p_n)},
	\end{split}
\end{equation}
where
\begin{equation}
	\mathbf{J}^\dagger_{(12)\mu}(q)\mathbf{J}^\mu_{(12)}(q) \simeq  \sum_{i,j=3}^n \mathbf{T}_i \scdot \mathbf{T}_j \ \mathcal{S}_{ij}(q)+2 \sum_{i=3}^n \mathbf{T}_{i}\scdot \mathbf{T}_{(12)}\mathcal{S}_{i(12)}(q),
\end{equation}
and
\begin{equation}
	\mathbf{T}_{(12)} = \mathbf{T}_{1} +\mathbf{T}_{2}.
\end{equation}
The Eikonal functions of the soft emissions are defined as
\begin{equation}
	\mathcal{S}_{ij}(q) =\frac{2 S_{ij}}{S_{iq} S_{jq}}, \quad 	\mathcal{S}_{i(12)}(q) = \frac{2 (S_{i1} + S_{i2})}{S_{iq}(S_{1q}+S_{2q})}.
\end{equation}
Note that the first factor in \eqref{eq:splitting-soft-collinear-2E}
exactly coincides with the single emission splitting function $\langle
\hat{P}_{q_i g_1}(z_i)\rangle/S_{i1}$ for
\begin{equation}
	z_i = \frac{n\scdot q_i}{n\scdot q_1 + n\scdot q_i}.
\end{equation} 
As in the double-soft case, we can now set $n=p_j$ in
\eqref{eq:splitting-soft-collinear-2E} to find
\begin{align}
	\left. \frac{\langle \hat{P}_{q_i g_1 g_2} \rangle}{ S_{i12}^2}\right\rvert_{S_2C_{i1}} 
	=
	\frac{\langle \hat{P}_{q_i g_1} (z_i)\rangle}{2 S_{i1} } 
	\left\{
	C_F \left[
	\mathcal{S}_{ij}(q_2) + \mathcal{S}_{j(i1)}(q_2)
	\right]
	-\frac{C_A}{2}\left[ \mathcal{S}_{ij}(q_2) - \mathcal{S}_{1j}(q_2)  \right]
	\right\}.
\end{align}
This shows how the genuine soft-collinear behaviour as described by
\eqref{eq:soft-col-factorisation} can be reproduced via the
two-emission splitting function (up to the correct reproduction of the
colour correlators).

In summary, we have shown that the two emission splitting function in
a physical gauge contains not only information on the triple
collinear, but also the double soft and soft-collinear limits.  The
only obvious difference between the genuine limiting expressions and
the splitting functions lies in the colour correlators and
combinatorial factors.  These results show the possibility to
construct an NNLO-subtraction scheme with multi-emission splitting
kernels in terms of a generalization of the NLO-dipole method
of~\cite{Catani:1996vz}.  Schematically, the procedure would be to
extract genuine limiting expressions from the results above, \eg a
purely collinear bit of the two emission splitting function via the
subtractions
\begin{equation}
	\left.\frac{\langle \hat{P}_{q_i g_1 g_2} \rangle}{S_{i12}^2}\right\rvert_{CC} =
	\frac{\langle \hat{P}_{q_i g_1 g_2} \rangle}{S_{i12}^2} 
	- \left.\frac{\langle \hat{P}_{q_i g_1 g_2} \rangle}{S_{i12}^2}\right\rvert_{S_1C_{i2}} 
	-  \left.\frac{\langle \hat{P}_{q_i g_1 g_2} \rangle}{S_{i12}^2}\right\rvert_{S_2C_{i1}} 
	- \left.\frac{\langle \hat{P}_{q_i g_1 g_2} \rangle}{S_{i12}^2}\right\rvert_{SS}.
\end{equation}
and then constructing a joint splitting operator as a linear
combination of such expressions together with appropriate mock-up
colour correlators and partitioning factors.  The result would be a
single operator which reproduces all limits correctly with a smooth
interpolation between them.  The downside to this procedure is the
reconstruction of colour correlators as in
\eqref{eq:double-soft-non-Ab} via reverse-engineering which does not
guarantee the correct inclusion of colour-correlations in intermediate
regions or \eg the large angle soft limit \cite{Dasgupta:2018nvj}.  In
contrast, the formalism outlined in the main part of this paper avoids
such features by leaving the subtractions mentioned above explicit
between all topologies contributing to a given limit at a given
scaling power (at the cost of having gauge-dependent intermediate
expressions).

\section{Two emission topologies}
\label{app:two-emission-diagrams}
In this section, we list the relevant topologies for two-emission
processes with one and two recoilers or spectators in terms of cut
diagrams.
This is helpful for a visual distinction of different types of contributions, yet for the actual use  in a splitting kernel we think of these diagrams in the operator-type picture from the main sections of this paper.
The set of topologies presented here represent a classification of all possible amplitudes squared which arise when adding two emissions to a set of hard legs.
Denoting the emissions by the fixed labels $j$ and $k$, we can write this generically as
\begin{align}
	|\mathcal{M}_{n+2}|^2 = &\sum_i \sum_\alpha \left(\tilde{E}^{(\alpha)}_{ijk} + (j \leftrightarrow k)\right) 
	\nonumber\\
	&+ \sum_i \sum_{l\neq i} \sum_\alpha \left(\tilde{A}^{(\alpha)}_{ijkl} + \tilde{B}^{(\alpha)}_{ijkl} + \tilde{X}^{(\alpha)}_{ijkl}  + (j \leftrightarrow k) \right)
	\nonumber \\
	&+ \sum_i \sum_{l\neq i} \sum_{m\neq l,i} \sum_\alpha \left( \tilde{F}^{(\alpha)}_{ijklm}  + (j \leftrightarrow k)\right),
\end{align}
where the respective topologies are defined below and the tilde refers to these topologies potentially being partitioned.

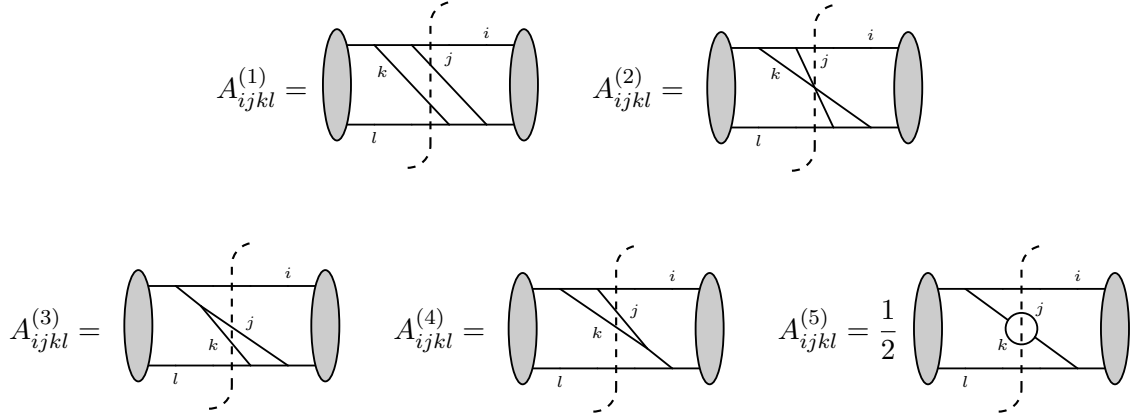
\begin{figure}
	\begin{fmffile}{double-emissionA1to5}
		\fmfset{thin}{.7pt}
		\fmfset{dash_len}{1.5mm}
		\fmfset{arrow_len}{2.5mm}
		\fmfset{wiggly_len}{2.2mm}
		\begin{align*}
			A^{(1)}_{ijkl} = \ \
			&  \begin{gathered}
				\begin{tikzpicture}
					\node (diagram) {%
						\begin{fmfgraph*}(70,30)
							\fmfstraight
							\fmfleft{ld,lm,lu}
							\fmfright{rd,rm,ru}
							\fmf{plain}{lu,vu1,vu2,vu3,vu4,ru}
							\fmf{plain}{ld,vd1,vd2,vd3,vd4,rd}
							\fmffreeze
							\fmfv{label=\tiny{$i$},label.angle=90,label.dist=1mm}{vu4}
							\fmfv{label=\tiny{$l$},label.angle=-90,label.dist=1mm}{vd1}
							\fmf{plain}{vu1,vd3}
							\fmf{plain}{vu2,vd4}
							\fmfv{label=\tiny{$j$},label.angle=-90,label.dist=1mm}{vu3}
							\fmfv{label=\tiny{$k$},label.angle=-70,label.dist=3mm}{vu1}
							\fmfovalblob{.6w}{.25}{lm}
							\fmfovalblob{.6w}{.25}{rm}
						\end{fmfgraph*}
					};
					\tikzset{shift={(0,0)}}
					\draw[thick, dashed] (-0.3,-1.1) arc(-90:0:0.3) (0,-0.8) -- (0,0.8) arc(180:90:0.3);
				\end{tikzpicture}
			\end{gathered}
			\qquad
			A^{(2)}_{ijkl} = \ \
			\begin{gathered}
				\begin{tikzpicture}
					\node (diagram) {%
						\begin{fmfgraph*}(70,30)
							\fmfstraight
							\fmfleft{ld,lm,lu}
							\fmfright{rd,rm,ru}
							\fmf{plain}{lu,vu1,vu2,vu3,vu4,ru}
							\fmf{plain}{ld,vd1,vd2,vd3,vd4,rd}
							\fmffreeze
							\fmfv{label=\tiny{$i$},label.angle=90,label.dist=1mm}{vu4}
							\fmfv{label=\tiny{$l$},label.angle=-90,label.dist=1mm}{vd1}
							\fmf{plain}{vu1,vd4}
							\fmf{plain}{vu2,vd3}
							\fmfv{label=\tiny{$j$},label.angle=-120,label.dist=1mm}{vu3}
							\fmfv{label=\tiny{$k$},label.angle=-60,label.dist=3mm}{vu1}
							\fmfovalblob{.6w}{.25}{lm}
							\fmfovalblob{.6w}{.25}{rm}
						\end{fmfgraph*}
					};
					\tikzset{shift={(0,0)}}
					\draw[thick, dashed] (-0.3,-1.1) arc(-90:0:0.3) (0,-0.8) -- (0,0.8) arc(180:90:0.3);
				\end{tikzpicture}
			\end{gathered}
		\end{align*}
		\begin{align*}
			A^{(3)}_{ijkl} = \ \
			\begin{gathered}
				\begin{tikzpicture}
					\node (diagram) {%
						\begin{fmfgraph*}(70,30)
							\fmfstraight
							\fmfleft{ld,lm1,lm2,lm3,lu}
							\fmfright{rd,rm1,rm2,rm3,ru}
							\fmf{plain}{lu,vu1,vu2,vu3,vu4,ru}
							\fmf{phantom}{lm3,vm1,vm2,rm3}
							\fmf{plain}{ld,vd1,vd2,vd3,vd4,rd}
							\fmffreeze
							\fmfv{label=\tiny{$i$},label.angle=90,label.dist=1mm}{vu4}
							\fmfv{label=\tiny{$l$},label.angle=-90,label.dist=1mm}{vd1}
							\fmf{plain}{vu1,vm1}
							\fmfv{label=\tiny{$k$},label.angle=90,label.dist=2mm}{vd2}
							\fmfv{label=\tiny{$j$},label.angle=-90,label.dist=4mm}{vu3}
							\fmf{plain}{vm1,vd4}
							\fmf{plain}{vm1,vd3}
							\fmfovalblob{.6w}{.25}{lm2}
							\fmfovalblob{.6w}{.25}{rm2}
						\end{fmfgraph*}
					};
					\tikzset{shift={(0,0)}}
					\draw[thick, dashed] (-0.3,-1.1) arc(-90:0:0.3) (0,-0.8) -- (0,0.8) arc(180:90:0.3);
				\end{tikzpicture}
			\end{gathered}
			\qquad
			A^{(4)}_{ijkl} = \ \
			\begin{gathered}
				\begin{tikzpicture}
					\node (diagram) {%
						\begin{fmfgraph*}(70,30)
							\fmfstraight
							\fmfleft{ld,lm1,lm2,lm3,lu}
							\fmfright{rd,rm1,rm2,rm3,ru}
							\fmf{plain}{lu,vu1,vu2,vu3,vu4,ru}
							\fmf{phantom}{lm1,vm1,vm2,rm1}
							\fmf{plain}{ld,vd1,vd2,vd3,vd4,rd}
							\fmffreeze
							\fmfv{label=\tiny{$i$},label.angle=90,label.dist=1mm}{vu4}
							\fmfv{label=\tiny{$l$},label.angle=-90,label.dist=1mm}{vd1}
							\fmfv{label=\tiny{$k$},label.angle=90,label.dist=4mm}{vd2}
							\fmfv{label=\tiny{$j$},label.angle=-90,label.dist=2mm}{vu3}
							\fmf{plain}{vd4,vm2}
							\fmf{plain}{vm2,vu1}
							\fmf{plain}{vm2,vu2}
							\fmf{phantom}{vu4,vd1}
							\fmfovalblob{.6w}{.25}{lm2}
							\fmfovalblob{.6w}{.25}{rm2}
						\end{fmfgraph*}
					};
					\tikzset{shift={(0,0)}}
					\draw[thick, dashed] (-0.3,-1.1) arc(-90:0:0.3) (0,-0.8) -- (0,0.8) arc(180:90:0.3);
				\end{tikzpicture}
			\end{gathered}
			\qquad 
			A^{(5)}_{ijkl} = \frac{1}{2} \ \ 
			\begin{gathered}
				\begin{tikzpicture}
					\node (diagram) {%
						\begin{fmfgraph*}(70,30)
							\fmfstraight
							\fmfleft{ld,lm1,lm2,lm3,lu}
							\fmfright{rd,rm1,rm2,rm3,ru}
							\fmf{plain}{lu,vu1,vu2,vu3,vu4,ru}
							\fmf{phantom}{lm1,vm1,vm2,rm1}
							\fmf{phantom}{lm3,vm31,vm32,rm3}
							\fmf{plain}{ld,vd1,vd2,vd3,vd4,rd}
							\fmf{phantom,tension=1.}{vm31,vm2}
							\fmffreeze
							\fmfv{label=\tiny{$i$},label.angle=90,label.dist=1mm}{vu4}
							\fmfv{label=\tiny{$l$},label.angle=-90,label.dist=1mm}{vd1}
							\fmfv{label=\tiny{$k$},label.angle=90,label.dist=2.5mm}{vd2}
							\fmfv{label=\tiny{$j$},label.angle=-90,label.dist=1.8mm}{vu3}
							\fmf{plain}{vu1,vm31}
							\fmf{plain}{vd4,vm2}
							\fmf{plain,right}{vm31,vm2}
							\fmf{plain,left}{vm31,vm2}
							\fmf{phantom}{vu4,vd1}
							\fmfovalblob{.6w}{.25}{lm2}
							\fmfovalblob{.6w}{.25}{rm2}
						\end{fmfgraph*}
					};
					\tikzset{shift={(0,0)}}
					\draw[thick, dashed] (-0.3,-1.1) arc(-90:0:0.3) (0,-0.8) -- (0,0.8) arc(180:90:0.3);
				\end{tikzpicture}
			\end{gathered}
		\end{align*}
		\captionof{figure}{Triplet-triplet exchange topologies.}
		\label{fig:triplet-triplet-diagrams}
	\end{fmffile}%
\end{figure}

\begin{figure}
	\begin{fmffile}{double-emissionB-diagrams}
		\fmfset{thin}{.7pt}
		\fmfset{dash_len}{1.5mm}
		\fmfset{arrow_len}{2.5mm}
		\fmfset{wiggly_len}{2.2mm}
		\begin{align*}
			B^{(1)}_{ijkl} = \ 
			\begin{gathered}
				\begin{tikzpicture}
					\node (diagram) {%
						\begin{fmfgraph*}(70,30)
							\fmfstraight
							\fmfleft{ld,lm,lu}
							\fmfright{rd,rm,ru}
							\fmf{plain}{lu,vu1,vu2,vu3,vu4,ru}
							\fmf{plain}{ld,vd1,vd2,vd3,vd4,rd}
							\fmf{phantom,tension=1.1}{vu1,vu2}
							\fmf{phantom,tension=1.1}{vu3,vu4}
							\fmf{phantom,tension=1.1}{vd3,vd4}
							\fmffreeze
							\fmfv{label=\tiny{$i$},label.angle=-45,label.dist=1.4mm}{vu2}
							\fmfv{label=\tiny{$l$},label.angle=-90,label.dist=1mm}{vd1}
							\fmfv{label=\tiny{$k$},label.angle=90,label.dist=1.6mm}{vd4}
							\fmf{plain,left}{vu2,vu3}
							\fmf{plain}{vu1,vd4}
							\fmfv{label=\tiny{$j$},label.angle=45,label.dist=1mm}{vu3}
							\fmf{phantom}{vu4,vd1}
							\fmfovalblob{.6w}{.25}{lm}
							\fmfovalblob{.6w}{.25}{rm}
						\end{fmfgraph*}
					};
					\tikzset{shift={(0,0)}}
					\draw[thick, dashed] (-0.3,-1.1) arc(-90:0:0.3) (0,-0.8) -- (0,0.8) arc(180:90:0.3);
				\end{tikzpicture}
			\end{gathered}
			\qquad
			B^{(2)}_{ijkl} = \ \
			\begin{gathered}
				\begin{tikzpicture}
					\node (diagram) {%
						\begin{fmfgraph*}(70,30)
							\fmfstraight
							\fmfleft{ld,lm,lu}
							\fmfright{rd,rm,ru}
							\fmf{plain}{lu,vu1,vu2,vu3,vu4,ru}
							\fmf{plain}{ld,vd1,vd2,vd3,vd4,rd}
							\fmf{phantom,tension=0.5}{vu1,vu4}
							\fmf{phantom,tension=0.5}{vu1,vu2}
							\fmffreeze
							\fmfv{label=\tiny{$i$},label.angle=-130,label.dist=1.4mm}{vu4}
							\fmfv{label=\tiny{$l$},label.angle=-90,label.dist=1mm}{vd1}
							\fmfv{label=\tiny{$j$},label.angle=60,label.dist=1.4mm}{vd3}
							\fmf{plain,left}{vu1,vu4}
							\fmf{plain}{vu2,vd3}
							\fmfv{label=\tiny{$k$},label.angle=135,label.dist=2mm}{vu1}
							\fmf{phantom}{vu4,vd1}
							\fmfovalblob{.6w}{.25}{lm}
							\fmfovalblob{.6w}{.25}{rm}
						\end{fmfgraph*}
					};
					\tikzset{shift={(0,0)}}
					\draw[thick, dashed] (-0.3,-1.1) arc(-90:0:0.3) (0,-0.8) -- (0,0.8) arc(180:90:0.3);
				\end{tikzpicture}
			\end{gathered}
			\qquad
			B^{(3)}_{ijkl} = \ \
			\begin{gathered}
				\begin{tikzpicture}
					\node (diagram) {%
						\begin{fmfgraph*}(70,30)
							\fmfstraight
							\fmfleft{ld,lm,lu}
							\fmfright{rd,rm,ru}
							\fmf{plain}{lu,vu1,vu2,vu3,vu4,ru}
							\fmf{plain}{ld,vd1,vd2,vd3,vd4,rd}
							\fmf{phantom,tension=1.1}{vu1,vu2}
							\fmf{phantom,tension=1.1}{vu3,vu4}
							\fmf{phantom,tension=1.1}{vd3,vd4}
							\fmffreeze
							\fmfv{label=\tiny{$i$},label.angle=45,label.dist=1.4mm}{vu2}
							\fmfv{label=\tiny{$l$},label.angle=-90,label.dist=1mm}{vd1}
							\fmfv{label=\tiny{$k$},label.angle=90,label.dist=1.6mm}{vd4}
							\fmf{phantom}{vu1,vx1,vx2,vd4}
							\fmf{plain,tension=4}{vu1,vx1}
							\fmf{plain}{vx1,vd4}
							\fmffreeze
							\fmf{plain,right=0.5}{vx1,vu3}
							\fmfv{label=\tiny{$j$},label.angle=-45,label.dist=1mm}{vu3}
							\fmf{phantom}{vu4,vd1}
							\fmfovalblob{.6w}{.25}{lm}
							\fmfovalblob{.6w}{.25}{rm}
						\end{fmfgraph*}
					};
					\tikzset{shift={(0,0)}}
					\draw[thick, dashed] (-0.3,-1.1) arc(-90:0:0.3) (0,-0.8) -- (0,0.8) arc(180:90:0.3);
				\end{tikzpicture}
			\end{gathered}\\
			B^{(4)}_{ijkl} = \
			\begin{gathered}
				\begin{tikzpicture}
					\node (diagram) {%
						\begin{fmfgraph*}(70,30)
							\fmfstraight
							\fmfright{ld,lm,lu}
							\fmfleft{rd,rm,ru}
							\fmf{plain}{lu,vu1,vu2,vu3,vu4,ru}
							\fmf{plain}{ld,vd1,vd2,vd3,vd4,rd}
							\fmf{phantom,tension=1.1}{vu1,vu2}
							\fmf{phantom,tension=1.1}{vu3,vu4}
							\fmf{phantom,tension=1.1}{vd3,vd4}
							\fmffreeze
							\fmfv{label=\tiny{$i$},label.angle=-135,label.dist=1mm}{vu2}
							\fmfv{label=\tiny{$l$},label.angle=-90,label.dist=1mm}{vd1}
							\fmfv{label=\tiny{$k$},label.angle=90,label.dist=1.6mm}{vd4}
							\fmf{plain,right}{vu2,vu3}
							\fmf{plain}{vu1,vd4}
							\fmfv{label=\tiny{$j$},label.angle=135,label.dist=1mm}{vu3}
							\fmf{phantom}{vu4,vd1}
							\fmfovalblob{.6w}{.25}{lm}
							\fmfovalblob{.6w}{.25}{rm}
						\end{fmfgraph*}
					};
					\tikzset{shift={(0,0)}}
					\draw[thick, dashed] (-0.3,-1.1) arc(-90:0:0.3) (0,-0.8) -- (0,0.8) arc(180:90:0.3);
				\end{tikzpicture}
			\end{gathered}
			\qquad
			B^{(5)}_{ijkl} = \ \
			\begin{gathered}
				\begin{tikzpicture}
					\node (diagram) {%
						\begin{fmfgraph*}(70,30)
							\fmfstraight
							\fmfright{ld,lm,lu}
							\fmfleft{rd,rm,ru}
							\fmf{plain}{lu,vu1,vu2,vu3,vu4,ru}
							\fmf{plain}{ld,vd1,vd2,vd3,vd4,rd}
							\fmf{phantom,tension=0.5}{vu1,vu4}
							\fmf{phantom,tension=0.5}{vu1,vu2}
							\fmffreeze
							\fmfv{label=\tiny{$i$},label.angle=-45,label.dist=1.4mm}{vu4}
							\fmfv{label=\tiny{$l$},label.angle=-90,label.dist=1mm}{vd1}
							\fmfv{label=\tiny{$j$},label.angle=135,label.dist=1.4mm}{vd3}
							\fmf{plain,right}{vu1,vu4}
							\fmf{plain}{vu2,vd3}
							\fmfv{label=\tiny{$k$},label.angle=45,label.dist=2mm}{vu1}
							\fmf{phantom}{vu4,vd1}
							\fmfovalblob{.6w}{.25}{lm}
							\fmfovalblob{.6w}{.25}{rm}
						\end{fmfgraph*}
					};
					\tikzset{shift={(0,0)}}
					\draw[thick, dashed] (-0.3,-1.1) arc(-90:0:0.3) (0,-0.8) -- (0,0.8) arc(180:90:0.3);
				\end{tikzpicture}
			\end{gathered}
			\qquad
			B^{(6)}_{ijkl} = \ \
			\begin{gathered}
				\begin{tikzpicture}
					\node (diagram) {%
						\begin{fmfgraph*}(70,30)
							\fmfstraight
							\fmfright{ld,lm,lu}
							\fmfleft{rd,rm,ru}
							\fmf{plain}{lu,vu1,vu2,vu3,vu4,ru}
							\fmf{plain}{ld,vd1,vd2,vd3,vd4,rd}
							\fmf{phantom,tension=1.1}{vu1,vu2}
							\fmf{phantom,tension=1.1}{vu3,vu4}
							\fmf{phantom,tension=1.1}{vd3,vd4}
							\fmffreeze
							\fmfv{label=\tiny{$i$},label.angle=135,label.dist=1.4mm}{vu2}
							\fmfv{label=\tiny{$l$},label.angle=-90,label.dist=1mm}{vd1}
							\fmfv{label=\tiny{$k$},label.angle=90,label.dist=1.6mm}{vd4}
							\fmf{phantom}{vu1,vx1,vx2,vd4}
							\fmf{plain,tension=4}{vu1,vx1}
							\fmf{plain}{vx1,vd4}
							\fmffreeze
							\fmf{plain,left=0.5}{vx1,vu3}
							\fmfv{label=\tiny{$j$},label.angle=-135,label.dist=1mm}{vu3}
							\fmf{phantom}{vu4,vd1}
							\fmfovalblob{.6w}{.25}{lm}
							\fmfovalblob{.6w}{.25}{rm}
						\end{fmfgraph*}
					};
					\tikzset{shift={(0,0)}}
					\draw[thick, dashed] (-0.3,-1.1) arc(-90:0:0.3) (0,-0.8) -- (0,0.8) arc(180:90:0.3);
				\end{tikzpicture}
			\end{gathered}
		\end{align*}
		\captionof{figure}{Triplet-pairs exchange topologies.}
		\label{fig:triplet-pairs-diagrams}
	\end{fmffile}%
\end{figure}
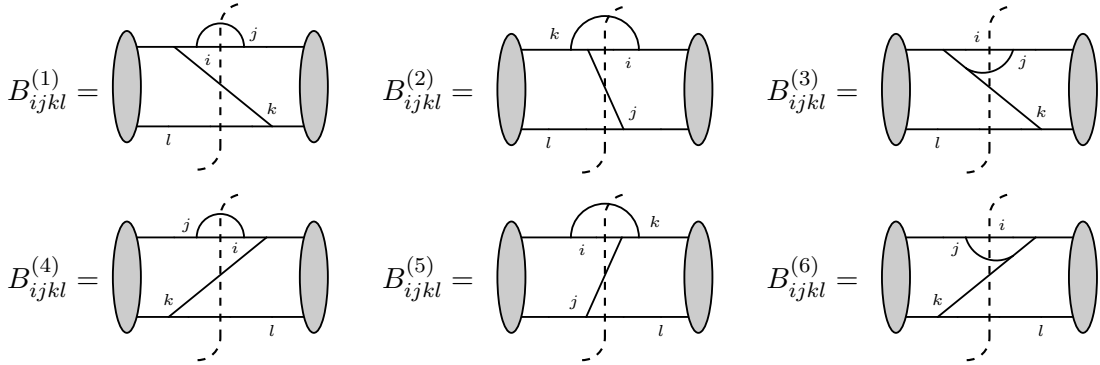

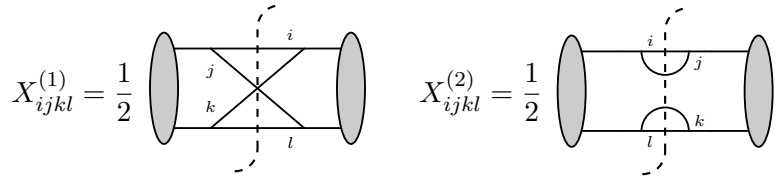
\begin{figure}
	\begin{fmffile}{double-emissionX}
		\fmfset{thin}{.7pt}
		\fmfset{dash_len}{1.5mm}
		\fmfset{arrow_len}{2.5mm}
		\fmfset{wiggly_len}{2.2mm}
		\begin{align*}
			X^{(1)}_{ijkl} = \frac{1}{2}\ \
			\begin{gathered}
				\begin{tikzpicture}
					\node (diagram) {%
						\begin{fmfgraph*}(70,30)
							\fmfstraight
							\fmfleft{ld,lm,lu}
							\fmfright{rd,rm,ru}
							\fmf{plain}{lu,vu1,vu2,vu3,ru}
							\fmf{plain}{ld,vd1,vd2,vd3,rd}
							\fmffreeze
							\fmfv{label=\tiny{$i$},label.angle=140,label.dist=1.8mm}{vu3}
							\fmfv{label=\tiny{$l$},label.angle=-140,label.dist=1.8mm}{vd3}
							\fmf{plain}{vu1,vd3}
							\fmf{phantom,label.side=right,label=\tiny$j$}{lu,vu2}
							\fmf{plain}{vd1,vu3}
							\fmf{phantom,label.side=right,label=\tiny$k$}{vd2,ld}
							\fmfovalblob{.6w}{.25}{lm}
							\fmfovalblob{.6w}{.25}{rm}
						\end{fmfgraph*}
					};
					\tikzset{shift={(0,0)}}
					\draw[thick, dashed] (-0.3,-1.1) arc(-90:0:0.3) (0,-0.8) -- (0,0.8) arc(180:90:0.3);
				\end{tikzpicture}
			\end{gathered}
			\qquad
			X^{(2)}_{ijkl} = \frac{1}{2}\ \
			\begin{gathered}
				\begin{tikzpicture}
					\node (diagram) {%
						\begin{fmfgraph*}(70,30)
							\fmfstraight
							\fmfleft{ld,lm,lu}
							\fmfright{rd,rm,ru}
							\fmf{plain}{lu,vu1,vu2,vu3,vu4,ru}
							\fmf{plain}{ld,vd1,vd2,vd3,vd4,rd}
							\fmf{phantom,tension=1.1}{vu1,vu2}
							\fmf{phantom,tension=1.1}{vu3,vu4}
							\fmf{phantom,tension=1.1}{vd1,vd2}
							\fmf{phantom,tension=1.1}{vd3,vd4}
							\fmffreeze
							\fmfv{label=\tiny{$i$},label.angle=45,label.dist=1mm}{vu2}
							\fmfv{label=\tiny{$j$},label.angle=-45,label.dist=1mm}{vu3}
							\fmf{plain,right}{vu2,vu3}
							\fmf{plain,left}{vd2,vd3}
							\fmfv{label=\tiny{$k$},label.angle=45,label.dist=1mm}{vd3}
							\fmfv{label=\tiny{$l$},label.angle=-45,label.dist=1mm}{vd2}
							\fmf{phantom}{vu4,vd1}
							\fmfovalblob{.6w}{.25}{lm}
							\fmfovalblob{.6w}{.25}{rm}
						\end{fmfgraph*}
					};
					\tikzset{shift={(0,0)}}
					\draw[thick, dashed] (-0.3,-1.1) arc(-90:0:0.3) (0,-0.8) -- (0,0.8) arc(180:90:0.3);
				\end{tikzpicture}
			\end{gathered}
		\end{align*}
		\captionof{figure}{Pairs-Pairs exchange topologies.}
		\label{fig:x-diagram}
	\end{fmffile}%
\end{figure}

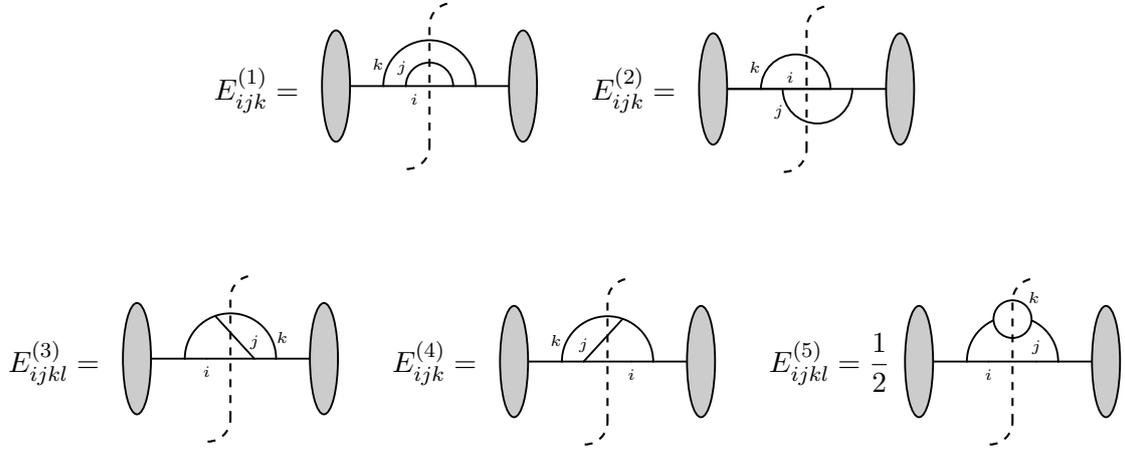
\begin{figure}
	\begin{fmffile}{self-energyE1to4}
		\fmfset{thin}{.7pt}
		\fmfset{dash_len}{1.5mm}
		\fmfset{arrow_len}{2.5mm}
		\fmfset{wiggly_len}{2.2mm}
		\begin{align*}
			E^{(1)}_{ijk} = \ \
			\begin{gathered}
				\begin{tikzpicture}
					\node (diagram) {%
						\begin{fmfgraph*}(70,80)
							\fmfleft{bl}
							\fmfright{br}
							\fmf{plain}{bl,v1,v2,v3,v4,br}
							\fmf{phantom,tension=1.1}{v1,v2}
							\fmf{phantom,tension=1.1}{v3,v4}
							\fmffreeze
							\fmf{plain,left}{v2,v3}
							\fmf{plain,left}{v1,v4}
							\fmfv{label=\tiny{$i$},label.angle=-60,label.dist=1.3mm}{v2}
							\fmf{phantom,label.side=left,label=\tiny$k$}{bl,v2}
							\fmfv{label=\tiny{$j$},label.angle=40,label.dist=2.4mm}{v1}
							\fmfovalblob{.6w}{.25}{bl}
							\fmfovalblob{.6w}{.25}{br}
						\end{fmfgraph*}
					};
					\tikzset{shift={(0,0)}}
					\draw[thick, dashed] (-0.3,-1.1) arc(-90:0:0.3) (0,-0.8) -- (0,0.8) arc(180:90:0.3);
				\end{tikzpicture}
			\end{gathered}
			\qquad
			E^{(2)}_{ijk} = \ \
			\begin{gathered}
				\begin{tikzpicture}
					\node (diagram) {%
						\begin{fmfgraph*}(70,80)
							\fmfleft{bl}
							\fmfright{br}
							\fmf{plain}{bl,v1,v2,v3,v4,br}
							\fmf{phantom,tension=1.2}{v1,v2}
							\fmf{phantom,tension=1.2}{v3,v4}
							\fmffreeze
							\fmf{plain,right}{v2,v4}
							\fmf{plain,left}{v1,v3}
							\fmfv{label=\tiny{$i$},label.angle=60,label.dist=1mm}{v2}
							\fmf{phantom,label.side=left,label=\tiny$k$}{bl,v2}
							\fmf{plain,label.side=right,label=\tiny$j$}{bl,v3}
							\fmfovalblob{.6w}{.25}{bl}
							\fmfovalblob{.6w}{.25}{br}
						\end{fmfgraph*}
					};
					\tikzset{shift={(0,0)}}
					\draw[thick, dashed] (-0.3,-1.1) arc(-90:0:0.3) (0,-0.8) -- (0,0.8) arc(180:90:0.3);
				\end{tikzpicture}
			\end{gathered}
		\end{align*}
		\begin{align*}
			E^{(3)}_{ijkl} = \ \
			\begin{gathered}
				\begin{tikzpicture}
					\node (diagram) {%
						\begin{fmfgraph*}(70,32)
							\fmfright{bl,ml,ul}
							\fmfleft{br,mr,ur}
							\fmf{plain}{ml,v1,v2,v3,v4,mr}
							\fmf{phantom}{ul,vu1,vu2,vu3,vu4,ur}
							\fmf{phantom,tension=1.2}{v1,v2}
							\fmf{phantom,tension=1.2}{v3,v4}
							\fmffreeze
							\fmf{plain}{v2,vu3}
							\fmf{plain,right}{v1,v4}
							\fmfv{label=\tiny{$i$},label.angle=-90,label.dist=1mm}{v3}
							\fmf{phantom,label.side=right,label=\tiny$k$}{ml,v2}
							\fmfv{label=\tiny{$j$},label.angle=90,label.dist=1.3mm}{v2}
							\fmfovalblob{.6w}{.25}{ml}
							\fmfovalblob{.6w}{.25}{mr}
						\end{fmfgraph*}
					};
					\tikzset{shift={(0,0)}}
					\draw[thick, dashed] (-0.3,-1.1) arc(-90:0:0.3) (0,-0.8) -- (0,0.8) arc(180:90:0.3);
				\end{tikzpicture}
			\end{gathered}
			\qquad
			E^{(4)}_{ijk} = \ \
			\begin{gathered}
				\begin{tikzpicture}
					\node (diagram) {%
						\begin{fmfgraph*}(70,32)
							\fmfleft{bl,ml,ul}
							\fmfright{br,mr,ur}
							\fmf{plain}{ml,v1,v2,v3,v4,mr}
							\fmf{phantom}{ul,vu1,vu2,vu3,vu4,ur}
							\fmf{phantom,tension=1.2}{v1,v2}
							\fmf{phantom,tension=1.2}{v3,v4}
							\fmffreeze
							\fmf{plain}{v2,vu3}
							\fmf{plain,left}{v1,v4}
							\fmfv{label=\tiny{$i$},label.angle=-90,label.dist=1mm}{v3}
							\fmf{phantom,label.side=left,label=\tiny$k$}{ml,v2}
							\fmfv{label=\tiny{$j$},label.angle=90,label.dist=1.3mm}{v2}
							\fmfovalblob{.6w}{.25}{ml}
							\fmfovalblob{.6w}{.25}{mr}
						\end{fmfgraph*}
					};
					\tikzset{shift={(0,0)}}
					\draw[thick, dashed] (-0.3,-1.1) arc(-90:0:0.3) (0,-0.8) -- (0,0.8) arc(180:90:0.3);
				\end{tikzpicture}
			\end{gathered}
			\qquad
			E^{(5)}_{ijkl} =  \frac{1}{2} \ \
			\begin{gathered}
				\begin{tikzpicture}
					\node (diagram) {%
						\begin{fmfgraph*}(70,32)
							\fmfright{bl,ml,ul}
							\fmfleft{br,mr,ur}
							\fmf{plain}{ml,v1,v2,v3,v4,mr}
							\fmf{phantom}{ul,vu1,vu2,vum,vu3,vu4,ur}
							\fmf{phantom,tension=1.2}{v1,v2}
							\fmf{phantom,tension=1.2}{v3,v4}
							\fmffreeze
							\fmfv{decor.shape=circle,decor.filled=empty,label=\tiny{$k$},label.angle=45,label.dist=3mm}{vum}
							\fmf{plain,right}{v1,v4}
							\fmfv{label=\tiny{$i$},label.angle=-90,label.dist=1mm}{v3}
							\fmfv{label=\tiny{$j$},label.angle=90,label.dist=1mm}{v2}
							\fmfovalblob{.6w}{.25}{ml}
							\fmfovalblob{.6w}{.25}{mr}
						\end{fmfgraph*}
					};
					\tikzset{shift={(0,0)}}
					\draw[thick, dashed] (-0.3,-1.1) arc(-90:0:0.3) (0,-0.8) -- (0,0.8) arc(180:90:0.3);
				\end{tikzpicture}
			\end{gathered}
		\end{align*}
	\end{fmffile}%
	\captionof{figure}{Self-energy like topologies.}
	\label{fig:self-energies}
\end{figure}

\begin{figure}
	\begin{fmffile}{two-spectatorF1to3}
		\fmfset{thin}{.7pt}
		\fmfset{dash_len}{1.5mm}
		\fmfset{arrow_len}{2.5mm}
		\fmfset{wiggly_len}{2.2mm}
		\begin{align*}
			&F^{(1)}_{ijklm} = \ 
			\begin{gathered}
				\begin{tikzpicture}
					\node (diagram) {%
						\begin{fmfgraph*}(70,40)
							\fmfstraight
							\fmfleft{ld,lm,lu}
							\fmfright{rd,rm,ru}
							\fmf{plain}{lu,vu1,vu2,vu3,vu4,ru}
							\fmf{plain}{lm,vm1,vm2,vm3,vm4,rm}
							\fmf{plain}{ld,vd1,vd2,vd3,vd4,rd}
							\fmf{phantom,tension=2}{rd,vd4}
							\fmffreeze
							\fmfv{label=\tiny{$i$},label.angle=90,label.dist=1mm}{vu4}
							\fmfv{label=\tiny{$l$},label.angle=90,label.dist=1mm}{vm1}
							\fmfv{label=\tiny{$m$},label.angle=-90,label.dist=1mm}{vd1}
							\fmf{plain,label.side=right,label=\tiny$k$,label.dist=0.5mm}{vu1,vm3}
							\fmf{plain}{vu2,vd4}
							\fmfv{label=\tiny{$j$},label.angle=-90,label.dist=1mm}{vu3}
							\fmf{phantom}{vu4,vd1}
							\fmfovalblob{.65w}{.2}{lm}
							\fmfovalblob{.65w}{.2}{rm}
						\end{fmfgraph*}
					};
					\tikzset{shift={(0,0)}}
					\draw[thick, dashed] (-0.3,-1.1) arc(-90:0:0.3) (0,-0.8) -- (0,0.8) arc(180:90:0.3);
				\end{tikzpicture}
			\end{gathered}
			\qquad
			F^{(2)}_{ijklm} = \frac{1}{2} \ 
			\begin{gathered}
				\begin{tikzpicture}
					\node (diagram) {%
						\begin{fmfgraph*}(70,40)
							\fmfstraight
							\fmfleft{ld,lm,lu}
							\fmfright{rd,rm,ru}
							\fmf{plain}{lu,vu1,vu2,vu3,vu4,ru}
							\fmf{plain}{lm,vm1,vm2,vmm,vm3,vm4,rm}
							\fmf{plain}{ld,vd1,vd2,vd3,vd4,rd}
							\fmffreeze
							\fmf{phantom}{vu1,vxx,vmm}
							\fmf{plain,tension=4}{vu1,vxx}
							\fmfv{label=\tiny{$i$},label.angle=90,label.dist=1mm}{vu4}
							\fmfv{label=\tiny{$l$},label.angle=90,label.dist=1mm}{vm1}
							\fmfv{label=\tiny{$m$},label.angle=-90,label.dist=1mm}{vd1}
							\fmf{plain,label.side=right,label=\tiny$k$,label.dist=1.5mm}{vxx,vd3}
							\fmf{plain,label.side=left,label=\tiny$j$,label.dist=0.5mm}{vxx,vm4}
							\fmfovalblob{.65w}{.2}{lm}
							\fmfovalblob{.65w}{.2}{rm}
						\end{fmfgraph*}
					};
					\tikzset{shift={(0,0)}}
					\draw[thick, dashed] (-0.3,-1.1) arc(-90:0:0.3) (0,-0.8) -- (0,0.8) arc(180:90:0.3);
				\end{tikzpicture}
			\end{gathered}
			\qquad
			F^{(3)}_{ijklm} = \frac{1}{2} \ 
			\begin{gathered}
				\begin{tikzpicture}
					\node (diagram) {%
						\begin{fmfgraph*}(70,40)
							\fmfstraight
							\fmfleft{ld,lm,lu}
							\fmfright{rd,rm,ru}
							\fmf{plain}{lu,vu1,vu2,vu3,vu4,ru}
							\fmf{plain}{lm,vm1,vm2,vm3,vm4,rm}
							\fmf{plain}{ld,vd1,vd2,vd3,vd4,rd}
							\fmffreeze
							\fmfv{label=\tiny{$i$},label.angle=90,label.dist=1mm}{vu4}
							\fmfv{label=\tiny{$l$},label.angle=90,label.dist=1mm}{vm1}
							\fmfv{label=\tiny{$m$},label.angle=-90,label.dist=1mm}{vd1}
							\fmf{plain,label.side=left,label=\tiny$j$,label.dist=0.5mm}{vu1,vm4}
							\fmf{plain,label.side=left,label=\tiny$k$,label.dist=0.5mm}{vm1,vd4}
							\fmfovalblob{.65w}{.2}{lm}
							\fmfovalblob{.65w}{.2}{rm}
						\end{fmfgraph*}
					};
					\tikzset{shift={(0,0)}}
					\draw[thick, dashed] (-0.3,-1.1) arc(-90:0:0.3) (0,-0.8) -- (0,0.8) arc(180:90:0.3);
				\end{tikzpicture}
			\end{gathered}
		\end{align*}
		\begin{align*}
			&F^{(4)}_{ijklm} =  \ 
			\begin{gathered}
				\begin{tikzpicture}
					\node (diagram) {%
						\begin{fmfgraph*}(70,40)
							\fmfstraight
							\fmfright{ld,lm,lu}
							\fmfleft{rd,rm,ru}
							\fmf{plain}{lu,vu1,vu2,vu3,vu4,ru}
							\fmf{plain}{lm,vm1,vm2,vm3,vm4,rm}
							\fmf{plain}{ld,vd1,vd2,vd3,vd4,rd}
							\fmf{phantom,tension=2}{rd,vd4}
							\fmffreeze
							\fmfv{label=\tiny{$i$},label.angle=90,label.dist=1mm}{vu4}
							\fmfv{label=\tiny{$l$},label.angle=90,label.dist=1mm}{vm1}
							\fmfv{label=\tiny{$m$},label.angle=-90,label.dist=1mm}{vd1}
							\fmf{plain,label.side=left,label=\tiny$k$,label.dist=0.5mm}{vu1,vm3}
							\fmf{plain}{vu2,vd4}
							\fmfv{label=\tiny{$j$},label.angle=-90,label.dist=1mm}{vu3}
							\fmf{phantom}{vu4,vd1}
							\fmfovalblob{.65w}{.2}{lm}
							\fmfovalblob{.65w}{.2}{rm}
						\end{fmfgraph*}
					};
					\tikzset{shift={(0,0)}}
					\draw[thick, dashed] (-0.3,-1.1) arc(-90:0:0.3) (0,-0.8) -- (0,0.8) arc(180:90:0.3);
				\end{tikzpicture}
			\end{gathered}
			\qquad
			F^{(5)}_{ijklm} = \frac{1}{2} \ 
			\begin{gathered}
				\begin{tikzpicture}
					\node (diagram) {%
						\begin{fmfgraph*}(70,40)
							\fmfstraight
							\fmfright{ld,lm,lu}
							\fmfleft{rd,rm,ru}
							\fmf{plain}{lu,vu1,vu2,vu3,vu4,ru}
							\fmf{plain}{lm,vm1,vm2,vmm,vm3,vm4,rm}
							\fmf{plain}{ld,vd1,vd2,vd3,vd4,rd}
							\fmffreeze
							\fmf{phantom}{vu1,vxx,vmm}
							\fmf{plain,tension=4}{vu1,vxx}
							\fmfv{label=\tiny{$i$},label.angle=90,label.dist=1mm}{vu4}
							\fmfv{label=\tiny{$l$},label.angle=90,label.dist=1mm}{vm1}
							\fmfv{label=\tiny{$m$},label.angle=-90,label.dist=1mm}{vd1}
							\fmf{plain,label.side=left,label=\tiny$k$,label.dist=1.5mm}{vxx,vd3}
							\fmf{plain,label.side=right,label=\tiny$j$,label.dist=0.5mm}{vxx,vm4}
							\fmfovalblob{.65w}{.2}{lm}
							\fmfovalblob{.65w}{.2}{rm}
						\end{fmfgraph*}
					};
					\tikzset{shift={(0,0)}}
					\draw[thick, dashed] (-0.3,-1.1) arc(-90:0:0.3) (0,-0.8) -- (0,0.8) arc(180:90:0.3);
				\end{tikzpicture}
			\end{gathered}
		\end{align*}
		\captionof{figure}{Two spectator topologies.}
		\label{fig:two-spectator-triplet-pairs-diagrams}
	\end{fmffile}%
\end{figure}
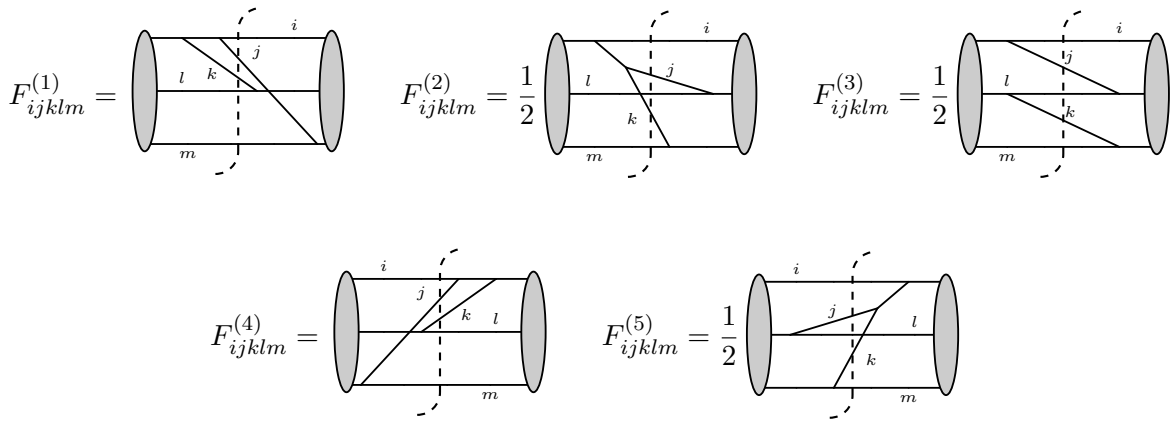

By the term `triplet', we denote emissions off of the same
emitter while `pairs' denotes the emission off of two separate
emitters.
We find five topologies for the triplet-triplet exchange case which are shown in
Fig.~\ref{fig:triplet-triplet-diagrams} and the six triplet-pairs in
Fig.~\ref{fig:triplet-pairs-diagrams}.
Moreover, we have two pairs-pairs topologies as shown
in Fig.~\ref{fig:x-diagram}.
In addition to the exchange ones, we have self energy-type diagrams in
both the triplet-triplet and pairs-pairs case, as shown in
Fig.~\ref{fig:self-energies}.
Tab.~\ref{tab:collinear-factors} shows the singular propagator factors for each topology in any of the possible collinearity settings.
\begin{table}
	\begin{center}
		\begin{tabular}{c|ccccc}
			& $A^{(1)}_{ijkl}$ & $A^{(2)}_{ijkl}$ &$A^{(3)}_{ijkl}$ &$A^{(4)}_{ijkl}$ &$A^{(5)}_{ijkl}$ \\
			&\tiny{$S_{ij}S_{kl}S_{ijk}S_{jkl}$} &\tiny{$S_{ij}S_{jl}S_{ijk}S_{jkl}$} &\tiny{$S_{jk}S_{kl}S_{ijk}S_{jkl}$} &\tiny{$S_{ij}S_{jk}S_{ijk}S_{jkl}$} &\tiny{$ S_{ijk}S_{jk}^2S_{jkl}$}  \\
			\hline
			$i \parallel j \parallel k$ & $S_{ij}S_{ijk}$ 	& $S_{ij}S_{ijk}$ & $S_{jk}S_{ijk}$ 	& $S_{ij}S_{jk}S_{ijk}$ & $S_{jk}^2 S_{ijk}$ \\
			$j \parallel k \parallel l$ & $S_{kl}S_{jkl}$ 	& $S_{jl}S_{jkl}$	& $S_{jk} S_{kl} S_{jkl}$ 	& $S_{jk}S_{jkl}$ 	& $S_{jk}^2 S_{jkl}$ \\
			$(i \parallel j), (k \parallel l)$ & $S_{ij}S_{kl}$ & $S_{ij}$ 		& $S_{kl}$ 	& $S_{ij}$ 			& \X \\
			$(i \parallel k), (j \parallel l)$ &  \X		& $S_{jl}$ 		& \X 	&  \X				& \X \\
		\end{tabular}
		
		\vspace{10pt}
		
		\begin{tabular}{c|ccc}
			& $B^{(1)}_{ijkl}$ & $B^{(2)}_{ijkl}$ & $B^{(3)}_{ijkl}$   \\
			&\tiny{$S_{ij}^2 S_{kl}S_{ijk}$} & \tiny{$S_{ij}S_{ik}S_{jl}S_{ijk}$} &\tiny{$S_{ij}S_{jk} S_{kl} S_{ijk} $} \\ \hline
			$i \parallel j \parallel k$  & $S_{ij}^2 S_{ijk}$	& $S_{ij}S_{ik}S_{ijk}$ & $S_{ij} S_{jk} S_{ijk}$		 \\
			$j \parallel k \parallel l$  & $S_{kl}$			& $S_{jl}$			&  $S_{jk} S_{kl}$\\
			$(i \parallel j), (k \parallel l)$  & $S_{ij}^2 S_{kl}$& $S_{ij}$			&  $S_{ij}S_{kl}$	 \\
			$(i \parallel k), (j \parallel l)$  & \X			& $S_{ik}S_{jl}$ 		&  \X		  \\
		\end{tabular}
		
		\vspace{10pt}
		
		\begin{tabular}{c|cc}
			& $X^{(1)}_{ijkl}$ & $X^{(2)}_{ijkl}$\\
			& \tiny{$S_{ij}S_{ik}S_{jl}S_{kl}$} & \tiny{$S_{ij}^2 S_{kl}^2$}\\ \hline
			$i \parallel j \parallel k$  	& $S_{ij}S_{ik}$ & $S_{ij}^2$\\
			$j \parallel k \parallel l$		&$S_{jl}S_{kl}$ & $S_{kl}^2$ \\
			$(i \parallel j), (k \parallel l)$ 		&$S_{ij}S_{kl}$ & $S_{ij}^2 S_{kl}^2$\\
			$(i \parallel k), (j \parallel l)$ 	&$S_{ik}S_{jl}$ & \X \\
		\end{tabular}
		
		\vspace{10pt}
		
		\begin{tabular}{c|ccccc}
			& $E^{(1)}_{ijk}$ & $E^{(2)}_{ijk}$ & $E^{(3)}_{ijk}$ & $E^{(4)}_{ijkl}$ & $E^{(5)}_{ijkl}$ \\
			&\tiny{$S_{ij}^2S_{ijk}^2$}&	 \tiny{$S_{ij}S_{ik}S_{ijk}^2$} & \tiny{$S_{ij}S_{jk}S_{ijk}^2$} & \tiny{$S_{ij}S_{jk}S_{ijk}^2$}  & \tiny{$S_{jk}^2 S_{ijk}^2$} \\
			\hline
			$i \parallel j \parallel k$  		&  $S_{ij}^2S_{ijk}^2$	&  $S_{ij}S_{ik}S_{ijk}^2$& $S_{ij}S_{jk}S_{ijk}^2$ & $S_{ij}S_{jk}S_{ijk}^2$ & $S_{jk}^2 S_{ijk}^2$\\
			$j \parallel k \parallel l$  		&  \X				& \X 				&$S_{jk}$ 			& $S_{jk}$  & $S_{jk}^2$\\
			$(i \parallel j), (k \parallel l)$  	& $S_{ij}^2$		& $S_{ij}$			&$S_{ij}$			& $S_{ij}$	 & \X\\
			$(i \parallel k), (j \parallel l)$  	&   \X			& $S_{ik}$			&\X 				&\X & \X \\
		\end{tabular}
	\end{center}
	\caption{Collection of propagator factors giving rise to singularities w.r.t.\ the set of collinearity structures for two emissions with one spectator. Note that $B^{(4)}$, $B^{(5)}$ and $B^{(6)}$ have the same propagators and therefore singular factors as $B^{(1)}$,$B^{(2)}$ and $B^{(3)}$, respectively}
	\label{tab:collinear-factors}
\end{table}
Lastly, we have five topologies with two spectators shown in Fig.~\ref{fig:two-spectator-triplet-pairs-diagrams}.
Just as for the one-spectator case, we show the singular propagator factors for the two-spectator diagrams in Tab.~\ref{tab:collinear-factors2}.
\begin{table}
	\begin{center}
		\begin{tabular}{c|ccc}
			& $F^{(1)}_{ijklm}$ & $F^{(2)}_{ijklm}$ & $F^{(3)}_{ijklm}$  \\
			&\tiny{$S_{ij}S_{kl}S_{jm}S_{ijk}$}&	 \tiny{$S_{jk}S_{jl}S_{km}S_{ijk}$} & \tiny{$S_{ij}S_{kl}S_{jl}S_{km}$} \\
			\hline
			$i \parallel j \parallel k$  	&  $S_{ij}S_{ijk}$ & $S_{jk} S_{ijk} $		& $S_{ij}$		 \\
			$j \parallel k \parallel l$  	&  $S_{kl}$		& $S_{jk}S_{jl}$ & $S_{kl} S_{jl}$		\\
			$j \parallel k \parallel m$ &  $S_{jm}$	& $S_{jk} S_{km}$	& $S_{km}$\\
			$(i \parallel j), (k \parallel l)$  	&   $S_{ij} S_{kl}$	& \X	& $S_{ij}S_{kl}$	\\
			$(i \parallel j), (k \parallel m)$ 	&$S_{ij}$ & $S_{km}$	& $S_{ij} S_{km}$		\\
			$(i \parallel k), (j \parallel l)$  	&   \X	& $S_{jl}$ 			& $S_{jl}$ 		\\
			$(i \parallel k), (j \parallel m)$   &  $S_{jm}$		& \X 			&\X		\\
			$(j \parallel l), (k \parallel m)$  	& \X	& $S_{jl}S_{km}$	& $S_{jl}S_{km}$			\\
			$(k \parallel l), (j \parallel m)$  	&  $S_{kl}S_{jm}$			& \X		& $S_{kl}$ \\
		\end{tabular}
	\end{center}
	\caption{Collection of propagator factors giving rise to singularities w.r.t.\ the set of collinearity structures for two emissions with three hard lines $i,l,m$. Note that $F^{(4)}$ and $F^{(5)}$ have the same singular factors as $F^{(1)}$ and $F^{(2)}$, respectively.}
	\label{tab:collinear-factors2}
\end{table}

\clearpage
\section{Removal of non-leading singular contributions}\label{sec:delta-tensor}
In order to construct the object $\hat{\Delta}^{r}_{\bar{r}}$ in \eqref{eqs:density-operator-general}, we start with a decomposition of one as in $1=\delta^i_{j}+\bar{\delta}^i_{j}$, using the usual Kronecker-delta and
\begin{equation}
	\bar{\delta}^i_{j} \equiv 1-\delta^i_{j}.
\end{equation}
The indices in $\delta^i_{j}$ and $\bar{\delta}^i_{j}$ refer to having external parton lines $i$ on the amplitude side and $j$ on the conjugate side of a diagram being connected (to either an emitter, emission or interferer line) or disconnected (meaning connected to a hard external line which does not affect any singular limit), respectively.
This decomposition reads
\begin{equation}\label{eq:decomp-one}
	\hat D^{r}_{\bar{r}} \equiv \frac{1}{n_\sigma} \sum_\sigma \prod\limits_{g=1}^{\mathrm{min}(p,\bar{p})} \prod \limits_{l=1}^{l_g} \left( \delta^{r_{gl}}_{\sigma(\bar{r}_{gl})} + b_{\sigma(g)}\bar{\delta}^{r_{gl}}_{\sigma(\bar{r}_{gl})} \right),
\end{equation}
where $\sigma$ stands for permutations of the set $\bar{S}_{n,p,k}$
and $n_\sigma$ is the number of permutations.\footnote{Note that in this definition, we have implied that $p\leq \bar{p}$ and that none of the legs on the amplitudes side are unconnected.} 
In this expression, the index $g$ numbers the splitting groups and $l_g$ represents the length of each group.
Then, \eqref{eq:decomp-one} serves as a generating function where the parameters $b_g$ are used to control the number of unconnected lines in each group of splitters (\ie between singular partons of the amplitude side complex conjugate one), \viz in
\begin{equation}
	\hat{\Delta}^{r}_{\bar{r}} \equiv \sum\limits_{n=0}^{\mathrm{min}(p,\bar{p})}\sum\limits_{\bar{g}=1}^{\bar{p}} \frac{\partial^n}{\partial b_{\bar{g}}^n} \hat{D}^r_{\bar{r}} \Big|_{b_{\bar{g}}=0}.
\end{equation}
Expanding out this objects can be used to generate the set of topologies corresponding to all diagrams of squared amplitudes that contribute in a singular setting with $k$ emissions.
Fully unconnected graphs can be neglected by setting $b_g^c = 0$ for $c>l_g$.

\newpage
\bibliography{multi-emissions}

\end{document}